DISS. ETH N° 23526

# Quantum Coarse-Graining: An Information-Theoretic Approach to Thermodynamics

A thesis submitted to obtain the degree of

DOCTOR OF SCIENCES of ETH ZURICH

(Dr. sc. ETH Zurich)

presented by

**Philippe Faist**

MSc. ETH in Physics

born on May 15, 1988

citizen of Montreux, VD

accepted on the recommendation of

| | |
|---|---|
| Prof. Renato Renner, | examiner |
| Prof. Seth Lloyd, | co-examiner |
| Prof. Terry Rudolph, | co-examiner |
| Prof. David Jennings, | co-examiner |

2016

*στην αγαπημένη μου Εβίτα*

*Penser importe plus que savoir ; combien de savants ne pensent pas ? Savoir du reste est nécessaire, mais savoir quoi ? ceci est la question.*

—Victor Hugo (1802–1885)

# Abstract

The present thesis investigates fundamental connections between thermodynamics and quantum information theory. The starting point is Landauer's principle: irreversible information processing cannot be carried out without some inevitable thermodynamical work cost.

First, existing frameworks for studying the thermodynamics of quantum systems in the finite-size regime are discussed. It is shown that two mainstream frameworks, an operational framework called "thermal operations" and the mathematically more convenient "Gibbs-preserving maps," are nonequivalent, and we comment on this gap.

These models serve as a basis to derive a new, fully information-theoretic framework which generalizes the above by making further abstraction of physical quantities such as energy. It is technically convenient to work with, and reproduces known results for finite-size quantum thermodynamics.

We apply our new framework to answer the question of determining the minimal work cost of implementing any logical process. The answer is given in terms of information-theoretic properties of the logical process. In the simpler case of information processing on memory registers with a degenerate Hamiltonian, the answer is given by the max-entropy, a measure of information known from quantum information theory. In the general case, we obtain a new information measure, the *coherent relative entropy*, which generalizes both the conditional entropy and the relative entropy. It satisfies a collection of properties which justifies its interpretation as a new kind of entropy measure and which connects it to known quantities.

Then, we turn to large systems and study how we can recover macroscopic thermodynamics. From our framework, macroscopic thermodynamics emerges by typicality, after singling out a class of thermodynamic states parametrized by thermodynamic variables taking values in a continuous range in an appropriate limit. These states are assumed to be singled out by some appropriate physical reason, such as equilibration or by symmetry properties. A natural thermodynamic potential emerges, dictating possible state transformations, and whose differential describes the physics of the system. The textbook thermodynamics of a gas is recovered as well as the form of the second law relating thermodynamic entropy and heat exchange.

Finally, noting that quantum states are relative to the observer, we point out that the procedure above gives rise to a natural form of coarse-graining in quantum mechanics. Several different quantum kets can be viewed as a single ket for a different observer. Such is the case of thermodynamics, for example, where the macroscopic observer's fundamental unit of information are the thermodynamic variables and not the individual microstates. The macroscopic observer only sees the traces left by the coarse-grained states in the *hidden information*, which describes which processes may spontaneously occur. A new picture of quantum information appears, where any observer can specify their fundamental units of information and describe them by quantum kets. In particular, it is possible in principle to observe quantum superpositions of thermodynamic states, provided a suitable reference frame is available.

# Résumé

La présente thèse étudie les liens fondamentaux entre la théorie de l'information quantique et la thermodynamique. Le principe de Landauer en constitue le point de départ : il est nécessaire de fournir du travail thermodynamique si l'on souhaite effacer de l'information stockée en mémoire.

Nous considérons en premier lieu des modèles existants permettant d'étudier la thermodynamique de systèmes quantiques dans le régime fini, c'est-à-dire permettant de décrire une seule exécution de l'expérience et non la moyenne de multiples répétitions. Il est démontré que le modèle connu sous le nom d'« opérations thermales » n'est pas équivalent aux « évolutions préservant l'état de Gibbs ». La différence fait l'objet d'une courte discussion.

Généralisant ces modèles, nous élaborons un nouveau cadre de travail entièrement formulé en termes d'information, et ne faisant référence à aucune quantité physique telle que l'énergie. Ce modèle possède des propriétés techniques agréables qui rendent son utilisation adaptée aux problèmes considérés, et permet de retrouver certains résultats connus au sujet de la thermodynamique de systèmes quantiques dans le régime fini.

Notre modèle est mis à l'épreuve pour déterminer le travail requis pour l'implémentation d'une fonction logique. La réponse est fournie en termes de propriétés logiques de la fonction sous la forme d'une mesure d'entropie. Dans le cas ou l'hamiltonien des registres mémoire considérés est complètement dégénéré, la réponse coïncide avec la « max-entropie », déjà connue dans le domaine de la théorie de l'information. Dans le cas général, la réponse nous fournit une nouvelle mesure d'entropie que nous nommons « entropie cohérente relative », généralisant à la fois l'entropie conditionnelle et l'entropie relative. Les propriétés de cette quantité en justifient l'interprétation comme une mesure d'information, et la relient à plusieurs mesures d'entropies connues.

Ensuite, nous étudions comment la thermodynamique de systèmes macroscopiques s'explique par le biais de notre approche. La thermodynamique émerge après avoir identifié une classe d'états d'un caractère typique. Ces états sont supposés sélectionnés suite à une raison physique, telle que dûs à un équilibre au contact d'un réservoir de chaleur, ou préférés pour des raisons de symétrie. Ces états constituent les « états thermodynamiques » du système, et sont supposés paramétrés par des variables prenant des valeurs continues dans une limite appropriée. Il en émerge un « potentiel thermodynamique naturel », qui détermine les évolutions spontanées possibles du système, et dont la différentielle décrit les propriétés physiques du système. Nous réexaminons l'exemple prototype d'un gaz, et obtenons une forme habituelle du deuxième principe de la thermodynamique reliant l'entropie à la chaleur dissipée.

Finalement, nous montrons que la procédure ci-dessus donne lieu à une forme de changement d'échelle intrinsèque à la mécanique quantique, permise grâce au fait que l'état quantique est relatif à l'observateur. En effet, plusieurs états distingués par un observateur peuvent être recombinés en un seul pour un observateur différent. Ceci est le cas, par exemple, en thermodynamique, où l'unité d'information de l'observateur macroscopique est l'état thermodynamique, déterminé par les grandeurs macroscopiques du système telles que la température, le volume, et la pression, alors qu'un observateur ayant accès aux degrés de liberté microscopiques pourrait décrire l'état individuel de chaque particule constituant le système. L'observateur macroscopique ne remarque la présence de degrés de libertés microscopiques que par l'« information cachée », qui détermine les évolutions possibles du système et donc ses propriétés thermodynamiques. Il en résulte un nouveau tableau pour la théorie de l'information quantique : chaque observateur peut préciser son unité d'information, et décrire le système en exprimant cette dernière en kets quantiques. En particulier, il est possible en principe d'observer des superpositions quantiques d'états thermodynamiques, pour autant qu'un système jouant le rôle de référentiel approprié soit disponible.

# Acknowledgments

I am deeply grateful to my thesis supervisor, Prof. Renato Renner, for successfully guiding my research with such a clear vision of quantum information and modern physics. I could always turn to you for much appreciated fresh ideas and counsel.

I would like to thank as well my colleagues and friends from the quantum information theory group, Lea Krämer Gabriel, Lídia del Rio, Rotem Arnon Friedman, Sandra Stupar, Daniela Frauchiger, David Sutter, Philipp Kammerlander, Elisa Bäumer, Christopher Portmann, and Joe Renes, who provide our daily enjoyable working atmosphere, as well as former members of the quantum information theory group Frédéric Dupuis, Cyril Stark, Mario Berta, Marco Tomamichel, Norm Beaudry, Aleksejs Fomins, Michael Walter, Christian Schilling, Felipe Lacerda, Fernando Brandão, Roger Colbeck, Omar Fawzi, Volkher Scholz, Yeong-Cherng Liang, Mário Ziman, Takanori Sugiyama, and Matthias Christandl. I am particularly indebted to my master's project supervisor and colleague Johan Åberg, for teaching me a great deal of quantum information and for entertaining discussions concerning many other areas of physics and chemistry.

It has equally been a pleasure to collaborate with wonderful people on several scientific projects. Mirjam Weilenmann, Nicole Yunger Halpern, Jonathan Oppenheim and Andreas Winter, thanks for your time and ideas.

I am very much grateful to David Jennings, Terry Rudolph and Seth Lloyd for joining my thesis committee, for useful feedback on this thesis and for enjoyable discussions. Thank you as well to Lea Krämer Gabriel and Evita Varela for feedback on an earlier version of this document.

My (mainly french-speaking) friends from the Institute of Theoretical Physics, Julián Cancino, Romain Müller, Vincent Beaud, Max Kelm, have always been a great company. I thank in particular Romain for beer and coffee time deep-reaching discussions on everything from programming languages to economics or emergent spacetime. Thank you as well Caterina Specchia and Gizem Öztürk for delightful discussions around coffee. I further wish to thank the wonderful team with whom I shared unforgettable studies at the University of Neuchâtel and who have remained great friends, Chahan Kropf, Aurélien Wyttenbach, Pierre-Nicolas Jolissaint, Mathieu Schopfer, and Jonas Geissbühler.

My family has constantly supported me throughout my studies and research. Jérôme, Corinne, Olivier, Daphné, thank you so much for your care.

I am lucky to have one very special person who has showed all her love and care, has constantly encouraged me, endured my lack of energy and presence at home, and provided the warm moral boost when I needed it. Evita, thank you from my heart for believing in me. My thesis is dedicated to you.

# Contents

# 1

# Introduction

With the advent of the digital age, information has increasingly established itself in our everyday lives: Computing devices accomplish essential tasks ranging from plain communication to complex calculations, information traffic has exploded with the growing popularity of the Internet, and credit card payments can be processed on the far side of the planet within seconds.

Equally revolutionary has information been to physics. The concepts and mathematical tools which were originally developed for cryptography and telecommunications have revealed themselves as fundamental building blocks underlying quantum theory and thermodynamics. The entropy of thermodynamics is none other than the information entropy describing our lack of knowledge of the microscopic state. The laws of quantum mechanics, which originally seemed to work only by spooky dark magic, acquire a sense of simplicity and inevitability when expressed in terms of information.

And as much as we have learned about thermodynamics and quantum mechanics, both still have secrets to reveal.

A surprising amount of confusion and litigious positions persists regarding fundamental conceptual elements of both theories. What is exactly the second law of thermodynamics? How fundamental is its status? How universal is thermodynamics? But also, what is a quantum measurement? What does it mean to have different observers in quantum mechanics? Is thermodynamics related any more closely to quantum information?

How can we attempt to answer these questions? These interrogations are of a conceptual nature, and cannot be deduced from a closed mathematical formulation of quantum mechanics or thermodynamics: The questions are precisely about those assumptions at the root of either theory. We thus turn to the physical concepts underlying those assumptions. As a starting point, we look at thought experiments which operate at the limit of the applicability of these frameworks. Such examples are, as we will see, Maxwell's demon, the Szilárd engine, Landauer's principle, Wigner's friend, and so on.

The first question we address is what happens to thermodynamics when it is taken to the limits of its intended domain of application, that is, macroscopic systems such as steam engines. In particular, what becomes of the second law of thermodynamics? The latter is often accepted with a universal sacro-saint status, even in as diverse contexts as electromagnetic radiation, chemical reactions, or even black hole physics (Bekenstein, 1973, 1974; Bousso, 2002). Yet the second law appears to break down in several examples. A prominent example is Maxwell's demon, followed by other setups in the same spirit such as Feynman's ratchet and pawl (Feynman, 1963). Paying special attention to how the second law is applied—including the demon or its memory explicitly in the picture, for example—restores its validity and lets everyone breathe a sigh of relief. Other situations are more challenging, including anomalous heat flows (Jennings and Rudolph, 2010; del Rio *et al.*, 2014), devices with feedback (Sagawa and Ueda, 2012) and thermodynamic manipulations with side information (del Rio *et al.*, 2011, 2014). There, concepts such as work, heat or even the thermal state may be ill-defined or may depend on the observer. Individually, the examples are well understood and present no serious challenge to the second law. Despite that, one strives for a global picture which clarifies these situations simultaneously.

Recently, a large amount of research has concentrated around *Landauer's principle* (Landauer, 1961). This is indeed a prototypical example of thermodynamics brought to its limits, in which an observer possesses some additional knowledge about the microscopic state of the system. Landauer argued that the erasure of one bit of information—for example, on a computer memory register—must be accompanied by the dissipation of an amount of heat of the order of $kT \ln 2$, where $k$ is Boltzmann's constant and $T$ is the temperature of the surroundings of the system. The required energy has to be brought in the form of work into the system, incurring an inevitable work cost. Bennett then significantly developed this argument, producing the new line of research of *information thermodynamics*, and finally exorcizing Maxwell's demon (Bennett, 1982, 2003). Many developments have since been achieved, models and frameworks proposed, and specific examples understood (see Section 1.2 for a brief review).

It might be argued that these problems have already been solved in the statistical mechanics community (Talkner and Hänggi, 2016; Campisi *et al.*, 2011; Anders *et al.*, 2010). There, the standard way of proceeding is to identify the thermodynamic internal energy with the expectation value of the Hamiltonian, the thermodynamic entropy with the von Neumann entropy, work and heat with the infinitesimal variation of the average energy with respect to changes of the Hamiltonian or the state respectively. However, our aim is to understand the regime where statements are to be made about a single instance of the process in question. Strictly speaking, and in the absence of further assumptions, average quantities are not appropriate in such a regime, as an average quantity would only gain an operational significance for large systems in a typical state, or if the experiment was repeated independently a large number of times.

In our approach, we develop a minimal model in which we specify a restriction on the possible operations the system can undergo. (The processes may either be thought of as occuring spontaneously, or as having been appropriately engineered.) The constraint is simple to formulate and is evidently true for most physical settings. Our model then specifies a notion of a *resource*. A resource is something valuable which can be used to enable the forbidden operations. In our model this resource is purity, that is, lack of randomness. Processes are then qualified by how much purity is required to lift the restriction, or how much purity they can yield on their own.

We purposely formulate our model solely in information-theoretic terms. There is no reference to any physical quantity such as energy. The only notions which appear are quantum states and an abstract object which encodes the microscopic restriction on the processes from an information-theoretic perspective. It is when we apply our model to physical systems that the abstract objects are determined by physical properties such as energy.

Our first main result is a second-law-type statement which quantifies how much purity (or thermodynamic work) has to be invested in order to carry out any given logical process. The result is first formulated in the simpler setting of information processing on memory registers whose Hamiltonian is completely degenerate. In this case, the amount of purity required is given by the max-entropy, a measure of information content of the system known from quantum information theory (Renner, 2005; Tomamichel, 2012). Eventually, we generalize this result to systems with any arbitrary Gibbs state. This gives rise to a novel measure of information which we term the *coherent relative entropy*.

Then, we turn our attention to how relevant this is for the original, macroscopic theory of thermodynamics as taught in textbooks. We spell out explicitly how macroscopic thermodynamics emerges from our model out of typicality of the underlying states. Indeed, we naturally recover thermodynamic variables and states, macroscopic notions of work and heat, as well as the thermodynamic potentials.

The picture presented here is more general than the usual paradigm taught in textbooks which is often formulated only for gases, and generally extended to other systems by sheer analogy. Our emergent picture of thermodynamics is automatically valid for any other system which is well modeled by our original microscopic information-theoretic constraint. As a rudimentary illustration, we present a purely information-theoretic example of information storage with varying degree of redundancy via a simple repetition code.

Our model may also be applied to different observers with varying degrees of knowledge. In summary, we have developed a theory generalizing thermodynamics, which can be applied to whichever observer is relevant, be it at the microscopic, mesoscopic or macroscopic scale (Figure 1.1).

What does this all tell us now about quantum mechanics itself? If we take quantum information seriously, an observer should be able to describe which are, for them, the fundamental units of information and correspondingly assign kets in a Hilbert space. A macroscopic observer's fundamental units of information are by definition the macroscopic states. It might seem contradictory at first sight to assign kets to macroscopic states, but we argue that this is fully compatible with both quantum mechanics and thermodynamics, owing to the fact that quantum states are relative to the observer. This gives us a natural notion of coarse-graining in quantum mechanics. Exploring this idea further, we argue that—in some specific sense—it is possible to prepare "quantum superpositions of thermodynamic states."

Now that must have sounded a bit wacky. Rest assured, we do not predict any different results for any typical tabletop experiment. Rather, it is the physical concepts such as "thermodynamic state" or "fundamental unit of information" which we suggest to be generalized in such a way to provide a consistent picture of both quantum information and thermodynamics.

## 1.1 Context: the landscape of quantum mechanics

The terminology *quantum information* may evoke very different meanings. Originally, quantum information was thought as a set of additional concepts and tools, of information-theoretic nature, which were formed by using the building blocks of quantum mechanics. By cleverly engineering interac-tions between atoms, photons and spin systems, it is possible to implement sequences of physical operations which serve for computational tasks or cryptography.

Ever more however, the opposite picture is presented. Quantum information is seen as the basic underlying framework generalizing probability theory, based on new notions of *information* and *events*. On top of this framework are then built the physical concepts completing the theory of quantum mechanics, such as the position and momentum operators, spin, time, and the Schrödinger equation. In the present thesis we adhere to the latter picture, which we find conceptually cleaner.

Figure 1.2 describes how our results fit in the landscape of quantum mechanics. Our finite-size information-theoretic framework is purely based on quantum information, and does not rely on any physical notion of position, momentum, or the Schrödinger equation. The emergent thermodynamic behavior is also formulated in purely information-theoretic terms. Once applied to physical systems, we recover traditional thermodynamics, such as for gases or magnetic systems (see, for instance, Callen 1985; Huang 1987).

## 1.2 Related work

The relation between thermodynamics and information has been extensively studied from various perspectives. We give a short overview in this section; for a rather comprehensive discussion we suggest Leff and Rex (2010). This section is adapted from Faist *et al.* (2015a).

The fundamental question we study was raised by Maxwell in the 19th century (Maxwell, 1871), who imagined a perpetuum mobile on a gas divided into two chambers, whose net effect was the reduction of entropy of an isolated system: A small being could have knowledge of the microscopic degrees of freedom of the gas and operate a trap door in the splitting wall, which he could use to filter the cold molecules from the hot molecules. Szilard (1929) realized that the crucial part of the problem was that the demon accessed microscopic degrees of freedom, which are not accessible normally in thermodynamics. He devised a thought experiment, the Szilárd engine, which illustrated the reversible conversion of $kT \ln 2$ work from or into well-defined accessible information. This suggested that the demon had to perform work to compensate for the entropy decrease of the gas. Scientists at the time were then led to believing that the measurement itself was a process that had to cost work, and some thought models were developed (Brillouin,

Figure 1.1 (on facing page): Main contributions of the thesis. A purely information-theoretic model implementing a simple restriction can be used, for example, to study finite-size thermodynamics. This model is solely based on the framework of quantum information. When the system is large and in a typical state, macroscopic thermodynamics emerges from our model complete with thermodynamic states, thermodynamic work and a natural thermodynamic potential. A macroscopic observer has only access to the thermodynamic variables and not to microscopic degrees of freedom. The thermodynamic states are this observer's fundamental unit of information. We show that the formalism of quantum information can be applied consistently by promoting these states to quantum kets. In particular, it is in principle possible to observe quantum superpositions of thermodynamic states provided an appropriate reference frame is available.

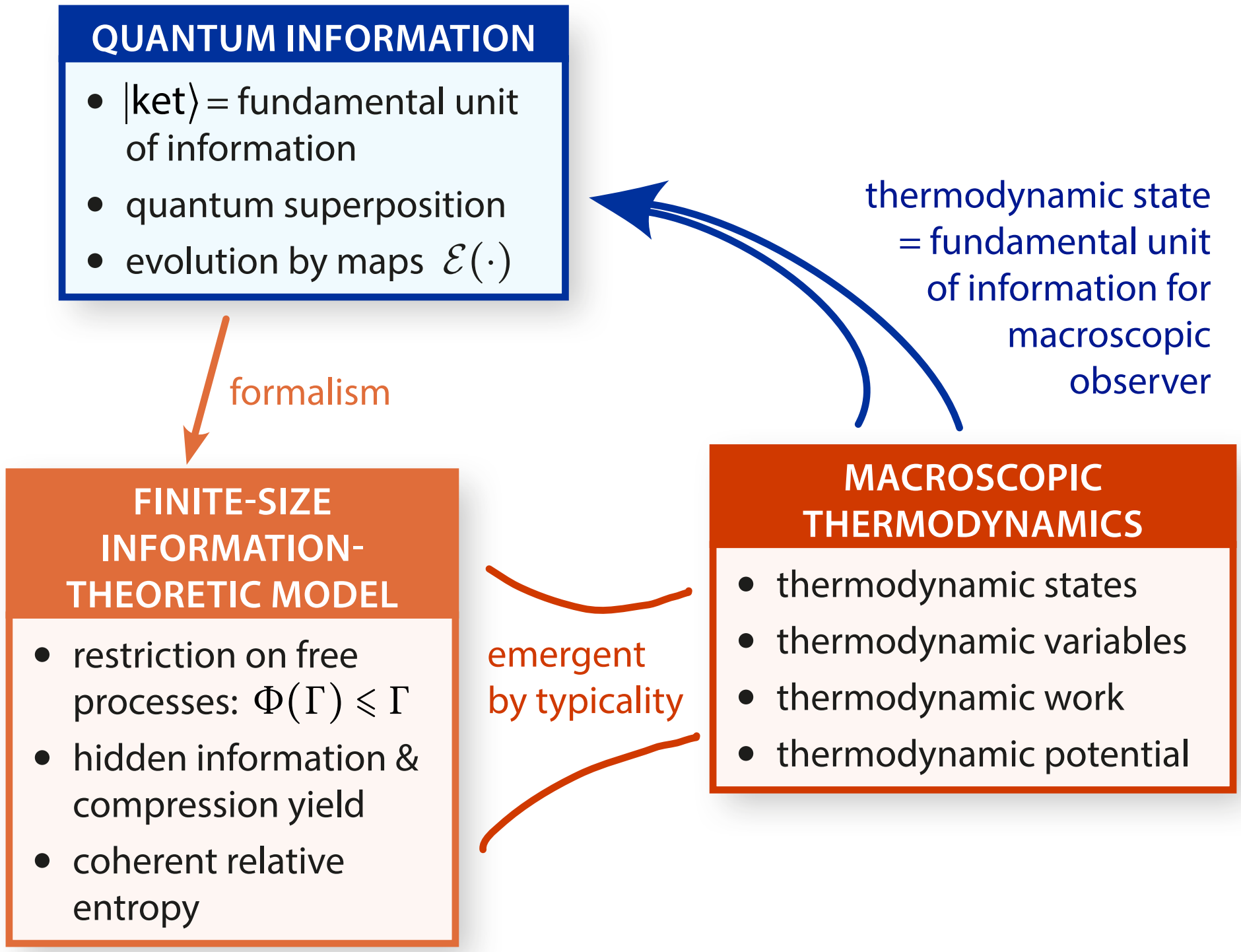
QUANTUM INFORMATION
|ket⟩ = fundamental unit of information
quantum superposition
evolution by maps $\mathcal{E}(\cdot)$
thermodynamic state = fundamental unit of information for macroscopic observer
formalism
FINITE-SIZE INFORMATION-THEORETIC MODEL
restriction on free processes: $\Phi(\Gamma) \leqslant \Gamma$
hidden information & compression yield
coherent relative entropy
emergent by typicality
MACROSCOPIC THERMODYNAMICS
thermodynamic states
thermodynamic variables
thermodynamic work
thermodynamic potential

1951; Gabor, 1961). It was Landauer (1961) who first associated work cost to the *logical irreversibility* of an operation, and who stated that the erasure of one bit of information had to cost $kT \ln 2$ work, and studied in particular the example of a particle in a double-V shaped potential (Landauer, 1961; Keyes and Landauer, 1970). Bennett showed that computations can be made completely reversible and devised an explicit measurement apparatus which required no work. On the other hand, resetting the demon's memory back to its original state does cost work, effectively exorcising Maxwell's demon (Bennett, 1982, 1973; Bub, 2001; Bennett, 2003, 1988; Maruyama *et al.*, 2009). Landauer's principle has been criticized (Earman and Norton, 1998, 1999), but became widely accepted as alternative proofs were proposed (Shizume, 1995; Piechocinska, 2000); cf. also Jacobs (2005). Its conceptual importance was further clarified (Bennett, 2003; Feynman, 1996). Various physical computational models were explored (Landauer and Woo, 1971; Likharev, 1982; Landauer, 1984, 1988). General considerations relating information and physics were further developed (Zurek, 1989; Landauer, 1996; Lloyd, 2000; Ladyman *et al.*, 2007). These efforts were lead in parallel to Jaynes showing the relevance of information theory for statistical mechanics and thermodynamics (Jaynes, 1957a,b, 1965, 1992).

With the development of quantum information in the last decades, importance was given to generalizing Landauer's principle to the quantum regime (Plenio and Vitelli, 2001; Alicki *et al.*, 2004; Oppenheim *et al.*, 2002; Janzing *et al.*, 2000; Janzing, 2006; Anders *et al.*, 2010; Reeb and Wolf, 2014), resulting in replacing the Gibbs, or Shannon, entropy with the quantum von Neumann entropy as relevant measure of information-theoretic entropy. These studies were motivated by the important technological advances making it possible to construct microscopic thermodevices (Scovil and Schulz-DuBois, 1959; Geusic *et al.*, 1967; Alicki, 1979; Howard, 1997; Lloyd, 1997; Geva and Kosloff, 1992; Hänggi and Marchesoni, 2009; Allahverdyan and Nieuwenhuizen, 2000; Feldmann and Kosloff, 2006; Baugh *et al.*, 2005; Lebedev *et al.*, 2016). Quantum heat engines were studied (Scully, 2001, 2002; Sagawa and Ueda, 2012, 2013; Park *et al.*, 2013) and experimental realizations of Maxwell demon-like setups carried out (Serreli *et al.*, 2007; Toyabe *et al.*, 2010; Koski *et al.*, 2015; Vidrighin *et al.*, 2016). Efforts were also undertaken to more generally understand the laws of thermodynamics from an information-theoretic point of view (Gemmer *et al.*, 2009; Popescu *et al.*, 2006; Jennings and Rudolph, 2010; Linden *et al.*, 2010b,a; Gogolin *et al.*, 2011; Jevtic *et al.*, 2015; Trotzky *et al.*, 2012; Hutter and Wehner, 2013; del Rio *et al.*, 2014; Gogolin and Eisert, 2016; Masanes and Oppenheim, 2014).

There followed a large number of contributions exploring various directions, which are collected in more detail in review works (Goold *et al.*, 2015a; Vinjanampathy and Anders, 2015; Parrondo *et al.*, 2015).

The frameworks that have been considered for the study of the thermodynamics of information processing are extremely varying. For example, systems have been considered as described by general Hamiltonians that are interacting, or for which we allow the modification of individual levels (Piechocinska, 2000; Alicki *et al.*, 2004; Sagawa and Ueda, 2009; Linden *et al.*, 2010b; del Rio *et al.*, 2011; Åberg, 2013; Skrzypczyk *et al.*, 2013, 2014; Åberg, 2014; Gallego *et al.*, 2014). Further studies are based on arrays of Szilárd engines (Dahlsten *et al.*, 2011; Egloff *et al.*, 2015). Another very promising approach is based on a resource theory of thermal operations, where the Gibbs states are for free (Horodecki *et al.*, 2003; Janzing, 2006; Brandão *et al.*, 2013; Horodecki and Oppenheim, 2013a; Brandão *et al.*, 2015; Buscemi, 2015). The two approaches are equivalent (Brandão *et al.*, 2013), and the set of operations don't in fact require fine control on the bath (Wilming *et al.*, 2016). The resource theory is tightly related to the notion of passivity (Pusz and Woronowicz, 1978; Lenard, 1978; Brandão *et al.*, 2013; Yunger Halpern and Renes, 2016; Brandão *et al.*, 2015; Perarnau-Llobet *et al.*, 2015a; Guryanova *et al.*, 2016; Yunger Halpern *et al.*, 2016). These resource theories are directly connected to macroscopic thermodynamics (Weilenmann *et al.*, 2015). More general resource theoretic frameworks have also been proposed (Brandão and Gour, 2015; del Rio, 2015; del Rio *et al.*, 2015; Kraemer and del Rio, 2016). Furthermore, effort was made to understand the role in thermodynamics of catalysis (Brandão *et al.*, 2015; Ng *et al.*, 2015) as well as coherence and the possible cost of time control (Malabarba *et al.*, 2015; Winter and Yang, 2016; Korzekwa *et al.*, 2016; Ćwikliński *et al.*, 2015; Lostaglio *et al.*, 2015a,c).

These frameworks have provided a solid set of tools to study the efficiency of nanoengines (Woods *et al.*, 2015), conserved charges which don't commute (Lostaglio *et al.*, 2015b; Guryanova *et al.*, 2016; Yunger Halpern *et al.*, 2016), relations with reversibility and recovery (Wehner *et al.*, 2015; Buscemi *et al.*, 2016) and to derive fully quantum fluctuation relations (Åberg, 2016; Alhambra *et al.*, 2016). The role of correlations has also been clarified (Friis *et al.*, 2016; Bruschi *et al.*, 2015; Hovhannisyan *et al.*, 2013).

An important contribution was to depart from the traditional view of studying averages over many independent repetitions of the same experiment, known as the i.i.d. regime. Rather, focusing on single instances of Figure 1.2 (on facing page): The landscape of quantum mechanics and thermodynamics. Quantum information is an abstract framework generalizing probability theory. A physical model explains how to apply this framework to physics, yielding the theory of quantum mechanics. Quantum Shannon theory deals with quantifying information and characterizing information-theoretic tasks such as information compression. Our finite-size information theoretic model generalizes this to units of information which may have different "values." On the other hand, our emergent thermodynamics generalizes usual thermodynamics of physical gases, which is usually derived directly from the physical model via statistical mechanics.

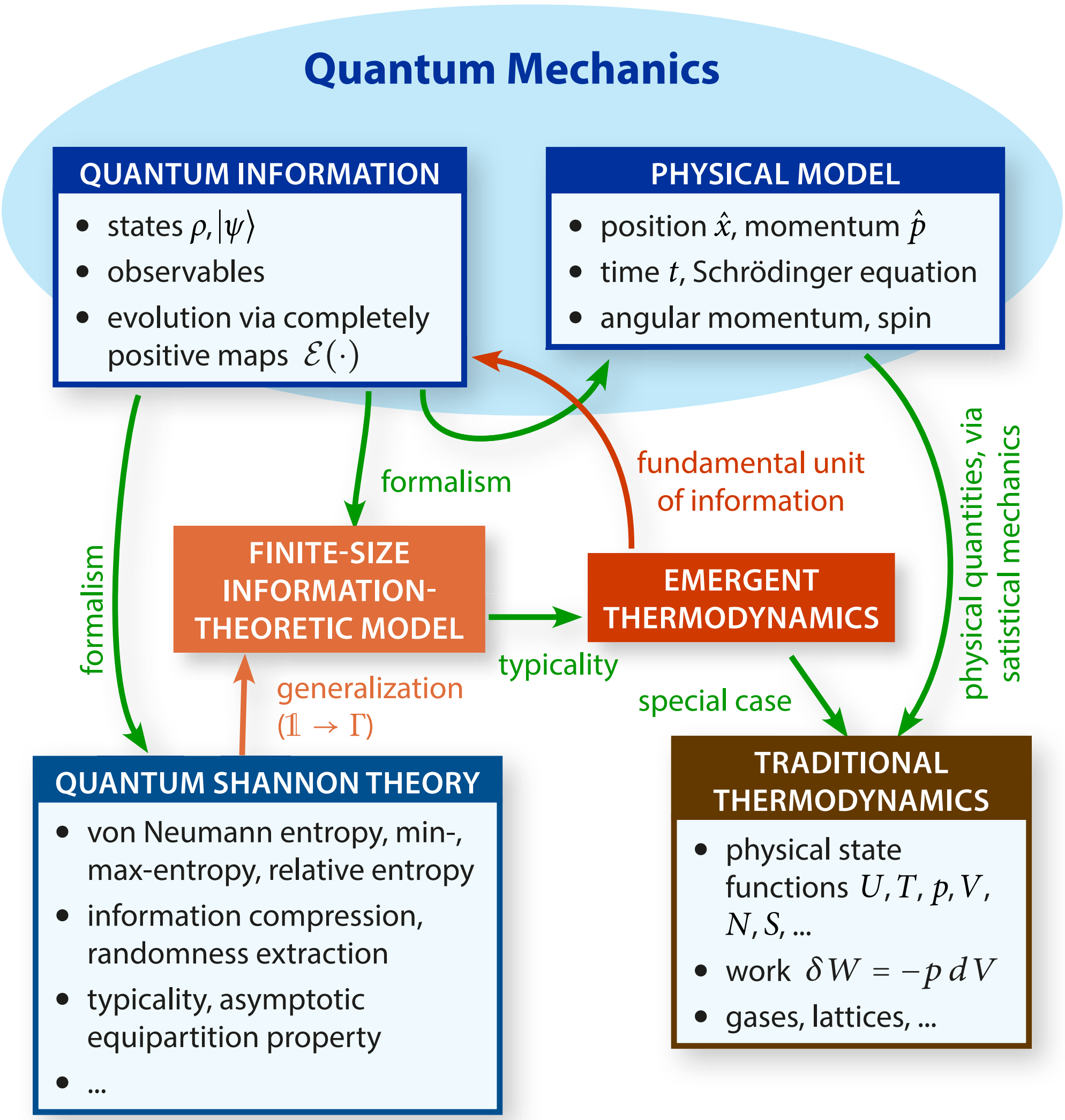
Quantum Mechanics
QUANTUM INFORMATION
• states ρ, |ψ⟩
• observables
• evolution via completely positive maps ℰ(·)
PHYSICAL MODEL
• position x̂, momentum p̂
• time t, Schrödinger equation
• angular momentum, spin
formalism
fundamental unit of information
formalism
FINITE-SIZE INFORMATION-THEORETIC MODEL
EMERGENT THERMODYNAMICS
physical quantities, via satistical mechanics
typicality
generalization (𝟙 → Γ)
special case
QUANTUM SHANNON THEORY
• von Neumann entropy, min-, max-entropy, relative entropy
• information compression, randomness extraction
• typicality, asymptotic equipartition property
• ...
TRADITIONAL THERMODYNAMICS
• physical state functions U, T, p, V, N, S, ...
• work δW = −p dV
• gases, lattices, ...

information-theoretic tasks (Renner, 2005) yields a new natural set of entropy measures to consider—the *smooth entropies* (Renner, 2005; Tomamichel *et al.*, 2009; Tomamichel, 2012). These entropies turn out to be just as relevant for quantum thermodynamics in the single instance regime (del Rio *et al.*, 2011; Dahlsten *et al.*, 2011; Åberg, 2013; Horodecki and Oppenheim, 2013a; Lloyd, 2013; Dahlsten, 2013; Egloff, 2010; Egloff *et al.*, 2015; Faist *et al.*, 2015a).

Further models were also proposed, with an explicitly modeled macroscopic storage system (Frenzel *et al.*, 2014), generalized operations (Binder *et al.*, 2015), spin models (Goold *et al.*, 2015b), but also in the context of generalized Gibbs ensembles (Perarnau-Llobet *et al.*, 2015b) as well as using measurement-based schemes (Tajima and Hayashi, 2015).

The field has also fueled conceptual contributions regarding the second law (Uffink, 2001; Bennett, 2008) as well as specifically studying how to define *work* (Gallego *et al.*, 2015; Talkner and Hänggi, 2016). Furthermore, other communities have expressed interest in the field, even for instance by developing a programming language for Szilárd engines (Abramsky and Horsman, 2015).

## 1.3 Assumed readership background

The present thesis is meant to be understandable as much for experts in the field as well as for students with basic knowledge of quantum information. The extended executive summary in Chapter 3 should present the essential concepts of quantum information which are needed and give additional pointers to more comprehensive references.

Several terms including the *second law*, *Landauer's principle*, or even *quantum information* may evoke surprisingly different meanings for different individuals in the community. The view adopted in this thesis is that quantum information is a framework generalizing classical probabilities; by adding the physical ingredients of position momentum operators, time and Schrödinger's equation etc., we obtain a full physical theory, quantum mechanics. The term *second law* in this thesis refers to the well-known law derived for macroscopic thermodynamics (in any of its equivalent formulations); its use will be restricted to systems which are clearly macroscopic and well described by textbook thermodynamics. *Landauer's principle* refers to the general principle as originally phrased (Landauer, 1961): Logically irreversible operations cannot be carried out without a corresponding dissipation of heat; in particular, the erasure of one bit of information in an environment at temperature $T$ must be accompanied by a dissipation of $kT \ln 2$ heat, where $k$ is Boltzmann's constant.

Further clarification of some common issues, such as the fact that entropy is a notion which is relative to an observer's knowledge, and the role of gambling in thermodynamics, are discussed in Chapter 2.

## 1.4 Contributions and outline of the thesis

This document covers the material of Faist *et al.* (2015a,b) as well as new results developed in the course of writing. Several other papers contributed to by the present author have not been included in this thesis (Dupuis *et al.*, 2013; Weilenmann *et al.*, 2015; Faist and Renner, 2015; Yunger Halpern *et al.*, 2016). Notes accompanying each chapter categorize its contents as REVIEW=review material, NOVEL=novel results which are as of yet unpublished, A=published in Faist *et al.* (2015a), and B=published in Faist *et al.* (2015b).

Chapter 2 provides some preliminary background on the concepts underlying the investigated connections between information theory and thermodynamics, mostly pioneered by Szilard (1929), Landauer (1961), and Bennett (1982, 2003). We further introduce more recent concepts underlying this thesis, such as the *logical* versus *physical* processes and the role of *gambling* in thermodynamics. (REVIEW)

Chapter 3 presents the formalism of quantum information we shall be working with, essentially as presented in textbooks such as those by Nielsen and Chuang (2000) and Wilde (2013). The information-theoretic nature of quantum theory is underscored, striving for a clearer separation of the different concepts which are usually presented bundled together in a quantum mechanics course. This chapter fixes along the way most of the notation used in the rest of this work; a summary of notation is also provided for convenience on page 253. (REVIEW)

Chapter 4 explores two frameworks which have been proposed to study quantum thermodynamics from the information theoretic perspective: thermal operations and Gibbs-preserving maps. We also prove that they differ (Faist *et al.*, 2015b), and comment on this gap. As a minor novel contribution we highlight the difference between frameworks which depend explicitly on the Hamiltonian and those which care only about the Gibbs state of the concerned systems. Finally, in order to clarify the notion of work, we review several models meant to represent work storage systems. (REVIEW, B, NOVEL)

Chapter 5 is devoted to the development of a novel, entirely information-theoretic framework generalizing the Gibbs-preserving maps. Classically, this framework may be interpreted dually as modeling thermodynamical operations, or as describing coarse-grained states on an underlying degenerate system (the *hidden information*). In its fully quantum version, this framework describes possible logical processes, and their *compression yield*, under the restriction that only the processes which preserve a particular operator—tentatively termed the *hidden information operator* by analogy—may be implemented for free. Basic properties of this framework are derived. The abstract version of the work storage models presented in the previous chapter may store compression yield, or provide the necessary resources to enable processes; furthermore they are shown to be equivalent in terms of how much resources they store or which logical processes they enable. (NOVEL)

Chapter 6 focuses on the application of the fully information-theoretic framework onto the special case of information processing, that is for operations which process information stored on memory registers described by completely degenerate Hamiltonians. This chapter is in most part reproduced from Faist *et al.* (2015a). The results include an explicit expression for the amount of work a logical process requires in terms of the smooth max-entropy. Several examples are presented, such as erasure with side information, generation of randomness, and quantum measurements. (A)

Chapter 7 considers the above framework in its full generality, and derives a novel quantity, the *coherent relative entropy*, as the optimal compression yield of a logical process implementing a given process matrix. This quantity generalizes both the relative entropy and the conditional entropy. Its properties are investigated, and it is shown to converge to the difference of input and output quantum relative entropies in the i.i.d. limit. (NOVEL)

Having studied the microscopic framework in detail, Chapter 8 turns to macroscopic thermodynamics. Thermodynamic variables and states, as well as thermodynamic potentials such as the free energy are shown to emerge naturally from our microscopic framework. These potentials, like the traditional thermodynamic potentials, exhibit a differential structure describing the physics of the system. The textbook example of a gas assists in illustrating this emergence picture. (NOVEL)

Chapter 9 is a necessary interlude, covering a modern—yet arguably not unanimously understood—picture of quantum mechanics with the notion of the quantum state being relative to the observer. Although to our knowledge never explicitly presented in this form, this understanding underlies most of the modern literature, including Everett's relative state interpretation, quantum Bayesianism, and most notably quantum reference frames. No novel technical results are presented in this chapter, yet this picture is essential for the final part of the thesis. (REVIEW)

In Chapter 10, the emergence picture of thermodynamics is combined with the observer-dependent nature of quantum states. The emergent thermodynamical theory can be interpreted as a new instance of a quantum system, where the thermodynamic variables are represented by quantum states, providing a fully quantum notion of *coarse-graining*. That is, the same framework—quantum information—may apply to different observers with varying degrees of knowledge, such as microscopic or macroscopic ones. The terminology *hidden information* is thus justified in the fully quantum context. This picture applies naturally to systems which are symmetric under the action of a group, connecting to the literature on reference frames. Furthermore, we argue that nothing prevents the preparation of a thermodynamical system in a quantum superposition of thermodynamical states, provided a suitable quantum reference frame is available. (NOVEL)

Finally, Chapter 11 wraps up the results presented in this thesis and provides an outlook for future developments and applications.

# 2

# Thermodynamics and Information

This chapter reviews some essential concepts connecting information theory to thermodynamics, providing an information-theoretic analysis of Landauer's principle, Maxwell's demon and further notions from "information thermodynamics."

An essential point to retain, understood by Landauer (1961) and Bennett (1982), is that information entropy and thermodynamic entropy are two facets of the same concept, which need to be placed on an equal footing in face of the second law of thermodynamics.

Furthermore, the notion of entropy—whether information-theoretic or thermodynamic—is *relative to an observer's knowledge*. In thermodynamics, the natural observer is the one who only has access to macroscopic physical quantities such as temperature and volume. The entropy state function of thermodynamics is then simply a characterization of the lack of information about the system from the perspective of the macroscopic observer.

Finally, we highlight the importance of gambling in thermodynamics: improbable events should be ignored. This is usually done implicitly in macroscopic thermodynamics and statistical mechanics. Treating this explicitly, however, will provide additional tools to formulate specific statements in the regime we are interested in.

## 2.1 Landauer's principle

The starting point of the field of information thermodynamics is Landauer's principle (Landauer, 1961). It asserts that the erasure of information, that is, the resetting of an unknown bit to a fixed known state, has an unavoidable change in the environment associated to it, typically the dissipation of heat.

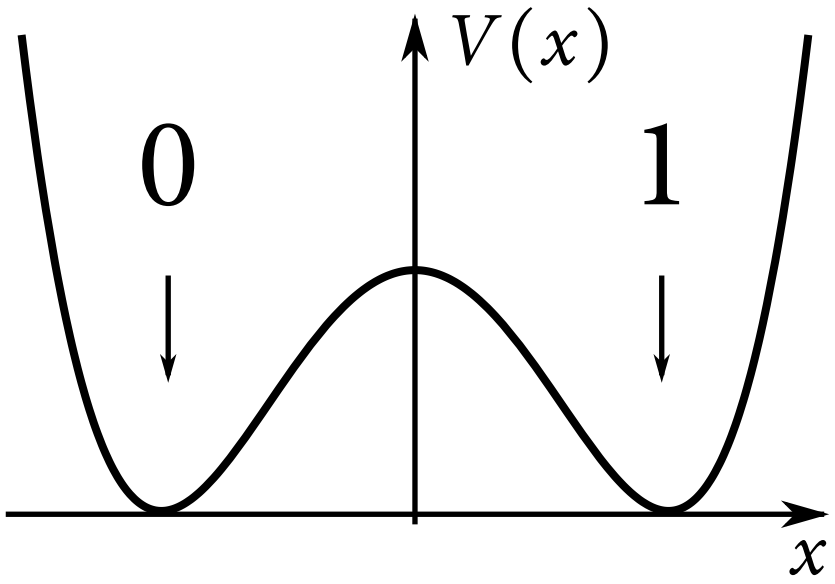

Figure 2.1: A particle trapped in a double-well potential as studied by Landauer (1961). The task is to reset the system to the state **0**, corresponding to bringing the particle in the left well. If the particle is known to be located in the left well, there is nothing to do; if it is known to be in the right well, it can be reversibly transferred to the opposite well. However, if its location is unknown, a dissipative force needs to be applied causing the dissipation of heat.

In his seminal paper, Landauer studied the example of a particle trapped in a double-well potential as depicted in Figure 2.1. Imagine that the particle occupies a ground state, and suppose that we wish to bring the particle in the left hand well. If the particle is already known to be on the left, then there is nothing to be done. If the particle is located in the right well, then we can bring the particle to the left reversibly: we may apply a force to the particle and supply work so that it climbs to the top of the barrier, and then reverse the force and collect back all the furnished work as the particle descends in the left well.

Suppose we initially don't know where the particle is. Because the laws of mechanics are completely reversible, it is clear that any action on the particle alone such as applying conservative forces will never have as net effect the mapping of the two distinct initial states onto the same final state. The solution is to introduce a form of dissipation: we apply a dissipative force on the particle towards the left for a long enough time, and then slowly let the particle relax to its ground state. This way, if the particle is in the right well, it crosses the barrier, and if it is on the left, it stays in the left well. Observe that work must be provided, and lost as heat, due to the dissipation.

Landauer's conclusion is the following: a reduction of entropy in the degrees of freedom which encode information (e.g., the particle being in the left well or right well), must be compensated by a corresponding increase of entropy of the remaining degrees of freedom (such as the surroundings of the system). This can be seen as a consequence of the second law of thermodynamics, provided we stick to the following rule: the *information entropy* should be counted as part of the *thermodynamic entropy* of a system when applying the second law of thermodynamics.

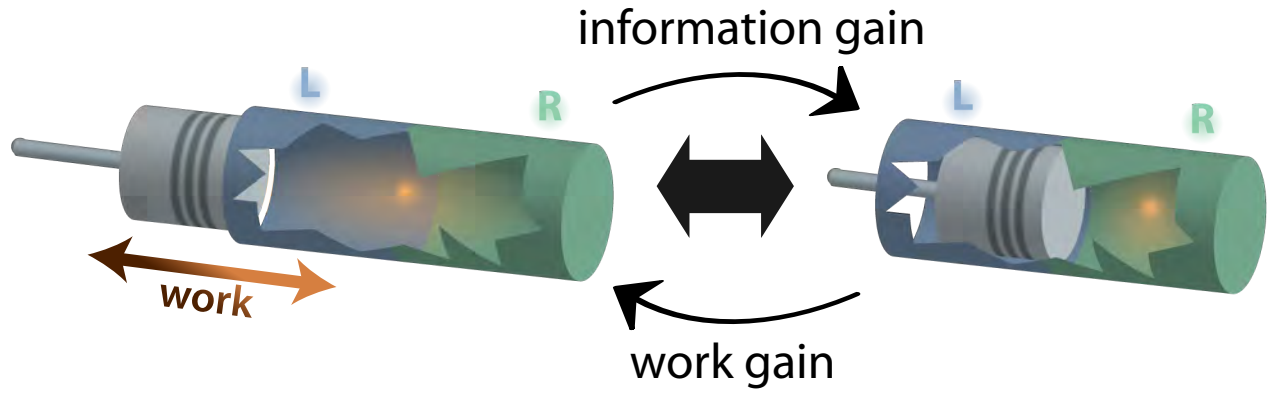

Figure 2.2: Szilárd engine. A single particle is thought of as a gas. If the particle occupies the full volume of the box, it can be brought to a definite side by compressing the gas, at the cost of supplying work. Conversely, work may be extracted if we know on which side the particle is, by letting it expand against a piston. Hence the information about which side the particle is on can be traded for work.

## 2.2 The Szilárd engine

A further minimal thought example dates back to Szilard (1929). Consider a box which contains a single particle. We will think of this single particle as an ideal gas; this approximation, however inaccurate, is sufficient to illustrate our considerations. The principles exposed here are further justified by more advanced and physically realistic models (Bennett, 1982; Alicki *et al.*, 2004; Brandão *et al.*, 2013; Frenzel *et al.*, 2014).

The box has a volume $2V$, and contains a one-particle gas at a temperature $T$. Imagine we wish to bring the particle to the right half of the box (Figure 2.2). First, assume that the gas occupies the whole box, and that we don't know where the particle is. Since it is an ideal gas, we may simply compress it isothermally. This can be done by setting up a piston on the left side of the box, setting the gas in contact with a heat bath at temperature $T$, and compressing the gas slowly. This thermodynamic process requires a work cost of $kT\ln(2V/V) = kT\ln 2$. So far, no surprises.

In contrast, if we know the position of the particle beforehand, the task can be accomplished at no work cost. Indeed, suppose that the two sides are separated by a membrane, and suppose that the particle is on the left side of the box. Then we could insert another separator on the left edge of the box, and then move slowly both separators to the left of the box while keeping the volume of the gas $V$ constant. This does not cost any work as any momentum or energy transfered by the particles to one membrane is compensated by a similar transfer on the other. In fact, the slightest kick on the separator on the left edge would suffice to send the whole volume gliding onto the right side, as pointed out by Feynman (1996).

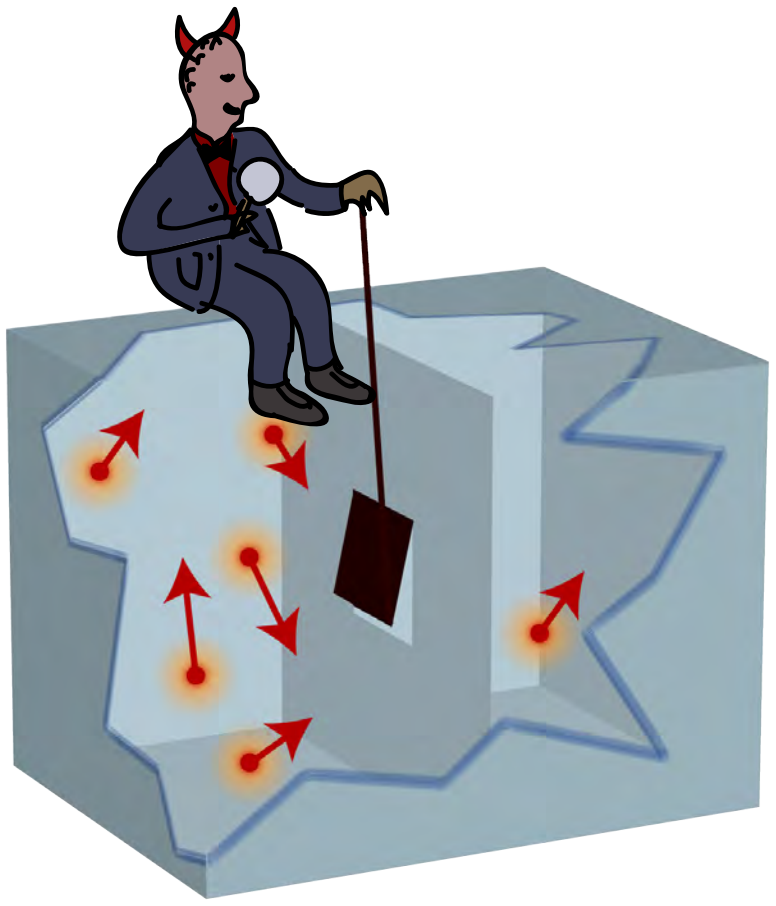

Figure 2.3: Maxwell's demon. A trap door separates two compartments and is operated by the demon. The latter brings all the particles onto one side by opening the trap door whenever a particle approaches from the right. This apparently violates the second law of thermodynamics, because the entropy of the gas decreases while no corresponding work is supplied.

For the Szilárd engine, we see that *the information about which side the particle is on can be traded for work*. Furthermore the simplicity of the setup suggests that this connection has a more fundamental character.

## 2.3 Maxwell's demon: entropy and observers

There are many ways of formulating the paradox of Maxwell's demon. The essence of its resolution is a confusion of statements made by different observers.

Consider a gas which initially occupies two chambers of equal volume. The two chambers may exchange particles through a trap door built into the wall separating the two compartments. An intelligent being, the demon, can operate this trap door without friction or any other form of energy dissipation (Figure 2.3).

The demon, we'll assume, can measure precisely the movement of the individual particles composing the gas. Each time a particle approaches the trap door from the right side, the demon lets it pass through the opening, but if a particle approaches from the left, then the demon maintains the door shut.[1] Ultimately, all the gas is confined within the left container, and so there is a net reduction of the entropy of the gas. Yet, no work is provided to the system. The second law of thermodynamics, which states that the entropy of an isolated system cannot decrease when no work is provided, has apparently been violated.

This paradox has intrigued many leading scientists and inspired numerous similar examples (Leff and Rex, 2010). It was originally thought that the process of acquiring information itself must require work, a statement which was further bolstered by studying explicit models of measurement processes (Brillouin, 1951; Gabor, 1961; Earman and Norton, 1998, 1999). Alternatively, the demon was to be considered as part of the thermodynamic system itself. Interactions with the gas then implied that the demon would thermalize, as illustrated by Feynman's ratchet and pawl (Feynman, 1963).

What if the demon itself is not well modeled as a thermodynamic system? It turns out that by invoking concepts of information theory, this example can be analyzed in a much cleaner picture. This thought experiment is in fact an ideal avenue to study the limits of traditional macroscopic thermodynamics in the case where one has additional information about the microstates of a thermodynamic system.

Let us start from the point of view of a macroscopic observer. Our macroscopic observer would gradually see the gas accumulating on the left side of the wall, in an apparent violation of the second law of thermodynamics (Figure 2.4**a**). There is, however, no reason to believe that the second law should apply to the gas. Indeed, thermodynamics is a theory about macroscopic systems: one of the founding assumptions of thermodynamics is a promise not to tamper with the microscopic degrees of freedom of a system. This is clearly not the case with the demon, so, strictly speaking, there is no surprise that the laws of thermodynamics don't accurately describe this situation.

Consider now the point of view of the demon. For him and his sharp eyesight, the particles are at definite locations with definite velocities, and it is just a matter of figuring out a schedule of opening and closing the trap door so that all the molecules composing the gas end up on the same side. The system is completely deterministic, and thermodynamics doesn't have to be invoked in any way (Figure 2.4**b**).

[1] Different formulations are found in the literature. This one is slightly simpler than the original one, in which the demon sorts out warm and cool particles.

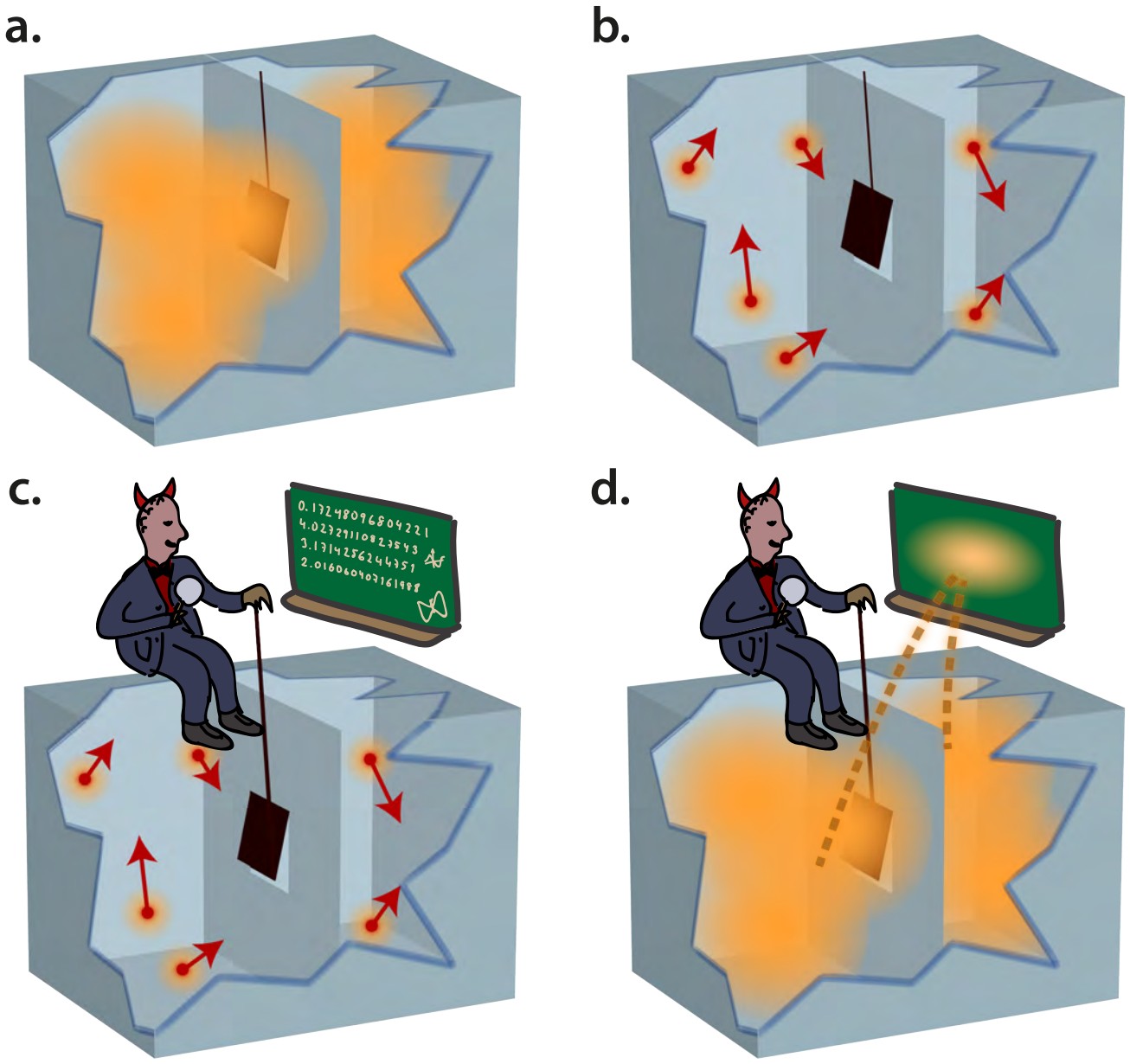

Figure 2.4: Maxwell's demon from different points of view. **a.** A macroscopic observer sees an apparent violation of the second law. This is no surprise, since microscopic degrees of freedom are being manipulated. **b.** The demon sees particles with definite positions and momentum, and operates the trap door as needed to organize them all on one side. The system is completely deterministic. **c.** We may explicitly model the demon's memory register which stores the measured information about the particles. Ultimately the memory register must be reset, incurring a work cost according to Landauer's principle. **d.** For a macroscopic observer, the demon's memory is perfectly correlated to the gas; effectively the demon's net accomplishment is to transfer entropy from the gas to his memory. Again, resetting this memory register costs work.

More interesting is the microscopic point of view where the demon is kept explicitly in the picture (Figure 2.4c). We need to model explicitly the demon's *memory*, which is the system in which the demon stores all the information about the particles which he has seen. Concretely, the demon must measure the position and velocity of each particle, and save the observed values in his memory. This information allows him to calculate the appropriate schedule for opening and closing the trap door. We require the demon to operate cyclically, meaning that after the demon has sorted the particles, then if we reset the state of the gas by refilling the two chambers as before, then the demon should be equally able to re-sort all the particles. This implies that the demon's memory must be reset to its initial state after each sorting process, or otherwise the memory would run out of free space. Crucially, the resetting operation may not be done for free according to Landauer's principle. This resolution is the currently widely accepted explanation of the paradox, and is largely due to Bennett (1982, 2003).

Let's now return to the macroscopic observer, and suppose that this observer knows of the demon's precise workings. The macroscopic observer knows that the demon performs measurements, stores the results in a memory, and operates the trap door. They do not know, however, what the actual state of the particles is: for them, the gas is in some thermal state. Also, the demon's memory is initially blank. When the demon performs measurements and records the results, it is effectively replicating the information about the positions and velocities of the particles into its memory register. Consequently, the demon's memory and the gas are described as being in a perfectly correlated state: even though the particles are in some random state, whenever the $n^{\text{th}}$ particle has a certain position and velocity, the memory's $n^{\text{th}}$ entry has those same values (Figure 2.4d). Then, the demon uses this side information to produce a operating schedule for the trap door, which places all the particles into one chamber. No information is lost, since the original positions and velocities of all the particles are still stored intact in the memory, and hence the operation is logically reversible. That is to say, all the demon has achieved up to this point is to transfer entropy from the gas to his memory. Eventually, if the demon is required to operate cyclically, then he must reset his memory; this requires a work cost.

Our final conclusion is that the thermodynamic entropy of the gas does indeed decrease, but this comes at the cost of either an entropy increase of the demon's memory, or a work cost for resetting the memory back to its initial state. This conclusion is independent of the point of view taken to describe the problem—whether microscopic or macroscopic, whether including the demon in the picture or not. So, when treated consistently with respect to any point of view, there is no contradiction to the laws of thermodynamics or information theory.

A particular emphasis should be put on one point: *the notion of entropy is relative to the observer, or point of view*. The entropy of the gas, or of the memory, may have differed in each of the pictures described above. According to the demon, after performing measurements on the particles the entropy of the gas is zero, because it is in a fully known state *according to him*. For the macroscopic observer, the entropy of the gas is that of the thermal state of the gas, which may be arbitrarily large. However, this remains consistent with the demon's description, because the macroscopic observer would still assert that the entropy of the gas is zero when conditioned on the demon's knowledge.

Notably, this departs from the traditional doctrine of thermodynamics where the entropy function is a physical property of the system. Both points of view are reconciled simply because, in thermodynamics, all statements are made with respect to a common, implicit observer: it is the macroscopic observer which has access to macroscopic degrees of freedom such as energy, temperature, volume, pressure, etc., but has maximal ignorance about the microscopic degrees of freedom such as the position and velocities of particles.

Formulated in this language, one of the goals of this thesis is to extend thermodynamics to more general observers.

## 2.4 Maxwell's demon with a Szilárd engine

While studying the role of Landauer's principle in Maxwell's demon, Bennett developed an analogous setup operating on a single Szilárd engine (Bennett, 1982, 2003), as depicted in Figure 2.5.

Initially, the demon does not know whether the particle is in the left or the right of the box. The demon inserts a separator in the middle of the box, and performs a measurement in order to determine the particle's state. If the state is found to be on the left of the box, then the demon attaches a piston on that side, and lets the one-particle gas expand in contact with a heat bath at temperature $T$. The demon extracts $kT\ln 2$ work in this way, and the gas fills again the full volume. If the particle was found on the other side, the demon proceeds analogously from the other side, also extracting $kT\ln 2$ work while also leaving the particle in an unknown state. In either case, the box is returned to its initial state, and the process may be repeated cyclically.

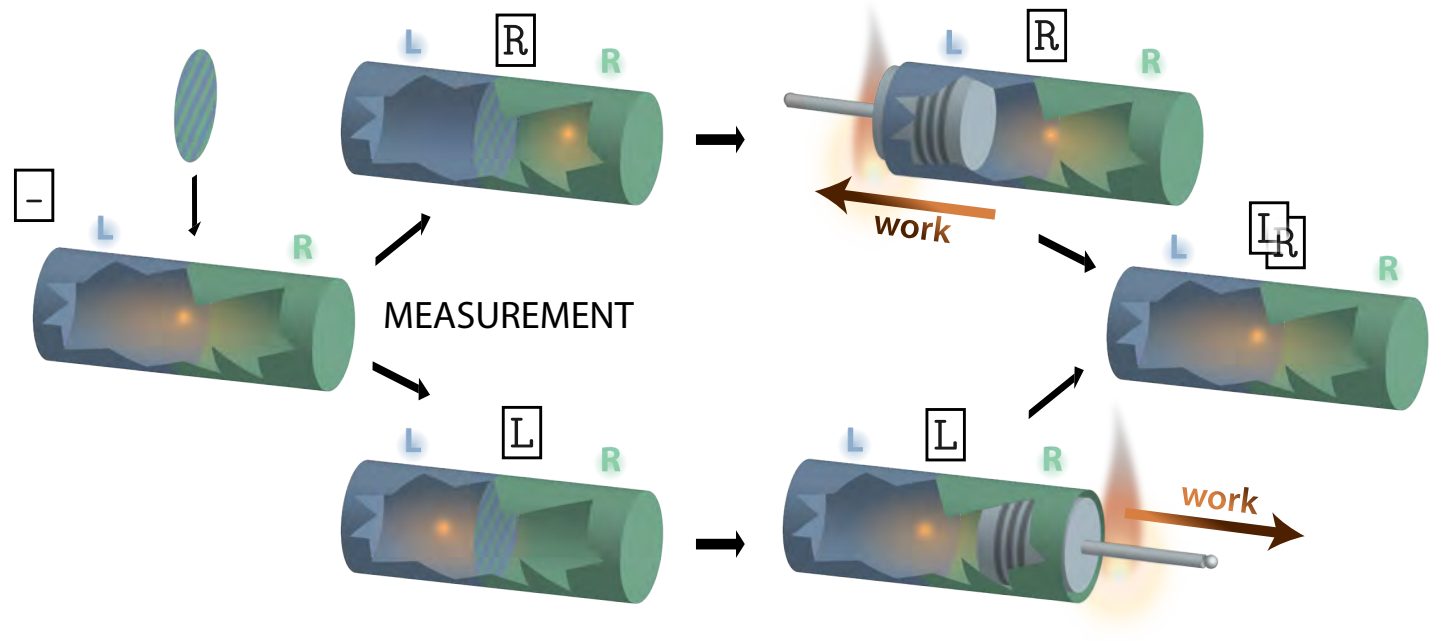

Figure 2.5: Maxwell's demon with a single Szilárd engine, proposed by Bennett (1982). A measurement detects on which side of the inserted separator the particle is, and extracts work with a piston in either case. The cylinder is left in its original state, apparently creating a perpetuum mobile with net work gain. However the measurement outcome (represented by "L" or "R") has to be stored in a memory register, which is initially in some pure state (represented by "–") and the work cost of resetting it to a pure state again compensates the work gain.

The demon appears to be able in this way to extract an arbitrary amount of work.

However, as Bennett notes, the demon must have recorded on which side of the box the particle is located. This is necessary in order to even remember the outcome of the measurement. In order to run the apparatus cyclically, all systems involved must be restored to their original state at the end of the process. This bit must then be reset, a process which costs at least $kT \ln 2$ according to Landauer's principle, compensating the work gain.

## 2.5 Logical processes and thermodynamical processes

Here we illustrate the conceptual differences between a *logical* and a *physical* or *thermodynamic process*, and explain how these notions interplay via some simple examples. Some content in this section is reproduced from (Faist *et al.*, 2015a).

A *logical process* is a balance sheet of inputs and outputs: It is an abstract rule that specifies which input should be mapped to which output, regardless of how the mapping is implemented. For example, an AND gate maps the logical states **00**, **01** and **10** to the logical state **0** and **11** to the logical state **1** (Figure 2.6a). This should be thought of as the specification of an abstract mathematical function, with no concern for how it is to be computed. Classically, a logical process may be given by a conditional probability distribution Pr[output | input] for each possible input and output states. In quantum mechanics, we will see that a logical process is represented by a so-called *quantum channel*.

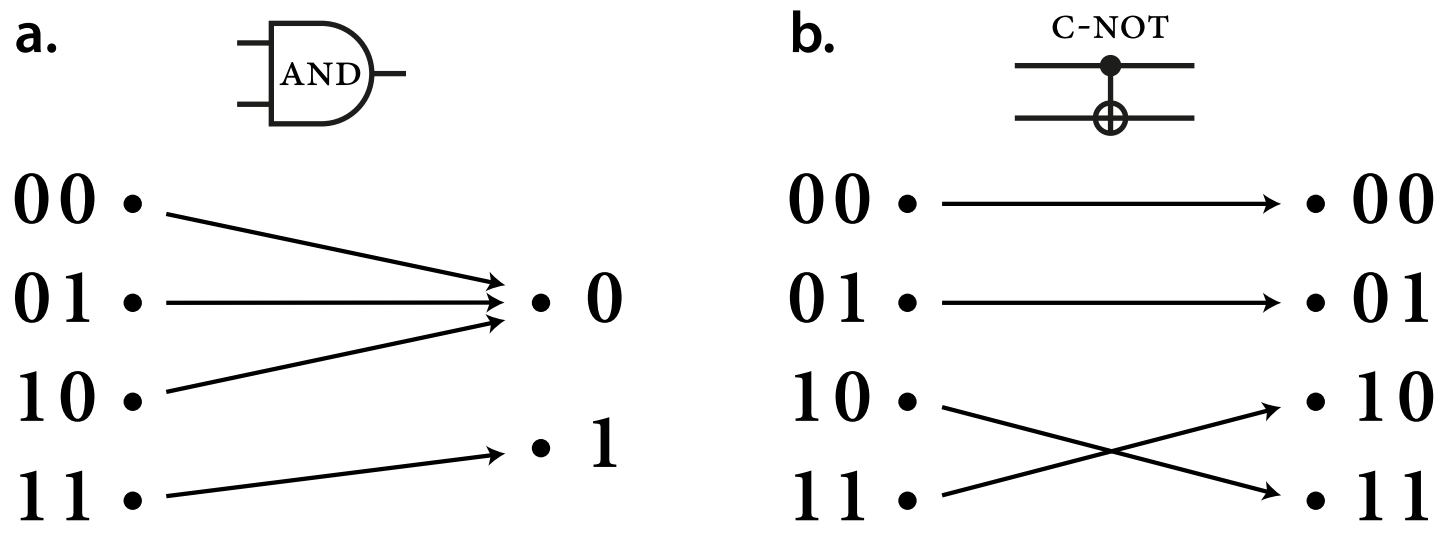

Figure 2.6: A logical process specifies which inputs are mapped to which outputs. **a.** The AND gate outputs '1' if both input bits are equal to '1,' and outputs '0' otherwise. **b.** The C-NOT gate flips the second bit if the first bit is equal to '1.' This gate is logically reversible, because the output allows to infer the input uniquely.

A *physical process* is a particular procedure involving physical components such as heat baths, ancillary systems, specific devices, a schedule of turning on and off interactions, and so on. In particular, a *thermodynamic process* is associated with the specific changes in the thermodynamic variables describing the system, making it a physical process. Any physical process results in some particular logical process if one establishes a balance sheet of the process and asks which inputs were effectively mapped to which outputs. Physical processes may thus be used to implement particular logical processes. There may be several physical processes which result in the same logical process.

### 2.5.1 Logical and thermodynamical reversibility

Both logical and thermodynamic processes have a well-defined notion of reversibility. A logical process is said to be *reversible* if the logical input state can be inferred uniquely from the logical output state. For example consider the C-NOT operation depicted in Figure 2.6**b**. It is logically reversible because by repeating the C-NOT operation we recover the initial state exactly. The notion of thermodynamical reversibility is defined in any thermodynamics textbook: A thermodynamic process is *thermodynamically reversible* if the same path in thermodynamic state space may be performed in reverse while recovering the full amount of work which was invested in the forward direction.

These two notions of reversibility are *a priori* unrelated. Indeed, logically irreversible processes can be implemented in a thermodynamically reversible way: For example, a Szilárd engine in a completely mixed state can be reset to the state "left" by a thermodynamically reversible isothermal process. While the thermodynamic transformation is reversible, the precise logical state the memory initially was in (that is, the information about whether the particle was on the left or the right side of the box) is irreversibly lost in the heat bath. On the other hand, one can well imagine a physical device implementing a logically reversible operation such as the C-NOT gate as operating either in the thermodynamically reversible or irreversible regimes.[2]

We also note that any irreversible logical operation may always be seen as part of a larger, reversible logical operation (Bennett, 1973, 1982, 1988). The idea is to keep a record of the input of all irreversible operations used during the computation in an extra memory system. Furthermore, this reversible operation can be designed such that the extra memory system is only left with a copy of the input. Ultimately, in order to reset this extra memory system, work is required according to Landauer's principle.[3]

*Logical Processes Corresponding to Thermodynamical Processes.*

Here are some further examples which might help clarify the difference between a logical and a thermodynamical process, as well as make the distinction between specifying a logical process and a pair of input and output states.

Consider a classical ideal gas of $N$ particles in a box of volume $2V$, at a temperature $T$. We wish to bring this gas to a new state with half the volume, given by the thermodynamic state $(T, V, N)$ (Figure 2.7). Up to this point, we have specified a *state transition*, that is, just a pair of input and output states.

A *thermodynamic process* implementing this transition would be for example a reversible isothermal compression of the gas. Other options are also possible, for example an irreversible fast compression of the gas, followed by a sustained contact with a heat bath at temperature $T$, letting the gas thermalize in the state $(T, V, N)$.

[2]These statements do not contradict Ladyman *et al.* (2007), because the thermodynamic processes they consider are those conditioned on the particular initial logical state of the device.

[3]Quantum mechanically, an analogous reasoning holds: any logical process can be seen as part of a larger unitary operation (known as the *Stinespring dilation*). Again, work may be required to reset the additional memory register to its original state.

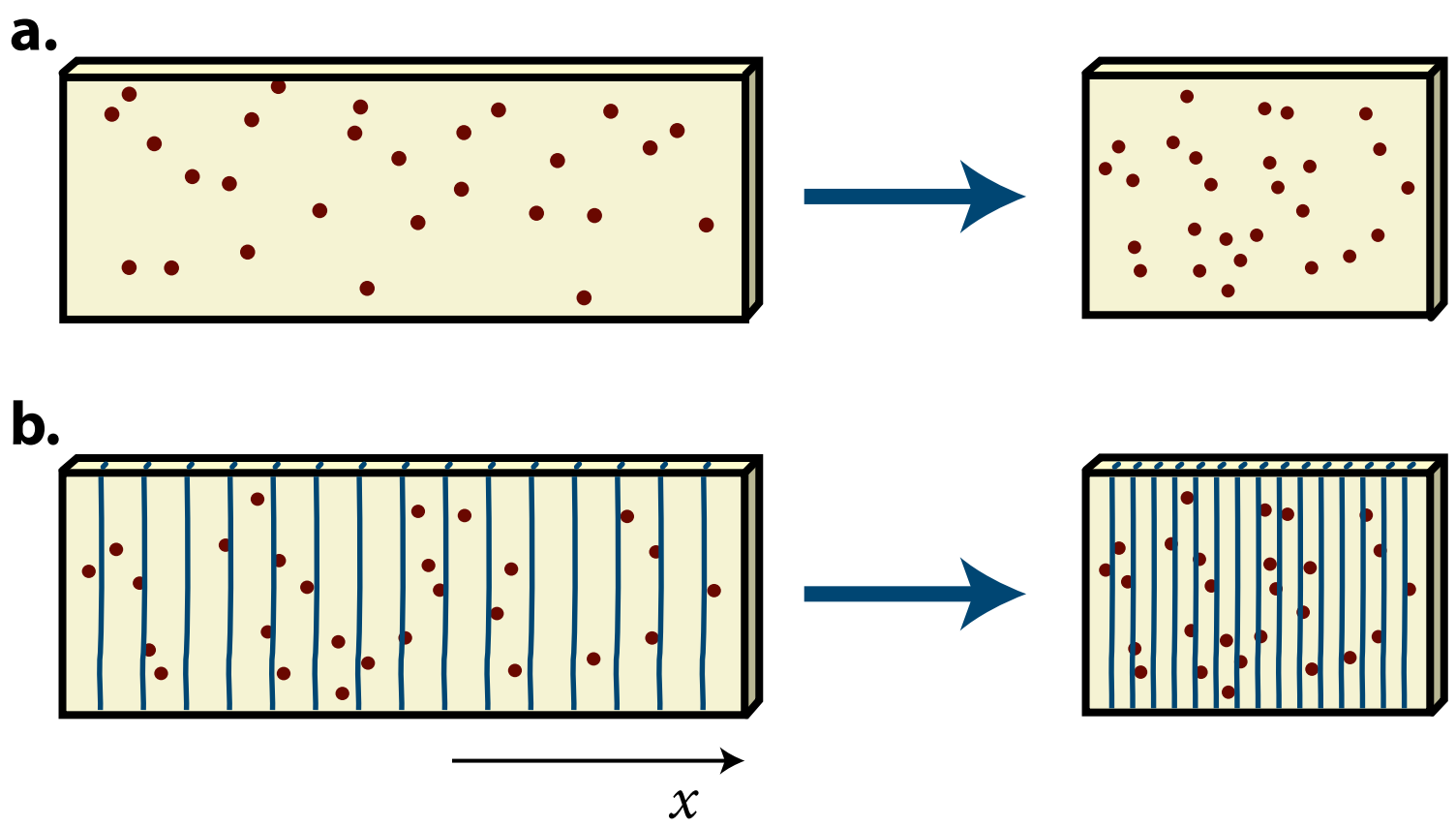

Figure 2.7: Examples of thermodynamic and logical processes. **a.** A reversible isothermal compression to half the original volume implements the logical process corresponding to randomizing the position of the particles within the new volume. **b.** Another logical process consists in mapping the $X$ positions of the particles to half their value. This can be implemented by introducing many separators, resolving the position of the particles to some acceptable precision, and then performing isothermal compression of each slice of gas.

The specification of a *logical process* goes in a different direction. For example, one can require that the position of each particle be completely randomized at the end of the process (Figure 2.7**a**), with no correlations between input and output; one can also require for example that a particle located at a position $(x, y, z)$ be located at $(x/2, y, z)$ at the output (Figure 2.7**b**). Both these logical processes have the same input and output state.[4]

The first logical process can be implemented by using either thermodynamical processes described above, i.e. the reversible isothermal process or the irreversible fast compression. It turns out that, in presence of only one heat bath at temperature $T$, the reversible isothermal process is the one which requires the least work, at the value $N \cdot kT \ln 2$.

The second logical process can be implemented similarly, by using the following trick: first, we insert many separators in the box, resolving the positions of the particles to some adequate precision, and then we perform a reversible isothermal compression of each of those slices of the gas independently. Then each particle originally located at position $x$ on the $X$-axis is now located at $x/2$ at the output, and $kT \ln 2$ work is expended per particle. (Should the $y$ and $z$ coordinates additionally be required to stay constant between the input and the output and not be randomized, a grid of separators along the other axes should also be inserted, while the compression is performed in the $X$ direction.) Again, it turns out that this work requirement of $N \cdot kT \ln 2$ is the optimal strategy to implement this logical process.

[4]The full specification of the logical process should also indicate how the momenta of the particles are mapped; we omit this for simplicity.

In this simple example both logical processes have the same minimal work requirement. This is in general not the case: We will see in the course of this thesis how the minimal work cost of implementing a specific logical process may be arbitrarily different from that of simply achieving the corresponding state transition.

## 2.6 Gambling against the second law

Finally, we introduce an additional essential concept: smoothing out improbable events in order to make physically relevant statements. This concept, in this form, originated from work in finite-size information theory (Renner, 2005; Tomamichel, 2012). From the statistical mechanical perspective, this is captured to some extent in fluctuation theorems (Jarzynski, 1997; Crooks, 1999), yet we prefer to introduce this notion from a more information-theoretic point of view, which is more general and flexible. Both approaches are connected (Åberg, 2016).

### 2.6.1 Gambling on a Szilárd engine

Consider the resetting scenario with a Szilárd engine: the particle is either on the left ("**0**") or right ("**1**") side of a Szilárd engine, and our task is to reset its state to the state **0**. Suppose now that we have prior knowledge about the particle: it is in the state **0** with probability $1 - p$, and in the state **1** with probability $p$. If $p$ is nonzero, then if we want to be certain that the particle is reset to the state **0**, we must run the Szilárd engine in "resetting mode," compressing the particle with the piston at the cost of $kT \ln 2$ work. Observe that $kT \ln 2$ work must be supplied even in this case, because when the box is put in contact with the heat bath the state is immediately randomized.

Now assume that $p$ is very small, say of the order of $10^{-23}$. Then, the particle will virtually never be observed in state **1**. So, we don't have to do anything in order to reset the engine, since from an operational point of view, no physical experiment on the box will ever be able to find the particle on the right side of the engine (except with disproportionately small probability). So, if $p$ is small enough, we may simply ignore this event and do nothing—evidently, this costs no work—instead of running the Szilárd engine which would cost $kT \ln 2$ work. So, *we can save work by ignoring events with disproportionately small probability of occurring.*

If $p$ is larger, then this event will be less negligible. If we choose to ignore this event anyway, then there is a probability $p$ that the particle is indeed observed in the state **1**; in this case the resetting procedure has failed and our "trick" has been discovered. At which point is it acceptable to ignore the event where the state of the particle is **1**? Say, if $p = 1/100$, may we still apply the same reasoning and leave the box as it is? This might depend on the application: for example, the implementation of a logical gate in a computer which is invoked millions of times may need to guarantee a very small failure probability.

The answer is to introduce an external parameter, $\epsilon$, representing the tolerated failure probability. The parameter $\epsilon$ may be chosen freely depending on the application. Our final statements will be given as a function of $\epsilon$. In the example above, the work requirement of resetting the particle is $W = kT \ln 2$ if $p > \epsilon$, but is $W = 0$ if $p \leqslant \epsilon$.

In the event of failure, which happens with probability at most $\epsilon$, the behavior is undefined and anything can happen. For an erasure process, in this case and depending on the implementation, typically more work is used than advertised, or the final state fails to be a pure state.

In conclusion, *gambling can save us work expenditure*: a higher $\epsilon$ implies more risk, but may allow us to reduce the work expenditure if successful.

### 2.6.2 Smoothing improbable events for physical statements

Introducing a failure probability is essential to recover physically relevant results, and is usually done implicitly in thermodynamics.

Consider a stone lying on the ground. There might be a very small chance that by a thermal fluctuation the stone spontaneously jumps in the air. However this event is so disproportionately unlikely that in a physical theory we may safely choose to ignore this possibility.

As another example, consider the worst-case work requirement of compressing a gas to half its volume. We then need to consider the situation where all the particles—by sheer luck—conspire to hit with high velocity against the piston. In this case much more work would be required than predicted by thermodynamics. However, this situation is so exceedingly improbable that we should ignore it in order to make a physically relevant statement.

These situations are treated explicitly with the parameter $\epsilon$ introduced above, which specifies the total probability of all events we want to exclude. In the case of compressing a gas, a reasonable $\epsilon$ to choose may be for example $\epsilon = 10^{-10}$. With this total failure probability we typically ignore all the unlucky events which would require more work, and as we will see, the minimal work requirement of compressing the gas is then exactly the one predicted by thermodynamics.

In the quantum regime, where events are generally not well-defined, this idea is captured by $\epsilon$-approximations: in our first example above, the stone has a very small amplitude of being found in the air, but its state is $\epsilon$-close to a state completely located on the ground. Analogously, we will study the work requirement of logical processes that are $\epsilon$-approximations of the desired logical process. This is a standard procedure in information theory (Renner, 2005; Tomamichel, 2012), and is justified by the fact that an $\epsilon$-approximation cannot be distinguished from the original logical process with a probability greater than $\epsilon$.

# 3

# The Formalism of Quantum Information

In this chapter, we build up the essential formalism of quantum information which forms the basis of our work.

In Section 3.1, we fix some notation and specify some concepts of linear algebra we use, and then proceed to present the elementary formalism of quantum information including quantum states, classical systems, evolutions, composition, correlations, purification, as well as distance measures between states. In Section 3.2, we use these notions to define *information entropy* as a measure of the "information content" of a system and present the *smooth entropy framework*. The toolbox of *semidefinite programming* is introduced in Section 3.3, which allows us to exploit additional structure in the optimization problems we study in later chapters. Eventually, we discuss several possible sources of confusion in Section 3.4, namely regarding the notion of a quantum system, the physical interpretation of probability as well as infinite-dimensional systems.

Along the way, we'll introduce some notation. For convenience, a summary of the notation is given on page 253.

## 3.1 Quantum information: elementary formalism

Here we present the essential formalism of quantum information, and how the individual concepts build up upon each other. We start by briefly recalling the necessary concepts from linear algebra and then proceed to introduce the elements of quantum information.

Most of the formalism presented here was principally developed by von

Neumann (1927), Fano (1957) and Uhlmann (1971, 1972, 1973, 1976, 1985), although in a somewhat different language. Statements are given for the most part without proofs, with a focus on the properties of the introduced concepts and how to apply them. For details, we refer to standard textbooks on quantum information theory such as Nielsen and Chuang (2000) and Wilde (2013). Useful references are also the lecture notes by Renner (2010) and Preskill (2015).

### 3.1.1 Necessary tools from linear algebra

Here are some basics about linear algebra on finite-dimensional spaces which we need in order to introduce the formalism of quantum information.

A Hilbert space $\mathscr{H}$ is a vector space equipped with an inner product, and which is complete, that is, in which all Cauchy sequences are convergent. All Hilbert spaces considered in quantum mechanics, and henceforth in this thesis, are complex. Also, in this thesis, we deal exclusively with Hilbert spaces which are of finite dimension.

The complex conjugate of a complex scalar $z$ is denoted as $z^*$. The inner product of the Hilbert space is denoted as $(\cdot,\cdot)$, although we solely make use of this notation in this section and otherwise faithfully stick to the bra-ket notation. The inner product is *Hermitian*, meaning that it is linear in the second variable and satisfies $(u,v) = (v,u)^*$ for all $u,v \in \mathscr{H}$. A natural norm $\|\cdot\|$ is defined on $\mathscr{H}$ as $\|u\| = \sqrt{(u,u)}$.

Extensive use will be made of Dirac's bra-ket notation. A vector in $\mathscr{H}$ is denoted by a *ket* $|\cdot\rangle \in \mathscr{H}$, where the label inside the ket notation may be anything which identifies which vector is meant. Also, an element of the dual space $\mathscr{H}^*$, the space of linear functionals on $\mathscr{H}$, is denoted by a *bra* $\langle\cdot|$, such that the *bra-ket* $\langle\phi|\psi\rangle$ is the scalar corresponding to applying the functional $\langle\phi|$ onto the vector $|\psi\rangle$. Because $\mathscr{H}$ is of finite dimension, $\mathscr{H}^*$ is canonically isomorphic to $\mathscr{H}$, and to each ket $|\phi\rangle$ corresponds a bra $\langle\phi|$ and *vice versa*; hence $\langle\phi|\psi\rangle$ can then be taken to be exactly the inner product $(|\phi\rangle, |\psi\rangle)$. Furthermore if $|\psi\rangle = \alpha|\psi_1\rangle + \beta|\psi_2\rangle$ then $\langle\psi| = \alpha^*\langle\psi_1| + \beta^*\langle\psi_2|$. Given two (finite dimensional) Hilbert spaces $\mathscr{H}$ and $\mathscr{H}'$, and given $\langle\phi| \in \mathscr{H}^*$ and $|\psi\rangle \in \mathscr{H}'$, we define the formal construct $|\psi\rangle\langle\phi|$ as the linear operator from $\mathscr{H}$ to $\mathscr{H}'$ which maps $|\phi\rangle$ onto $|\psi\rangle$ and which maps the orthogonal space to $|\phi\rangle$ onto the zero vector. This ensures a consistency in the notation in that for all $|\phi\rangle, |\psi\rangle, |\chi\rangle, |\zeta\rangle \in \mathscr{H}$, we have $\langle\chi|\left(|\psi\rangle\langle\phi|\right)|\zeta\rangle = \left(\langle\chi|\psi\rangle\right)\left(\langle\phi|\zeta\rangle\right) = \langle\chi|\psi\rangle\langle\phi|\zeta\rangle$. The bra-ket notation is designed such that all expressions involving kets and bras are linear in each

ket and bra: for example, if we have $\langle\phi| = \alpha\langle\phi_1| + \beta\langle\phi_2|$, then $|\psi\rangle\langle\phi|\chi\rangle\langle\zeta| = \alpha\,|\psi\rangle\langle\phi_1|\chi\rangle\langle\zeta| + \beta\,|\psi\rangle\langle\phi_2|\chi\rangle\langle\zeta|$.

The vector in $\mathscr{H}$ is identified by whatever is written inside the ket notation. Generally we use upright letters or Greek letters to denote a particular, chosen ket: For example, $|\mathrm{f}\rangle$ or $|\mathrm{final}\rangle$ might denote a ket corresponding to a final state of some process, and otherwise typical choices of symbols to identify particular vectors are $\psi$, $\phi$, $\chi$, etc. On the other hand, italic latin letters are used both to identify an orthonormal basis and to range over its items. For example, it is common to define a basis of $\mathscr{H}$ using the simple notation $\{|k\rangle\}$ with a single index $k$ which identifies this choice of a basis, and at the same time may be used to range over all the basis elements. This allows us to use the lighter notation, and yet still define for example a different basis, say $\{|j\rangle\}$, whose elements are solely be referenced using the other index $j$. If we need to range over the same basis several times independently, we use derived index names $k$, $k'$, $k''$, $\bar{k}$, etc. Whether the indexing starts at zero or one is greatly a matter of taste and convenience, and the exact range of the indexing is made unambiguous only whenever necessary.

Consider now an operator $A$ acting on $\mathscr{H}$, that is, a linear map $\mathscr{H} \to \mathscr{H}$. Given a basis $\{|k\rangle\}$ of $\mathscr{H}$ we refer to the scalar quantity $A_{kk'} = \langle k|A|k'\rangle$ as a *matrix element* of $A$ in that basis, and is indeed the entry in the $k^{\mathrm{th}}$ row and the $k'^{\mathrm{th}}$ column of the matrix representation of $A$ in that basis $\{|k\rangle\}$. Note consequently that $A = \sum_{kk'} A_{kk'}|k\rangle\langle k'|$.

A *projector* is an operator $P$ satisfying $P^2 = P$. In the remainder of this thesis, we exclusively consider Hermitian projectors; this is henceforth implied. For any $A$, we denote by $\Pi^A$ the projector onto the support of $A$.

The identity operator is denoted by $\mathbb{1}$, and is defined via $\mathbb{1}|\psi\rangle = |\psi\rangle$ for all $|\psi\rangle \in \mathscr{H}$. For any basis $\{|k\rangle\}$, the *closure relation* reads $\sum_k |k\rangle\langle k| = \mathbb{1}$.

The *trace* of an operator $A$ is defined as $\operatorname{tr} A = \sum_k \langle k|A|k\rangle$ for any basis $\{|k\rangle\}$; the choice of basis is irrelevant. The trace is cyclic, $\operatorname{tr} AB = \operatorname{tr} BA$. It also intuitively transforms outer products into an inner product: $\operatorname{tr}|\psi\rangle\langle\phi| = \langle\phi|\psi\rangle$.

The *adjoint*, or *Hermitian conjugate* of an operator $A$ is defined as the unique operator $A^\dagger$ satisfying the condition $(A^\dagger|\phi\rangle, |\psi\rangle) = (|\phi\rangle, A|\psi\rangle)$, or equivalently, $\langle\phi|A^\dagger|\psi\rangle = \langle\psi|A|\phi\rangle^*$, for all $|\phi\rangle, |\psi\rangle$.

A *Hermitian* operator $A$ is one that satisfies $A = A^\dagger$. All Hermitian operators have real diagonal entries, are diagonalizable in an orthogonal basis, and have real eigenvalues. An operator $A$ is *positive definite* if it is Hermitian and all its eigenvalues are strictly positive. An operator $A$ is *positive semidefinite* if it is Hermitian and all its eigenvalues are greater or equal to zero. Equivalently, $A$ is positive semidefinite if for all $|\psi\rangle$, we have $\langle\psi|A|\psi\rangle \geqslant 0$. In general, it may not be obvious to recognize whether a matrix is positive or not; we present some useful tools in Section A.1.

A *density operator* is a positive semidefinite operator of trace unity. The terms *density operator* and *density matrix* are used interchangeably. A density matrix is necessarily Hermitian, and has eigenvalues between zero and one inclusive. The sum of the eigenvalues is equal to one. We usually denote density operators with the Greek letters $\rho$, $\sigma$, and $\tau$. Density operators display a convex structure: A convex combination of any collection of density operators $\{\rho_j\}$, that is, an object of the form $\sum_j q_j\rho_j$ with $\sum_j q_j = 1$ and $0 \leqslant q_j \leqslant 1$, is again a density operator.

A *unitary operator* is an operator $U$ which satisfies $U^\dagger U = UU^\dagger = \mathbb{1}$. Equivalently, $U$ maps an orthonormal basis onto an orthonormal basis. An *isometry* is a linear map $V : \mathscr{H} \to \mathscr{H}'$ with $V^\dagger V = \mathbb{1}$ and where $VV^\dagger$ is a projector; $V$ effectively embeds $\mathscr{H}$ isometrically into a subspace of $\mathscr{H}'$.

A real function $f : \mathbb{R} \to \mathbb{R}$ is extended to accept a Hermitian operator $X$ as argument, making it a *matrix function*, by diagonalizing $X$ as $X = U\,\mathrm{diag}(x_1, \ldots, x_d)\,U^\dagger$ for some unitary matrix $U$ and with some real eigenvalues $x_1 \ldots x_d$, and by defining $f(X) := U\,\mathrm{diag}\,(f(x_1), \ldots, f(x_d))\,U^\dagger$. Alternatively, if the original $f$ has a Taylor expansion around $x = x_0$ as $f(x) = \sum a_k(x - x_0)^k$, then we may define $f(X) = \sum a_k(X - x_0\mathbb{1})^k$; these definitions are equivalent. A useful example is the *matrix square root* of a positive semidefinite operator $A$, denoted by $\sqrt{A}$ or $A^{1/2}$, which is defined in this way by taking $f(x) = \sqrt{x}$. The absolute value $|A|$ is defined analogously, and note as well that $|A| = \sqrt{AA^\dagger}$. In the same spirit, we define for $A$ its *Moore-Penrose pseudoinverse* $A^{-1}$, with $f(x) = 1/x$ for $x \neq 0$ and $f(0) = 0$, as well as its *matrix logarithm* $\log A$, by taking the function $f(x) = \log x$ for $x \neq 0$ and $f(0) = 0$. The Moore-Penrose pseudoinverse is equivalent of taking the matrix inverse of $A$ restricted to its support, and is equal to the regular inverse when $A$ is invertible.

For any operator $A$, both $AA^\dagger$ and $A^\dagger A$ are positive semidefinite, and have the same nonzero eigenvalues with matching multiplicity. The square root of these eigenvalues are called the *singular values* of $A$. For any $A$, there exists unitaries $U$, $V$ such that $A = USV^\dagger$ with a diagonal matrix $S$ whose nonzero diagonal entries are the singular values of $A$: this is called the *singular value decomposition*; the unitaries $U$ and $V$ are not unique. If $A$ is Hermitian, then the singular values are the absolute values of the eigenvalues of $A$.

For any operator $A$, we define the *trace norm* or *Schatten-1 norm*, denoted by $\|A\|_1$, as the sum of the singular values of $A$, or, equivalently, $\|A\|_1 = \operatorname{tr}|A| = \operatorname{tr}\sqrt{AA^\dagger}$. The *infinity norm* or *operator norm* $\|A\|_\infty$ of $A$ is the greatest singular value of $A$. If $A$ is positive semidefinite, this is equal to the maximal eigenvalue of $A$. Both these norms are unitarily invariant and obey $\|AB\|_\bullet \leqslant \|A\|_\bullet \|B\|_\bullet$.

### 3.1.2 Quantum states

A quantum system is always described as being in a particular *quantum state*. The quantum state should be thought of as the best possible description we can attribute to a system, in the sense that it contains everything we know about future outcomes of manipulations on this system.

***Quantum states.*** *A 'quantum state' is a density operator acting on a complex Hilbert space $\mathscr{H}$. The latter is called the 'quantum state space' associated with the quantum system.*

*The quantum state contains all the necessary information to form the most accurate predictions about the system which the observer may formulate.*

As mentioned before, we concentrate our attention solely on quantum state spaces of finite dimension. We briefly comment on infinite-dimensional quantum systems in a further section (Section 3.4.3).

A particular case of a quantum system, which is used in many examples throughout the literature in quantum information, is the *quantum bit*, or *qubit*. This is the quantum system with a two-dimensional state space $\mathscr{H} = \mathbb{C}^2$, and its basis is often denoted by $\{|0\rangle, |1\rangle\}$.

It will furthermore prove useful to note that all information contained in a quantum state $\rho$ can be retrieved via linear operators $X$ applied onto $\rho$ via constructs of the form

$$\operatorname{tr}(X\rho)\ . \tag{3.1}$$

This is simply because, for example, one can get all matrix elements of $\rho$ in this way by choosing $X$ appropriately. At this point, this observation is not useful, however later when dealing with composite systems we'll need this.

For a system of dimension $d$, the state $\mathbb{1}/d$ is called the *maximally mixed state* or *fully mixed state*. This is because it is the state of the system which is "maximally uncertain," in the sense that it has maximal entropy. It is also the state obtained by statistical mixture of all elements in any chosen basis.

There is a class of states which play a special role: those which are of the form $|\psi\rangle\langle\psi|$ for some $|\psi\rangle \in \mathscr{H}$ with $\langle\psi|\psi\rangle = 1$. These are known as the *pure*

*states*. These are states which are "maximally known," in the sense that they have zero entropy.

***Pure states***. *A quantum state $\rho$ is called 'pure' if it is rank-one. Equivalently, there exists a vector $|\psi\rangle \in \mathscr{H}$ with $\langle\psi|\psi\rangle = 1$ such that $\rho = |\psi\rangle\langle\psi|$. Also equivalently,* $\operatorname{tr}\rho^2 = 1$.

Also, a quantum state is pure if and only if it has one eigenvalue equal to one with a one-dimensional corresponding eigenspace, and all the other eigenvalues are equal to zero. Pure states are represented one-to-one by normalized kets $|\psi\rangle \in \mathscr{H}$, up to the global phase of $|\phi\rangle$. We may thus refer to a ket $|\psi\rangle$ as a *quantum state*, in place of the pure state $|\psi\rangle\langle\psi|$.

***Mixed states***. *A state which is not pure is called 'mixed.'*

The quantum states, being density operators, have a convex structure. This represents a physical *statistical mixture*.

***Statistical mixture***. *For any quantum states $\rho$, $\sigma$, and for any* $0 \leqslant \lambda \leqslant 1$, *the operator $\lambda\rho + (1-\lambda)\sigma$ is a quantum state.*

Any density matrix can be written as a convex combination of only pure states. This can be done for example by writing the density matrix in its eigenbasis as $\rho = \sum_k p_k |k\rangle\langle k|$. In fact, the pure states are the extremal points of the convex space of all density matrices on $\mathscr{H}$.

Let's restrict for a moment our consideration to pure states, represented by kets in the Hilbert space. We see that if a system can be described in either states $|\phi\rangle$ or $|\psi\rangle$, then necessarily it is in principle possible to describe it in the state $\alpha|\phi\rangle + \beta|\psi\rangle$ (with $|\alpha|^2 + |\beta|^2 = 1$ to ensure normalization). This is known as the *superposition principle*, and is a fundamental principle underlying all of quantum mechanics.

***Superposition principle***. *If a quantum system can be described as being in either state from a set of pure states $\{|\phi_i\rangle\}$, then it may also in principle be described as being in a state corresponding to any linear combination of the $\{|\phi_i\rangle\}$.*

The superposition principle allows a system to be described by infinitely more states than classical systems. Consider, for example the classical bit: it may be in only two different states **0** or **1**. However, in quantum mechanics the qubit can, thanks to the superposition principle, reach any state of the form $\alpha|0\rangle + \beta|1\rangle$ where $|0\rangle$ and $|1\rangle$ are the two basis elements. The superposition principle is simply a consequence of the linearity of quantum mechanics.

### 3.1.3 Classical states: connection to observations

A special class of quantum states are those which are diagonal with respect to a fixed, predefined basis. These are the *classical states*. They are classical because they reproduce exactly probability theory the way we've always known it.

***Classical states.*** *A quantum state $\rho$ is said to be 'classical with respect to the basis $\{|k\rangle\}$' if it is diagonal in that basis.*

Of course, any state is classical with respect to its eigenbasis. But the interest of classical states is if a system inherently carries a preferred basis due to some external physical reason. This may be caused, for example, by a physical restriction on possible manipulations or measurements on the system. A *classical system* or *classical register* is a system which has such an external restriction causing the possible states of the system to be classical with respect to a fixed basis. This fixed basis is referred to as the *classical basis* of the system.

The interest of classical states is that they can be connected to something concrete: they are the systems which behave exactly like we're classically used to. The elements $\{|k\rangle\}$ of the classical basis are those exclusive possible states the system can be in. The system may not be in any linear combination of those pure states. The eigenvalues of the density matrix are real values $\{p_k\}$ which satisfy $p_k \geqslant 0$ and $\sum_k p_k = 1$. This is precisely a probability distribution.

For classical systems, we require that the density operator act as a probability distribution, in the sense that it gives the probabilities of possible outcomes of measurements. If the system is measured, one of the possible outcomes $\{k\}$ is observed, with each $k$ appearing with probability $p_k$.

***Measurement of a classical system.*** *Suppose a classical system with classical basis $\{|k\rangle\}$, is in the state $\rho = \sum p_k |k\rangle\langle k|$. Then a 'measurement of the system in its classical basis' yields as random outcome one of the possible k labels, and where the outcome k is observed with probability $p_k$. Immediately after the measurement, the system remains in the pure state $|k\rangle$ it was observed in.*

A measurement of the classical system in its classical basis just means looking at the outcome of a random experiment. For example, if a coin is tossed without looking at the result, then the state of the coin can be described as being "heads" with probability $p$, and "tails" with probability $1 - p$ where $p$ is determined by the coin and the tossing procedure and the properties of the coin. Once we look at the coin in order to learn the result, one of the two outcomes is observed with probability $p$ or $1 - p$ respectively. Clearly, the coin remains in that state until it is tossed again. That is, if we look at the coin a second time again right after looking at once, obviously it is observed in the exact same state as the first observation.

The collapse of the state after the measurement is a phenomenon which is related to the observer: nothing happens to the system itself, but our knowledge about the system is updated. Formally, the collapse is represented by projecting the quantum state with the as $(|k\rangle\langle k|\,\rho\,|k\rangle\langle k|)/\langle k|\rho|k\rangle$ with the projector $|k\rangle\langle k|$, where the denominator ensures the normalization of the new density matrix.

Hence, standard probability theory can be fully recovered from quantum information by embedding classical distributions onto the diagonals of quantum states. This implies, in turn, that classical systems store information in Shannon's sense, and we may consider such concepts such as the Shannon entropy, the amount of classical information contained in the system, and so on.

### 3.1.4 Composite systems

Now, we need to describe the possible states which a system composed of several subsystems can take.

Here, we introduce a notation in which systems are denoted by with capital letters $A$, $B$, ..., and their corresponding Hilbert spaces are denoted by $\mathscr{H}_A$, $\mathscr{H}_B$, and so on.

***Composition of systems.*** *The Hilbert space $\mathscr{H}_{AB}$ corresponding to the joint system composed of a system A and a system B, described individually with respective Hilbert spaces $\mathscr{H}_A$ and $\mathscr{H}_B$, is given by the tensor product space $\mathscr{H}_{AB} = \mathscr{H}_A \otimes \mathscr{H}_B$. If the systems A and B are individually prepared in states $\rho_A$ and $\sigma_B$ respectively, then the joint state of the composite system is $\rho_A \otimes \sigma_B$.*

Recall that the tensor product of Hilbert spaces may be defined in the following way. If $\{|i\rangle_A\}$ and $\{|j\rangle_B\}$ are bases of $\mathscr{H}_A$ and $\mathscr{H}_B$ respectively, then $\mathscr{H}_{AB}$ is defined as the (formal) linear span of a collection of orthogonal basis elements which we denote by $|i\rangle_A \otimes |j\rangle_B$ for all possible combinations of $i$ and $j$:

$$\mathscr{H}_A \otimes \mathscr{H}_B = \mathrm{span}\,\{|i\rangle_A \otimes |j\rangle_B : i, j\} \;, \tag{3.2}$$

where $i$ and $j$ range independently on all possible elements of the basis $\{|i\rangle_A\}$ and $\{|j\rangle_B\}$. In addition, we extend the notation '⊗' from a formal definition on the basis elements to a full operation on all kets of $\mathscr{H}_A$ and $\mathscr{H}_B$, subject to the following rules for all $|a\rangle_A, |a'\rangle_A \in \mathscr{H}_A$, $|b\rangle_B, |b'\rangle_B \in \mathscr{H}_B$ and $\lambda \in \mathbb{C}$:

$$(\lambda\,|a\rangle_A) \otimes |b\rangle_B = |a\rangle_A \otimes (\lambda\,|b\rangle_B) = \lambda\,(|a\rangle_A \otimes |b\rangle_B) \;; \tag{3.3-a}$$

$$\big(|a\rangle_A + |a'\rangle_A\big) \otimes |b\rangle_B = |a\rangle_A \otimes |b\rangle_B + |a'\rangle_A \otimes |b\rangle_B \;; \tag{3.3-b}$$

$$|a\rangle_A \otimes \big(|b\rangle_B + |b'\rangle_B\big) = |a\rangle_A \otimes |b\rangle_B + |a\rangle_A \otimes |b'\rangle_B \;. \tag{3.3-c}$$

These rules ensure that we can consistently embed arbitrary individual kets $|a\rangle_A \in \mathscr{H}_A$ and $|b\rangle_B \in \mathscr{H}_B$ as a combined ket $|a\rangle_A \otimes |b\rangle_B \in \mathscr{H}_{AB}$, by writing those kets in components of their respective bases. These rules also ensure that the whole definition does not depend on the original choice of the bases $\{|i\rangle_A\}$ and $\{|j\rangle_B\}$.

Often the '⊗' symbol may be omitted when the tensor product is clear: thus $|a\rangle_A \otimes |b\rangle_B$ becomes $|a\rangle_A|b\rangle_B$ or even $|a, b\rangle_{AB}$ or $|a\,b\rangle_{AB}$. We also often refer to the composite system as $A \otimes B$ or $AB$.

Systems in the tensor product are identified by an explicit label. In this case the ordering of terms does not matter, $|a\rangle_A \otimes |b\rangle_B = |b\rangle_B \otimes |a\rangle_A$. However, in order to represent the tensor product ket in terms of a vector of components, a conventional ordering must be fixed.

Observe first that the tensor product space is the natural generalization of the classical product of probability distributions. Indeed, given two random variables $X$ and $Y$ which may take the values $\{i\}_{i=1}^{d_X}$ and $\{j\}_{j=1}^{d_Y}$ respectively, then the joint random variable describing both together is itself a random variable which may take any of the values $\{(i, j)\}_{i=1\ldots d_X, j=1\ldots d_Y}$. This is reflected in the definition of the tensor product, since we have defined the basis of the composed space as all the elements of the form $|i\rangle_A \otimes |j\rangle_B$ for all combinations of $i, j$. Secondly, our definition is compatible with the superposition postulate. Indeed, we have imposed that the elements $|i\rangle_A \otimes |j\rangle_B$ be in the joint space for all combinations of $i$ and $j$, but also, we have allowed all linear combinations of these elements by decreeing that they form a basis of the new space. This means that by representing the composition of quantum systems by a tensor product space, we have satisfied two of our ground principles from above: the relation to classical probability distributions via representation on the diagonal, and the superposition principle.

The tensor product of vectors can be lifted to tensor product of operators in the following way: for an operator $X_A$ acting on $\mathscr{H}_A$, and an operator $Y_B$ acting on $\mathscr{H}_B$, the tensor product operator $X_A \otimes Y_B$ is given by

$$X_A \otimes Y_B = \sum_{i,i',k,k'} \langle i|X|i'\rangle_A \langle j|Y|j'\rangle_B \cdot |i,j\rangle\langle i',j'|_{AB} \,. \tag{3.4}$$

In particular observe that $|i\rangle\langle i'|_A \otimes |j\rangle\langle j'|_B = |i,j\rangle\langle i',j'|_{AB}$. Notably, the tensor product is compatible with the structure of density operators: if the system $A$ is in the state $\rho_A$, and independent from system $B$ which is in the state $\sigma_B$, then the composite system is described as in the state

$$\rho_A \otimes \sigma_B = \sum_{i,i',k,k'} \langle i|\rho|i'\rangle_A \langle j|\sigma|j'\rangle_B \cdot |i,j\rangle\langle i',j'|_{AB} \,. \tag{3.5}$$

which is a well-defined density operator on the state space $\mathscr{H}_A \otimes \mathscr{H}_B$, i.e. it is both positive semidefinite and has unit trace.

Now, consider two classical, independent random variables $X$ and $Y$. If the random variables have probability distributions $p_X(i)$ and $p_Y(j)$, then the joint probability distribution is given by

$$p_{XY}(i,j) = p_X(i) \cdot p_Y(j) \,, \tag{3.6}$$

since they are independent. On the other hand if we insert two classical states $\rho_A = \sum p_i|i\rangle\langle i|_A$ and $\sigma_B = \sum q_j|j\rangle\langle j|_B$ into (3.5), then the joint density operator is given by

$$\rho_A \otimes \sigma_B = \sum_{i,j} p_i q_j \, |i\, j\rangle\langle i\, j|_{AB} \,, \tag{3.7}$$

which is again diagonal in the basis $|i,j\rangle_{AB} = |i\rangle_A \otimes |j\rangle_B$ and thus classical with respect to that basis. Furthermore, the entries have precisely the same form as in the classical situation (3.6): again, we can embed classical probabilities into quantum states naturally onto the diagonal of the state, as we did in the last section.

As mentioned above, any physical property, evolution, observation, or experiment outcome should depend only on quantities of the form $\operatorname{tr}(X_A \rho_A)$ for appropriate choices of $X_A$, where $\rho_A$ is the state of the system $A$ alone: the operators $X_A$ serve to "extract information" from system $A$. Suppose now that instead of being given $\rho_A$, we are given the description $\rho_{AB}$ of the joint system. How do we then describe the same "information extraction" on $A$ only, but now given $\rho_{AB}$? The answer is first, to stick to the tensor product for extending "information extraction" operators as well, just as we did for the density matrices themselves, and second, to identify the operator which "extracts no information at all" on $B$. The latter is simply the identity operator, because $\mathrm{tr}\,(\mathbb{1}\,\rho) = \mathrm{tr}\;\rho = 1$ for any $\rho$—the identity operator extracts nothing we didn't know already. Hence, we define the *extension of operators on additional systems* by taking the tensor product with the identity operator,

$$X_A \to X_A \otimes \mathbb{1}_B\,. \tag{3.8}$$

In compound expressions, we may often omit the identity operator and let it be implied by the context. For example, $\mathrm{tr}\,(X_A \rho_{AB}) := \mathrm{tr}\,((X_A \otimes \mathbb{1}_B)\,\rho_{AB})$.

Given a joint description $\rho_{AB}$ of the two systems, it is possible to find a corresponding description $\rho_A$ of $A$ which fully describes all information which can be "extracted" from $A$ only. Specifically, there is a $\rho_A$ such that for all $X_A$,

$$\mathrm{tr}\,(X_A \rho_A) = \mathrm{tr}\,((X_A \otimes \mathbb{1}_A)\,\rho_{AB})\,. \tag{3.9}$$

This $\rho_A$ is called the *reduced state* or *marginal state*, and is given by the *partial trace*.

***Reduced state and partial trace****. Assume that a composite system AB, with Hilbert space $\mathscr{H}_A \otimes \mathscr{H}_B$, is described as being in the state $\rho_{AB}$. Then the 'reduced state' $\rho_A$ on A is given by the 'partial trace'* $\mathrm{tr}_B(\cdot)$*:*

$$\mathrm{tr}_B\,\rho_{AB} = \sum_j (\mathbb{1}_A \otimes \langle j|_B)\,\rho_{AB}\,(\mathbb{1}_A \otimes |j\rangle_B)\;, \tag{3.10}$$

*for any basis* $\{|j\rangle_B\}$ *of* $\mathscr{H}_B$*, the choice of which is irrelevant.*

Physically, *there is no need to know the global state of a compound system to deduce the information available locally on system A* in terms of "extracting information" with operators $X_A$. This is remarkable, and even desired: it ensures that the physical theory is "local," in the sense that all information which can be "extracted" from $A$ is represented in the local object $\rho_A$. It is indeed expected that the presence, or absence of system $B$ shouldn't influence any actions performed solely on system $A$.

Often global states are defined on several systems, say $\rho_{ABC}$; afterwards, if we refer to reduced states $\rho_A$, $\rho_{AB}$, $\rho_C$ etc., they are always implicitly defined via the partial trace.

If several systems are traced out, one simply composes partial traces over each of the systems traced out or treats the ignored systems as a single system, all of these being equivalent: $\rho_C = \mathrm{tr}_A\,(\mathrm{tr}_B\,(\rho_{ABC})) = \mathrm{tr}_B\,(\mathrm{tr}_A\,(\rho_{ABC})) = \mathrm{tr}_{(AB)}\,(\rho_{ABC})$.

Again, the partial trace transforms outer products to inner products on the systems being ignored. For example,

$$\mathrm{tr}_B(|a\rangle_A |b\rangle_B \langle a'|_A \langle b'|_B) = \langle b'|b\rangle_B |a\rangle\langle a|_A \,. \tag{3.11}$$

The partial trace is compatible with the marginal of a classical probability distribution if we encode the distribution on the diagonal of a quantum state. Consider the quantum state representing the classical probability distribution $\{p_{XY}(i,j)\}$ as

$$\rho_{AB} = \sum_{ij} p_{XY}(i,j)\, |i,j\rangle\langle i,j|_{AB} \,; \tag{3.12}$$

then taking the partial trace on system $B$ yields the corresponding reduced state on $A$:

$$\rho_A = \mathrm{tr}_B\, \rho_{AB} = \sum_{ij} p_{XY}(i,j)\ \mathrm{tr}_B |i,j\rangle\langle i,j|_{AB} = \sum_i p_X(i)\, |i\rangle\langle i|_A \,, \tag{3.13}$$

where we recognize the quantum state corresponding to the marginal distribution $p_X(i) = \sum_j p_{XY}(i,j)$ of the joint distribution.

### 3.1.5 Correlations

Let's now get some intuition about the possible states on $\mathscr{H}_A \otimes \mathscr{H}_B$. These are called *bipartite states*. First of all, there are quantum states on $AB$ which cannot be written in the form $\rho_A \otimes \sigma_B$. This is clear because for classical systems such states must encode uncorrelated probability distributions, whereas $\rho_{AB}$ may encode correlated distributions. For example, the state $\rho_{AB} = \frac{1}{2}|00\rangle\langle 00|_{AB} + \frac{1}{2}|11\rangle\langle 11|_{AB}$ describes two bits which are classically perfectly correlated.

We call states which can be written as a tensor product *product states*; they are completely uncorrelated. Observe that a system in a pure state is always completely uncorrelated to any other system.

Some correlations are of classical nature: they can be explained by a joint random choice of density matrices for each system. These states are called *separable*. The joint random choice is often called a *hidden variable*.

***Separable states.*** *A state $\rho_{AB}$ on $AB$ is 'separable' if it can be written in the*

*form*

$$\rho_{AB} = \sum_i p_i \, \rho_A^{(i)} \otimes \rho_B^{(i)} \tag{3.14}$$

*for a set of pairs of density matrices* $\rho_A^{(i)}, \rho_B^{(i)}$ *and a corresponding probability distribution* $\{p_i\}$.

Among the separable states, let's distinguish those which are classical on one or the other system in a fixed, predefined basis. These naturally embed classical distributions and correlations into quantum systems.

***Classical-classical states.*** *A state* $\rho_{AB}$ *is 'classical-classical,' or simply 'classical,' with respect to bases* $\{|i\rangle_A\}$ *and* $\{|j\rangle_B\}$ *if* $\rho_{AB}$ *can be written in the form*

$$\rho_{AB} = \sum_{ij} p_{ij} \, |i\rangle\langle i|_A \otimes |j\rangle\langle j|_B \,, \tag{3.15}$$

*where* $\{p_{ij}\}_{ij}$ *is a probability distribution.*

***Classical-quantum states.*** *A state* $\rho_{AB}$ *is 'classical-quantum' (often abbreviated 'c-q' in the literature) with respect to the basis* $\{|i\rangle_A\}$ *if* $\rho_{AB}$ *can be written in the form*

$$\rho_{AB} = \sum_i p_i \, |i\rangle\langle i|_A \otimes \rho_B^{(i)} \,, \tag{3.16}$$

*for a set of states* $\{\rho_B^{(i)}\}$ *and a corresponding probability distribution* $\{p_i\}$.

*By extension, a 'classical-classical-quantum' state is one which can be written in the form*

$$\rho_{ABC} = \sum_{ij} p_{ij} \, |i\rangle\langle i|_A \otimes |j\rangle\langle j|_B \otimes \rho_C^{(ij)} \,, \tag{3.17}$$

*and so on.*

Evidently, a classical-classical state is in particular classical-quantum, and any classical-quantum state is separable.

Not all bipartite states are separable. Those states which are not separable exhibit forms of correlations which cannot be explained classically. Those states are called *entangled*.

***Entangled states.*** *A state* $\rho_{AB}$ *on AB is 'entangled' if it is not separable.*

A particularly famous class of entangled states for two qubits are the so-called *Bell states* or the *Bell basis*. These four ket vectors form indeed a basis of the two-qubit system, and are widely used in protocols which exploit entanglement, such as teleportation (Bennett *et al.*, 1993).

$$\begin{aligned} |\Phi^+\rangle &= \frac{1}{\sqrt{2}}\left[|00\rangle + |11\rangle\right]; \quad |\Phi^-\rangle = \frac{1}{\sqrt{2}}\left[|00\rangle - |11\rangle\right]; \\ |\Psi^+\rangle &= \frac{1}{\sqrt{2}}\left[|01\rangle + |10\rangle\right]; \quad |\Psi^-\rangle = \frac{1}{\sqrt{2}}\left[|01\rangle - |10\rangle\right]. \end{aligned} \tag{3.18}$$

If the qubits are realized as physical spin-½ systems as $|0\rangle \to |\uparrow\rangle$ and $|1\rangle \to |\downarrow\rangle$, then $|\Psi^-\rangle$ is the spin singlet state. This one is also called the *EPR pair*, as Einstein, Podolsky and Rosen used it to derive their paradox (Einstein *et al.*, 1935). The other three states span the spin triplet subspace. The Bell states are named after John Bell, who famously proved that there cannot be local hidden variables as Einstein, Podolsky and Rosen had suggested (Bell, 1964).

The Bell states are examples of *maximally entangled states.*

***Maximally entangled state.*** *Given two systems A and B which have the same dimension d, a 'maximally entangled state' is a state on AB of the form*

$$|\phi\rangle_{A:B} = \frac{1}{\sqrt{d}}\sum_i |i\rangle_A \otimes |i\rangle_B \,, \tag{3.19}$$

*where $\{|i\rangle_A\}$ and $\{|i\rangle_B\}$ are bases of A and B respectively.*

A maximally entangled state has the following property: it is pure as a global state, but the reduced state on either system is maximally mixed. Any state with this property is in fact a maximally entangled state.

Any pure bipartite state $\rho_{AB} = |\psi\rangle\langle\psi|_{AB}$ is either a tensor product of pure states, or it is entangled and its reduced state on either system is mixed. Both reduced states on $A$ and $B$ are closely related by the *Schmidt decomposition* of $|\psi\rangle_{AB}$:

***Schmidt decomposition.*** *For any pure bipartite state $\rho_{AB} = |\psi\rangle\langle\psi|_{AB}$, there exists two collections of orthonormal kets $\{|i\rangle_A\}$ and $\{|i\rangle_B\}$ with the same number of elements, such that*

$$|\psi\rangle_{AB} = \sum_i \gamma_i\, |i\rangle_A \otimes |i\rangle_B \,, \tag{3.20}$$

*for some strictly positive coefficients $\gamma_i$. These coefficients are the square roots of the nonzero eigenvalues of both $\rho_A$ and $\rho_B$, which have the same spectrum and multiplicities. The number of vectors is equal to the rank of both $\rho_A$ and $\rho_B$. The vectors $\{|i\rangle_A\}$ (resp. $\{|i\rangle_B\}$) are eigenvectors of $\rho_A$ (resp. $\rho_B$) corresponding* *to the eigenvalue* $\gamma_i^2$.

It will prove handy to "store" the Schmidt decomposition of $|\psi\rangle\langle\psi|_{AB} = \rho_{AB}$ by defining the following ket:

$$|\Phi^{\rho}\rangle_{AB} = \sum_i |i\rangle_A |i\rangle_B \,. \tag{3.21}$$

Its norm is given by $\langle\Phi^{\rho}|\Phi^{\rho}\rangle_{AB} = \operatorname{rank}\rho_A = \operatorname{rank}\rho_B$. It has the convenient property that

$$|\psi\rangle_{AB} = \rho_A^{1/2}|\Phi^{\rho}\rangle_{AB} = \rho_B^{1/2}|\Phi^{\rho}\rangle_{AB} \,, \tag{3.22}$$

which also gives us a way to determine $|\Phi^{\rho}\rangle_{AB}$:

$$|\Phi^{\rho}\rangle_{AB} = \rho_A^{-1/2}|\psi\rangle_{AB} = \rho_B^{-1/2}|\psi\rangle_{AB} \,. \tag{3.23}$$

If the systems $A$ and $B$ have the same dimension, it is useful to introduce the *(unnormalized) maximally entangled ket* with respect to bases $\{|i\rangle_A\}$ and $\{|i\rangle_B\}$. This ket is denoted by $|\Phi\rangle_{A:B}$ and defined as

$$|\Phi\rangle_{A:B} = \sum_i |i\rangle_A \otimes |i\rangle_B \,. \tag{3.24}$$

It satisfies $\operatorname{tr}_B|\Phi\rangle\langle\Phi|_{A:B} = \mathbb{1}_A$ and $\operatorname{tr}_A|\Phi\rangle\langle\Phi|_{A:B} = \mathbb{1}_B$, and hence $\langle\Phi|\Phi\rangle_{A:B} = \dim\mathscr{H}_A = \dim\mathscr{H}_B$. The ket $|\Phi\rangle_{A:B}$ provides us some interesting properties for pure states, which follow from the Schmidt decomposition: for any pure $\rho_{AB} = |\psi\rangle\langle\psi|_{AB}$, we have

$$|\psi\rangle_{AB} = (\rho_A^{1/2} \otimes \mathbb{1}_B)\,|\Phi\rangle_{A:B} = (\mathbb{1}_A \otimes \rho_B^{1/2})\,|\Phi\rangle_{A:B} \,, \tag{3.25}$$

from which we see that

$$|\Phi^{\rho}\rangle_{AB} = \Pi_A^{\rho_A}|\Phi\rangle_{A:B} = \Pi_B^{\rho_B}|\Phi\rangle_{A:B} \,, \tag{3.26}$$

where $\Pi_A^{\rho_A}$ (resp. $\Pi_B^{\rho_B}$) is the projector onto the support of $\rho_A$ (resp. $\rho_B$).

More generally, the unnormalized maximally entangled state $|\Phi\rangle_{A:B}$ has the property that, for any $X_A$,

$$(X_A \otimes \mathbb{1}_B)\,|\Phi\rangle_{A:B} = \left(\mathbb{1}_B \otimes X'_B\right)|\Phi\rangle_{A:B} \,, \tag{3.27}$$

where $X'_B$ is defined such that $\langle i|X'_B|j\rangle_B = \langle j|X_A|i\rangle_A$ for the same bases as those used to define $|\Phi\rangle_{A:B}$.[1]

### 3.1.6 Evolution: completely positive maps and unitaries

On the level of quantum information, we only care about evolution in terms of mapping one state to another in a discrete step. It turns out that it is not necessary to invoke any continuous *time variable*, and that the formalism can be formulated solely in terms of mappings. In practice though, a discrete mapping would have to be implemented via time evolution with a properly engineered Hamiltonian.[2]

Evolution in quantum information is described by a linear mapping $\mathcal{E}_{A\to B}$ from operators on $\mathscr{H}_A$ to operators on $\mathscr{H}_B$. Since this operator acts itself on operators, it is called a *superoperator*.

First, we need some properties of superoperators. A *superoperator* $\mathcal{E}_{A\to B}$ is a linear mapping which maps linear operators on $\mathscr{H}_A$ to linear operators on $\mathscr{H}_B$. It is *Hermiticity-preserving* if it maps Hermitian operators to Hermitian operators, *positive* if it maps positive semidefinite operators to positive semidefinite operators, and *trace-preserving* (respectively *trace-nonincreasing*) if the trace of the argument is preserved for all inputs (respectively preserved or reduced for all inputs).

The *identity process*, denoted by $\mathrm{id}_{A\to A}$ or $\mathrm{id}_A$, maps an operator on $\mathscr{H}_A$ onto itself. For two isometric spaces $\mathscr{H}_A$ and $\mathscr{H}_B$, the identity process $\mathrm{id}_{A\to B}$ maps operators on $A$ to their counterparts on $B$ (defined by a conventional choice of standard bases, or by the isometry itself). Another useful superoperator, defined for two isometric spaces $A$ and $B$ with respective bases $\{|i\rangle_A\}$ and $\{|i\rangle_B\}$, is the *partial transpose* defined as

$$t_{A\to B}(\cdot) = \sum_{i\,i'} \langle i|\cdot|i'\rangle_A |i'\rangle\langle i|_B \,. \qquad (3.28)$$

The partial transpose acting on an operator $X_{AC}$ may be either denoted by $t_{A\to B}(X_{AC})$ or by $X_{AC}^{t_{A\to B}}$, whichever is more convenient. Both the identity process and the partial transpose are positive superoperators.

Superoperators acting on different systems may be composed by tensor product, just like kets and operators. The combined action of two superoper-ators $\mathcal{E}_{A\to B}$ and $\mathcal{F}_{C\to D}$ mapping operators on $AC$ to operators on $BD$ can be described by the joint superoperator

$$(\mathcal{E}_{A\to B}\otimes\mathcal{F}_{C\to D})(\cdot)=\sum_{ii'jj'}\langle i\,j|\cdot|i'\,j'\rangle_{AC}\cdot\mathcal{E}_{A\to B}\left(|i\rangle\langle i'|_A\right)\otimes\mathcal{F}_{C\to D}\left(|j\rangle\langle j'|_C\right)\ , \tag{3.29}$$

where $\{|i\rangle_A\}$ and $\{|j\rangle_C\}$ are bases of $A$ and $C$, the choices of which are irrelevant.

[1]This is nothing else than the *(partial) transpose* operation which shall be defined in a moment, $X'_B = t_{A\to B}(X_A)$.

[2]We note that the fundamental nature of time in quantum physics has been questioned (Page and Wootters, 1983; Vedral, 2014; Giovannetti *et al.*, 2015; Ranković *et al.*, 2015, and references therein). Indeed, this external continuous parameter which has to be introduced by hand doesn't seem to fit with the rest of the formalism.

Perhaps surprisingly, the positivity of superoperators is not stable under taking a tensor product. For example, the partial transpose is positive, because it maps any operator to the operator described by the matrix transpose when represented in the corresponding basis. However, if we add a system $R$, it is straightforward to check that the mapping $t_{A\to B}\otimes\mathrm{id}_C$ applied to a maximally entangled state results in an operator which is not positive. Hence there are positive superoperators which are nonetheless nonphysical because they may output objects which are not density operators.

In order to identify the class of physical mappings, we look at those positive superoperators which always map positive operators to positive operators, even if tensor products with other systems are taken into account. Such superoperators are called *completely positive*. More precisely, a superoperator $\mathcal{E}_{A\to B}$ is *completely positive* if for any system $C$, the joint mapping $\mathcal{E}_{A\to B}\otimes\mathrm{id}_C$ is positive.

We can now formulate how quantum systems can evolve. It is still useful to describe the evolution operation as acting on two distinct systems $A$ and $B$, even if both may designate in fact the same system at different times.

***Quantum evolution***. *The most general evolution of a system A into a system B possible in quantum information is given by a trace-preserving, completely positive map which maps operators on $\mathscr{H}_A$ to operators on $\mathscr{H}_B$. Such superoperators are also called 'quantum operation' or 'channel.' They are also further on referred to as 'logical processes.'*

The assumption of linearity is in line with the postulate that the quantum state space is a linear vector space. Furthermore, this assumption makes the mapping $\mathcal{E}_{A\to B}$ automatically compatible with the structure of statistical mixture of quantum states, since it preserves convex combinations of states:

$$\mathcal{E}_{A\to B}\left(\lambda\,\rho+(1-\lambda)\,\sigma\right)=\lambda\,\mathcal{E}_{A\to B}\left(\rho\right)+(1-\lambda)\,\mathcal{E}_{A\to B}\left(\sigma\right)\ . \tag{3.30}$$

A more practical characterization of completely positive maps is that a map $\mathcal{E}_{A\to B}$ is completely positive if and only if it can be written in the form

$$\mathcal{E}_{A\to B}(\cdot) = \sum M_j (\cdot) M_j^\dagger , \tag{3.31}$$

for a collection of complex matrices $\{M_j\}$. The $M_j$'s have no constraints other than taking an input in $\mathscr{H}_A$ and yielding a vector in $\mathscr{H}_B$. This representation is called an *operator-sum representation* of $\mathcal{E}$. It is not unique.

Furthermore, a completely positive map $\mathcal{E}_{A\to B}$ is trace-preserving if and only if an operator-sum representation of $\mathcal{E}_{A\to B}$ satisfies

$$\sum M_j^\dagger M_j = \mathbb{1}_A . \tag{3.32}$$

(If one operator-sum representation satisfies (3.32), then all operator-sum representations do, so we don't have to worry about the representation not being unique.)

The systems we've been designating by $A$ and $B$ don't, in fact, have to be the same system; they don't even need to have the same dimension. If they are not the same system, this corresponds to describing part of a more global evolution. For example, if four systems $ABCD$ evolve as a global isolated system, then we can come up with a description of the evolution of $A$ onto $CD$ while ignoring $BCD$ at the input and ignoring $AB$ at the output. This "part of the global evolution" can perfectly well be itself represented by a trace-preserving, completely positive map.

There is a special class of trace-preserving, completely positive maps which play an important role: *unitary mappings*. These are the maps which preserve exactly the eigenvalues of its input for all input density matrices.

***Unitary evolution.*** *A completely positive map $\mathcal{E}_{A\to B}$ is a 'unitary mapping' if there is a unitary operator $U$ such that*

$$\mathcal{E}_{A\to B}(\cdot) = U (\cdot) U^\dagger . \tag{3.33}$$

*In particular, the spaces $A$ and $B$ must have the same dimension.*

The operator-sum representation of a unitary evolution is unique and given by (3.33).

Unitary maps play a particular role in quantum mechanics, as it turns out that the time evolution of isolated quantum systems, as given by Schrödinger's equation, is of this type. Also, as we will see, there is a duality between evolution and bipartite states, in which unitary evolutions correspond to pure states.

The fact that any evolution in quantum mechanics has to be represented by a trace-preserving, completely positive map imposes constraints on the possible processes which can happen. For example, an important result in quantum information is the *no-cloning theorem*, which asserts that it is impossible to "clone" a quantum state. Specifically, there exists no quantum mechanical evolution $\mathcal{E}_{A\to AA'}$, mapping a system $A$ to two copies of $A$, the effect of which is to clone quantum states: $\mathcal{E}(\rho) = \rho \otimes \rho$ for all $\rho$. Indeed, this mapping is not linear. Note that this does not mean that we can't create twice the same quantum state, it just means that in order to create twice the same state we need to know what the quantum state is *a priori*, that is, we must have a full classical representation of the density operator. What the no-cloning theorem asserts is that there exists no machine which is able to duplicate any quantum state one feeds into it.

### 3.1.7 Quantum measurements

Now that we have established the possible evolutions of the quantum state and that we understand how a classical memory register works, we may turn to more general forms of extracting information from quantum systems, known as *quantum measurements*.

The possible quantum mechanical evolutions severely limit the classical information one can extract from a quantum state. Consider a quantum process $\mathcal{E}_{A\to C}$ which maps a quantum system $A$ onto a classical register $C$. We require the map to always produce a classical state on $C$ with respect to a fixed basis $\{|k\rangle_C\}$. This map is the most generic form of a quantum measurement: it extracts classical information from a quantum system. Because the map always produces a classical system, we may write the output of the map for all $\rho$ as

$$\mathcal{E}_{A\to C}(\rho) = \sum f_k(\rho)\, |k\rangle\langle k|_C \,, \tag{3.34}$$

for some real functions $\{f_k(\rho)\}$. Because $\mathcal{E}$ is linear, these functions necessarily have to be linear, and hence must have the form $f_k(\rho) = \operatorname{tr}(Q_k\rho)$ for some operators $\{Q_k\}$. The output of $\mathcal{E}$ must be a classical distribution embedded onto the diagonal of the output quantum state, and hence $\{f_k(\rho)\}_k$ must form a probability distribution for all $\rho$. One can verify that this implies that the $Q_k$ must be positive semidefinite, and that $\sum Q_k = \mathbb{1}$. We obtain the following important result: a quantum measurement is necessarily described by a quantum evolution of the form

$$\mathcal{E}_{A\to C}(\rho) = \sum \mathrm{tr}\left(Q_k \rho\right) |k\rangle\langle k|_C \,, \tag{3.35}$$

where the $\{Q_k\}$ form what we call a *positive operator-valued measure*, or *POVM*: this is for our purposes any collection of positive semidefinite operators which sum to the identity.[3] An individual element $Q_k$ of a POVM is referred to as a *POVM element* or *POVM effect*. Now recall our interpretation of classical states above: a classical state of the form $\sum q_k |k\rangle\langle k|_C$ is observed as being in state "$k$" with probability $q_k$. This is the "measurement outcome." So finally, we arrive at the general result for quantum measurements:

***Quantum measurement.*** *A 'quantum measurement' is any attempt to retrieve classical information from a quantum system, which is necessarily described by a mapping from the quantum system to a classical register. This mapping is uniquely characterized by a 'POVM' $\{Q_k\}$, and if the quantum system is in the state $\rho$, then the classical register is left in the state $k$ (the 'measurement outcome') with probability* $\mathrm{tr}(Q_k \rho)$.

Let's emphasize that in this construction of quantum information theory, we have derived this result for general quantum measurements based solely on the interpretation above of classical systems, and we didn't need to introduce any *ad hoc* measurement postulate as it is common in quantum mechanics textbooks. Of course, both approaches are perfectly equivalent and the difference is essentially a matter of taste of how we want to "see things." The approach presented here might however prove more intuitive for the rest of this thesis.

Now, what is the state of the system after we've performed a quantum measurement? Let us analyze this question again using our formalism: describe the measurement process more explicitly as a quantum evolution $\mathcal{E}_{A\to CA}$ which maps the quantum system $A$ onto the classical register $C$ as well as the quantum system $A$ which contains the post-measurement state, in the form of a classical-quantum state. The most general form this mapping

[3] This terminology is often found to be confusing: yes, the "M" stands for "measure" and not "measurement." This is because the most general definition, for instance for a real-valued outcome, is of the form of a mathematical measure: it is a function which assigns a "weight" to measurable outcome sets; yet, instead of the measure taking values between zero and one, it takes as values *positive operators* which allow to calculate probabilities according to the input state. Hence the 'measure' is 'positive operator-valued.'

is then

$$\mathcal{E}_{A\to CA}(\rho) = \sum_k |k\rangle\langle k|_C \otimes \mathcal{C}^{(k)}_{A\to A}(\rho) , \tag{3.36}$$

for some trace-nonincreasing completely positive maps $\mathcal{C}^{(k)}_{A\to A}$ which satisfy $\sum_k \operatorname{tr} \mathcal{C}^{(k)}_{A\to A}(\rho) = 1$ for all $\rho$. The probability of observing each outcome $k$ is given by $f_k(\rho) = \operatorname{tr} \mathcal{C}^{(k)}_{A\to A}(\rho)$. On the classical register, the post-measurement state is given by projecting the state onto $|k\rangle\langle k|_C$ corresponding to the observed outcome; extending this projector as per (3.8) requires us to project the outcome state on $AC$ with the projector $|k\rangle\langle k|_C \otimes \mathbb{1}_A$. Hence, the state which is left on the output $A$ after observing the outcome $k$ is $\mathcal{C}^{(k)}_{A\to A}(\rho)/f_k(\rho)$, where the denominator ensures the proper normalization of the density operator as in the classical case. Furthermore, the POVM effects $\{Q_k\}$ corresponding to this measurement are fixed by the choice of the $\{\mathcal{C}^{(k)}_{A\to A}\}$.[4]

An important class of measurements are those known as *projective measurements*. These are measurements whose POVM effects are projectors, and whose post-measurement states are the input state projected using that projector corresponding to the measurement outcome.

***Projective measurement.*** *A 'projective measurement' is described by a set of mutually orthogonal projectors $\{P_k\}$ whose combined support spans all state space. Each projector corresponds to a possible measurement outcome. If the state of the quantum system to be measured is $\rho$, then the outcome labeled by "k" is observed with probability* $\operatorname{tr} P_k\rho$, *and the system is left in the post-measurement state $P_k\rho P_k$.*

For projective measurements, the explicit completely positive map (3.36) representing the evolution of the quantum state during the measurement takes the form

$$\mathcal{E}_{A\to CA}(\rho) = \sum_k |k\rangle\langle k|_C \otimes (P_k\, \rho\, P_k) . \tag{3.37}$$

A projective measurement can be represented by what is called an *observable*. An observable is any Hermitian matrix. Recall that any Hermitian matrix can be diagonalized in an orthogonal basis, and hence an observable $A$ defines a set of mutually orthogonal projectors $\{P_a\}$, labeled by eigenval-ues $a$ of $A$, with $P_a$ projecting onto the eigenspace of $A$ corresponding to the eigenvalue $a$. These projectors define a projective measurement as introduced above. This representation is the one often encountered in textbooks on quantum mechanics, and is indeed particularly useful when physical quantities are involved: the eigenvalue $a$ may then be given a physical meaning such as energy, component of momentum, spin, etc.

[4]For completeness, $Q_k$ is given by $Q_k = \mathcal{C}^{(k)\dagger}_{A\leftarrow A}(\mathbb{1}_A)$, where the adjoint of a superoperator is to be defined later in Section 3.3.1.

### 3.1.8 Purification of quantum states and maps

Purification is a striking and unique feature of quantum information. It asserts that any quantum state can be seen as part of a pure state on a larger system.

***Purification.*** *For any quantum state $\rho_A$ on a system $A$, there exists a system $R$ such that there exists a pure state $|\psi\rangle_{AR}$ with the property that*

$$\mathrm{tr}_R |\psi\rangle\langle\psi|_{AR} = \rho_A \,. \tag{3.38}$$

*Furthermore, the system $R$ may be chosen to have any dimension greater or equal to* $\mathrm{rank}\,\rho_A$.

The purification is unique up to a partial isometry on the purifying system: if $|\psi\rangle_{AR}$ and $|\phi\rangle_{AR'}$ are two pure states with $\rho_A = \mathrm{tr}_R|\psi\rangle\langle\psi|_{AR} = \mathrm{tr}_{R'}|\phi\rangle\langle\phi|_{AR'}$, then there exists a partial isometry $V_{R\to R'}$ such that $|\phi\rangle_{AR'} = (\mathbb{1}_A \otimes V_{R\to R'})|\psi\rangle_{AR}$.

For any $\rho_A$, we may explicitly construct a purification by considering a system $R$ of same dimension as $A$. A purification of $\rho_A$ is then given by $|\psi\rangle_{AR} = (\rho_A^{1/2} \otimes \mathbb{1}_R)\,|\Phi\rangle_{A:R}$, where $|\Phi\rangle_{A:R}$ is the (unnormalized) maximally entangled state between $A$ and $R$ for some choices of bases for $A$ and $R$.

It turns out that for quantum channels a similar statement can be formulated. Any channel can be seen as part of a unitary evolution.

***Stinespring dilation.*** *For any trace-preserving completely positive map $\mathcal{E}_{A\to B}$, there exists a system $E$ and an isometry $V_{A\to EB}$ such that for all $\rho$,*

$$\mathcal{E}_{A\to B}(\rho) = \mathrm{tr}_E\big[V_{A\to EB}\,\rho\,(V_{A\to EB})^\dagger\big] \,. \tag{3.39}$$

*This is called the 'Stinespring dilation' of $\mathcal{E}_{A\to B}$.*

Another form of the Stinespring dilation can be given if $B$ and $A$ are of the same dimension: for any quantum channel $\mathcal{E}_{A\to A}$ there exists a system $E$, a unitary $U_{EA}$ and a pure state $|\psi\rangle_E$ such that $\mathcal{E}_{A\to A}$ can be seen as part of the unitary process starting with a pure state on $E$:

$$\mathcal{E}_{A\to A}(\rho_A) = \mathrm{tr}_E\big[U_{EA}\,(|\psi\rangle\langle\psi|_E \otimes \rho_A)\,U_{EA}^\dagger\big]\,. \tag{3.40}$$

The latter form is physically more intuitive, as we can interpret $E$ a part of the environment of the system $A$ which $A$ interacts unitarily with.

### 3.1.9 States-mappings duality and the process matrix

A very elegant feature of quantum information is the duality between quantum states and quantum channels, which is given by the *Choi-Jamiołkowski isomorphism.*

Consider two systems $X$ and $X'$, and let a system $R$ be of same dimension as $X$. To any superoperator $\mathcal{E}_{X\to X'}(\cdot)$ corresponds an operator $E_{X'R}$, the *Choi matrix*, defined as

$$E_{X'R} = (\mathcal{E}_{X\to X'} \otimes \mathrm{id}_R)(|\Phi\rangle\langle\Phi|_{X:R})\,, \tag{3.41}$$

where $|\Phi\rangle_{X:R}$ is the (unnormalized) maximally entangled state between $X$ and $R$ in some fixed bases. The superoperator is recovered with

$$\mathcal{E}_{X\to X'}(\cdot) = \mathrm{tr}_R\left[E_{X'R}\,(\cdot)^{t_{X\to R}}\right], \tag{3.42}$$

where $t_{X\to R}$ denotes the partial transpose operation in the same bases as those used to define $|\Phi\rangle_{X:R}$. This correspondence between the superoperator $\mathcal{E}_{X\to X'}$ and the operator $E_{X'R}$ is the *Choi-Jamiołkowski isomorphism.*

Notably, the map $\mathcal{E}_{X\to X'}$ is completely positive if and only if $E_{X'R}$ is positive semidefinite. This is finally a more elegant reason why it is the completely positive maps that are physically relevant, other than by stability through tensor products.

Also, the map $\mathcal{E}_{X\to X'}$ is trace-preserving if and only if $\mathrm{tr}_{X'}\,E_{X'R} = \mathbb{1}_R$. This means, by including a factor $(\dim\mathscr{H}_R)^{-1}$ in (3.41), that *all quantum channels $X \to X'$ are in one-to-one correspondence with all bipartite density operators on $X \otimes X'$ with a fully mixed reduced state on $X$.*

Unitary evolutions correspond to pure states: for any unitary $U_X$, then the Choi matrix corresponding to the mapping $U_X(\cdot)U_X^\dagger$ is the unitary applied locally to the (unnormalized) maximally entangled state, $E_{X'R} = U_X\Phi_{X:R}U_X^\dagger$. The Choi matrix corresponding to the identity channel $\mathrm{id}_{X\to X'}$ is simply $E_{X'R} = \Phi_{X':R}$, and the Choi matrix corresponding to the partial transpose $t_{X\to X'}(\cdot)$ is $E_{X'R} = t_{X\to X'}(\Phi_{X:R}) = \sum_{i\,i'}|i\rangle\langle i'|_{X'} \otimes |i'\rangle\langle i|_R$.

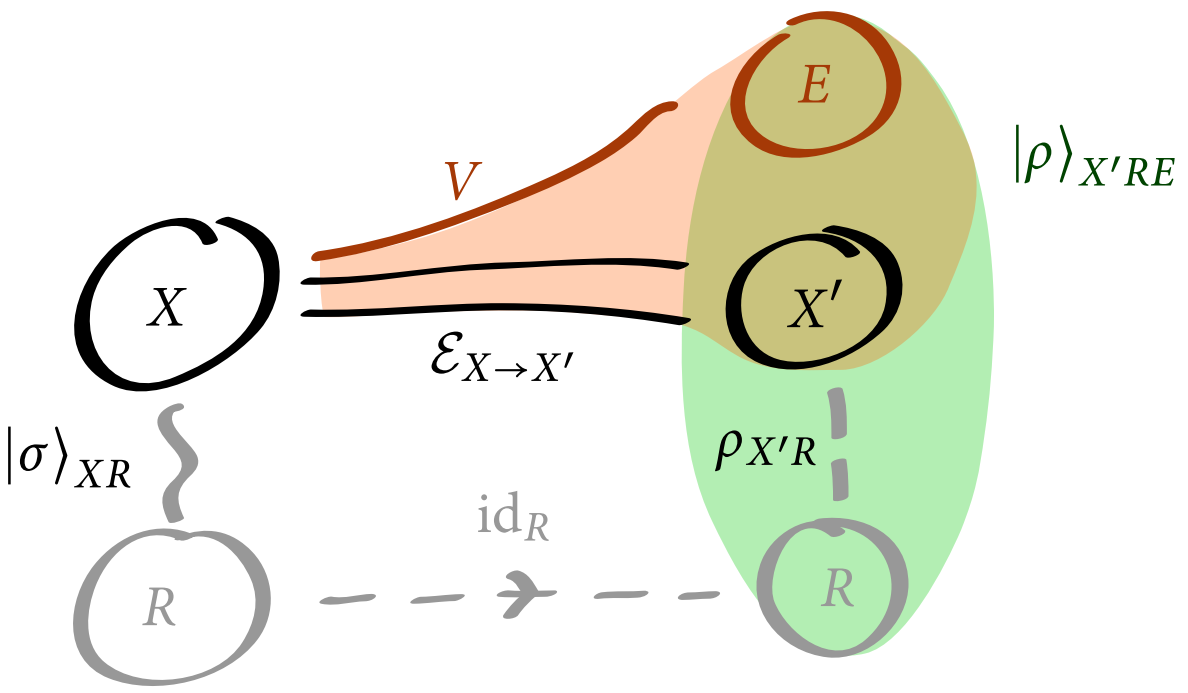

Figure 3.1: Stinespring dilation and process matrix. A quantum channel $\mathcal{E}_{X\to X'}$ maps a system $X$ to another system $X'$. Suppose the input state $\sigma_A$ is purified in a system $R$ as $|\sigma\rangle_{XR}$. Then the joint output state $\rho_{X'R}$ is the corresponding *process matrix*. Bipartite states $\rho_{X'R}$ are in one-to-one correspondence with all pairs $(\mathcal{E}_{X\to X'}, \sigma_X)$. The output of the Stinespring dilation $V$ of $\mathcal{E}$ is a purification $\rho_{X'RE}$ of $\rho_{X'R}$.

Also, for any classical process, then the corresponding Choi matrix is diagonal, and encodes all possible conditional probabilities of output states conditioned on each input.

There is more: consider again a quantum channel $\mathcal{E}_{X\to X'}$ and suppose we fix an input state $\sigma_X$ (Figure 3.1). This input state may be purified by a pure state $|\sigma\rangle_{XR} = \sigma_X^{1/2}|\Phi\rangle_{X:R}$ on a reference system $R$. Now define the state

$$\rho_{X'R} = (\mathcal{E}_{X\to X'} \otimes \mathbb{1}_R)(|\sigma\rangle\langle\sigma|_{XR}) \,. \tag{3.43}$$

The state $\rho_{X'R}$ then contains all the information about both the input state and the channel on the support of the input state. Indeed, the input state is recovered via

$$\sigma_R = \rho_R\,; \qquad \text{and} \qquad |\sigma\rangle_{XR} = \sigma_R^{1/2}|\Phi\rangle_{X:R}\,, \tag{3.44}$$

and the channel can be recovered on the support of the input state via

$$\mathcal{E}_{X\to X'}(\cdot) = \mathrm{tr}_R\Big[\rho_R^{-1/2}\rho_{X'R}\rho_R^{-1/2}\,(\cdot)^{t_{X\to R}}\Big]\,. \tag{3.45}$$

The physical meaning is remarkable: if we keep a purification of the input state to a channel on a reference system $R$, then the bipartite state remaining on the output and on $R$ contains all the information about the channel on the support of the input in the form of correlations between $X'$ and $R$.

We refer to the state $\rho_{X'R}$ as the *process matrix* of the mapping $\mathcal{E}_{X\to X'}$ with input state $\sigma_X$. For a classical process, $\rho_{X'R}$ is classical and encodes the joint probability distribution of having a certain pair of an input and an output.

This result allows us furthermore to specify an explicit form for the Stinespring dilation of a channel $\mathcal{E}_{X\to X'}$ restricted to the support of $\sigma_X$. Let $\rho_{X'R}$ be defined as above, and consider a purification $|\rho\rangle_{X'RE}$ of $\rho_{X'R}$ on an additional system $E$. As two purifications of the same reduced state $\sigma_R = \rho_R$, both states necessarily have to be related by a partial isometry, $|\rho\rangle_{X'RE} = V_{X\to X'E}\,|\sigma\rangle_{XR}$. This is precisely the Stinespring dilation on the support of $\sigma_X$:

$$\mathcal{E}_{X\to X'}(\sigma_{XR}) = \mathrm{tr}_E\Big[V_{X\to X'E}\,\sigma_{XR}\,(V_{X\to X'E})^\dagger\Big]\,. \tag{3.46}$$

### 3.1.10 Distance measures and distinguishability

Some pairs of state are easily distinguishable and others aren't. For instance, imagine that one are provided with a qubit which is prepared either in the state $\rho = |0\rangle\langle 0|$ or in the state $\sigma = |1\rangle\langle 1|$, each with probability ½. Then a simple projective measurement in the basis $\{|0\rangle, |1\rangle\}$ yields an outcome which indicates exactly whether it was $\rho$ or $\sigma$ which was prepared. In contrast, if $\rho = |0\rangle\langle 0|$ and $\sigma = (1-\epsilon)|0\rangle\langle 0| + \epsilon|1\rangle\langle 1|$ for a small $\epsilon$, the task is more difficult. Indeed, the same projective measurement yields the outcome $|0\rangle$ with high probability and thus won't be informative about whether it was $\rho$ or $\sigma$ which had been prepared originally. It turns out that no measurement would do a better job. Intuitively, these states are "close to each other." This brings us to the *trace distance*. As the name indicates, it is a distance measure between quantum states which somehow relates to the trace.

***Trace distance.*** *The trace distance between two states $\rho$ and $\sigma$ on a quantum system $S$ is defined by*

$$D(\rho,\sigma) = \frac{1}{2}\,\mathrm{tr}\,|\rho - \sigma|\,, \tag{3.47}$$

*with $|A| = \sqrt{AA^\dagger}$.*

As a distance measure, the trace distance is positive, is equal to zero only if both states are equal, is symmetric in its arguments, and obeys the triangle inequality $D(\rho,\tau) \leqslant D(\rho,\sigma) + D(\sigma,\tau)$. For pure states $|\psi\rangle$,$|\phi\rangle$, the trace distance is given by $D(|\psi\rangle\langle\psi|, |\phi\rangle\langle\phi|) = \sqrt{1 - \langle\psi|\phi\rangle^2}$. We note also a convenient expression for the trace distance given in Proposition A.5 in the appendix.

The trace distance between two states can only decrease if the states are evolved with a same quantum channel $\mathcal{E}$:

$$D(\rho, \sigma) \geqslant D(\mathcal{E}(\rho), \mathcal{E}(\sigma)) . \tag{3.48}$$

Hence, processing the state won't make the states any further apart. In particular, this holds for the partial trace:

$$D(\rho_{AB}, \sigma_{AB}) \geqslant D(\rho_A, \sigma_A) . \tag{3.49}$$

The trace distance is related to distinguishability of quantum states via the following property: if two states $\rho$ or $\sigma$ are prepared each with probability ½, then any physical process which attempts to distinguish correctly the two has a probability of success bounded by

$$\Pr[\text{success}] \leqslant \frac{1}{2}\big(1 + D(\rho, \sigma)\big) . \tag{3.50}$$

The trace distance is particularly useful because it allows us to extend operational properties of a state to states close to it at the expense of a *failure probability*. If we prove an operational statement for a state $\rho$, then that statement must hold for the state $\sigma$ except with probability at most $D(\rho, \sigma)$; this idea underlies for example the framework of universal composability in cryptography (Canetti, 2001; Ben-Or *et al.*, 2005; Renner, 2005).

Consider now the following variant of the above distinguishability problem. Assume that we are faced with the task of distinguishing the two states $\rho$ and $\sigma$, but we have potentially access to purifications of these states. The question is now slightly different: how well are we sure we can distinguish $|\psi\rangle_{AB}$ from $|\phi\rangle_{AB}$, if $|\psi\rangle_{AB}$ is a purification of $\rho_A$ and $|\phi\rangle_{AB}$ is a purification of $\sigma_A$? There is again a distance measure-like quantity which describes this setting: the *fidelity*.

***Fidelity.*** *The 'fidelity' between $\rho$ and $\sigma$ on a system $A$ is defined as the maximum overlap of the purifications of these states on any reference system $B$:*[5]

$$F(\rho, \sigma) = \max_{|\psi\rangle_{AB}, |\phi\rangle_{AB}} |\langle\psi|\phi\rangle| , \tag{3.51}$$

[5]Some authors prefer to refer to this quantity as the 'root fidelity' while reserving the term 'fidelity' for its square. The convention used here follows Nielsen and Chuang (2000).

*where* $\mathrm{tr}_B|\psi\rangle\langle\psi|_{AB} = \rho_A$ *and* $\mathrm{tr}_B|\phi\rangle\langle\phi|_{AB} = \sigma_B$.

Uhlmann's theorem relates this definition to the following closed form for the fidelity,

$$F(\rho,\sigma) = \|\sqrt{\rho}\sqrt{\sigma}\|_1 = \mathrm{tr}\sqrt{\sqrt{\rho}\sigma\sqrt{\rho}}\ . \tag{3.52}$$

The fidelity is not a distance measure *per se*, but enjoys analogous properties. It satisfies $0 \leqslant F(\rho,\sigma) \leqslant 1$, is symmetric in both arguments, and $F(\rho,\sigma) = 1$ if and only if $\rho = \sigma$. It may never decrease under simultaneous application of a channel $\mathcal{E}$ on both arguments:

$$F(\rho,\sigma) \leqslant F(\mathcal{E}(\rho),\mathcal{E}(\sigma))\ . \tag{3.53}$$

The *purified distance* is a proper distance measure which is derived from the fidelity (Tomamichel *et al.*, 2010; Tomamichel, 2012).

***Purified distance***. *The 'purified distance' is defined as*

$$P(\rho,\sigma) = \sqrt{1 - F^2(\rho,\sigma)}\ . \tag{3.54}$$

It is symmetric in its arguments, equal to zero only if both states are equal, and may never increase under quantum channels. In particular, for the partial trace we have $P(\rho_{AB},\sigma_{AB}) \geqslant P(\rho_A,\sigma_A)$.

Thanks to Uhlmann's theorem, the purified distance enjoys the useful property that, for any bipartite state $\rho_{AB}$ and for any $\sigma_A$, there exists a state $\sigma_{AB}$ such that $P(\rho_{AB},\sigma_{AB}) = P(\rho_A,\sigma_A)$. Hence, given two states $\rho_A$ and $\sigma_A$ which are close, we can "follow" an extension $\rho_{AB}$ of the state $\rho_A$ with an extension $\sigma_{AB}$ of $\sigma_A$ which is again close to it. This is in general not possible with the trace distance.

There is also a useful relation between the trace distance and the purified distance. For any $\rho$, $\sigma$, we have

$$D(\rho,\sigma) \leqslant P(\rho,\sigma) \leqslant \sqrt{2D(\rho,\sigma)}\ . \tag{3.55}$$

## 3.2 Entropy and information

In classical information theory as originally developed by Shannon (Shannon, 1948), a source emitting random symbols contains a certain amount of information as quantified by the *information entropy*. Here, we extend this to quantum systems.

The kets in a quantum system may be understood as a fundamental *unit*

*of information*: when a system is prepared in a pure state $|\psi\rangle$, it cannot be correlated with anything else, and a measurement which commutes with $|\psi\rangle\langle\psi|$ yields a single certain outcome. In this sense, the pure state is the "certain" or "definite" state. If a system is in a mixed state, then it may be correlated to other systems, and thus contain useful information about possible measurement outcomes of these other systems. In other words, it contains information we might be interested in.

To quantify the amount of information present in a quantum system, we consider *information-theoretic tasks*. These are particular tasks relating to the processing of information. One well-known task is *information compression*: given a system $X$ in a state $\rho_X$ which might be correlated to other systems, what is the smallest system size in which we can transform $\rho_X$ such that it retains its correlations with the other systems? Another task, a useful primitive in cryptography, is *randomness extraction*: given a system in the state $\rho_X$, how much uniform randomness can be extracted from it?

Historically, information theory has mostly concentrated in proving statements in the so-called *i.i.d. limit*: this is the asymptotic limit in which an average of the resource of interest is counted over many independent repetitions of the task. This situation is where the *von Neumann entropy* is most relevant.

Lately however, there has been renewed interest in studying single instances of information-theoretic tasks. Here, the von Neumann entropy is to be replaced by different measures of information, representing different information-theoretic tasks. These single instance entropy measures are more precise, and generalize the von Neumann entropy in that they all converge to the von Neumann entropy when considered in the i.i.d. limit.

### 3.2.1 Information theory in the i.i.d. limit

On average, it turns out that both tasks of information compression and randomness extraction are characterized by the same quantity. Both the average amount of storage space needed to compress $\rho_X$, as well as the amount of uniform randomness which can be extracted from $\rho_X$, are given by the *von Neumann entropy*,

$$H(X)_{\rho_X} = -\operatorname{tr}\rho_X \log\rho_X \; . \tag{3.56}$$

We emphasize that this is the case only when the average is considered over many independent repetitions of the task. This is called the *i.i.d. limit*.

The von Neumann entropy reduces to Shannon's original entropy mea-sure for classical states, the *Shannon entropy* of a random variable $X$ distributed according to $p_X(x)$:

$$H(X)_{p_X} = -\sum p_X(x) \log p_X(x) \,. \tag{3.57}$$

Furthermore, the von Neumann entropy of a quantum state is the minimal Shannon entropy of the distribution of outcomes of any possible measurement on the quantum system.

The basis of the logarithm in (3.56) and (3.57) defines the unit of entropy. In any case, the entropy quantifies an amount of information. A usual choice in information theory is to choose the logarithm in base 2, in which case the unit of entropy is called the *bit*. In physics, one often prefers to choose the natural logarithm in base $e$, and the corresponding unit of the entropy is then the *nat*. The conversion from bits to nats is given as (# of nats) = (# of bits) × ln 2.

### 3.2.2 Venturing beyond the i.i.d. limit

If the information-theoretic tasks described above are to be carried out once only, then we need to resort to different entropy measures. This is often called the *single-shot regime* in the literature. Information-theoretic tasks can no longer all essentially be described by the von Neumann entropy, is the case in the i.i.d. limit. Since the early concepts of so-called "single-shot" information theory were introduced (Renner and Wolf, 2004; Renner, 2005), which built on earlier work by Rényi (1960), the field has seen the production of a myriad of specialized entropy measures.

There are two main types of single-shot entropy measures. First, there are the *smooth entropies*. These are usually based on the smoothing of some worst-case single-shot operational measure. The *smooth min-entropy* and *smooth max-entropy*, for example, are "smoothed" versions of the min- and max-entropy respectively (Renner, 2005; Tomamichel, 2012). The latter were defined based on single instances of the operational tasks of randomness extraction and information reconciliation (Renner, 2005; König *et al.*, 2009).

Second, there are various generalizations of the Rényi entropies. Rényi originally introduced this family of entropies as a generalization of the Shannon entropy (Rényi, 1960). However, further generalizations have proven very useful in quantum information, especially in proof techniques (Renner and Wolf, 2004; Tomamichel *et al.*, 2009; Datta, 2009; Müller-Lennert *et al.*, 2013; Audenaert and Datta, 2015; Tomamichel *et al.*, 2014; Wilde, 2015). They have found an operational interpretation for error rates in hypothesis testing (Mosonyi and Ogawa, 2014).

Both of these categories of quantum information entropy measures provide generalizations with different "flavors," such as *relative entropies* and *conditional entropies*.

In addition, there are the central *von Neumann entropy* and the *quantum relative entropy*, which are recovered by essentially all other entropy measures in the i.i.d. limit. There is also an additional, curious lone wolf which does not belong directly to one of the categories above: the *hypothesis testing entropy*. It is defined via the operational task of single-shot hypothesis testing (Dupuis *et al.*, 2013). This maverick smoothly interpolates between the min- and max-entropy in a very different fashion as the $\alpha$-Rényi entropies do.

There is some evident structure in this zoo of entropy measures. We have overviewed above the main categories in which these entropies lie. However there are also additional relations between different measures: some are recovered by others in certain limits, while others act as "parents" to form "descendant entropies." The precise relations are detailed in the mentioned literature.

We present our own perspective on this flock, summarized in the form of a diagram in Figure 3.2.

### 3.2.3 The smooth entropy framework

Here we focus on a class of entropy measures characterizing information-theoretic tasks beyond the i.i.d. limit. In the course of this thesis, we shall establish strong connections between these entropy measures and thermodynamics.

The amount of quantum information required to compress the state $\rho_X$ in a single instance of the process, and be certain to succeed in doing so, is given by the *max-entropy*

$$H_{\max,0}(X)_{\rho_X} = \log \operatorname{rank} \rho_X \, . \tag{3.58}$$

The amount of uniform randomness which can be extracted on a single instance from $\rho_X$, and be certain to succeed in doing so, is given by the *min-entropy*

$$H_{\min}(X)_{\rho_X} = -\log \|\rho_X\|_\infty \, . \tag{3.59}$$

The units of entropy are again determined by the basis of the logarithms in (3.58) and in (3.59).

If side information is available, then these information-theoretic tasks may be performed more efficiently in general. For example, if we want to compress a state $\rho_X$ but we have access to a register which is perfectly correlated to it, then the state is effectively entirely reflected by our side information and there is nothing left to compress. This is given by the *conditional entropy*. Each of the entropy measures defined above has a conditional variant:

$$H(X\,|\,M)_{\rho_{XM}} = H(XM)_\rho - H(M)_\rho \; ; \tag{3.60-a}$$

$$H_{\min}(X\,|\,M)_{\rho_{XM}} = -\log \min\{\operatorname{tr}\sigma_M : \mathbb{1}_X \otimes \sigma_M \geqslant \rho_{XM}\} \; ; \tag{3.60-b}$$

$$H_{\max,0}(X\,|\,M)_{\rho_{XM}} = \log \left\| \operatorname{tr}_X[\Pi^{\rho_{XM}}_{XM}] \right\|_\infty \, . \tag{3.60-c}$$

These entropy measures are positive for classical distributions and for separable states. They may be negative for entangled states. In the case of information compression, for example, this corresponds to cases where not only the state is successfully compressed, but we even have potential for future compression for free (Horodecki *et al.*, 2005).

The min-entropy and the max-entropy can be discontinuous as a function of the quantum state. This is worrisome as the entropy measure is supposed to represent an operational quantity, yet two nearby quantum states cannot be distinguished. The answer is to *smooth the quantum states*. As mentioned in Section 3.1.10, if we design an information compression machine which works for a state, it will also work for a state which is close to that state with high probability. Hence, instead of designing an information compression machine compressing $\rho$ into a system of size $H_{\max,0}(X)_\rho$, we may design a different machine compressing a state $\hat{\rho}$ into a system of size $H_{\max,0}(X)_{\hat{\rho}}$ where we choose $\hat{\rho}$ such that $H_{\max,0}(X)_{\hat{\rho}}$ is as small as possible with $\hat{\rho}$ close to $\rho$, and the machine will still work well for $\rho$ with high probability. How close the state $\hat{\rho}$ has to be chosen to $\rho$, and hence the eventual failure probability, is specified freely by a parameter $\epsilon$. We may then define the *smooth max-entropy*, which represents the optimal compression size if a failure probability is tolerated:

$$H^\epsilon_{\max,0}(X\,|\,M)_{\rho_{XM}} = \min_{\hat{\rho}_{XM} \approx_\epsilon \rho_{XM}} H_{\max,0}(X\,|\,M)_{\hat{\rho}_{XM}} \, , \tag{3.61}$$

where the optimization is performed over all quantum states $\hat{\rho}_{XM}$ which

Figure 3.2 (on next page): The Entropy Zoo.

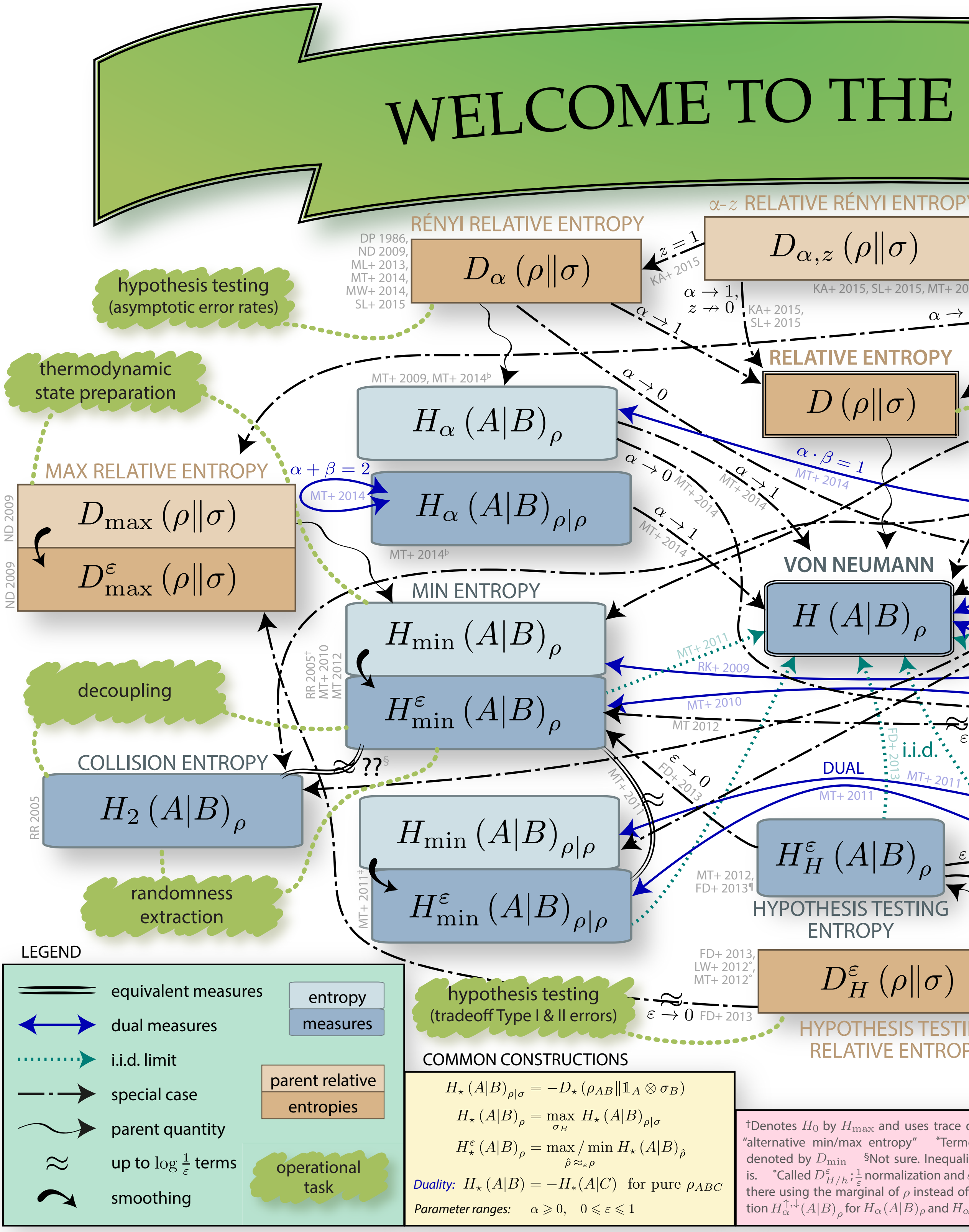

WELCOME TO THE
RÉNYI RELATIVE ENTROPY
$D_\alpha(\rho\|\sigma)$
DP 1986, ND 2009, ML+ 2013, MT+ 2014, MW+ 2014, SL+ 2015
α-z RELATIVE RÉNYI ENTROPY
$D_{\alpha,z}(\rho\|\sigma)$
KA+ 2015, SL+ 2015, MT+ 20
$z=1$
KA+ 2015
$\alpha\to 1$, $z\nrightarrow 0$
KA+ 2015, SL+ 2015
$\alpha\to 1$
RELATIVE ENTROPY
$D(\rho\|\sigma)$
hypothesis testing (asymptotic error rates)
thermodynamic state preparation
MT+ 2009, MT+ 2014♭
$H_\alpha(A|B)_\rho$
$\alpha\to 0$
$\alpha\cdot\beta=1$
MT+ 2014
$\alpha+\beta=2$
MT+ 2014
$H_\alpha(A|B)_{\rho|\rho}$
MT+ 2014♭
$\alpha\to 0$ MT+ 2014
$\alpha\to 1$ MT+ 2014
MAX RELATIVE ENTROPY
$D_{\max}(\rho\|\sigma)$
$D^\varepsilon_{\max}(\rho\|\sigma)$
ND 2009
VON NEUMANN
$H(A|B)_\rho$
MIN ENTROPY
$H_{\min}(A|B)_\rho$
$H^\varepsilon_{\min}(A|B)_\rho$
RR 2005†, MT+ 2010, MT 2012
MT+ 2011
RK+ 2009
MT+ 2010
MT 2012
decoupling
$\approx$ ??§
COLLISION ENTROPY
$H_2(A|B)_\rho$
RR 2005
$\varepsilon\to 0$
FD+ 2013
MT+ 2011
FD+ 2013
i.i.d.
DUAL
MT+ 2011
$H_{\min}(A|B)_{\rho|\rho}$
$H^\varepsilon_{\min}(A|B)_{\rho|\rho}$
MT+ 2011‡
randomness extraction
$H^\varepsilon_H(A|B)_\rho$
MT+ 2012, FD+ 2013¶
HYPOTHESIS TESTING ENTROPY
FD+ 2013, LW+ 2012°, MT+ 2012°
$D^\varepsilon_H(\rho\|\sigma)$
hypothesis testing (tradeoff Type I & II errors)
$\varepsilon\to 0$ FD+ 2013
HYPOTHESIS TESTI
RELATIVE ENTROP
LEGEND
equivalent measures
dual measures
i.i.d. limit
special case
parent quantity
$\approx$ up to $\log\frac{1}{\varepsilon}$ terms
smoothing
entropy
measures
parent relative
entropies
operational task
COMMON CONSTRUCTIONS
$H_\star(A|B)_{\rho|\sigma} = -D_\star(\rho_{AB}\|\mathbb{1}_A\otimes\sigma_B)$
$H_\star(A|B)_\rho = \max_{\sigma_B} H_\star(A|B)_{\rho|\sigma}$
$H^\varepsilon_\star(A|B)_\rho = \max/\min_{\hat\rho\approx_\varepsilon\rho} H_\star(A|B)_{\hat\rho}$
Duality: $H_\star(A|B) = -H_*(A|C)$ for pure $\rho_{ABC}$
Parameter ranges: $\alpha\geqslant 0$, $0\leqslant\varepsilon\leqslant 1$
†Denotes $H_0$ by $H_{\max}$ and uses trace c
"alternative min/max entropy" *Term
denoted by $D_{\min}$ §Not sure. Inequali
is. °Called $D^\varepsilon_{H/h}$; $\frac{1}{\varepsilon}$ normalization and
there using the marginal of $\rho$ instead of
tion $H^{\uparrow,\downarrow}_\alpha(A|B)_\rho$ for $H_\alpha(A|B)_\rho$ and $H_\alpha$
No guarantee of correctness or completeness

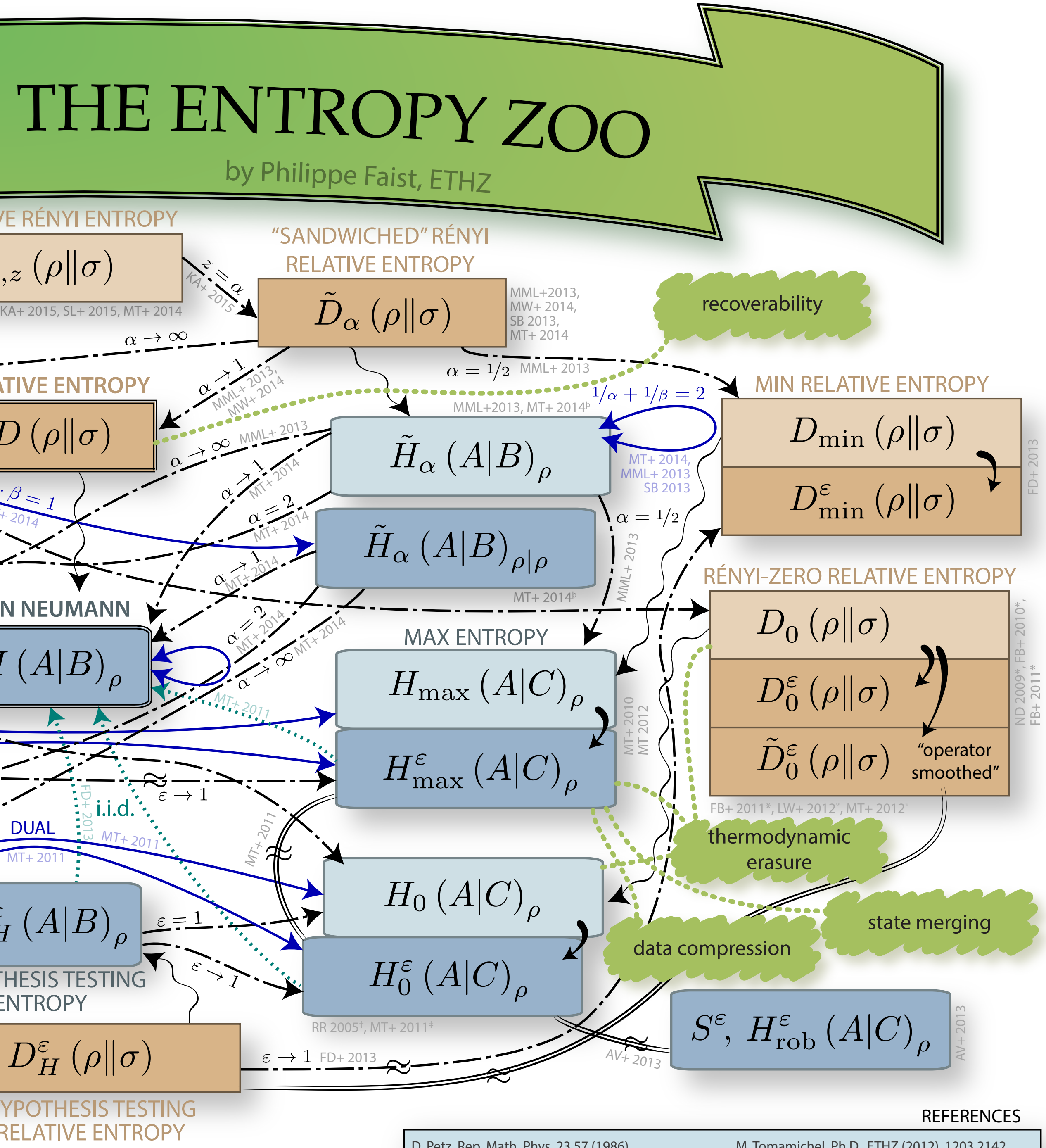

## FOOTNOTES

₀ by $H_{\max}$ and uses trace distance smoothing ‡Termed there "min/max entropy" *Termed there "min-relative entropy" and $D_{\min}$ §Not sure. Inequality at least I think. Ask Frédéric Dupu- $D^\varepsilon_{H/h}$; $\frac{1}{\varepsilon}$ normalization and $\varepsilon$ parameter meaning vary ¶Defined the marginal of $\rho$ instead of the optimization on $\sigma$. ♭Uses nota- $A|B)_\rho$ for $H_\alpha(A|B)_\rho$ and $H_\alpha(A|B)_{\rho|\rho}$ (same for $\tilde{H}_\alpha$) satisfy $P(\hat\rho_{XM}, \rho_{XM}) \leqslant \epsilon$.[6] We may define analogously the *smooth min-entropy*:

$$H^{\epsilon}_{\mathrm{min}}(X\,|\,M)_{\rho_{XM}} = \max_{\hat\rho_{XM}\approx_\epsilon \rho_{XM}} H_{\mathrm{min}}(X\,|\,M)_{\hat\rho_{XM}} \,. \tag{3.62}$$

For technical reasons, it is convenient to consider an alternative version to $H_{\mathrm{max},0}(\cdot)$, which we define as

$$H_{\mathrm{max}}(X\,|\,M)_{\rho_{XM}} = \log \max_{\substack{\sigma_M\geqslant 0\\ \mathrm{tr}\,\sigma_M\leqslant 1}} \left\| \sqrt{\rho_{XM}}\sqrt{\mathbb{1}_X\otimes\sigma_M} \right\|_1^2 ; \tag{3.63-a}$$

$$H^{\epsilon}_{\mathrm{max}}(X\,|\,M)_{\rho_{XM}} = \min_{\hat\rho_{XM}\approx_\epsilon \rho_{XM}} H_{\mathrm{max}}(X\,|\,M)_{\hat\rho_{XM}} \,, \tag{3.63-b}$$

which is also called the *max-entropy*. The two smooth measures are in fact equivalent up to factors in $\log\epsilon$, justifying their shared name (Tomamichel *et al.*, 2011, Lemma 18). The two variants $H_{\mathrm{max}}$ and $H_{\mathrm{max},0}$ of the max-entropy have also been called *max-entropy* and *alternative max-entropy*, respectively (Tomamichel *et al.*, 2011; Tomamichel, 2012).

This collection of min-entropy and max-entropy variants, which characterize the resource costs of many information-theoretic tasks in the single instance regime, forms the *smooth entropy framework* (Renner, 2005; Tomamichel, 2012).

The smooth min-entropy and the max-entropy are dual to each other: for any pure state $\rho_{ABC}$,

$$H^{\epsilon}_{\mathrm{max}}(A\,|\,B)_\rho = -H^{\epsilon}_{\mathrm{min}}(A\,|\,C)_\rho \,. \tag{3.64}$$

The two entropy measures also obey a property called strong subadditivity: for any $\rho_{ABC}$,

$$H^{\epsilon}_{\mathrm{max}}(A\,|\,BC)_\rho \leqslant H^{\epsilon}_{\mathrm{max}}(A\,|\,B)_\rho \,; \tag{3.65-a}$$

$$H^{\epsilon}_{\mathrm{min}}(A\,|\,BC)_\rho \leqslant H^{\epsilon}_{\mathrm{min}}(A\,|\,B)_\rho \,. \tag{3.65-b}$$

Physically, this property ensures that if part of the side information is ignored, then one can only have more ignorance about the system $A$.

Furthermore, in the limit of many independent copies of the quantum state (the *i.i.d. limit*), the min- and the max-entropy obey the asymptotic

[6] The choice of using the trace distance or purified distance is largely conventional, as the two are related by (3.55). We follow the convention used in the more recent literature.

equipartition property,

$$\lim_{n\to\infty} \frac{1}{n} H^{\epsilon}_{\max}(X\,|\,M)_{\rho} = H(X\,|\,M)_{\rho} \; ; \tag{3.66}$$

$$\lim_{n\to\infty} \frac{1}{n} H^{\epsilon}_{\min}(X\,|\,M)_{\rho} = H(X\,|\,M)_{\rho} \; . \tag{3.67}$$

### 3.2.4 Relative entropy

Another useful entropy measure is the *quantum relative entropy*, which describes the relative amount of uncertainty a state $\rho_X$ contains with respect to another reference operator $\Gamma_X \geqslant 0$. It is defined as

$$D(\rho_X \,\|\, \Gamma_X) = \operatorname{tr}[\rho_X(\log\rho_X - \log\Gamma_X)] \; . \tag{3.68}$$

The von Neumann entropy is recovered if we take $\Gamma = \mathbb{1}$:

$$H(X)_{\rho} = -D(\rho_X \,\|\, \mathbb{1}_X) \; . \tag{3.69}$$

This relation can be generalized for the conditional entropy as

$$\begin{aligned} H(X\,|\,M)_{\rho} &= -D(\rho_{XM} \,\|\, \mathbb{1}_X \otimes \rho_M) \\ &= -\min_{\substack{\sigma_M \geqslant 0 \\ \operatorname{tr}\sigma_M = 1}} D(\rho_{XM} \,\|\, \mathbb{1}_X \otimes \sigma_M) \; . \end{aligned} \tag{3.70}$$

In this sense, the quantum relative entropy acts as a *parent relative entropy* for the von Neumann entropy.

The quantum relative entropy, as well as its generalizations which are presented below, is only well defined if $\rho$ lies within the support of $\Gamma$.

There are also parent relative entropies for the min- and max-entropies. These are the *min-relative entropy* and the *max-relative entropy* (Datta, 2009; Dupuis *et al.*, 2013):

$$D_{\min}(\rho \,\|\, \Gamma) = -\log \|\sqrt{\rho}\sqrt{\Gamma}\|_1^2 \; ; \tag{3.71-a}$$

$$D^{\epsilon}_{\min}(\rho \,\|\, \Gamma) = \max_{\hat{\rho} \approx_{\epsilon} \rho} D_{\min}(\hat{\rho} \,\|\, \Gamma) \; , \tag{3.71-b}$$

and

$$D_{\max}(\rho \,\|\, \Gamma) = \log \|\Gamma^{-1/2}\,\rho\,\Gamma^{-1/2}\|_{\infty} \; ; \tag{3.72-a}$$

$$D^{\epsilon}_{\max}(\rho \,\|\, \Gamma) = \min_{\hat{\rho} \approx_{\epsilon} \rho} D_{\max}(\hat{\rho} \,\|\, \Gamma) \; . \tag{3.72-b}$$

These quantities come again each with both a nonsmooth and a smooth version. The expression inside the logarithm in (3.71-a) coincides with the squared fidelity if both $\operatorname{tr}\Gamma = 1$. As before, the distance measure used for smoothing is the purified distance. The min- and max-entropies may be recovered as

$$H_{\min}(X\,|\,M)_\rho = -\min_{\substack{\sigma_M\geqslant 0\\ \operatorname{tr}\sigma_M=1}} D_{\max}(\rho_{XM}\,\|\,\mathbb{1}_X\otimes\sigma_M)\,; \tag{3.73-a}$$

$$H_{\max}(X\,|\,M)_\rho = -\min_{\substack{\sigma_M\geqslant 0\\ \operatorname{tr}\sigma_M=1}} D_{\min}(\rho_{XM}\,\|\,\mathbb{1}_X\otimes\sigma_M)\,. \tag{3.73-b}$$

A similar parent measure also exists for $H_{\max,0}$ (Datta, 2009):

$$D_{\min,0}(\rho\,\|\,\Gamma) = -\log\operatorname{tr}(\Pi^\rho\Gamma)\,; \tag{3.74-a}$$

$$D^\epsilon_{\min,0}(\rho\,\|\,\Gamma) = \max_{\hat\rho\approx_\epsilon\rho} D_{\min,0}(\hat\rho\,\|\,\Gamma)\,, \tag{3.74-b}$$

which leads to

$$H_{\max,0}(X\,|\,M)_\rho = -\min_{\substack{\sigma_M\geqslant 0\\ \operatorname{tr}\sigma_M=1}} D_{\min,0}(\rho_{XM}\,\|\,\mathbb{1}_X\otimes\sigma_M)\,. \tag{3.75}$$

Let's also introduce the *hypothesis testing relative entropy* (Buscemi and Datta, 2010; Wang and Renner, 2012; Tomamichel and Hayashi, 2013; Dupuis *et al.*, 2013), and which is defined as

$$D^\eta_{\mathrm{H}}(\rho\,\|\,\Gamma) = -\frac{1}{\eta}\log\min_{\substack{0\leqslant Q\leqslant\mathbb{1}\\ \operatorname{tr}[Q\rho]\geqslant\eta}} \operatorname{tr}[Q\Gamma]\,. \tag{3.76}$$

An interesting property this quantity is that it interpolates between min- and max-relative entropy measures (Dupuis *et al.*, 2013), as

$$D^{\eta\approx 1-\epsilon}_{\mathrm{H}}(\rho\,\|\,\Gamma)\approx D^\epsilon_{\min}(\rho\,\|\,\Gamma)\,;\qquad D^{\eta\approx\epsilon}_{\mathrm{H}}(\rho\,\|\,\Gamma)\approx D^\epsilon_{\max}(\rho\,\|\,\Gamma)\,. \tag{3.77}$$

The relative entropy measures presented above all satisfy the *data processing inequality*, which states that processing of information cannot increase the relative entropy: for any completely positive, trace-nonincreasing map $\mathcal{E}$,

$$D^{\epsilon}_{\min}(\rho \,\|\, \Gamma) \geqslant D^{\epsilon}_{\min}(\mathcal{E}(\rho) \,\|\, \mathcal{E}(\Gamma))\ ; \tag{3.78-a}$$

$$D^{\epsilon}_{\max}(\rho \,\|\, \Gamma) \geqslant D^{\epsilon}_{\max}(\mathcal{E}(\rho) \,\|\, \mathcal{E}(\Gamma))\ ; \tag{3.78-b}$$

$$D^{\eta}_{\mathrm{H}}(\rho \,\|\, \Gamma) \geqslant D^{\eta}_{\mathrm{H}}(\mathcal{E}(\rho) \,\|\, \mathcal{E}(\Gamma))\ . \tag{3.78-c}$$

Furthermore all of the generalized relative entropy measures obey the *asymptotic equipartition property* in the i.i.d. limit,

$$\lim_{n\to\infty} \frac{1}{n} D^{\epsilon}_{\min}(\rho^{\otimes n} \,\|\, \Gamma^{\otimes n}) = D(\rho \,\|\, \Gamma)\ ; \tag{3.79-a}$$

$$\lim_{n\to\infty} \frac{1}{n} D^{\epsilon}_{\max}(\rho^{\otimes n} \,\|\, \Gamma^{\otimes n}) = D(\rho \,\|\, \Gamma)\ ; \tag{3.79-b}$$

$$\lim_{n\to\infty} \frac{1}{n} D^{\eta}_{\mathrm{H}}(\rho^{\otimes n} \,\|\, \Gamma^{\otimes n}) = D(\rho \,\|\, \Gamma)\ , \tag{3.79-c}$$

which hold for any $0 < \epsilon < 1$ and $0 < \eta < 1$.

## 3.3 Semidefinite calculus and optimization

This section serves to introduce some notation and simple definitions which prove useful when working with positive semidefinite operators and completely positive maps. They are useful technical tools for proving properties of entropy measures.

### 3.3.1 Operator inequalities; adjoint of superoperator; unital maps; and subnormalized states

If $A$ is a positive semidefinite operator we write $A \geqslant 0$. For two Hermitian operators $A$ and $B$ we write $A \geqslant B$ to signify that $A - B \geqslant 0$.

In this thesis, there appear only Hermiticity-preserving superoperators, which are in most cases completely positive. Given a Hermiticity-preserving superoperator $\Phi(\cdot)$, the *adjoint* $\Phi^\dagger$ of $\Phi$ is defined as the superoperator which satisfies $\mathrm{tr}(\Phi^\dagger(X)\,Y) = \mathrm{tr}(X\,\Phi(Y))$ for all Hermitian $X, Y$. The adjoint of a completely positive map is completely positive.

A superoperator which satisfies $\Phi(\mathbb{1}_X) = \mathbb{1}_Y$ is *unital*. The adjoint of a trace-preserving, completely positive map is a unital, completely positive map and vice versa.

When studying smooth entropy measures it is useful to consider not only quantum states as positive semidefinite operators $\rho$ which are normalized to $\mathrm{tr}\,\rho = 1$, but also which are *subnormalized*, i.e. which obey $\mathrm{tr}\,\rho \leqslant 1$. By abuse of terminology we also call such objects *subnormalized quantum states*. If a quantum state is assumed to be subnormalized rather than normalized, this shall always be stated explicitly.

Subnormalized quantum states have a simple physical interpretation: They correspond to states which do not account for all possible outcomes of an experiment. For example, consider a qubit defined as two selected electronic levels of an ion. If the qubit is described by the subnormalized quantum state $\rho = \mathrm{diag}(0.8, 0.1)$, this means that there is 80% probability of finding the qubit in the state $|0\rangle$, 10% probability of finding it in the state $|1\rangle$, and 10% probability that the qubit is an inaccurate representation of the ion, and that its state was in a different, unaccounted for, electronic state. In general, it is always possible to see a subnormalized state as a projected version of a normalized state on a larger system.

Similarly, it is convenient to consider logical processes which are trace nonincreasing instead of trace-preserving. A completely positive map $\mathcal{E}_{X\to X'}$ is trace-nonincreasing if and only of the Choi matrix $E_{X'R}$ of (3.41) satisfies $\mathrm{tr}_{X'} E_{X'R} \leqslant \mathbb{1}_R$. A subnormalized process matrix $\rho_{X'R}$ may be interpreted as a trace-nonincreasing logical process on a subnormalized state.

It is useful to extend the definitions of the trace distance and the purified distance to subnormalized states (Tomamichel *et al.*, 2010; Tomamichel, 2012). The definitions are chosen such that the generalized distance measures correspond to the usual distance measures for extensions of the subnormalized states to normalized states. The *generalized trace distance* is defined for two subnormalized states $\rho, \sigma$ as

$$D(\rho, \sigma) = \frac{1}{2}\|\rho - \sigma\|_1 + \frac{1}{2}|\mathrm{tr}\,\rho - \mathrm{tr}\,\sigma|\,, \tag{3.80}$$

and reduces to the usual trace distance (3.47) for normalized quantum states. The *generalized fidelity* is defined as

$$F(\rho, \sigma) = \|\sqrt{\rho}\sqrt{\sigma}\|_1 + \sqrt{(1 - \mathrm{tr}\,\rho)(1 - \mathrm{tr}\,\sigma)}\,, \tag{3.81}$$

which also reduces to the fidelity (3.52) for normalized states. Finally, the *generalized purified distance* (or also simply *purified distance* in the literature) is defined as

$$P(\rho, \sigma) = \sqrt{1 - F^2(\rho, \sigma)}\,, \tag{3.82}$$

where we have extended the fidelity in (3.54) with the generalized fidelity.

### 3.3.2 Semidefinite programming

Semidefinite programming is a useful toolbox which brings a rich structure to a certain class of optimization problems. We follow the notation of Watrous (2009, 2011), where proofs to the statements given here may also be found.

Let $A$ and $B$ be Hermitian matrices, let $\Phi(\cdot)$ be a Hermiticity-preserving superoperator, and let $X \geqslant 0$ be the optimization variable, which is a Hermitian matrix constrained to the cone of positive semidefinite matrices. The prototypical *semidefinite program* is an optimization problem of the following form:[7]

$$\text{minimize}: \quad \operatorname{tr}(A X) \tag{3.83-a}$$
$$\text{subject to}: \quad \Phi(X) \geqslant B\,. \tag{3.83-b}$$

To any such problem corresponds another, related problem in terms of a different variable $Y \geqslant 0$:

$$\text{maximize}: \quad \operatorname{tr}(B Y) \tag{3.84-a}$$
$$\text{subject to}: \quad \Phi^{\dagger}(Y) \leqslant A\,. \tag{3.84-b}$$

The first problem is called the *primal problem*, and the second, *dual problem*. Either problem is deemed *feasible* if there exists a valid choice of the optimization variable satisfying the corresponding constraint. If there exists a $X \geqslant 0$ such that $\Phi(X) - B$ is positive definite, the primal problem is said to be *strictly feasible*; the dual is *strictly feasible* if there is a $Y \geqslant 0$ such that $A - \Phi^{\dagger}(Y)$ is positive definite. For these two problems, we define their optimal attained values

$$\alpha = \inf\left\{\operatorname{tr}(A X) : \Phi(X) \geqslant B, X \geqslant 0\right\}; \tag{3.85-a}$$
$$\beta = \sup\left\{\operatorname{tr}(B Y) : \Phi^{\dagger}(Y) \leqslant A, Y \geqslant 0\right\}, \tag{3.85-b}$$

with the convention that $\alpha = -\infty$ if the primal problem is not feasible and $\beta = +\infty$ if the dual problem is not feasible.

The first important relation between the primal and dual problems is *weak duality*.

***Weak duality***. *For any semidefinite program, we have*

$$\alpha \geqslant \beta\,. \tag{3.86}$$

[7] Several equivalent prototypical forms for semidefinite programs exist in the literature.

This convenient relation allows us to immediately bound the optimal attained value of one of the two problems by picking any valid candidate in the other.

For some pairs of problems, we may have $\alpha = \beta$. In those cases we speak of *strong duality*. This is often the case in practice. A useful result here is Slater's theorem, providing sufficient conditions for strong duality (Watrous, 2009, Theorem 2.2).

***Slater's conditions for strong duality***. *Consider any semidefinite program written in the form* (3.83), *and let its dual problem be given by* (3.84). *Then:*

*(i) if the primal problem is feasible and the dual is strictly feasible, then strong duality holds and there exists a valid choice $X$ for the primal problem with* $\operatorname{tr}(AX) = \alpha$*;*

*(ii) if the dual problem is feasible and the primal is strictly feasible, then strong duality holds and there exists a valid choice $Y$ for the dual problem with* $\operatorname{tr}(BY) = \beta$*;*

*(iii) if both problems are strictly feasible, then strong duality holds and there exist valid choices of $X$ and $Y$ with* $\alpha = \beta = \operatorname{tr}(AX) = \operatorname{tr}(BY)$.

We note that strong duality in itself doesn't necessarily imply the existence of an optimal choice of variables attaining the infimum or supremum. In practice, the existence of optimal primal or dual choices may be explicitly stated by Slater's conditions. Alternatively, it may be deduced by straightforward arguments if, for example, the constraints force the integration region to be compact.

Finally, there are many optimization problems which do not immediately ressemble (3.83) and yet which can be recast in that form. Examples include the fidelity between to states, the square of the fidelity, the trace distance, the infinity norm, as well as most of the smooth entropies (Tomamichel, 2012). Some tricks include casting a maximization problem into a minimization with a minus sign; several variables or constraints can be grouped together into a block-diagonal matrix or superoperator; equality constraints can be accommodated by imposing inequality constraints in both directions; and finally variables can be Hermitian instead of positive semidefinite, by decomposing them into positive and negative parts. We may also optimize over quantum channels by representing them via their Choi state (3.41). Furthermore, we note that *strict feasibility* only refers to inequality constraints and in this sense is not automatically killed by equality constraints (Watrous, 2009).

## 3.4 Conceptual remarks

### 3.4.1 Physical systems and logical systems

Compared to more traditional quantum mechanics, the power of quantum information theory lies in the handling of information as an abstract concept, regardless of the physical system the information is encoded on. This coincides with Shannon's original ideas: describe information in an abstract mathematical framework regardless of its physical support.

In quantum information, a quantum system may be any abstract place where quantum information is stored, and does not have to correspond exactly to a particular physical entity such as an atom or an electron. A quantum system may be a part of a physical system. For example, in typical implementations of a qubit, two well isolated energy levels are singled out to form a two-dimensional subsystem of a more complex physical system. We use the term *logical system* to refer to such quantum systems, that is, abstract quantum systems describing quantum information which may be realized in any way we would like. A logical system is usually realized as a part of a physical system; a logical system could even be an entangled subspace of several physical systems.

The distinction between a logical system and a physical system is for us artificial and serves only to hint at how the system may be thought of. A physical system may be associated with one of (or a composition of) the basic building blocks the physical theory provides us—atoms, electrons, photons, etc., and inherits additional properties such as a Hamiltonian and ways to interact with it which are relevant for experiments. By contrast, the term *logical system* serves to emphasize that the quantum system does not need to be associated with a clearly defined physical entity such as an atom or an electron. A logical system is typically a virtual quantum system storing information which we choose to treat explicitly, and which in practice is a part of a larger physical system.

Consider a system $\mathscr{H}_A$ described with a basis $\{|k, l, \ldots\rangle_A\}$ with quantum numbers $k, l, \ldots$, and consider systems $\mathscr{H}_K, \mathscr{H}_L, \ldots$ with bases $\{|k\rangle_K\}$, $\{|l\rangle_L\}$ and so on. The state space $\mathscr{H}_A$ can be embedded isometrically into the tensor product space $\mathscr{H}_K \otimes \mathscr{H}_L \otimes \cdots$ with the help of the isometry $|k, l, \ldots\rangle_A \to |k\rangle_K \otimes |l\rangle_L \otimes \cdots$. The systems $\mathscr{H}_K, \mathscr{H}_L, \ldots$ are called *logical systems*. They describe explicitly information which is represented implicitly as part of the state of the full physical system. The interest in doing so is that we can now describe the information about only part of the system, say $\mathscr{H}_K$, as we would do with a proper subsystem; we may for instance trace out the systems $L, \ldots$ from a general state on $A$ to obtain information that this state contains about the quantum number $k$.

Consider for example the electronic states of the hydrogen atom with a single electron, which are derived in any quantum mechanical textbook (Cohen-Tannoudji *et al.*, 1977). A basis of the state space $\mathscr{H}_A$ is given by the states $|k\, l\, m\rangle_A$, where the quantum numbers $k$, $l$ and $m$ relate to the eigenvalues of the energy, total angular momentum and Z angular momentum observables. These quantum numbers obey the relations $0 \leqslant l < k$ and $-l \leqslant m \leqslant l$. We may embed these basis states into logical systems $\mathscr{H}_K \otimes \mathscr{H}_L \otimes \mathscr{H}_M$ as described above. Suppose for example that for some physical reason the state of the atom is invariant under rotations: then the information this state carries about the quantum number $k$ is readily given by tracing out from the global state on $A$ the systems $L$ and $M$. Furthermore, a time evolution which commutes with rotations must act locally on the $K$ register.

While logical systems in the sense above may be seen as "parts" of a physical system, they may also be seen as the result of a quantum computation. In this picture, the isometrical embedding above is the quantum computation itself, and the input and output systems have to be padded with zero states appropriately to make the computation unitary. This is the case for example of the Schur transform, and in particular the quantum Fourier transform (Harrow, 2005).

### 3.4.2 Probabilities and repetitions of the experiment

Here we briefly recall a basic interpretation of probability and its relevance in our context. We adhere to the view that probabilities are associated with an *experiment*, that is, a particular procedure which can be repeated in principle an arbitrary number of times, and not a *particular instance* of an experiment. The same is true for the quantum state.

Suppose an experiment yields random outcomes, and assume, for simplicity, that there is a finite amount of possibilities. Assigning probabilities to outcomes amounts to assigning a value between zero and one to each individual outcome, where the sum of all the attributed values must be equal to one.

These values—*probabilities*—can be used to characterize a classical physical system about which we don't know some information, for example a coin which is tossed. But what do these values physically mean? An unambiguous meaning is given if we consider the experiment as being repeated independently an asymptotically large number of times. We don't actually have to physically repeat the experiment that many times, we just need to hypothetically be able to do so in principle. For example, the experiment consisting in a coin toss is definitely an experiment we could repeat in principle an arbitrary large number of times. Then, we would empirically observe that the fraction of times we see the outcome "heads" converges to certain particular value; this value is the *probability* of the outcome "heads." Thus, the probability of an event has the interpretation of the empirical fraction of trials in which that event is observed if the experiment were to be repeated asymptotically many times. Note that there, the probabilities refer to a *particular experiment* or *experimental procedure*, and not necessarily to a single instance of an experiment.

The same reasoning applies in quantum information. The calculated probabilities of the measurement outcomes, and hence the quantum state, can be given a clear, unambiguous physical meaning when associated with a *situation* or *experiment* which can be in principle repeated an arbitrary number of times.

The strict view presented here suffices for the purposes of this thesis. It is also representative of the standard setting in physics: by nature, physics deals with the description of experiments which can *in principle* be carried out an arbitrary number of times.

We note that the present view can be extended to give a well defined meaning to probabilities also for finite instances of an experiment, for example, by adopting a Bayesian point of view (de Finetti, 1974; Jaynes, 1986; Caves *et al.*, 2002a; Fuchs and Schack, 2015). Using de Finetti techniques, it is furthermore known that the Bayesian point of view coincides with the description given above in the asymptotic limit of many independent repetitions (de Finetti, 1937; Caves *et al.*, 2002b; König and Renner, 2005).

### 3.4.3 Infinite-dimensional quantum systems

As stated above, the systems considered in this thesis have finite-dimensional state spaces. There are two main reasons for this restriction. First, it greatly simplifies the technical machinery which needs to be put in place, and avoids issues with divergent quantities. Second, usual physical systems which are represented by infinite-dimensional systems can be approximated to arbitrary precision by finite-dimensional descriptions. What's more, finite-dimensional approximations are arguably closer to a physical observer's actual description than the infinite-dimensional one, because of the limited precision of the observer's measurements.

We do note that several of the tools used in this thesis have been successfully adapted to the language of infinite-dimensional quantum systems, in particular the smooth entropy framework (Furrer *et al.*, 2011; Berta *et al.*, 2016). Furthermore the idea of approximating infinite-dimensional systems with finite-dimensional approximations is not novel, and has been shown to apply even to quantum field theory (König and Scholz, 2015, 2016). These results suggest that our results can be suitably generalized to infinite-dimensional systems, too.

# 4

# Frameworks for Finite-Size Thermodynamics

In this chapter we concentrate on how to extend the concepts of thermodynamics to micro and nano scales.

Even though virtually every paper in the area has its own framework, each one differing in subtle ways, these frameworks are largely equivalent when it comes to recovering the second law of thermodynamics, for example. However, the question remains whether a single framework is capable of providing a unified picture of thermodynamics of quantum systems.

There are two main lines of thought for developing such a framework: the *operational approach*, in which one tries to characterize all operations which can be performed by an agent without any resource cost, and the *fundamental limits* approach, in which one rather characterizes which operations can under no circumstance be performed by the agent. The first approach is better suited for proving achievability of a particular transformation or operation: by definition there exists an explicit procedure to carry out the transformation. The second approach is better suited for second-law type limitation statements: in any case the agent could not do better. The second approach is impermeable to the possible criticism arguing that we could have forgotten to include some further possible operations.

Thermal operations is a framework of the first kind. This framework allows only operations which can be performed with energy-preserving unitaries and ancillas initialized in thermal states. This framework characterizes which state transitions are possible on the the a system, making it a so-called *resource theory* (Janzing *et al.*, 2000; Janzing, 2006; Brandão *et al.*, 2013, 2015). Thermal operations have been extensively exploited to understand thermodynamics at the quantum level (Horodecki *et al.*, 2003; Janzing, 2006; Brandão *et al.*, 2013; Horodecki and Oppenheim, 2013a; Gour *et al.*, 2015).

Gibbs-preserving maps, on the other hand, is a framework of the second kind. In this framework, all operations are allowed for free except those which do not preserve the Gibbs state on the system. This is the most general free operation one can allow without making the theory trivial.

It turns out that these two frameworks do not allow the same state transformations. We show that there are state transformations which are allowed by Gibbs-preserving maps which cannot be performed by a thermal operation. The difference may be surprising since the two frameworks are equivalent classically. The cause of this gap could be thought to be a different treatment of *time control*. However we conjecture that the gap in the frameworks can be explained by the difference in the framework depending on the Hamiltonian directly or only through Gibbs states.

The present chapter is organized as follows. First, we introduce the Gibbs state and its essential properties, as it forms the cornerstone of the frameworks we introduce later. Then, we expose and comment on the gap between thermal operations and Gibbs-preserving maps. We then investigate so-called *batteries* or *work storage systems*, which serve to define the concept of *work*. Several models used in the literature are presented, namely the *work bit*, or *wit*, the *ladder of energy levels* meant to emulate a weight, and the *information battery* which stores work in the form of purity.

In this chapter we focus only on those frameworks which are relevant to our results. We point to more comprehensive reviews for further discussion of various other models (Goold *et al.*, 2015a; Vinjanampathy and Anders, 2015).

## 4.1 The Gibbs state

In thermodynamics and statistical mechanics, a particular class of microscopic states are singled out: these are the states which are *at thermal equilibrium*, that is, which are left invariant by the stochastic nature of the system's time evolution when in contact with a thermal bath. This state takes the form of the so-called *Gibbs state*, and simultaneously enjoys several defining properties. It is the state which a system in contact with a bath will tend to for long times under suitable conditions on the dynamics. It is also the state which, in a statistical inference picture, maximizes the von Neumann entropy subject to relevant constraints. Finally, it is the only state from which no work can be extracted. Crucially, these states correspond exactly to the *thermodynamic states* in macroscopic thermodynamics—that is, the states described with thermodynamic variables such as $U, T, S, p, V, \mu$ and $N$.

Let any system interact suitably with a bath, and its state will approach an equilibrium state as time evolves. Under typical assumptions, such as weak coupling to a very large heat bath, a quantum system $S$ described by a Hamiltonian $H_S$ approaches the *Gibbs state* (Huang, 1987; Gemmer *et al.*, 2009)

$$\gamma_S = \frac{1}{Z} e^{-\beta H_S} , \tag{4.1}$$

where $Z = \operatorname{tr} e^{-\beta H_S}$ is the *partition function* and where $\beta$ is the inverse temperature of the heat bath. The system is typically well described after large times by the state (4.1) regardless of its initial state before contact with the bath. The more specific conditions under which systems equilibrate, and at which speed they do so, may be highly nontrivial to determine, and constitute an active field of research (Gemmer *et al.*, 2009; Riera *et al.*, 2012; Gogolin and Eisert, 2016).

Similarly, the Gibbs state is well defined if the system interacts with several other baths, such as a particle reservoir (cf. any textbook). If a system exchanges energy and particles with a bath, then its Gibbs state is given by

$$\gamma_S = \frac{1}{Z} e^{-\beta(H_S + \mu N_S)} , \tag{4.2}$$

with the partition function $Z = \operatorname{tr} e^{-\beta(H_S+\mu N_S)}$, and where $\beta$ is the inverse temperature and $\mu$ is the chemical potential of the bath. More generally, if the physical quantities designated by observables $\{Z_S^{(j)}\}$ are exchanged each with a bath of that particular type, then the Gibbs state is simply given as

$$\gamma_S = \frac{1}{Z} e^{-\sum_j \mu_j Z_S^{(j)}} , \tag{4.3}$$

where $Z = \operatorname{tr} e^{-\sum_j \mu_j Z_S^{(j)}}$ is the partition function and where $\mu_j$ is the intensive variable corresponding to the bath $j$ which generalizes temperature and the chemical potential. Equation (4.3) is also valid if the baths exchange charges $Z_S^{(j)}$ which don't commute (Guryanova *et al.*, 2016; Yunger Halpern *et al.*, 2016).

The role that time evolution plays in macroscopic thermodynamics is relatively minor: it only serves to guarantee that the set of possible states is restricted to the class of thermal states. Hence, one can wonder whether time evolution is needed at all in order to single out this special class of states.

Indeed, the Gibbs state also appears as a statistical inference problem, independently of any explicit dynamical argument. It is, following Jaynes (1957a,b), the state with maximal uncertainty subject to constraints involving physical quantities (such as energy) which are exchanged with a bath. Specifically, Jaynes points out that (4.3) is precisely the solution to the optimization problem

$$\text{maximize:} \quad H(\rho) \tag{4.4-a}$$

$$\text{subject to:} \quad \operatorname{tr}\left[Z_S^{(j)}\rho\right] = z_j \quad (\forall\, j)\,, \tag{4.4-b}$$

where $H(\rho)$ is the von Neumann entropy, the $z_j$ are some fixed values which constrain the expectation value of $Z_S^{(j)}$, and where $\rho$ ranges over all quantum states. This form of statistical inference is commonly called *Max-Ent*.

It might seem unnatural in this approach to constrain the expectation value of the observables $Z_S^{(j)}$, rather than impose that the state lives within an eigenspace of $Z_S^{(j)}$ of corresponding eigenvalue $z_j$ in the spirit of a microcanonical ensemble. However this is justified if a large number $n$ of independent copies of the system are considered. In that case, by typicality, the average value of $Z_S^{(j)}$ for a single system is directly related to the total, peaked value of the $Z_{S^n}^{(j),n}$ quantity (Yunger Halpern *et al.*, 2016).

There is a further defining property of the class of Gibbs states: they are the only states from which no work can be extracted. This notion is made formal via what is known as *passivity* and *complete passivity*. A state $\rho$ is *passive* if for any unitary $U$, the expectation value of the Hamiltonian $H$ decreases or stays constant if the unitary is applied onto $\rho$: $\operatorname{tr}(U\rho U^\dagger H) \leqslant \operatorname{tr}(\rho H)$. The state $\rho$ is *completely passive* if $\rho^{\otimes n}$ is passive for all $n > 0$. The class of states which are completely passive is exactly the family of Gibbs states of the form (4.1) with $\beta$ which may be chosen freely (Pusz and Woronowicz, 1978; Lenard, 1978). The result also extents to multiple baths exchanging heat, particles, or other physical quantities which don't even have to commute (Yunger Halpern and Renes, 2016; Yunger Halpern *et al.*, 2016). An important implication is in the context of resource theories of thermodynamics (Janzing *et al.*, 2000; Janzing, 2006; Brandão *et al.*, 2013; Horodecki and Oppenheim, 2013a,b; Brandão *et al.*, 2015; Goold *et al.*, 2015a): the thermal states are the only states which may be allowed for free in the resource theory without rendering the theory trivial.

The frameworks we will now present belong to a more resource theory approach, and heavily rely on the notion of a Gibbs state.

## 4.2 Thermal operations

Thermal operations is a resource theory, that is, a set of rules specifying which operations can be performed on a system. Consider a system $S$ which we would like to apply transformations on. The system has a well-defined Hamiltonian, denoted by $H_S$, and does not interact with any other system. The set of rules are then given as follows.

First, one is allowed to bring in any arbitrary additional systems, as long as they are in a thermal state. The Hamiltonians of the additional systems may be chosen freely. Let's denote by $A$ the joint system consisting of all the added ancillary systems, and let $H_A$ be their joint Hamiltonian.

Second, we'll allow to carry out any global unitary which preserves the total energy. Energy conservation is required in a strong sense, namely that the unitary $U_{SA}$ commutes with the total Hamiltonian $H_S + H_A$.

Third, we may discard any system, in whatever state it is. Here we'll only be discarding the ancillas $A$ and keeping the system $S$, although in principle the framework allows more generally to trace out any system; for example, one could keep instead an ancilla system only and discard the rest.

We assume in particular that in the second step, the unitary can be carried out at no energy cost and to arbitrary good precision. This strong assumption is motivated by explicit models which can carry out such an operation: we can, for example, switch on an interaction term which generates the global unitary for a well defined duration; the interaction term may be made arbitrarily small, and in that limit doesn't require any energy to turn on or off. Similarly time control is not an issue (Horodecki and Oppenheim, 2013a; Brandão *et al.*, 2015; Malabarba *et al.*, 2015).

In the case of a trivial Hamiltonian on $S$ and $A$, the ancilla is necessarily initialized in a fully mixed state and the energy constraints are superfluous. In this case, the operations are called *noisy operations* (Horodecki *et al.*, 2003).

Let's now study the possible state transitions which can be achieved with these operations. Suppose that the system $S$, with Hamiltonian $H_S$, is initialized in a state $\rho_S$. Then Horodecki and Oppenheim (2013a) show that, if $\rho_S$ is block-diagonal in the eigenbasis of the Hamiltonian, then the set of states $\sigma_S$ which can be reached by a thermal operation is given by the notion of *thermo-majorization*, a partial order which is a generalization of the regular majorization relation (Hardy *et al.*, 1952; Bhatia, 1997; Uhlmann, 1971; Marshall *et al.*, 2010). We shall give a definition which is equivalent to the original one (Horodecki and Oppenheim, 2013a, Figure 2), but which is slightly shorter to formulate.

***Thermo-majorization.*** *Let $\rho_S$ and $\sigma_S$ be block-diagonal in the eigenbasis of $H_S$. Let $\gamma_S = e^{-\beta H_S}/Z$ denote the Gibbs state on the system S, with $Z = \operatorname{tr} e^{-\beta H_S}$. Let $p(E,i)$ (resp. $q(E,i)$) be the $i^{th}$ eigenvalue of $\rho_S$ (resp. of $\sigma_S$) inside the eigenspace of $H_S$ corresponding to the eigenvalue $E$. Then the state $\rho_S$ is said to 'thermo-majorize' $\sigma_S$ if there exists a matrix $D_{E,i}^{E',i'}$ such that for all $E$, $i$, the matrix satisfies*

$$\sum_{E',i'} D_{E,i}^{E',i'} \frac{e^{-\beta E'}}{Z} = \frac{e^{-\beta E}}{Z} \; ; \tag{4.5-a}$$

$$\sum_{E',i'} D_{E',i'}^{E,i} = 1 \; ; \; \textit{as well as} \tag{4.5-b}$$

$$\sum_{E',i'} D_{E,i}^{E',i'} p(E',i') = q(E,i) \; . \tag{4.5-c}$$

Such a majorization relation was introduced by Ruch and Mead (1976); Ruch *et al.* (1980); Joe (1990), and presented as *d-majorization* in Marshall *et al.* (2010). The latter reference also provides a proof that the above definition is equivalent to the one presented in Horodecki and Oppenheim (2013a), which is based on the appropriate ordering of eigenvalues and domination of the so-called *Lorentz curves*. In the case where the Hamiltonian on $S$ is trivial, that is in the case of noisy operations, thermo-majorization coincides with the standard notion of majorization, and the matrix $D_{E,i}^{E',i'}$ above is doubly stochastic (Bhatia, 1997; Marshall *et al.*, 2010).

We note in passing that if catalysis is in addition allowed in the transformation, then the thermo-majorization condition can be replaced by the condition that all the Rényi-$\alpha$ relative entropies to the Gibbs state must decrease, providing a set of "second laws" (Brandão *et al.*, 2015).

It is clear from the definition of thermal operations that it is impossible from a block-diagonal state to generate any superposition of different energy levels. This is simply because the all thermal operations commute with the total Hamiltonian: they are so-called *time covariant*.

If, however, the initial state $\rho_S$ is not block-diagonal in the energy eigenbasis, no general criteria are known which provide both a necessary and sufficient condition for state transformations, with the exception of the fully worked-out case of a single qubit (Ćwikliński *et al.*, 2015). General necessary conditions have been given in terms of decreasing relative entropy laws and decreasing coherence laws (Lostaglio *et al.*, 2015c).

## 4.3 Gibbs-preserving maps

In the framework of Gibbs-preserving maps, one allows to carry out any completely positive, trace-preserving map on a system which preserves the Gibbs state at a given temperature $T$ (or *Gibbs-preserving map*, for short). These maps are a natural quantum-mechanical generalization of the stochastic matrix $D_{E,i}^{E',i'}$ used in our definition of thermo-majorization.

***Gibbs-preserving map.*** *Given a quantum system S with corresponding Gibbs states $\gamma_S$, a completely positive map $\Phi_S$ is said to be 'Gibbs-preserving' if it is trace-preserving and maps the Gibbs state onto itself, that is, if $\Phi_S(\gamma_S) = \gamma_S$.*

Note that we include in the definition of a Gibbs-preserving map that it be also trace-preserving, for pure convenience.

The main motivation for studying the framework of Gibbs-preserving maps is that it provides the most permissive framework one could imagine, in the sense that if any non-Gibbs-preserving operation is allowed for free, then one could construct *perpetuum mobile* violating the macroscopic second law of thermodynamics. Indeed, if we allowed any operation $\mathcal{E}$ for free for which $\mathcal{E}(\gamma_S) = \tau_S \neq \gamma_S$, then we could repeat for free this operation on many copies of the Gibbs state to create the state $\tau_S^{\otimes n}$ for free. However, with any reasonable model for thermodynamics (such as thermal operations), one can extract an arbitrarily large amount of work from $\tau_S^{\otimes n}$ as $n \to \infty$. Hence one could extract an arbitrarily large amount of work by using only free operations, and the theory would be trivial.

This means that the framework of Gibbs-preserving maps is a conservative choice which is ideal for proving fundamental limits on possible transformations. The framework of Gibbs-preserving maps is also technically convenient to work with: Indeed, the condition of being a Gibbs-preserving map is a semidefinite constraint. One may thus resort to the toolbox of semidefinite programming in order to, for example, decide whether a transformation from a particular initial state to a particular final state is possible or not.

## 4.4 Gibbs-preserving maps outperform thermal operations in the quantum regime

The results in this section have been reported in Faist *et al.* (2015b), from which the content of this section has been adapted.

### 4.4.1 Initial remarks

Observe that thermal operations cannot change the Gibbs state into any other state (Janzing, 2006; Brandão *et al.*, 2013; Horodecki and Oppenheim, 2013a). Also, we have seen above that thermal operations are not capable of generating coherent superpositions of energy levels.

Since a thermal operation preserves the Gibbs state, the state transformations possible with thermal operations are necessarily included in those achievable with Gibbs-preserving maps. Is the converse true? It is in the classical case, i.e. for states which are block diagonal in their energy eigenbasis. This can be seen as follows. As mentioned above, transformations by thermal operations for block-diagonal input states are completely characterized in terms of thermo-majorization of the initial and final states' spectrum with respect to the Gibbs state. Now, if there exists a Gibbs-preserving map which relates a state $\rho_S$ to another state $\sigma_S$, then classic results about majorization (Marshall *et al.*, 2010) ensure that $\rho_S$ thermo-majorizes $\sigma_S$, meaning there exists also a thermal operation performing the transformation.

On a side note, this does not imply that Gibbs-preserving maps are equivalent to thermal operations as channels even when acting on block diagonal states. Rather, they are only equivalent in terms of state transitions. In other words, while the same pair (*input state*, *output state*) can be achieved in both frameworks for block-diagonal states, the actual channels that one can perform, differ. Note also that even for a given fixed input state the actual channel performed is in general relevant, and not only the input and output state, as the full information about the channel can be obtained by keeping a purification of the input (Faist *et al.*, 2015a). Additionally, a classic example (for the trivial Hamiltonian $H = 0$) of a map preserving the fully mixed state but which is not a thermal operation is the Choi-Jamiolkowski map of the two-party reduced state of the Aharonov or determinant state (Fitzi *et al.*, 2001, Note [9])

$$|\mathcal{A}\rangle_{ABC} = \frac{1}{\sqrt{6}} \left[ |012\rangle + |120\rangle + |201\rangle - |210\rangle - |102\rangle - |021\rangle \right] , \qquad (4.6)$$

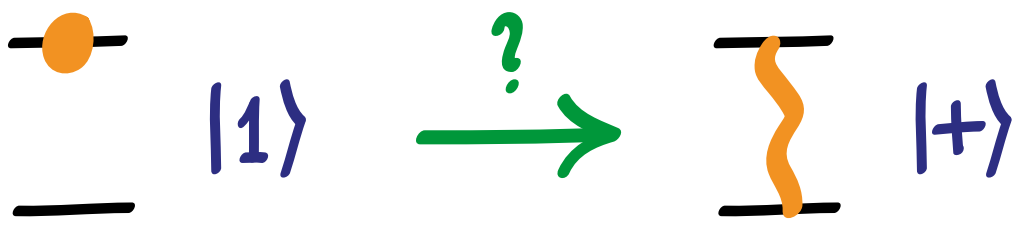

Figure 4.1: Problematic state transformation: If a qubit system is in a pure excited energy eigenstate $|1\rangle$, one would expect it is possible to bring it into any other state at no work cost, in particular in the coherent superposition of energy eigenstates $|+\rangle = \frac{1}{\sqrt{2}}[|0\rangle + |1\rangle]$. This is indeed possible with a Gibbs-preserving map, however thermal operations forbid this transition because it requires nontrivial time control.

which is up to a local unitary the same example as in Landau and Streater (1993).

### 4.4.2 Gibbs-preserving maps are more powerful

We now address the question of whether Gibbs-preserving maps are strictly more powerful than thermal operations in terms of state transitions, on arbitrary, quantum, input states. We show that this is the case, by exhibiting an example of a Gibbs-preserving map that performs a transformation forbidden by thermal operations.

Consider a two-level system with an energy gap $\Delta E$. We denote the ground state by $|0\rangle$ and the excited state by $|1\rangle$. Consider now the transformation:

$$|1\rangle \to \rho \,, \tag{4.7}$$

where $\rho$ is any pure or mixed state. Depending on $\rho$, in particular if $|\rho\rangle = |+\rangle := \frac{1}{2}\left[|0\rangle + |1\rangle\right]$, this transformation may need to "build" coherence between the energy levels, which, as noted above, cannot be achieved with thermal operations (Figure 4.1). Yet we'll see that for any $\rho$, there exists nevertheless a Gibbs-preserving map performing this transition. Let $\beta$ be a fixed inverse temperature, and denote the Gibbs state on the system by $\gamma = p_0|0\rangle\langle 0| + p_1|1\rangle\langle 1|$ with $p_0 = 1/Z$, $p_1 = e^{-\beta\Delta E}/Z$ and $Z = 1 + e^{-\beta\Delta E}$. Let $\Phi$ be defined as

$$\Phi\left(\cdot\right) = \langle 0|\cdot|0\rangle\, \sigma + \langle 1|\cdot|1\rangle\, \rho \,, \tag{4.8}$$

for some state $\sigma$ which we have not yet fixed. Note that $\Phi$ is completely positive and trace-preserving. We also have $\Phi\left(|1\rangle\langle 1|\right) = \rho$ by construction.

The condition that $\Phi$ be Gibbs-preserving, $\Phi(\gamma) = \gamma$, gives us

$$p_0\sigma + p_1\rho = \gamma \,,$$

which implies

$$\sigma = p_0^{-1}(\gamma - p_1\rho) \ . \tag{4.9}$$

This choice of $\sigma$ has unit trace, and is positive semidefinite; indeed, as $\gamma \geqslant p_1\mathbb{1}$ (since $p_1$ is the smallest eigenvalue of $\gamma$) and $\rho \leqslant \mathbb{1}$, we have $\gamma - p_1\rho \geqslant 0$. This means that, with this choice of $\sigma$, $\Phi$ is precisely a completely positive, trace-preserving, Gibbs-preserving channel which maps $|1\rangle$ to $\rho$. This map is forbidden by thermal operations if $\rho$ contains a coherent superposition over energy levels, and we have the desired counter-example.

This example can easily be generalized to a system of $n$ arbitrary energy levels: if $|n\rangle$, of energy $E_n$, is such that no other state has higher energy, a Gibbs-preserving map $\Phi$ transforming $|n\rangle$ into any $\rho$ is given by

$$\Phi(\cdot) = \mathrm{tr}\left[(\mathbb{1} - |n\rangle\langle n|)\,(\cdot)\right]\sigma + \mathrm{tr}\left[|n\rangle\langle n|\,(\cdot)\right]\rho \,, \tag{4.10}$$

where $\sigma = (\gamma - p_n\rho)/(1-p_n)$ and where the Gibbs state is $\gamma = \sum p_i|i\rangle\langle i|$ with $p_i = e^{-\beta E_i}/Z$ and $Z = \sum e^{-\beta E_i}$.

### 4.4.3 The problem of time control

The observation of a gap between these two classes of operations leaves open the question which of the two captures the actual physical situation. The Gibbs-preserving maps are useful as the most permissive framework that is nontrivial; there is however no known explicit microscopic model which corresponds to these operations. Furthermore, to observe any coherence between energy levels one needs a time reference frame (Bartlett *et al.*, 2006; Mølmer, 1997; Aharonov and Susskind, 1967; Bartlett *et al.*, 2007), which might cost work to produce and eventually get degraded. Allowing the use of such a resource catalytically enables operations that were otherwise forbidden (Brandão *et al.*, 2013; Åberg, 2014; Brandão *et al.*, 2015; Ng *et al.*, 2015), yet if the catalyst may be returned only approximately in its original state, then work can be embezzled and all transformations are possible, rendering the framework trivial (Åberg, 2014; Brandão *et al.*, 2015; Ng *et al.*, 2015). Also, one usually expects from a physical theory that one can ignore very unlikely events; this is by definition not possible in the framework of exact catalysis.

However, this does not yet conclusively show that the Gibbs-preserving maps are physically irrelevant. Intuitively, one could have argued from the start that the transition (4.7) should have been possible for any $\rho$: indeed, the initial state has both maximal purity and highest possible energy. In fact, it is not uncommon to assume that some form of coherence is available, for example in the context of quantum computation, or, more generally, whenever a quantum system interacts with a macroscopic system such as a detector or a laser field. Indeed, the latter are usually modeled in a coherent state.

### 4.4.4 Hamiltonians vs. Gibbs states

In the definitions of thermal operations and of Gibbs-preserving maps, whether we refer to the Gibbs state $\gamma_S$ of the system $S$ or to its Hamiltonian $H_S$ seems at first sight to be an incidental detail of presentation, as one directly relates to the other via the relation $\gamma_S = e^{-\beta H_S}/Z$. However, the Gibbs state inherits the structure of quantum states: from a given Gibbs state on two systems $A \otimes B$, we may calculate its reduced state on a single system $A$ via the partial trace. This is not possible for a joint Hamiltonian: there is no analogous simple concept for a "reduced Hamiltonian."

In fact, we demonstrate that it is possible to perform the transition described in Figure 4.1 without having to create any coherent superposition of energy levels, by interpreting the states $|0\rangle$ and $|1\rangle$ differently. This is done by leaving the output qubit interacting with another ancillary qubit, so that the Gibbs state on the output qubit is well defined, but not its Hamiltonian. Then, the states $|0\rangle$ and $|1\rangle$ are defined to be the eigenstates of the reduced Gibbs state; these are not energy levels because the system is interacting.

The statement is formulated more specifically as follows. The task consists in implementing the mapping $|1\rangle_S \to |+\rangle_S$ on a qubit system $S$. We require that the qubit be described at the input and at the output by the Gibbs state $\gamma_S = g_1|1\rangle\langle 1|_S + g_0|0\rangle\langle 0|_S$ with fixed given $g_0$ and $g_1$ satisfying the constraints $0 < g_1 \leqslant g_0$ and $g_1 + g_0 = 1$. Then, we claim that this transition can be carried out by an energy-preserving unitary involving a larger system. Crucially, no restriction is formulated about the Hamiltonian of the system, nor about its interactions with other systems. We may, in particular, leave the qubit strongly interacting with another ancillary system. This means that the states $|0\rangle_S$ and $|1\rangle_S$ now refer to Gibbs state eigenstates, which don't need to coincide with well-defined energy levels.

In the remainder of this section, we prove this claim.

Let's start with the qubit $S$ isolated from other systems, and described by the Hamiltonian $H_S^{(i)} = \Delta E\,|1\rangle\langle 1|_S$, with the energy gap $\Delta E$ chosen such that $g_1 = e^{-\beta\Delta E}/\bigl(1 + e^{-\beta\Delta E}\bigr)$ and $g_0 = 1/\bigl(1 + e^{-\beta\Delta E}\bigr)$, that is, with $\Delta E$ chosen exactly such that $e^{-\beta H_S} = \gamma_S$.

Now let's put to work our modest engineering skills to design the output system. Take two qubits $A$ and $B$, and define the state

$$\gamma_{AB}^{(f)} = \delta\,|0\rangle\langle 0|_A \otimes \mathbb{1}_B + |1\rangle\langle 1|_A \otimes \left[(g_0 - \delta)\,|0\rangle\langle 0|_B + (g_1 - \delta)\,|1\rangle\langle 1|_B\right], \quad (4.11)$$

with $0 < \delta < g_0, g_1$, such that $\gamma_{AB}^{(f)} \geqslant 0$ and $\gamma_{AB}^{(f)}$ has full rank. Observe that the reduced state on $B$ is the same as the required $\gamma_S$:

$$\gamma_B^{(f)} = \delta\,\mathbb{1}_B + (g_0 - \delta)\,|0\rangle\langle 0|_B + (g_1 - \delta)\,|1\rangle\langle 1|_B = g_0\,|0\rangle\langle 0|_B + g_1\,|1\rangle\langle 1|_B\,. \quad (4.12)$$

We can arrange for $\gamma_{AB}^{(f)}$ to coincide with the Gibbs state of the joint $AB$ system by engineering a joint interacting Hamiltonian as $H_{AB}^{(f)} = -\beta^{-1}\ln\gamma_{AB}$:

$$H_{AB}^{(f)} = -\frac{1}{\beta}\left[(\ln\delta)|0\rangle\langle 0|_A \otimes \mathbb{1}_B + |1\rangle\langle 1|_A \otimes \begin{pmatrix}\ln(g_0 - \delta) & \\ & \ln(g_1 - \delta)\end{pmatrix}\right]. \quad (4.13)$$

Note that if the qubit $A$ is the state $|0\rangle_A$, then the Hamiltonian state on $B$ is completely degenerate, with energy $\delta$.

Lastly, we need to agree on the fact that ancillas can be brought in prepared in a pure energy eigenstate, if we account for their energy. Also, we may dispose of an ancillary system; If it is not in a pure energy eigenstate, we may not recover any energy stored within. This gross model is sufficient for our purposes, and agrees with virtually any existing reasonable framework.

All right, now let's put all the puzzle pieces together. Here are the operations we perform, in sequence, where $\delta$ is still to be fixed, and defining $\varepsilon = -\beta^{-1}\ln\delta$:

1. Start with the system $S$, with Hamiltonian $H_S^{(i)}$ (defined above), initialized in the $|1\rangle_S$ state with energy $\Delta E$;

2. Bring in an ancillary composite system $AB$ consisting of two qubits, with the Hamiltonian $H_{AB}^{(f)}$ defined above, and initialized in the state $|0\rangle_A \otimes |0\rangle_B$. It has energy $\varepsilon$;

At this point, the joint state is $|1\rangle_S \otimes |0\rangle_A \otimes |0\rangle_B$, which is an eigenstate of the joint Hamiltonian $H_{SAB} = H_S^{(\mathrm{i})} + H_{AB}^{(\mathrm{f})}$ for the energy $\Delta E + \varepsilon$. Then:

3. Perform a strictly energy-preserving unitary which maps the state $|1\rangle_S \otimes |0\rangle_A \otimes |1\rangle_B$ onto the state $|1\rangle_S \otimes |0\rangle_A \otimes |+\rangle_B$, which also has the same energy $\Delta E + \varepsilon$;

4. Discard the $S$ ancilla, which is left in the pure energy eigenstate of energy $\Delta E$, and hence from which we can recover this energy;

5. Discard the $A$ ancilla, which is not left in any energy eigenstate, so from which *a priori* we don't know how to recover any energy.

In total, we have invested an energy $\Delta E$ and have recovered the energy $\varepsilon$. By choosing $\varepsilon \to \Delta E$ with $\varepsilon > \Delta E$, we have $0 < \delta < g_0, g_1$ as required, and we may perform the transition with arbitrary small energy cost.

There are of course several variants of this procedure. For example, we may have directly assumed the input qubit was in fact the system $B$ itself, and equally assumed that the qubit $A$ was initialized in the $|0\rangle_A$ state. Then an energy-conserving unitary would have sufficed to bring the joint state to $|0\rangle_A \otimes |+\rangle_B$, which has same energy, after which we could have discarded $A$. The example above avoids however the assumption that the input state itself is interacting already with another system we have control over.

### 4.4.5 Discussion and interpretation of Gibbs-preserving maps

The question as to whether Gibbs-preserving maps can be implemented with processes which can be operationally justified, such a thermal operations with additional resources, remains open.

We have seen that the central example demonstrating the gap between thermal operations and Gibbs-preserving maps may be carried out with thermal operations if the transition is not required to be performed between isolated systems, but only systems whose Gibbs state are given. This may suggest that any Gibbs-preserving map may be implemented with such operations under this loosened condition, possibly by appropriately generalizing the decoupling techniques used in (Dupuis *et al.*, 2014; del Rio *et al.*, 2011). We leave this as an open question.

## 4.5 Battery models for storing extracted work

How should we define *work* when studying finite-size thermodynamics? It is not clear if it is possible to identify individual energy exchanges into well defined notions of *work* and *heat*. Indeed, the concept of *work* should embody some notion of a "clean" form of energy (Åberg, 2013). However, in finite-size thermodynamics, energy exchange is usually characterized by a random variable which might have more or less spread and might thus have varying degrees of "cleanliness."

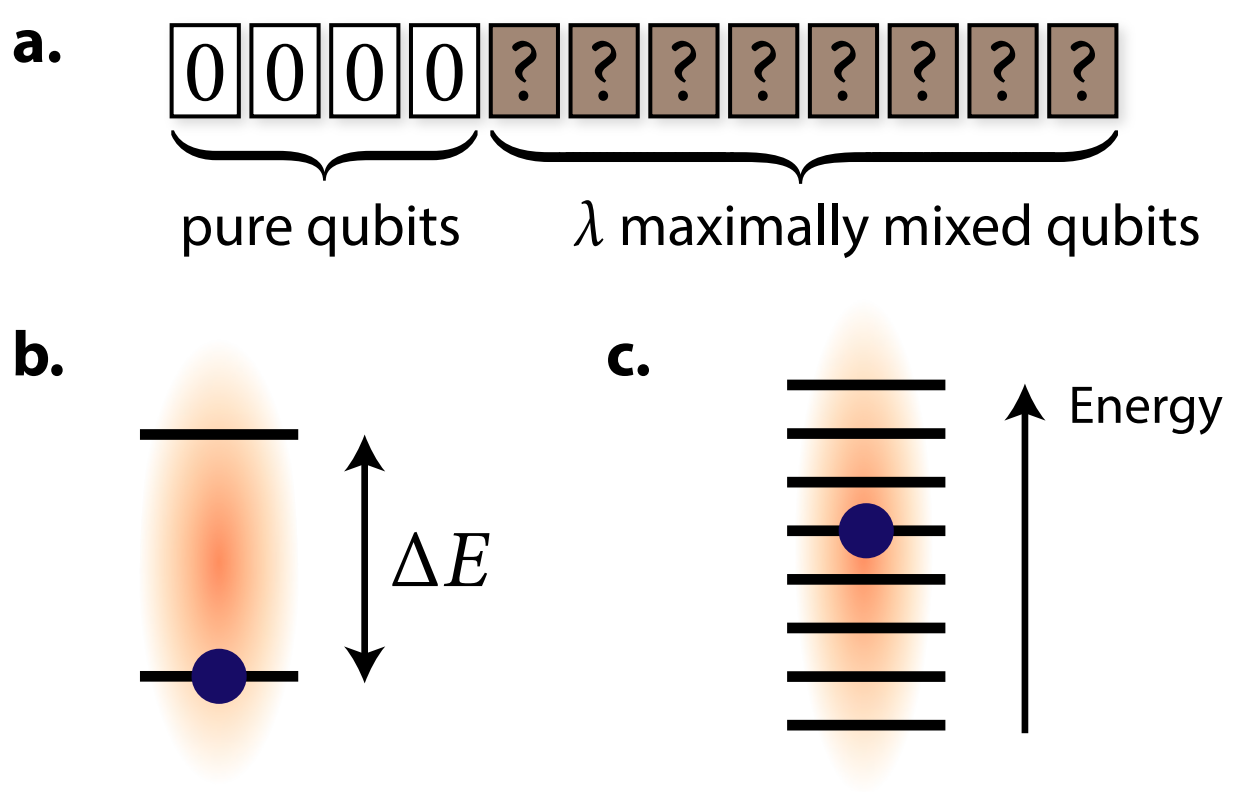

Figure 4.2: Explicit Battery Models. **a.** The *information battery* is a memory register with a degenerate Hamiltonian, in which a certain number of qubits are in a fully mixed state ('?') and the rest are in a pure state ('0'). The number $\lambda$ of mixed qubits determines the state of depletion of the battery. **b.** The *work bit*, or *wit* is a two-level system which stores the energy $\Delta E$ when it is excited to the higher energy level. **c.** The *weight system* consists of many evenly-spaced energy levels in a pure energy eigenstate, mimicking a weight which can be raised or lowered. Energy can be stored by raising or lowering the state.

One compelling solution is to model explicitly the device which is meant to store the extracted work. In thermodynamics indeed, work is often defined as a form of energy which can be converted into standard mechanical or electrical work. For example, thermodynamic work might be defined as raising or lowering a weight by a certain difference in height (Lieb and Yngvason, 1999, 2004, 2013).

### 4.5.1 The information battery

A simple work storage model is the *information battery* (Bennett, 1982; Feynman, 1996). Consider a large number of qubits, each described by a degenerate Hamiltonian $H = 0$, and which may each be prepared either in a predefined pure state $|0\rangle$ or the maximally mixed state $\mathbb{1}_2/2$ (Figure 4.2a).

The idea of this battery model is the fact that a pure qubit can in principle be traded for a maximally mixed qubit at a work expense or yield of $kT \ln 2$, by using a heat bath at temperature $T$. This is justified by a number of explicit models which achieve this, for example the Szilárd engine (Section 2.2), but also more generic models where the Hamiltonian is manipulated directly (Alicki *et al.*, 2004; Åberg, 2013), in resource theories (Horodecki and Oppenheim, 2013a; Brandão *et al.*, 2015) or with other explicit storage systems (Barnett and Vaccaro, 2013; Frenzel *et al.*, 2014).

Consider a free process, which may be for example a thermal operation or a Gibbs-preserving map, that simultaneously maps a state $\rho$ to a state $\sigma$ as well as the information battery from a state with $\lambda_1$ maximally mixed qubits at the beginning to a state of $\lambda_2$ maximally mixed qubits (Figure 4.3). Then the amount of work extracted can be defined by to the difference $\kappa \cdot (\lambda_1 - \lambda_2)$, where $\kappa$ is a proportionality factor which is to be fixed. If $\lambda_1 < \lambda_2$, then the free process has increased the purity of our battery, which now has more potential to enable other forbidden transitions; work was extracted. If $\lambda_2 < \lambda_1$, then we end up with less purity than at the beginning, and work was invested; this is represented by a negative amount of extracted work.

The proportionality factor $\kappa$ is given by the way we choose to make the correspondence between information and any form of standard work, such as mechanical work. By invoking the models cited above with a heat bath at temperature $T$, the proportionality factor is $\kappa = kT \ln 2$. If work is not provided in the form of energy, but spin for example (Barnett and Vaccaro, 2013; Frenzel *et al.*, 2014; Weilenmann *et al.*, 2015), the proportionality factor should be adapted appropriately.

In order to account more precisely for amounts of work which are not given by an integer, we need to allow $\lambda_1$ and $\lambda_2$ to take real values for which $2^{\lambda_1}, 2^{\lambda_2}$ are still integers. This corresponds to having the total battery system in a state with a flat spectrum of a given rank $2^{\lambda_1}$ or $2^{\lambda_2}$. Then, by choosing a large enough battery, any specific work amount can be approximated a difference $\lambda_1 - \lambda_2$ of such $\lambda$'s to arbitrary good precision.

### 4.5.2 The work bit, or 'wit'

The *work bit*, or *wit* (Horodecki and Oppenheim, 2013a; Brandão *et al.*, 2015), is a two-level system with two energy levels spaced by a gap $\Delta E$ (Figure 4.2**b**). The energy gap may be chosen freely, and usually has to be tuned to the value required to enable a particular forbidden process or to extract a given amount of work.

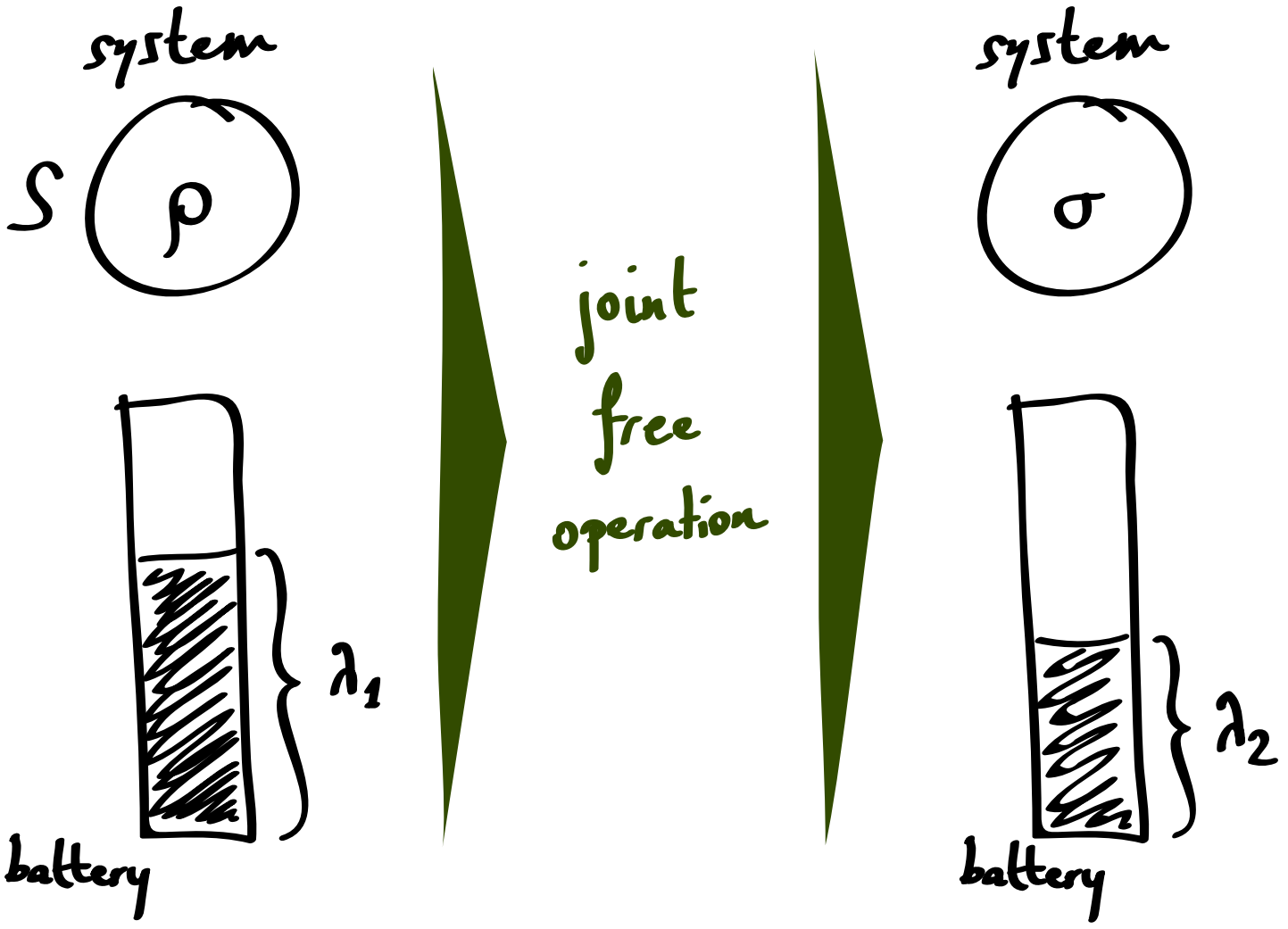

Figure 4.3: Work is extracted with a battery. If there is a free transformation (such as a thermal operation) which transforms $\rho$ to $\sigma$ along with changing the battery state from $\lambda_1$ mixed qubits to $\lambda_2$ mixed qubits, then this may be interpreted as extracting an amount of work proportional to $\lambda_1 - \lambda_2$. This can be used to define work in our framework. The interpretation as work is justified by a variety of explicit models which perform this conversion, such as with a Szilárd engine (Szilard, 1929; Dahlsten *et al.*, 2011), or by raising or lowering Hamiltonian levels in contact with a heat bath (Alicki *et al.*, 2004; Åberg, 2013). Work may also be defined in the same way by using a different kind of battery instead of the information battery, such as a *wit* or a *weight system*.

We proceed as before by replacing the information battery in Figure 4.3 by a wit. We say that an amount $\Delta E$ of work is extracted if the battery starts in the state $|0\rangle$ and is brought to the state $|1\rangle$ during a free operation (such as a thermal operation). Similarly, $\Delta E$ work is supplied if the battery starts in the state $|1\rangle$ and a free operation causes the battery to end up in the state $|0\rangle$.

The fact that the battery is required to remain in a pure state justifies interpreting the corresponding difference in energy of the system as *work* rather than *heat*, since the energy does not have any fluctuating nature.

### 4.5.3 A weight modeled as a ladder of energy levels

Another battery model, often called *weight system*, is defined as a doubly-infinite ladder of evenly-spaced energy levels (Skrzypczyk *et al.*, 2014, 2013). The system is required to remain in a pure energy eigenstate. The state may be raised or lowered to other energy eigenstates, a process which is analogous to raising or lowering a weight in a gravitational field. The difference between the energy of the system before and after the process determines the amount of work extracted or supplied to the system.

This model has the advantage that it may be used as is for any particular transition with any required work (up to a precision given by the energy spacing). In contrast, the wit's gap $\Delta E$ must be tuned for any specific process to the amount of work one needs to extract or provide.

# 5

# A Fully Information-Theoretic Model for Finite-Size Thermodynamics

Here, we draw inspiration from the frameworks presented in Chapter 4 to set up a purely information-theoretic approach to study possible transformations under a simple information-theoretic restriction. With the appropriate associations this model corresponds to Gibbs-preserving maps with an information battery as a work storage system.

The underlying idea is based on the following observation: even though the frameworks presented above have been developed for the specific situation of finite-size thermodynamics, the concepts on which some of these models depend are in fact entirely information-theoretic. For example, Gibbs-preserving maps make absolutely no explicit reference to the Hamiltonian of the system and only care about its Gibbs state. Similarly, an information battery makes no reference to physical quantities such as energy or heat, as it solely cares about keeping track of purity.

In fact, we point out through simple classical examples how the Gibbs state may be associated to counting of degrees of freedom of a heat bath—in other words, it is a witness of "hidden" degrees of freedom one doesn't have access to, which we refer to as *hidden information*. Particularly, the model of finite-size thermodynamics based on thermo-majorization, taken in a fully classical context, can be mapped one-to-one onto a purely information-theoretic problem which includes a coarse-graining. This mapping essentially corresponds to the standard derivation in statistical mechanics of the canonical ensemble from the microcanonical ensemble. In a second, more information-theoretic example, we study a simple coding scheme between symbols with abstract "values." We reach similar conclusions, in that the "values" may either represent the energy required to send the symbol through a channel, or the length of an underlying codeword which was coarse-grained.

The purpose of this chapter is to take this observation seriously and formulate our framework entirely in information-theoretic terms. By analogy with the classical setting, we continue using the terminology *hidden information*, and introduce the notion of *compression yield*. In the context of thermodynamics, our model corresponds effectively to Gibbs-preserving maps combined with an information battery, where the hidden information operator is taken to be the Gibbs state, and where the compression yield measures an amount of work required by a logical process.

A first technical result is that a logical process $\Phi$ which "sub-preserves" the hidden information operator $\Gamma$, that is, satisfies $\Phi(\Gamma) \leqslant \Gamma$, may be dilated to a fully $\Gamma$-preserving channel on a larger system, generalizing a similar result for subunital and unital maps (Faist *et al.*, 2015a).

A second result is that if we generalize the work storage system models of Section 4.5 to our fully information-theoretic setting, these models are all equivalent, in that they all enable the exact same *a priori* forbidden logical processes.

It is also possible to define the *reverse process* of a logical process. It is based on the Petz recovery map (Petz, 1986b; Hayden *et al.*, 2004), known in the context of the data processing inequality for the relative entropy (Petz, 2003; Wilde, 2015) and quantum Markov chains (Fawzi and Renner, 2015), and with known connections to thermodynamics (Wehner *et al.*, 2015; Åberg, 2016; Alhambra *et al.*, 2016). We present some examples which motivate this definition and which will serve later in the context of coarse-graining.

This chapter is organized as follows. In Section 5.1, two simple classical settings are presented which illustrate how the Gibbs state can, in certain contexts, be interpreted as counting underlying information which was coarse-grained. In Section 5.2, we present our fully quantum model, phrased in purely information-theoretic terms. Several properties of the framework are shown in Section 5.3, including the equivalence of several generalizations of work storage models. In Section 5.4, we define the notion of a *reverse process* and explore some examples. The proofs are presented separately in Section 5.5.

## 5.1 Classical examples

### 5.1.1 Classical thermodynamics and degrees of freedom of the bath

Classically, it is possible to map thermal operations one-to-one onto a purely information-theoretic situation with noisy operations. This is performed via the operation known as Gibbs-rescaling (Egloff *et al.*, 2015; Brandão *et al.*, 2015).

Consider a classical system $S$ and let the free operations on that system be represented by thermal operations at a temperature $T$. To avoid introducing a new notation and terminology, we just consider quantum states which are diagonal in the energy eigenbasis. Suppose there are $m$ energy eigenspaces, and that the $i^{\text{th}}$ energy eigenspace has a degeneracy $d_i$. The Gibbs state is then

$$\gamma_\beta = \operatorname{diag}\Big(\underbrace{e^{-\beta E_1}/Z, e^{-\beta E_1}/Z, \ldots,}_{d_1} \underbrace{e^{-\beta E_2}/Z, \ldots,}_{d_2} \ldots\Big), \tag{5.1}$$

where $Z = \sum_i d_i\, e^{-\beta E_i}$.

The *Gibbs-rescaling* operation $G_\beta(\cdot)$ is defined by mapping a state $\rho$ onto a new state $G_\beta(\rho)$ living in a different space, as follows:

$$\rho = \operatorname{diag}\left(p_{ij}\right) \;\to\; G_\beta(\rho) = \operatorname{diag}\Big(\underbrace{e^{\beta E_1} p_{1,1}, \ldots,}_{e^{-\beta E_1}} \underbrace{e^{\beta E_1} p_{1,2}, \ldots,}_{e^{-\beta E_1}} \ldots\Big). \tag{5.2}$$

That is, each eigenvalue $p_{ij}$ in the energy block $i$ is mapped to a block of dimension $e^{\beta E_i}$ with constant eigenvalue $e^{-\beta E_i} p_{ij}$. The space of the Gibbs-rescaled states has dimension $Z$. If the $e^{-\beta E_i}$'s are not integers, then decrease all the energies by a constant shift such that all the $e^{-\beta E_i}$ are close enough to integers. The Gibbs state itself is mapped by Gibbs-rescaling to the maximally mixed state: $G_\beta(\gamma_\beta) = \mathbb{1}/Z$.

One may interpret the Gibbs-rescaled state as representing the degrees of freedom of the heat bath, much like the canonical ensemble is derived from an application of the microcanonical ensemble jointly on the system and a reservoir. Indeed, each microstate of energy $E$ on the system corresponds to $\sim e^{-\beta E}$ microstates of the heat bath. This may follow from a simple counting argument under the constraint that the total energy stays fixed, and is presented in standard statistical mechanics textbooks (Huang, 1987).

The Gibbs-rescaling operation allows us to map thermo-majorization to regular majorization: $\rho$ thermo-majorizes $\sigma$ if and only if $G_\beta(\rho)$ majorizes $G_\beta(\sigma)$. This is proven in Horodecki and Oppenheim (2013a, Theorem 2).

Let us represent the system as interacting with a large heat bath, and suppose that the total energy is fixed. It is well known from statistical mechanics that the microcanonical state on the joint system, when reduced onto the system, gives the Gibbs state. In this picture, the Gibbs-rescaling operation can be interpreted as listing explicitly the individual degrees of freedom of the heat bath which correspond to the system being in a particular energy level. The eigenvalues of the Gibbs state are then a witness for degrees of freedom which are inaccessible, or hidden.

In the following, we'll take this point of view at a more fundamental level. The framework which we present will be based on an operator which generalizes this idea. Its eigenvalues are interpreted, like the Gibbs state, as witnessing degrees of freedom which are inaccessible. Long live the *hidden information operator*.

### 5.1.2 Classical information theory with "intrinsic values" of states

Before setting off for the actual model, it might be useful to consider a purely information-theoretic example. This toy example revisits the basic information-theoretic protocol of source coding, which was the original setting studied by Shannon in his seminal paper (Shannon, 1948) and is covered in any standard textbook on information theory (Cover and Thomas, 2006). We consider a particular variant of the problem, one for which the encoded symbols may have different "costs" or "payoffs," in some similarity to the gambling setting presented in Cover and Thomas (2006, Chapter 6).

The resources required to decode or encode a symbol are typically ignored in textbook information theory, where the focus is kept in the mathematical aspect of information. Similar considerations are necessary, though, in continuous variable coding (Cover and Thomas, 2006, Chapter 8), where there may not be any natural equivalent of the maximally mixed state acting as reference for entropy.

Suppose that a discrete memoryless source $X$ emits symbols in an alphabet $\mathcal{X}$, each with a specified probability $p_X(x)$. These symbols come on some classical physical support, and our task is to transcode these messages onto some other physical support. The new physical support has possible classical states in some abstract set $\mathcal{Y}$. There is no need *a priori* for $\mathcal{Y}$ and $\mathcal{X}$ to coincide.

Now here's the twist: suppose that each physical state $x \in \mathcal{X}$ and $z \in \mathcal{Z}$ carries some form of *intrinsic value*. This may correspond to the energy of the state, the length of a code word, the amount of ink required to write the encoded symbol on a piece of paper (assuming, e.g., that the ink can be recovered upon decoding), etc. Denote by $g_X(x)$ the intrinsic value corresponding to the physical state $x$ on the input, and define similarly $g_Y(y)$. We require this "intrinsic value" to be multiplicative upon composition, a property which will prove useful later:

$$g_{X^2}(x_1, x_2) = g_X(x_1) \cdot g_X(x_2) \; . \tag{5.3}$$

For any other measure which should be additive instead, we define $g_X(x)$ and $g_Y(y)$ to be the exponential of the corresponding quantities. For example, if each symbol $x$ carries an energy $E_x$, then we take $g_X(x) = e^{-\beta E_x}$, where $\beta$ is some constant which makes the argument of the exponential dimensionless.

Suppose that the "intrinsic values" of the input symbols can be recovered and stored, and suppose that we need to invest these resources into producing the output. There might be a surplus of "intrinsic value" if the symbols we output cost less to produce, or we might need to invest additional "intrinsic value" if the symbols we output are expensive.

Assume that the transcription is to be performed symbol by symbol. Suppose also for simplicity that the transcription has to be deterministic, assigning a unique $y$ for every input $x$ written as $y(x)$. The inverse function, defined on the range of $y(x)$, is denoted by $x(y)$. The question we ask here is, what is the average (log) intrinsic value gain in the transcription? We ask about the minus logarithm of the quantity, because that one is additive. The minus sign follows the convention that a larger $g_X(x)$ indicates a state with less desirable value, which is the case if $g_X(x)$ is the dimension of the system necessary to store the symbol. So we are interested in the quantity

$$\big\langle \text{log-“intrinsic value” gain} \big\rangle = \big\langle -\log g_X(x) + \log g_Y(y) \big\rangle \; , \tag{5.4}$$

where the average is to be taken over the input symbols $x$, and where $y$ is determined from the transcription function $y(x)$. A simple calculation gives

$$\begin{aligned}
(5.4) &= \sum_x p_X(x) \left[ -\log g_X(x) + \log g_Y(y(x)) \right] \\
&= \sum_x p_X(x) \log \frac{p_X(x)}{g_X(x)} - \sum_y p_Y(y) \log \frac{p_Y(y)}{g_Y(y)} \\
&= D(p_X \,\|\, g_X) - D(p_Y \,\|\, g_Y) \; ,
\end{aligned} \tag{5.5}$$

recalling that the distributions $p_X(x)$ and $p_Y(y)$ are the same up to a per-mutation and possibly additional zero-probability events in $Y$. Hence the relative entropy to the distribution of "intrinsic values" can be seen as the "average value potential" of a state. Note that the second argument to the relative entropy is not a normalized distribution: rather, it's just a vector of positive values which act as "weights" for each state.

What is instructive here is that the "value" may be anything which is multiplicative, and may describe in particular any additive quantity via its exponential. It describes in particular states with energies. But it describes equally as well lengths of code words: suppose that what is described as one "state" or "symbol" $x$ is in fact a string of "sub-symbols," for example the letters which form a word. The state $x$ is then a coarse-grained version of the resolved states of the sub-symbols. The "intrinsic value" $g(x)$ of a symbol $x$ is then simply the dimension of the state space of the corresponding sub-state or microstate. Then a good transcription procedure simply consists in substituting states $x$ which are long words with states with a shorter length: this is simply a form of information compression. These sub-symbols can be viewed as information which is not being accessed by the transcription procedure, or *hidden* from the procedure.

As before, the same unified picture describes at the same time states with energies as well as states which are in fact coarse-grained versions of higher-dimensional states which are only being accessed via their coarse-grained properties. This inaccessible information is what we refer to as *hidden information*.

Observe that the hidden degrees of freedom witnessed by $g_X(x)$ don't have to actually exist. Instead, the two problems where $g_X(x)$ corresponds to the energy of a symbol or the number of letters required to write it, are formally equivalent. In other words, we may well study either situation *as if* the hidden degrees of freedom existed.

## 5.2 The model

Time has come to present the framework itself. The model is specified via a set of *free operations*, as well as an explicit system, the *battery*, which stores the resources necessary to implement non-free operations.

As much as this setting resembles a standard resource theory, there are some key differences with respect to the usual resource theories of entanglement or thermodynamics in the literature (Horodecki *et al.*, 2009, 2003; Brandão *et al.*, 2013; Horodecki and Oppenheim, 2013b; Brandão and Gour, 2015; Gour *et al.*, 2015). The differences are mostly in the spirit in which the model is used. First, this model is developed in the aim of proving fundamental limits to possible operations. Instead of postulating free operations which are explicitly allowed, we single out those operations which are admissible—they are simply not *a priori* excluded. Our model does treat these admissible operations as if they could be carried out, even if they may be impossible to implement physically. This results in stronger statements because the proven limitation is valid even for the larger class of operations. Second, we will not consider state transitions as has been typically done with previous resource theories, but rather fully specified *logical processes* (Section 2.5).

### 5.2.1 Hidden information

Before defining the admissible operations, we must define the *hidden information operator* in terms of which the former are defined.

To each system $S$ corresponds a *hidden information operator* $\Gamma_S$, which may be any positive semidefinite operator. This operator should intuitively represent "intrinsic values or costs of states" in a certain basis, by analogy to the classical examples presented in Section 5.1.

Conventionally, if $\Gamma_S$ has zero eigenvalues, then the corresponding states are considered to be impossible to prepare. These states will never be observed. (They would correspond to an infinite "intrinsic value.")

There is no reason *a priori* to think that the classical equivalence of the Gibbs states with a coarse-grained picture directly generalizes to the quantum realm. Here, we stick to the terminology *hidden information* only by analogy to the classical case. The fully quantum justification is deferred to Chapter 10 which is devoted to coarse-graining in a fully quantum context.

### 5.2.2 Admissible operations

In our model, the *admissible operations*, or *free operations*, are those quantum operations which "don't leak hidden information." The quantum operation is given as a completely positive, trace-nonincreasing map $\Phi_{A\to B}$ mapping a state on $A$ to a state on $B$. Formally, the map $\Phi_{A\to B}$ is defined to be an *admissible operation* if it satisfies

$$\Phi_{A\to B}(\Gamma_A) \leqslant \Gamma_B \,. \tag{5.6}$$

The restriction we consider here generalizes that of Gibbs-preserving maps. The definition with an operator inequality instead of an equality is meant to deal with the situation in which $\operatorname{tr}\Gamma_A \neq \operatorname{tr}\Gamma_B$. Furthermore the condition of being trace-nonincreasing, rather than trace-preserving, is purely for convenience. Both these choices are justified by the fact that any completely positive, trace-nonincreasing map $\Phi_{A\to B}$ which satisfies (5.6) may be dilated to a map on a larger system which is trace-preserving and preserves the hidden information operator exactly. The details and proof of this claim will be provided later in Section 5.3.1.

Let's fix some terminology for the sake of clarity.

***Γ-sub-preserving map and Γ-preserving map.*** *We say that a completely positive map* $\Phi$ *is a 'Γ-sub-preserving map' if it is is trace nonincreasing and satisfies* $\Phi(\Gamma) \leqslant \Gamma$. *Similarly,* $\Phi$ *is a 'Γ-preserving map' if it is trace preserving and satisfies* $\Phi(\Gamma) = \Gamma$.

The constraint of being a Γ-sub-preserving map is also technically very useful to work with, because it is a semidefinite constraint. (The same applies to Γ-preserving maps.) Indeed, a completely positive map $\Phi_{A\to B}$ is Γ-sub-preserving by definition if both

$$\Phi_{A\to B}\left(\Gamma_A\right) \leqslant \Gamma_B\ ; \quad \text{and} \tag{5.7-a}$$

$$\Phi^{\dagger}_{A\leftarrow B}\left(\mathbb{1}_B\right) \leqslant \mathbb{1}_A\ . \tag{5.7-b}$$

Consequently, a number of optimization problems for admissible operations may be formulated as semidefinite programs.

### 5.2.3 Lifting the restriction with a battery

The battery is an explicit system which stores the necessary resources to enable operations which are *a priori* forbidden. In the same spirit as above, it does not guarantee the operation may be carried out, but merely lifts the restriction and transforms a forbidden operation into an admissible one.

The battery model we use is the information battery presented in Section 4.5.1. It is a quantum system $A$ consisting in a large number of qubits, for which $\Gamma_A = \mathbb{1}_A$. The individual qubits should be thought of as either in a pure state or completely mixed. Technically, we require the battery to be in a state $2^{-\lambda_1}\mathbb{1}_{2^{\lambda_1}}$ at the beginning of the process, and to finish in a state $2^{-\lambda_2}\mathbb{1}_{2^{\lambda_2}}$ at the end, where both states are simply a state with a flat spectrum of rank $2^{\lambda_1}$ or $2^{\lambda_2}$. If the $\lambda_i$'s are integers, this corresponds exactly to having $\lambda_i$ qubits in a fully mixed state and the rest in a pure state.

How many pure qubits do we need to sacrifice to make a forbidden transition admissible? It turns out that for any transition, we can always turn it into an admissible one by investing a sufficient number of pure qubits.

Suppose that a logical process $\mathcal{E}_{X\to Y}(\cdot)$ is forbidden. By involving a battery prepared in a state with appropriately chosen $\lambda_1$ and $\lambda_2$, we will show that there is an admissible operation $\Phi_{AX\to AY}$ such that

$$\Phi_{AX\to AY}(2^{-\lambda_1}\mathbb{1}_{2^{\lambda_1}} \otimes (\cdot)) = 2^{-\lambda_2}\mathbb{1}_{2^{\lambda_2}} \otimes \mathcal{E}_{X\to Y}(\cdot)\ , \tag{5.8}$$

i.e. which reproduces exactly the channel $\mathcal{E}$ on $X$, while taking the battery from the state $2^{-\lambda_1}\mathbb{1}_{2^{\lambda_1}}$ to the state $2^{-\lambda_2}\mathbb{1}_{2^{\lambda_2}}$. The amount of pure qubits which we have invested to make $\mathcal{E}$ to an admissible process is given by $\lambda_2 - \lambda_1$.

On the other hand, there may be admissible operations that are "so admissible" that we could be charging our battery by using them. Indeed, there are logical processes $\mathcal{E}_{X\to Y}$ for which an admissible channel $\Phi_{AX\to AY}$ exists satisfying (5.8) also for $\lambda_1 > \lambda_2$. In that case, we gain pure qubits for free. The amount of pure qubits which we have gained on the battery is simply given by $\lambda_1 - \lambda_2$.

The difference $\lambda_1 - \lambda_2$ is called the *compression yield* of an implementation of the logical process. It is, as we've said, the number of pure qubits we can extract by running the process cleverly with admissible operations, or the negative of the number of qubits we've invested in order to make the logical process admissible.

We also note that the use of an information battery in our model serves to highlight the pure information-theoretical aspect of all the concepts involved. In fact, it is a simple choice of presentation: we will later show that other choices of battery models lead to equivalent claims.

In the case of thermodynamics as above, the compression yield corresponds to the amount of work that we may extract from a logical process. We shall discuss this in reasonable, but hopefully not exceedingly boring, detail in the following chapters.

### 5.2.4 Special case: standard information processing

Traditionally, in information theory the individual states are all equivalent. This means that $\Gamma_S = \mathbb{1}_S$, and the free operations are the subunital operations, which may always be dilated to unital operations on a larger system.

The resulting theory which is described in this case is simply one of purity and entropy. Indeed, the unital operations are precisely those which never decrease the entropy, and are often used to model operations which are noisy.

This simpler case will be studied in more detail in the next chapter, where we draw the full connection with the smooth max-entropy.

### 5.2.5 Special case: quantum thermodynamics of a system in contact with a heat bath

In the standard setting of quantum thermodynamics, to each system $S$ corresponds a Hamiltonian $H_S$. The hidden information operator further depends on whether contact with reservoirs are in principle allowed as well as possible conservation laws. If we are allowed a heat bath at an inverse temperature $\beta$, then the hidden information is given by $\Gamma_S = e^{-\beta H_S}$. If no contact with any heat bath is allowed, and a strict energy conservation law is enforced at a given energy $E$,

Note that this operator does not have to be normalized because it does not directly represent a quantum state, in contrast to the Gibbs state. The latter indeed is defined as the quantum state to which thermalizing dynamics of the system tends to.

The admissible operations are consistent with the usual thermodynamic models—in fact, they essentially correspond to Gibbs-preserving maps. The difference is the normalization of the Gibbs state. In fact, we find the unnormalized version more consistent to work with when different systems are involved. In the case of Gibbs-preserving maps, we didn't have to worry about this, because we required the input and output Hamiltonian to be the same. However, consider the following example. Suppose that we shift all energy levels by a constant $\Delta E$, which means that $\Delta E$ work has to be supplied. The Gibbs state remains the same, so Gibbs-preserving maps wouldn't notice the difference. However, the operator $\Gamma_S$ is multiplied by a factor $e^{-\beta\Delta E}$. Because of this, Condition (5.6) does notice the difference, and it forbids the new transition without any external battery supplying the required work.

## 5.3 Properties of the model

### 5.3.1 Dilation of Γ-sub-preserving maps to Γ-preserving maps

Our first result here is that we may be somewhat flexible in our choice of how strictly the admissible operations should preserve the Γ operator.

From a technical point of view the Γ-preserving maps don't handle nicely systems of varying sizes or with different Γ operators. For example, if $X$ and $Y$ are systems with $\operatorname{tr}\Gamma_X \neq \operatorname{tr}\Gamma_Y$, there may clearly be no Γ-preserving map from $X$ to $Y$ (because it has to be also trace preserving).

It turns out that the difference is a mere question of convenience. A $\Gamma$-sub-preserving map can always be seen as a restriction of a $\Gamma$-preserving map on a larger system. Furthermore, the systems we have to include may be taken to be prepared in, or finishing up in, eigenstates of the respective $\Gamma$ operators.

***Proposition 5.1** (Dilation of $\Gamma$-sub-preserving maps). Let $K$ and $L$ be quantum systems with corresponding $\Gamma_K$ and $\Gamma_L$. Let $\tilde{\Phi}_{K\to L}$ be a $\Gamma$-sub-preserving map. Choose two arbitrary eigenvectors $|\mathrm{k}\rangle_K$ and $|\mathrm{l}\rangle_L$ of $\Gamma_K$ and $\Gamma_L$, respectively. Then there exists a qubit system $\mathscr{H}_Q$ with corresponding $\Gamma_Q$ diagonal in a basis composed of two orthogonal states $\{|\mathrm{i}\rangle_Q, |\mathrm{f}\rangle_Q\}$, such that there exists a $\Gamma$-preserving map $\Phi_{KLQ\to KLQ}$ satisfying*

$$\tilde{\Phi}_{K\to L}(\cdot) = \langle \mathrm{k}\,\mathrm{f}|\ \Phi_{KLQ\to KLQ}\big((\cdot)\otimes|\mathrm{l}\,\mathrm{i}\rangle\langle\mathrm{l}\,\mathrm{i}|_{LQ}\big)\ |\mathrm{k}\,\mathrm{f}\rangle_{KQ}\ . \tag{5.9}$$

*Here, the joint hidden information operator on $K, L, Q$ is $\Gamma_{KLQ} = \Gamma_K\otimes\Gamma_L\otimes\Gamma_Q$. Furthermore, the corresponding eigenvalues satisfy*

$$\langle \mathrm{l}|\Gamma_L|\mathrm{l}\rangle_L\langle \mathrm{i}|\Gamma_Q|\mathrm{i}\rangle_Q = \langle \mathrm{k}|\Gamma_K|\mathrm{k}\rangle_K\langle \mathrm{f}|\Gamma_Q|\mathrm{f}\rangle_Q\ . \tag{5.10}$$

*(Proof on page 110.)*

This means that for any $\Gamma$-sub-preserving map $\tilde{\Phi}_{K\to L}$, we may find a larger system and a $\Gamma$-preserving map $\Phi_{KLQ}$ such that $\tilde{\Phi}_{K\to L}$ is seen as the restriction of $\Phi_{KLQ}$ to the case where the input is fixed to $|\mathrm{l}\,\mathrm{i}\rangle_{LQ}$ on $LQ$ and where the output is post-selected with $|\mathrm{k}\,\mathrm{f}\rangle_{KQ}$ on $KQ$.

If the operators $\Gamma_K, \Gamma_L, \Gamma_Q$ come from Hamiltonians $H_K, H_L, H_Q$ as $\Gamma_i = e^{-\beta H_i}$ for a fixed inverse temperature $\beta$, then the ancillas are prepared and left in pure energy eigenstates, specifically $|\mathrm{l}\,\mathrm{i}\rangle_{LQ}$ for the input and $|\mathrm{k}\,\mathrm{f}\rangle_{KQ}$ for the output. Furthermore condition (5.10) ensures that the total energy of the ancillas remains the same:

$$\langle \mathrm{l}|H_L|\mathrm{l}\rangle_L + \langle \mathrm{i}|H_Q|\mathrm{i}\rangle_Q = \langle \mathrm{k}|H_K|\mathrm{k}\rangle_K + \langle \mathrm{f}|H_Q|\mathrm{f}\rangle_Q\ . \tag{5.11}$$

### 5.3.2 Logical processes and equivalence of a class of battery models

One of the goals of the approach described here is to determine how much compression yield is possible to extract (or has to be invested) while performing an arbitrary trace-preserving completely positive map on the system. The setup is the one described above: a joint admissible operation is to be performed over the system and an information battery. The difference in "charge status" of the battery is the compression yield.

We approach the problem by proving a first result which provides a necessary and sufficient criterion on the logical process for the admissibility of extracting a certain amount of compression yield. A logical process $\mathcal{E}_{X\to X'}$ can give a compression yield $y$ by using an information battery if and only if $\mathcal{E}_{X\to X'}(\Gamma_X) \leqslant 2^{-y}\Gamma_{X'}$. This condition is simpler to check rather than explicitly investigating the existence of a joint admissible operation on the system and battery. A detailed statement is formulated as part of the proposition below.

One may also criticize the choice of the specific type of battery model used, rather than other possible models developed in the context of finite-size thermodynamics like the wit or the weight (Section 4.5). The choice of an information battery highlights the information-theoretic character of our model. However, we may use other battery models also phrased in this abstract terminology, and it turns out—unsurprisingly—that they are equivalent. This second result, formulated below in the same proposition, generalizes to full logical processes the equivalence of the different work storage systems shown in quantum thermodynamics (Brandão *et al.*, 2015).

Perhaps the most intuitive alternative battery model is the *wit*, that is, a simple qubit system $Q$, with $\Gamma_Q = g_1|1\rangle\langle 1|_Q + g_2|2\rangle\langle 2|_Q$. It is required to start in the state $|1\rangle_Q$ and finish in the state $|2\rangle_Q$. By tuning the values of $g_1$ and $g_2$, this battery system enables the exact same logical processes as an information battery with $2^{\lambda_1} = g_1$ and $2^{\lambda_2} = g_2$.

If we add many other states to a wit, and have all the eigenvalues of $\Gamma$ with a log-constant spacing, then we get an analogous to the weight battery model seen in the previous chapter. It also enables the exact same logical processes as the "properly tuned" wit.

In fact, the most general form of a battery we have managed to show is equivalent to the others is simply one with an arbitrary $\Gamma$ operator, and which is in states that are "a part of the $\Gamma$ operator" in the sense that they can be written in the form $\sigma = (P\Gamma P)/(\operatorname{tr} P\Gamma)$ for a projector $P$ which commutes with $\Gamma$. The "value" or "uselessness" or "amount of hidden information" of this state is then simply $\operatorname{tr}(P\Gamma)$. The wit, the weight, as well as the information battery are all special cases of this general model.

***Proposition 5.2*** *(Equivalence of various battery models). Let $\mathcal{E}_{X\to X'}$ be a completely positive, trace-nonincreasing map. Let $y \in \mathbb{R}$. Then, the following are equivalent:*

*(i) The map $\mathcal{E}_{X\to X'}$ satisfies*

$$\mathcal{E}_{X\to X'}(\Gamma_X) \leqslant 2^{-y}\,\Gamma_{X'}\ ; \tag{5.12}$$

*(ii)* *For any* $\lambda_1, \lambda_2 \geqslant 0$ *such that* $2^{\lambda_1}, 2^{\lambda_2}$ *are integers and* $\lambda_1 - \lambda_2 \leqslant y$, *there exists a large enough system* $A$ *with* $\Gamma_A = \mathbb{1}_A$ *as well as a trace-nonincreasing,* $\Gamma$*-sub-preserving map* $\Phi_{XA\to X'A}$ *satisfying for all* $\omega_X$,

$$\Phi_{XA\to X'A}\big((2^{-\lambda_1}\mathbb{1}_{2^{\lambda_1}}) \otimes \omega_X\big) = (2^{-\lambda_2}\mathbb{1}_{2^{\lambda_2}}) \otimes \mathcal{E}_{X\to X'}(\omega_X)\ ; \quad (5.13)$$

*(iii)* *For a two-level system* $Q$ *with two orthonormal states* $|1\rangle_Q, |2\rangle_Q$, *and with* $\Gamma_Q = g_1|1\rangle\langle 1|_Q + g_2|2\rangle\langle 2|_Q$ *chosen such that* $g_2/g_1 \geqslant 2^{-y}$, *there exists a trace-nonincreasing,* $\Gamma$*-sub-preserving map* $\Phi'_{XQ\to X'Q}$ *satisfying for all* $\omega_X$,

$$\Phi'_{XQ\to X'Q}\big(\omega_X \otimes |1\rangle\langle 1|_Q\big) = \mathcal{E}_{X\to X'}(\omega_X) \otimes |2\rangle\langle 2|_Q\ ; \quad (5.14)$$

*(iv)* *Let* $\tilde{Q}$ *be any system and choose two orthogonal states* $|1\rangle_{\tilde{Q}}, |2\rangle_{\tilde{Q}}$ *which are eigenstates of* $\Gamma_{\tilde{Q}}$ *corresponding to respective eigenvalues* $g_1, g_2$ *which satisfy* $g_2/g_1 \geqslant 2^{-y}$. *Then there exists a trace-nonincreasing,* $\Gamma$*-sub-preserving map* $\Phi'_{X\tilde{Q}\to X'\tilde{Q}}$ *satisfying for all* $\omega_X$,

$$\Phi'_{X\tilde{Q}\to X'\tilde{Q}}\big(\omega_X \otimes |1\rangle\langle 1|_{\tilde{Q}}\big) = \mathcal{E}_{X\to X'}(\omega_X) \otimes |2\rangle\langle 2|_{\tilde{Q}}\ ; \quad (5.15)$$

*(v)* *Let* $W_1, W_2$ *be quantum systems with respective corresponding hidden information operators* $\Gamma_{W_1}, \Gamma_{W_2}$, *and let* $P_{W_1}, P'_{W_2}$ *be projectors satisfying* $[P_{W_1}, \Gamma_{W_1}] = 0$ *and* $[P'_{W_2}, \Gamma_{W_2}] = 0$, *such that*

$$\frac{\operatorname{tr} P'_{W_2}\Gamma_{W_2}}{\operatorname{tr} P_{W_1}\Gamma_{W_1}} \geqslant 2^{-y}\ . \quad (5.16)$$

*Then there exists a* $\Gamma$*-sub-preserving, trace-nonincreasing map* $\Phi''_{XW_1\to X'W_2}$ *such that for all* $\omega_X$,

$$\Phi''_{XW_1\to X'W_2}\left(\frac{P_{W_1}\Gamma_{W_1}P_{W_1}}{\operatorname{tr}(P_{W_1}\Gamma_{W_1})} \otimes \omega_X\right) = \frac{P'_{W_2}\Gamma_{W_2}P'_{W_2}}{\operatorname{tr}(P'_{W_2}\Gamma_{W_2})} \otimes \mathcal{E}_{X\to X'}(\omega_X)\ . \quad (5.17)$$

*(Proof on page 114.)*

## 5.4 Reverse of a logical process and recoverability

Here we introduce the notion of a *reverse process* for a given logical process $\mathcal{E}_{X\to X'}$, and for given hidden information operators $\Gamma_X, \Gamma_{X'}$. The reverse process is a logical process which has the property of "undoing" the process $\mathcal{E}$ "as best as we can," where the information erased by $\mathcal{E}$ is essentially replaced by a state derived from $\Gamma_X$. In a way, the reverse process gives an accurate description of the knowledge of the input state if we know only the output of the process.

A similar reverse process has been introduced in (Åberg, 2016) as *time-reversal process* of a energy-preserving unitary between a system and an ancilla to derive fully quantum fluctuation relations. It is based on the *Petz recovery map*, known in the context of the data processing inequality for the relative entropy (Petz, 1986b, 2003; Hayden *et al.*, 2004; Wilde, 2015) and quantum Markov chains (Fawzi and Renner, 2015), and whose connection to reversibility in thermodynamics was already pointed out in (Wehner *et al.*, 2015). A similar expression has also been used in (Alhambra *et al.*, 2016) to derive a precise second law in the form of an equality.

***Reverse process.*** *Let $X, X'$ be two systems described by $\Gamma_X, \Gamma_{X'}$. Let $\mathcal{E}_{X\to X'}$ be a trace-nonincreasing, completely positive map and let $y$ be such that*

$$\mathcal{E}_{X\to X'}(\Gamma_X) \leqslant 2^{-y}\Gamma_{X'}\ . \tag{5.18}$$

*Then the 'reverse process,' or 'y-reverse process,' of $\mathcal{E}_{X\to X'}$ is*

$$\mathcal{R}^{\mathcal{E},y}_{X'\to X}(\cdot) = 2^y\,\Gamma_X^{1/2}\,\mathcal{E}^\dagger_{X\leftarrow X'}\Big(\Gamma_{X'}^{-1/2}\,(\cdot)\,\Gamma_{X'}^{-1/2}\Big)\,\Gamma_X^{1/2}\ . \tag{5.19}$$

We note that for $y = 0$, expression (5.19) coincides with the Petz recovery map (Petz, 1986b, 2003; Wilde, 2015; Fawzi and Renner, 2015). Also, the coefficient $2^y$ can be seen as treating implicitly the battery which is considered explicitly in (Alhambra *et al.*, 2016).

Crucially, the reverse process is a physical process, and its compression yield is the negative of that of $\mathcal{E}$.

***Proposition 5.3.*** *The reverse process map $\mathcal{R}^{\mathcal{E},y}_{X'\to X}$ is completely positive and trace nonincreasing, and it satisfies $\mathcal{R}^{\mathcal{E},y}_{X'\to X}(\Gamma_{X'}) \leqslant 2^y\,\Gamma_X$.* *(Proof on page 115.)*

Because the reverse process is defined *a priori* for any $y$ satisfying (5.18), and not only the optimal such $y$, one may wonder in which respect the reverse process changes for different $y$'s. In fact, it only changes the normalization of the output, as is clear from the definition (5.19). Hence, if a non-optimal $y$ is used, then the output of the reverse process is systematically subnormalized. The interpretation of this is a systematic nonzero probability of failing to prepare the correct state on $X$.

It may not appear immediately clear in which respect the reverse process is useful. In the present thesis, the reverse process will find its use only in the much later chapter on coarse-graining. For now, let us consider a few simple examples to forge some intuition and to motivate our above definition.

### 5.4.1 Trivial example: the identity channel

As an obvious first example, the reverse process corresponding to the identity channel is again the identity channel, assuming that $\Gamma_{X'} = \Gamma_X$ and provided we consider the $(y=0)$-reverse process. Indeed,

$$\mathcal{R}^{\mathrm{id},y=0}_{X'\to X}(\cdot) = \Gamma_X^{1/2}\ \mathrm{id}_{X'\to X}\big(\Gamma_{X'}^{-1/2}(\cdot)\Gamma_{X'}^{-1/2}\big)\ \Gamma_X^{1/2} = \mathrm{id}(\cdot)\ . \tag{5.20}$$

### 5.4.2 Example: the erasure channel

The simple erasure channel is the logical process that forgets its input entirely, and whose output $X'$ is a trivial system $X' \simeq \mathbb{C}$. It is given by

$$\mathcal{E}_{X\to X'}(\cdot) = \mathrm{tr}(\cdot)\ . \tag{5.21}$$

Clearly, $\mathcal{E}_{X\to X'}(\Gamma_X) = \mathrm{tr}(\Gamma_X) = 2^{-y}\,\Gamma_{X'}$ with $y = -\log\mathrm{tr}(\Gamma_X)$ if we suppose that $\Gamma_{X'} = 1$.

The reverse process of $\mathcal{E}_{X\to X'}$ is a channel which prepares the state proportional to $\Gamma_X$ on $X$ (note the input is a scalar, which we can take to be one if the input is normalized):

$$\mathcal{R}^{\mathcal{E},y=-\log\mathrm{tr}\,\Gamma_X}_{X'\leftarrow X}(1) = \frac{\Gamma_X}{\mathrm{tr}\,\Gamma_X}\ . \tag{5.22}$$

This is consistent: Because the output contains no information, we have no information about the input; this is represented by a state proportional to the hidden information.

### 5.4.3 Example: erasure channel on subsystem only

In this example, the logical process $\mathcal{E}$ applies the erasure channel on a subsystem of $X$ only. Suppose that the system $X$ consists of two subsystems, $X = A \otimes B$, and that the output $X'$ is a copy of $A$. The operator $\Gamma_X = \Gamma_{AB}$ may be correlated between $A$ and $B$, and let $\Gamma_A = \mathrm{tr}_B\,\Gamma_{AB}$ be defined by the partial trace. Consider the logical process $\mathcal{E}_{AB\to A}$ defined by erasing the system $B$,

$$\mathcal{E}_{AB\to A}(\cdot) = \mathrm{tr}_B(\cdot)\ . \tag{5.23}$$

The compression yield of this process is clearly $y = 0$. Indeed,

$$\mathcal{E}_{AB\to A}(\Gamma_{AB}) = \mathrm{tr}_B(\Gamma_{AB}) = \Gamma_A \,. \tag{5.24}$$

The reverse process of $\mathcal{E}_{AB\to A}$ is given by

$$\mathcal{R}^{\mathcal{E},y=0}_{A\to AB}(\cdot) = \Gamma_{AB}^{1/2}\left(\left(\Gamma_A^{-1/2}\,(\cdot)\,\Gamma_A^{-1/2}\right)\otimes \mathbb{1}_B\right)\Gamma_{AB}^{1/2}\,, \tag{5.25}$$

since the adjoint map of the partial trace is tensoring with the identity operator.

Observe that the reverse process does not always successfully undo the partial trace operation. Namely, if we start with a bipartite state $\sigma_{AB}$, and apply the logical process $\mathcal{E}_{AB\to A}$ resulting in the state $\sigma_A$, then an application of the reverse process $\mathcal{R}^{\mathcal{E},y=0}_{A\to AB}$ may not yield $\sigma_{AB}$. Again, this is because the state on $B$ has been forgotten, and cannot be recovered, as illustrated in what follows.

It is instructive to consider the special case in which $\Gamma_{AB} = \Gamma_A^{(0)} \otimes \Gamma_B^{(0)}$. This would be the case for noninteracting systems in contact with a heat bath. The reverse process simply becomes

$$\mathcal{R}^{\mathcal{E},y=0}_{A\to AB}(\cdot) = (\cdot)\otimes \frac{\Gamma_B^{(0)}}{\mathrm{tr}\,\Gamma_B^{(0)}}\,, \tag{5.26}$$

recalling that $\Gamma_A = \mathrm{tr}_B\,\Gamma_{AB} = \mathrm{tr}\big(\Gamma_B^{(0)}\big)\cdot\Gamma_A^{(0)}$. Hence, the reverse process prepares the state proportional to the hidden information operator on the system which was forgotten, as in the previous example.

In thermodynamics, this corresponds to assigning a Gibbs state to the bath. Indeed, if we describe the bath as an explicit system, we may associate to it a hidden information operator describing which operations are possible. For a heat bath, the hidden information operator is simply given by the Hamiltonian of the bath and its temperature. If we want to assign a state describing our level of knowledge of the bath, then we should consider the reverse process of that corresponding to forgetting the bath: this is the process considered here, and consists in preparing the Gibbs state on the bath.

## 5.5 Proofs

*Proof of Proposition 5.1.* By definition of a $\Gamma$-sub-preserving map, $\tilde{\Phi}_{K\to L}$ satisfies both (5.7-a) and (5.7-b). Hence let $F_K, G_L \geqslant 0$ such that

$$\tilde{\Phi}_{K\to L}\left(\Gamma_K\right) = \Gamma_L - G_L\,; \tag{5.27-a}$$

$$\tilde{\Phi}^{\dagger}_{K\leftarrow L}\left(\mathbb{1}_L\right) = \mathbb{1}_K - F_K\,. \tag{5.27-b}$$

Let $\Pi^{\Gamma}_L$ be the projector onto the support of $\Gamma_L$. We have $\Pi^{\Gamma}_L \leqslant \mathbb{1}_L$ and thus $\tilde{\Phi}^{\dagger}_{K\leftarrow L}\left(\Pi^{\Gamma}_L\right) \leqslant \tilde{\Phi}^{\dagger}_{K\leftarrow L}\left(\mathbb{1}_L\right) \leqslant \mathbb{1}_K$. So define $F'_K \geqslant 0$ such that

$$\tilde{\Phi}^{\dagger}_{K\leftarrow L}\left(\Pi^{\Gamma}_L\right) = \mathbb{1}_K - F'_K\,. \tag{5.27-c}$$

Let the system $Q$ be as in the claim, with $\Gamma_Q$ diagonal in the basis $\{\lvert\mathrm{i}\rangle_Q, \lvert\mathrm{f}\rangle_Q\}$. Define now the completely positive map

$$\begin{aligned}\Phi_{KLQ\to KLQ}\left(\cdot\right) = {}& \tilde{\Phi}_{K\to L}\left(\langle \mathrm{l\,i}\rvert \cdot \lvert \mathrm{l\,i}\rangle_{LQ}\right) \otimes \lvert\mathrm{k\,f}\rangle\langle\mathrm{k\,f}\rvert_{KQ} \\ & + \Gamma_K^{1/2}\tilde{\Phi}^{\dagger}_{K\leftarrow L}\left(\left(\Gamma_L^{-1/2}\langle\mathrm{k\,f}\rvert_{KQ}\right)\left(\cdot\right)\left(\Gamma_L^{-1/2}\lvert\mathrm{k\,f}\rangle_{KQ}\right)\right)\Gamma_K^{1/2} \otimes \lvert\mathrm{l\,i}\rangle\langle\mathrm{l\,i}\rvert_{LQ} \\ & + \Xi_{KL\to KL}\left(\langle\mathrm{i}\rvert\cdot\lvert\mathrm{i}\rangle_Q\right) \otimes \lvert\mathrm{i}\rangle\langle\mathrm{i}\rvert_Q \\ & + \Omega_{KL\to KL}\left(\langle\mathrm{f}\rvert\cdot\lvert\mathrm{f}\rangle_Q\right) \otimes \lvert\mathrm{f}\rangle\langle\mathrm{f}\rvert_Q\,,\end{aligned} \tag{5.28}$$

with some completely positive maps $\Xi_{KL\to KL}$ and $\Omega_{KL\to KL}$ yet to be determined.

First, note that the property (5.9) is obvious for this $\Phi_{KLQ}$, simply because $\lvert\mathrm{i}\rangle_Q$ and $\lvert\mathrm{f}\rangle_Q$ are orthogonal. It remains to exhibit explicit $\Xi_{KL\to KL}$ and $\Omega_{KL\to KL}$ such that $\Phi_{KLQ}$ is trace-preserving and $\Gamma$-preserving. Define as shorthands

$$\begin{aligned} g_{\mathrm{k}} &= \langle\mathrm{k}\rvert\Gamma_K\lvert\mathrm{k}\rangle_K\,; & g_{\mathrm{l}} &= \langle\mathrm{l}\rvert\Gamma_L\lvert\mathrm{l}\rangle_L\,; \\ g_{\mathrm{i}} &= \langle\mathrm{i}\rvert\Gamma_Q\lvert\mathrm{i}\rangle_Q\,; & g_{\mathrm{f}} &= \langle\mathrm{f}\rvert\Gamma_Q\lvert\mathrm{f}\rangle_Q\,. \end{aligned} \tag{5.29}$$

Note that Condition (5.10) is then equivalent to

$$g_{\mathrm{l}}\cdot g_{\mathrm{i}} = g_{\mathrm{k}}\cdot g_{\mathrm{f}}\,, \tag{5.30}$$

and that this is straightforwardly satisfied for an appropriate choice of $\Gamma_Q$ (and hence of $g_{\mathrm{i}}, g_{\mathrm{f}}$).

At this point, we'll derive conditions that $\Xi_{KL\to KL}$ and $\Omega_{KL\to KL}$ need to satisfy in order for $\Phi_{KLQ\to KLQ}$ to map $\Gamma_{KLQ}$ onto itself and to be trace-preserving. Calculate

$$\begin{aligned}\Phi_{KLQ\to KLQ}&\left(\Gamma_{KLQ}\right) \\ &= g_{\mathrm{l}}g_{\mathrm{i}}\,\tilde{\Phi}_{K\to L}\left(\Gamma_K\right)\otimes\lvert\mathrm{k\,f}\rangle\langle\mathrm{k\,f}\rvert_{KQ} \\ &\quad + g_{\mathrm{k}}g_{\mathrm{f}}\,\Gamma_K^{1/2}\tilde{\Phi}^{\dagger}_{K\leftarrow L}\left(\Pi^{\Gamma}_L\right)\Gamma_K^{1/2}\otimes\lvert\mathrm{l\,i}\rangle\langle\mathrm{l\,i}\rvert_{LQ} \\ &\quad + g_{\mathrm{i}}\Xi_{KL\to KL}\left(\Gamma_{KL}\right)\otimes\lvert\mathrm{i}\rangle\langle\mathrm{i}\rvert_Q + g_{\mathrm{f}}\Omega_{KL\to KL}\left(\Gamma_{KL}\right)\otimes\lvert\mathrm{f}\rangle\langle\mathrm{f}\rvert_Q \\ &= \lvert\mathrm{f}\rangle\langle\mathrm{f}\rvert_Q\otimes\left[g_{\mathrm{l}}g_{\mathrm{i}}\left(\Gamma_L - G_L\right)\otimes\lvert\mathrm{k}\rangle\langle\mathrm{k}\rvert_K + g_{\mathrm{f}}\Omega_{KL\to KL}\left(\Gamma_{KL}\right)\right] \\ &\quad + \lvert\mathrm{i}\rangle\langle\mathrm{i}\rvert_Q\otimes\left[g_{\mathrm{k}}g_{\mathrm{f}}\Gamma_K^{1/2}\left(\mathbb{1}_K - F'_K\right)\Gamma_K^{1/2}\otimes\lvert\mathrm{l}\rangle\langle\mathrm{l}\rvert_{KQ} + g_{\mathrm{i}}\Xi_{KL\to KL}\left(\Gamma_{KL}\right)\right]\,.\end{aligned} \tag{5.31}$$

We see that in order for this last expression to equal $\Gamma_{KLQ} = g_{\mathrm{f}}\lvert\mathrm{f}\rangle\langle\mathrm{f}\rvert_Q\otimes\Gamma_{KL} + g_{\mathrm{i}}\lvert\mathrm{i}\rangle\langle\mathrm{i}\rvert_Q\otimes\Gamma_{KL}$,

we need that the terms in square brackets above obey

$$g_{\mathrm{l}} g_{\mathrm{i}} \left(\Gamma_L - G_L\right) \otimes |\mathrm{k}\rangle\langle\mathrm{k}|_K + g_{\mathrm{f}} \Omega_{KL\to KL}\left(\Gamma_{KL}\right) = g_{\mathrm{f}}\, \Gamma_{KL} \; ; \tag{5.32-a}$$

$$g_{\mathrm{k}} g_{\mathrm{f}} \Gamma_K^{1/2} \left(\mathbb{1}_K - F_K'\right) \Gamma_K^{1/2} \otimes |\mathrm{l}\rangle\langle\mathrm{l}|_{KQ} + g_{\mathrm{i}} \Xi_{KL\to KL}\left(\Gamma_{KL}\right) = g_{\mathrm{i}}\, \Gamma_{KL} \; . \tag{5.32-b}$$

On the other hand, the adjoint map of $\Phi_{KLQ\to KLQ}$ is relatively straightforward to identify as

$$\begin{aligned}
&\Phi^\dagger_{KLQ\leftarrow KLQ}\left(\cdot\right) \\
&\quad = \tilde\Phi^\dagger_{K\leftarrow L}\left(\langle\mathrm{k\,f}|\cdot|\mathrm{k\,f}\rangle_{KQ}\right) \otimes |\mathrm{l\,i}\rangle\langle\mathrm{l\,i}|_{LQ} \\
&\qquad + \Gamma_L^{-1/2} \tilde\Phi_{K\to L}\left(\left(\Gamma_K^{1/2}\langle\mathrm{l\,i}|_{LQ}\right)\left(\cdot\right)\left(\Gamma_K^{1/2}|\mathrm{l\,i}\rangle_{LQ}\right)\right)\Gamma_L^{-1/2} \otimes |\mathrm{k\,f}\rangle\langle\mathrm{k\,f}|_{KQ} \\
&\qquad + \Xi^\dagger_{KL\leftarrow KL}\left(\langle\mathrm{i}|\cdot|\mathrm{i}\rangle_Q\right) \otimes |\mathrm{i}\rangle\langle\mathrm{i}|_Q \\
&\qquad + \Omega^\dagger_{KL\leftarrow KL}\left(\langle\mathrm{f}|\cdot|\mathrm{f}\rangle_Q\right) \otimes |\mathrm{f}\rangle\langle\mathrm{f}|_Q \; .
\end{aligned} \tag{5.33}$$

We may thus now derive the conditions on $\Xi_{KL\to KL}$ and $\Omega_{KL\to KL}$ for $\Phi_{KLQ\to KLQ}$ to be trace-preserving. Specifically, we need to ensure that $\Phi^\dagger_{KLQ\leftarrow KLQ}\left(\mathbb{1}_{KLQ}\right) = \mathbb{1}_{KLQ}$. A calculation gives us

$$\begin{aligned}
&\Phi^\dagger_{KLQ\leftarrow KLQ}\left(\mathbb{1}_{KLQ}\right) \\
&\quad = \tilde\Phi^\dagger_{K\leftarrow L}\left(\mathbb{1}_L\right) \otimes |\mathrm{l\,i}\rangle\langle\mathrm{l\,i}|_{LQ} + \Gamma_L^{-1/2}\tilde\Phi_{K\to L}\left(\Gamma_K\right)\Gamma_L^{-1/2} \otimes |\mathrm{k\,f}\rangle\langle\mathrm{k\,f}|_{KQ} \\
&\qquad + \Xi^\dagger_{KL\leftarrow KL}\left(\mathbb{1}_{KL}\right) \otimes |\mathrm{i}\rangle\langle\mathrm{i}|_Q + \Omega^\dagger_{KL\leftarrow KL}\left(\mathbb{1}_{KL}\right) \otimes |\mathrm{f}\rangle\langle\mathrm{f}|_Q \; . \\
&\quad = |\mathrm{f}\rangle\langle\mathrm{f}|_Q \otimes \left[\Gamma_L^{-1/2}\left(\Gamma_L - G_L\right)\Gamma_L^{-1/2} \otimes |\mathrm{k}\rangle\langle\mathrm{k}|_K + \Omega^\dagger_{KL\leftarrow KL}\left(\mathbb{1}_{KL}\right)\right] \\
&\qquad + |\mathrm{i}\rangle\langle\mathrm{i}|_Q \otimes \left[\left(\mathbb{1}_K - F_K\right) \otimes |\mathrm{l}\rangle\langle\mathrm{l}|_L + \Xi^\dagger_{KL\leftarrow KL}\left(\mathbb{1}_{KL}\right)\right] \; .
\end{aligned} \tag{5.34}$$

Thus, for $\Phi_{KLQ\to KLQ}$ to be trace-preserving we must have

$$\Gamma_L^{-1/2}\left(\Gamma_L - G_L\right)\Gamma_L^{-1/2} \otimes |\mathrm{k}\rangle\langle\mathrm{k}|_K + \Omega^\dagger_{KL\leftarrow KL}\left(\mathbb{1}_{KL}\right) = \mathbb{1}_{KL} \; ; \tag{5.35-a}$$

$$\left(\mathbb{1}_K - F_K\right) \otimes |\mathrm{l}\rangle\langle\mathrm{l}|_L + \Xi^\dagger_{KL\leftarrow KL}\left(\mathbb{1}_{KL}\right) = \mathbb{1}_{KL} \; . \tag{5.35-b}$$

Let us now explicitly construct an $\Xi_{KL\to KL}$ which satisfies both (5.32-b) and (5.35-b). These conditions may be written as

$$\Xi_{KL\to KL}\left(\Gamma_{KL}\right) = \Gamma_{KL} - g_{\mathrm{l}}\, \Gamma_K^{1/2}\left(\mathbb{1}_K - F_K'\right)\Gamma_K^{1/2} \otimes |\mathrm{l}\rangle\langle\mathrm{l}|_L =: A_{KL} \; ; \tag{5.36-a}$$

$$\Xi^\dagger_{KL\leftarrow KL}\left(\mathbb{1}_{KL}\right) = \mathbb{1}_{KL} - \left(\mathbb{1}_K - F_K\right) \otimes |\mathrm{l}\rangle\langle\mathrm{l}|_L =: B_{KL} \tag{5.36-b}$$

where we have used (5.29) and defined two new operators $A_{KL}$ and $B_{KL}$. Observe now that since $g_{\mathrm{l}}\, \Gamma_K^{1/2}\left(\mathbb{1}_K - F_K'\right)\Gamma_K^{1/2} \otimes |\mathrm{l}\rangle\langle\mathrm{l}|_L \leqslant \Gamma_K \otimes \left(g_{\mathrm{l}}\, |\mathrm{l}\rangle\langle\mathrm{l}|_L\right) \leqslant \Gamma_{KL}$, we have that $A_{KL} \geqslant 0$. Similarly, $\left(\mathbb{1}_K - F_K\right) \otimes |\mathrm{l}\rangle\langle\mathrm{l}|_L \leqslant \mathbb{1}_{KL}$ and hence $B_{KL} \geqslant 0$. Now, let $\xi = \operatorname{tr} A_{KL}$ and define

$$\Xi_{KL\to KL}\left(\cdot\right) = \xi^{-1} \operatorname{tr}\left(B_{KL}\left(\cdot\right)\right) A_{KL} \; . \tag{5.37}$$

We then have

$$\Xi^{\dagger}_{KL\leftarrow KL}\left(\mathbb{1}_{KL}\right) = \xi^{-1}\,\mathrm{tr}\left(A_{KL}\,\mathbb{1}_{KL}\right)\,B_{KL} = B_{KL}\,, \tag{5.38}$$

thus satisfying condition (5.36-b). On the other hand we have

$$\Xi_{KL\to KL}\left(\Gamma_{KL}\right) = \xi^{-1}\,\mathrm{tr}\left[B_{KL}\,\Gamma_{KL}\right]\,A_{KL}\,, \tag{5.39}$$

which would trivially satisfy condition (5.36-a) if we had $\xi = \mathrm{tr}\left(B_{KL}\Gamma_{KL}\right)$. Let's verify that this is the case. We have

$$\begin{aligned}\xi = \mathrm{tr}\,A_{KL} = \mathrm{tr}\,\Gamma_{KL} - g_{\mathrm{l}}\,\mathrm{tr}\left[\left(\mathbb{1}_K - F'_K\right)\Gamma_K\right] &= \mathrm{tr}\,\Gamma_{KL} - g_{\mathrm{l}}\,\mathrm{tr}\left[\tilde{\Phi}^{\dagger}_{K\leftarrow L}\left(\Pi^{\Gamma}_L\right)\,\Gamma_K\right]\\ &= \mathrm{tr}\,\Gamma_{KL} - g_{\mathrm{l}}\,\mathrm{tr}\left[\Pi^{\Gamma}_L\,\tilde{\Phi}_{K\to L}\left(\Gamma_K\right)\right]\,.\end{aligned} \tag{5.40}$$

Now, because $\tilde{\Phi}_{K\to L}\left(\Gamma_K\right) \leqslant \Gamma_L$, the operator $\tilde{\Phi}_{K\to L}\left(\Gamma_K\right)$ must lie within the support of $\Gamma_L$. Thus the projector in the last term of (5.40) has no effect and can be replaced by an identity operator. We then have

$$\begin{aligned}(5.40) &= \mathrm{tr}\,\Gamma_{KL} - g_{\mathrm{l}}\,\mathrm{tr}\left[\mathbb{1}_L\tilde{\Phi}_{K\to L}\left(\Gamma_K\right)\right] = \mathrm{tr}\,\Gamma_{KL} - g_{\mathrm{l}}\,\mathrm{tr}\left[\tilde{\Phi}^{\dagger}_{K\leftarrow L}\left(\mathbb{1}_L\right)\Gamma_K\right]\\ &= \mathrm{tr}\,\Gamma_{KL} - g_{\mathrm{l}}\,\mathrm{tr}\left[\left(\mathbb{1}_K - F_K\right)\Gamma_K\right] = \mathrm{tr}\,\Gamma_{KL} - \mathrm{tr}\left[\left(\mathbb{1}_K - F_K\right)\otimes |\mathrm{l}\rangle\langle\mathrm{l}|_L\Gamma_{KL}\right]\\ &= \mathrm{tr}\left(B_{KL}\Gamma_{KL}\right)\,.\end{aligned} \tag{5.41}$$

We have thus constructed $\Xi_{KL\to KL}$ such that it satisfies conditions (5.32-b) and (5.35-b). Let's now proceed analogously for $\Omega_{KL\to KL}$. We can rewrite conditions (5.32-a) and (5.35-a) as

$$\Omega_{KL\to KL}\left(\Gamma_{KL}\right) = \Gamma_{KL} - g_{\mathrm{k}}|\mathrm{k}\rangle\langle\mathrm{k}|_K \otimes \left(\Gamma_L - G_L\right) =: C_{KL}\,; \tag{5.42}$$

$$\Omega^{\dagger}_{KL\leftarrow KL}\left(\mathbb{1}_{KL}\right) = \mathbb{1}_{KL} - |\mathrm{k}\rangle\langle\mathrm{k}|_K \otimes \Gamma_L^{-1/2}\left(\Gamma_L - G_L\right)\Gamma_L^{-1/2} =: D_{KL}\,, \tag{5.43}$$

defining the operators $C_{KL}$ and $D_{KL}$. We have $g_{\mathrm{k}}|\mathrm{k}\rangle\langle\mathrm{k}|_K \otimes \left(\Gamma_L - G_L\right) \leqslant \Gamma_{KL}$ and thus $C_{KL} \geqslant 0$. Also $\Gamma_L^{-1/2}\left(\Gamma_L - G_L\right)\Gamma_L^{-1/2} \leqslant \mathbb{1}_L$ and thus $D_{KL} \geqslant 0$. Proceeding as for $\Xi_{KL\to KL}$, let $\omega = \mathrm{tr}\,C_{KL}$ and define

$$\Omega_{KL\to KL}\left(\cdot\right) = \omega^{-1}\,\mathrm{tr}\left(D_{KL}\,\left(\cdot\right)\right)C_{KL}\,. \tag{5.44}$$

Then

$$\Omega^{\dagger}_{KL\leftarrow KL}\left(\mathbb{1}_{KL}\right) = \omega^{-1}\,\mathrm{tr}\left(C_{KL}\mathbb{1}_{KL}\right)\,D_{KL}\,, \tag{5.45}$$

which satisfies (5.43). On the other hand, we have

$$\Omega_{KL\to KL}\left(\Gamma_{KL}\right) = \omega^{-1}\,\mathrm{tr}\left(D_{KL}\Gamma_{KL}\right)\,C_{KL}\,. \tag{5.46}$$

So it remains to show that $\omega = \mathrm{tr}\left(D_{KL}\Gamma_{KL}\right)$. Indeed,

$$\mathrm{tr}\left(D_{KL}\Gamma_{KL}\right) = \mathrm{tr}\,\Gamma_{KL} - g_{\mathrm{k}}\,\mathrm{tr}\left(\left(\Gamma_L - G_L\right)\Pi^{\Gamma}_L\right) = \mathrm{tr}\,\Gamma_{KL} - g_{\mathrm{k}}\,\mathrm{tr}\left(\Gamma_L - G_L\right) = \mathrm{tr}\,C_{KL} = \omega\,, \tag{5.47}$$

where the projector $\Pi_L^\Gamma$ has no effect in the second expression since $\Gamma_L - G_L$ is entirely contained within the support of $\Gamma_L$.

We have thus constructed a completely positive, trace preserving map $\Phi_{KLQ\to KLQ}$ which maps $\Gamma_{KLQ}$ onto itself and which satisfies (5.9). This concludes the proof. ∎

*Proof of Proposition 5.2.* The proof consists in showing (i) ⇒(v) ⇒(iv) ⇒(iii) ⇒(i) as well as ((i) ⇒)(v) ⇒(ii) ⇒(i).

(i) ⇒(v): By assumption we have $\mathcal{E}_{X\to X'}(\Gamma_X) \leqslant 2^{-y}\Gamma_{X'}$. Let $\Gamma_{W_1}, \Gamma_{W_2}$ and $P_{W_1}, P'_{W_2}$ satisfy the assumptions in the claim (v), and define the shorthands

$$\sigma_{W_1}^{(1)} = \frac{P_{W_1}\Gamma_{W_1}P_{W_1}}{\mathrm{tr}(P_{W_1}\Gamma_{W_1})}\ ; \qquad\qquad \sigma_{W_2}^{(2)} = \frac{P'_{W_2}\Gamma_{W_2}P'_{W_2}}{\mathrm{tr}(P'_{W_2}\Gamma_{W_2})}\ . \tag{5.48}$$

Define the map

$$\Phi''_{XW_1\to X'W_2}(\cdot) = \sigma_{W_2}^{(2)} \otimes \mathcal{E}_{X\to X'}[\mathrm{tr}_{W_1}(P_{W_1}(\cdot))]\ . \tag{5.49}$$

This map is completely positive by construction, and is trace nonincreasing because it is a composition of trace nonincreasing maps. We need to show that it is Γ-sub-preserving. We have

$$\begin{aligned}\Phi''_{XW_1\to X'W_2}(\Gamma_X \otimes \Gamma_{W_1}) &= \left(\mathrm{tr}\, P_{W_1}\Gamma_{W_1}\right)\cdot \sigma_{W_2}^{(2)} \otimes \mathcal{E}_{X\to X'}(\Gamma_X)\\ &\leqslant 2^{-y}\frac{\mathrm{tr}\, P_{W_1}\Gamma_{W_1}}{\mathrm{tr}\, P'_{W_2}\Gamma_{W_2}}\cdot\left(P'_{W_2}\Gamma_{W_2}P'_{W_2}\right)\otimes\Gamma_{X'}\\ &\leqslant \Gamma_{W_2}\otimes\Gamma_{X'}\ ,\end{aligned} \tag{5.50}$$

using the fact that $P'_{W_2}\Gamma_{W_2}P'_{W_2} \leqslant \Gamma_{W_2}$ since $\Gamma_{W_2}$ commutes with the projector $P'_{W_2}$.

(v) ⇒(iv): This special case follows directly from (v) with $W_1 = W_2 = \tilde{Q}$, $\Gamma_{W_1} = \Gamma_{W_2} = \Gamma_{\tilde{Q}}$ and by choosing $P_{W_1} = |1\rangle\langle 1|_{\tilde{Q}}$, $P'_{W_2} = |2\rangle\langle 2|_{\tilde{Q}}$. Note that $g_1 = \mathrm{tr}\, P_{W_1}\Gamma_{W_1}$ and $g_2 = \mathrm{tr}\, P'_{W_2}\Gamma_{W_2}$ and hence indeed $\left(\mathrm{tr}\, P'_{W_2}\Gamma_{W_2}\right)/(\mathrm{tr}\, P_{W_1}\Gamma_{W_1}) = g_2/g_1 \geqslant 2^{-y}$.

(iv) ⇒(iii): This is a trivial special case of (iv).

(iii) ⇒(i): Let $\Gamma_Q, |1\rangle_Q, |2\rangle_Q, g_1, g_2$ and $\Phi'_{XQ\to X'Q}$ be any choices which satisfy the assumptions of (iii) and which also satisfy the choice $g_2/g_1 = 2^{-y}$. Observe that for any $\omega_X$

$$\mathcal{E}_{X\to X'}(\omega_X) = \left\langle 2\right|\Phi'_{XQ\to X'Q}\left(\omega_X \otimes |1\rangle\langle 1|_Q\right)\left|2\right\rangle_Q\ . \tag{5.51}$$

Plugging in $\omega_X = \Gamma_X$, and using the fact that $g_1|1\rangle\langle 1|_Q \leqslant \Gamma_Q$ and that $\Phi'_{XQ\to X'Q}$ is Γ-sub-preserving,

$$\begin{aligned}\mathcal{E}_{X\to X'}(\Gamma_X) &\leqslant \left\langle 2\right| g_1^{-1}\cdot\Phi'_{XQ\to X'Q}\left(\Gamma_X\otimes\Gamma_Q\right)\left|2\right\rangle_Q\ .\\ &\leqslant \left\langle 2\right| g_1^{-1}\cdot\Gamma_{X'}\otimes\Gamma_Q\left|2\right\rangle_Q\ .\\ &= \frac{g_2}{g_1}\cdot\Gamma_{X'} = 2^{-y}\,\Gamma_{X'}\ .\end{aligned} \tag{5.52}$$

(v) ⇒(ii): This is in fact another special case of (v). Let $\lambda_1, \lambda_2$ such that $\lambda_1 - \lambda_2 \leqslant y$ and that $2^{\lambda_1}, 2^{\lambda_2}$ are integers. Let $A$ be any quantum system of dimension at least $\max\{2^{\lambda_1}, 2^{\lambda_2}\}$ and with $\Gamma_A = \mathbb{1}_A$. Now we'll use our assumption that (v) holds. Choose $W_1 = W_2 = A$, $P_{W_1} = \mathbb{1}_{2^{\lambda_1}}$, $P'_{W_2} = \mathbb{1}_{2^{\lambda_2}}$. Observe that $\operatorname{tr} P_{W_1}\Gamma_{W_1} = \operatorname{tr} P_{W_1} = 2^{\lambda_1}$ and $\operatorname{tr} P'_{W_2}\Gamma_{W_2} = \operatorname{tr} P'_{W_2} = 2^{\lambda_2}$, and hence the assumptions of the choices in (v) are satisfied. Then we know that there must exist a $\Gamma$-sub-preserving, trace-nonincreasing map $\Phi''_{XA\to X'A}$ obeying (5.17). The latter condition reads by plugging in our choices

$$\Phi''_{XA\to X'A}\big((2^{-\lambda_1}\mathbb{1}_{2^{\lambda_1}}) \otimes \omega_X\big) = \big(2^{-\lambda_2}\mathbb{1}_{2^{\lambda_2}}\big) \otimes \mathcal{E}_{X\to X'}(\omega_X) \tag{5.53}$$

for all $\omega_X$. This is exactly the condition that $\Phi$ has to fulfill, and hence $\Phi$ may be taken equal to the map $\Phi''$. It follows that (ii) is true.

(ii) ⇒(i): Consider any $\lambda_1, \lambda_2 \geqslant 0$ with $\lambda_1 - \lambda_2 \leqslant y$. Let $\Phi_{XA\to X'A}$ be the corresponding $\Gamma$-sub-preserving map given by the assumption that (ii) holds. Observe that for all $\omega_X$,

$$\mathcal{E}_{X\to X'}(\omega_X) = \operatorname{tr}_A\Big\{\mathbb{1}_{2^{\lambda_2}}\ \Phi_{XA\to X'A}\big((2^{-\lambda_1}\mathbb{1}_{2^{\lambda_1}}) \otimes \omega_X\big)\Big\}\,. \tag{5.54}$$

Plugging in $\omega_X = \Gamma_X$, and using the fact that $\Phi$ is $\Gamma$-sub-preserving,

$$\begin{aligned}\mathcal{E}_{X\to X'}(\Gamma_X) &\leqslant \operatorname{tr}_A\Big\{\mathbb{1}_{2^{\lambda_2}}\ \Phi_{XA\to X'A}\big(2^{-\lambda_1}\cdot\Gamma_A\otimes\Gamma_X\big)\Big\}\\ &\leqslant 2^{-\lambda_1}\cdot\operatorname{tr}_A\{\mathbb{1}_{2^{\lambda_2}}\ \Gamma_A\otimes\Gamma_{X'}\}\\ &= 2^{-(\lambda_1-\lambda_2)}\,\Gamma_{X'}\end{aligned} \tag{5.55}$$

The statement (i) follows by choosing a sequence of $(\lambda_1, \lambda_2)$ with $\lambda_1 - \lambda_2 \to y$. ∎

*Proof of Proposition 5.3.* These properties are straightforward to very. The map $\mathcal{R}^{\mathcal{E},y}_{X'\to X}$ is clearly completely positive, as a composition of completely positive maps. We further have

$$\mathcal{R}^{\mathcal{E},y\,\dagger}_{X'\leftarrow X}(\mathbb{1}_X) = 2^y\,\Gamma_{X'}^{-1/2}\,\mathcal{E}_{X\to X'}\Big(\Gamma_X^{1/2}\ \mathbb{1}_X\ \Gamma_X^{1/2}\Big)\,\Gamma_{X'}^{-1/2} \leqslant \mathbb{1}_{X'}\,, \tag{5.56}$$

using the fact that $\mathcal{E}_{X\to X'}(\Gamma_X) \leqslant 2^{-y}\Gamma_{X'}$. Then,

$$\mathcal{R}^{\mathcal{E},y}_{X'\to X}(\Gamma_{X'}) \leqslant 2^y\,\Gamma_X^{1/2}\,\mathcal{E}^\dagger_{X\leftarrow X'}(\mathbb{1}_{X'})\,\Gamma_X^{1/2} \leqslant 2^y\,\Gamma_X\,.$$ ∎

# 6

# The Minimal Work Cost of Information Processing

In this chapter, we'll study the minimal amount of work needed to perform a quantum logical operation using memory registers.[1]

This chapter concerns itself solely with input and output states encoded into systems with completely degenerate Hamiltonians. This may be viewed as a warm-up for the next chapter, where we consider general systems described by general Gibbs states.

Suppose a system $X$ is in a state $\sigma_X$. Suppose its Hamiltonian is completely degenerate. The goal is to perform a given completely positive, trace preserving map $\mathcal{E}_{X\to X'}$ on the system. The output system (of possibly different dimension) is described also by a trivial Hamiltonian. The question is, how much work is required?

We refer to the model developed in the previous chapter, applied to quantum thermodynamics. The admissible operations (or free operations) are those which are $\Gamma$-sub-preserving, that is, which obey the operator inequality (5.6). In this chapter we restrict ourselves to the special case $\Gamma = \mathbb{1}$, corresponding to information processing on memory registers with a completely degenerate Hamiltonian. The battery is thought of as an information battery for concreteness; we've seen however in the last chapter the information battery model is equivalent to other standard models. The compression yield $y$ is the number of pure qubits which can be extracted for free in principle while performing a logical process, and we have seen in Section 4.5.1 that this corresponds to extracted work $W_{\text{extracted}} = kT \ln 2 \cdot y$. The work cost is then simply the negative of the extracted work, $W_{\text{cost}} = -W_{\text{extracted}}$.

[1]These results are reported in Faist *et al.* (2015a), and this chapter reproduces in large part verbatim content from that reference.

## 6.1 The minimal work cost of a logical process

### 6.1.1 Statement of the problem

The problem is the following. We are given a logical process $\mathcal{E}_{X \to X'}$, that is, any completely positive, trace-preserving map. We are given an input state $\sigma_X$. What is the minimal work cost of a physical implementation of this process?

A physical implementation only has to implement the mapping $\mathcal{E}_{X \to X'}$ correctly on the support of $\sigma_X$, as other states will never be observed. Hence the full information about the problem is encoded in the process matrix $\rho_{X'R}$ following the construction presented in Section 3.1.9.

So the reformulated problem is: given a process matrix $\rho_{X'R}$, what is the work cost of implementing the corresponding logical process and input state?

### 6.1.2 A result with zero error tolerance

The answer to our question is essentially given by Proposition 5.2. The latter gives us a criterion indicating exactly how much work is needed for a specific logical process. The difference is that in that case, we didn't worry about the input state. So we need to reformulate the problem again as follows: we want to find the mapping $\mathcal{T}_{X \to X'}$ which costs the least work, and which is compatible with the given process matrix $\rho_{X'R}$.

Let us formulate the technical statement as a proposition for clarity. It follows directly from Proposition 5.2 along with Proposition 5.1.

***Proposition 6.1*** *(Minimal work cost of a logical process). Let $X, X'$ be quantum systems, of possibly different dimensions. Let $|\sigma\rangle_{XR}$ be a pure state on $X$ and a reference system $R$. Let $\mathcal{E}_{X \to X'}$ be a completely positive, trace-preserving map. Then, for any $\lambda_1, \lambda_2 \geqslant 0$ such that $2^{\lambda_1}, 2^{\lambda_2}$ are integers, the following are equivalent:*

*(i) There exists an information battery system $A$ with $\Gamma_A = \mathbb{1}_A$, as well as a*

*map $\Phi_{XA\to X'A}$ which is an admissible operation, such that:*

$$\Phi_{XA\to X'A}\left[\mathbb{1}_{XA}\right] = \mathbb{1}_{X'A}\ ; \tag{6.1-a}$$

$$\Phi_{XA\to X'A}\left[\sigma_{XR}\otimes\left(2^{-\lambda_1}\mathbb{1}_{2^{\lambda_1}}\right)\right] = \rho_{X'R}\otimes\left(2^{-\lambda_2}\mathbb{1}_{2^{\lambda_2}}\right)\ , \tag{6.1-b}$$

*where the states $\left(2^{-\lambda_i}\mathbb{1}_{2^{\lambda_i}}\right)$ denote any uniform quantum state of rank $2^{\lambda_i}$ on $A$.*

*(ii) The following condition holds:*

$$\lambda_1 - \lambda_2 \leqslant -\log\left\|\mathcal{E}_{X\to X'}(\Pi_X^\sigma)\right\|_\infty = -H_{\max,0}(E\,|\,X')_\rho\ , \tag{6.2}$$

*where $\Pi_X^\sigma$ denotes the projector onto the support of $\sigma_X$, and where $|\rho\rangle_{X'RE}$ is any purification of $\rho_{X'R}$.*

Recall that $kT\ln(2)\cdot(\lambda_1-\lambda_2)$ is interpreted as the amount of work extracted from the process and stored into the information battery. Hence, the proposition tells us that in our framework, the optimal work cost (the negative of the extracted work) is then given by the quantity

$$W^{\epsilon=0}(\rho_{X'R}) = kT\ln(2)\cdot\log\left\|\mathcal{E}_{X\to X'}(\Pi_X)\right\|_\infty \tag{6.3}$$

$$= kT\ln(2)\cdot H_{\max,0}(E\,|\,X')_\rho\ , \tag{6.4}$$

recalling that $\sigma_X$ and $\mathcal{E}_{X\to X'}(\Pi_X)$ are determined unambiguously from $\rho_{X'R}$.

### 6.1.3 A physically robust result

As seen in Section 2.6.2, it is necessary to ensure the robustness of the bound (6.3) when low probability events are ignored. It is well known that the quantity $H_{\max,0}(E\,|\,X')_\rho$ is not robust: for instance, if $X'$ is a trivial system, this corresponds to the rank of $\rho_E$ which can be clearly discontinuous.

A smooth version of (6.4) is straightforward to obtain. In this case, we allow the actual process to not implement precisely $\mathcal{E}$, but only approximate it well via an $\epsilon$-approximation (and hence allowing for a small failure probability). The best strategy to detect this inexactness is to prepare $|\sigma\rangle_{XR}$ and send $\sigma_X$ into the process, and then perform a measurement on $\rho_{X'R}$. To ensure that the approximate process is not distinguishable from the ideal process with probability greater than $\epsilon$, we require that the trace distance between the ideal output of the process $\rho_{X'R}$ and the actual output $\hat{\rho}_{X'R}$ must not exceed $\epsilon$.

We can apply our main result to the approximate process that brings $\sigma_{XR}$ to $\hat{\rho}_{X'R}$, and lower bound the work cost of that process by

$$\begin{aligned} W^{\epsilon}(\hat{\rho}_{X'R}) &\geqslant kT\ln(2)\cdot H_{\max,0}(E\,|\,X')_{\hat{\rho}} \\ &\geqslant kT\ln(2)\cdot H_{\max}(E\,|\,X')_{\hat{\rho}}\ , \end{aligned} \tag{6.5}$$

where the second inequality is shown in Tomamichel *et al.* (2011). This relaxation of $H_{\max,0}$ to $H_{\max}$ is done for the sake of presentation and consistency with other results within the smooth entropy framework. With a smoothing parameter $\epsilon > 0$, there is no significant difference with this relaxation: indeed, the two quantities are equivalent up to adjustment of the $\epsilon$ parameter and up to a logarithmic term in $\epsilon$ (Tomamichel *et al.*, 2011, Lemma 18).

If we optimize (6.5) over all possible maps $\mathcal{T}$ that implement a process matrix $\hat{\rho}_{X'R}$, we obtain the work requirement of the best $\epsilon$-approximation,

$$\begin{aligned} W^{\epsilon}(\rho_{X'R}) &\geqslant \min_{\hat{\rho}_{X'R}\approx_{\epsilon}\rho_{X'R}} kT\ln(2)\cdot H_{\max}(E\,|\,X')_{\hat{\rho}} \\ &\geqslant \min_{\hat{\rho}_{X'RE}\approx_{\bar{\epsilon}}\rho_{X'RE}} kT\ln(2)\cdot H_{\max}(E\,|\,X')_{\hat{\rho}} \end{aligned} \tag{6.6}$$

where the first optimization ranges over all quantum states $\hat{\rho}_{X'R}$ such that the trace distance $D(\hat{\rho}_{X'R}, \rho_{X'R}) \leqslant \epsilon$, and where the second optimization ranges over all quantum states $\hat{\rho}_{X'RE}$ such that $P(\rho_{X'RE}, \hat{\rho}_{X'RE}) \leqslant \bar{\epsilon}$, with the purified distance defined in Section 3.1.10 and where $\bar{\epsilon} = \sqrt{2\epsilon}$. Finally, recalling the definition of $H_{\max}$ (Section 3.2.3), we obtain

$$W^{\epsilon}(\rho_{X'R}) \geqslant kT\ln(2)\cdot H_{\max}^{\bar{\epsilon}}(E\,|\,X')_{\rho}\ . \tag{6.7}$$

## 6.2 Work cost and discarded information

The $E$ system is introduced in (6.2) as a purifying system for $\rho_{X'R}$. It is however also the ancillary system necessary for the Stinespring dilation of $\mathcal{E}_{X\to X'}$ (see Section 3.1.8). This is a hypothetical system which represents the information which is discarded by the process.

Thus, the entropic term $H_{\max,0}(E\,|\,X')$ in (6.3) simply measures the amount of discarded information, from the point of view of the observer which has completed the process and has access to the output $X'$.

In a practical implementation of the process, the Stinespring $E$ system is a subsystem of the actual environment of the system, such as the heat bath and any control apparatuses.

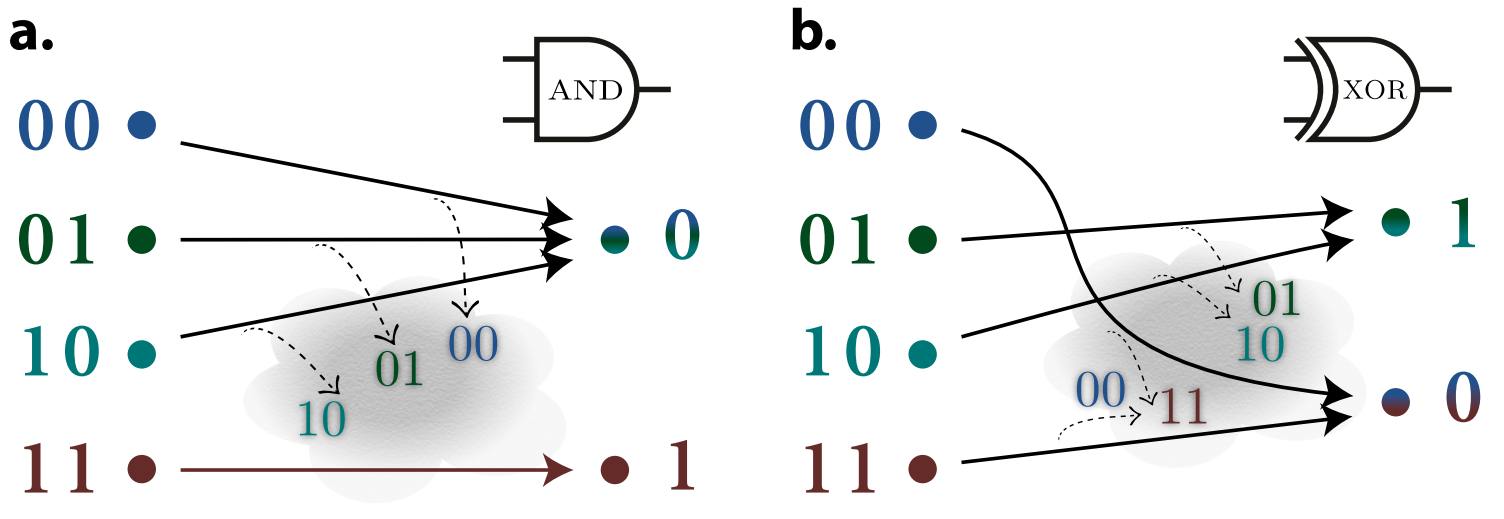

Figure 6.1: Examples of logical processes. **a.** The AND gate is one of the building blocks of computers. Our result implies that any successful implementation of this logically irreversible gate requires at least work $\log_2(3) \cdot kT \ln 2 \approx 1.6\, kT \ln 2$ due to the entropy of the discarded information (dotted arrows). **b.** The XOR gate only requires $kT \ln 2$ work, as it discards less entropy per output event than the AND gate.

## 6.3 Classical mappings; the AND gate

Our result, which is applicable to arbitrary quantum processes, applies to all classical computations as a special case. Classically, logical processes correspond to stochastic maps, of which deterministic functions are a special case. As a simple example, consider the AND gate. This is one of the elementary operations computing devices can perform, from which more complex circuits can be designed. The gate takes two bits as input, and outputs a single bit that is set to **1** exactly when both input bits are **1**, as illustrated in Figure 6.1a. The logical process is manifestly irreversible, as the output alone doesn't allow to infer the input uniquely. Intuitively, if one of the inputs is zero, then the logical process effectively has to reset a three-level system to zero, forgetting which of the three possible inputs **00**, **01**, or **10** was given; this information can be viewed as being discarded, and hence dumped into the environment.

We can use our main result to study this situation, because the formalism of quantum information includes that of classical information as a special case. A classical logical process is specified by a conditional probability distribution $p(x'|x)$ describing the probability of observing $x'$ at the output if the input was $x$. Let us write this classical process as a quantum logical process which measures its input and selectively produces the appropriate classical output state:[2]

$$\mathcal{E}_{X\to X'}(\cdot) = \sum_{x,x'} \langle x|\cdot|x\rangle_X \cdot p(x'|x)\, |x'\rangle\langle x'|_{X'} \,. \tag{6.8}$$

Our main result (6.4) then simply reduces to

$$W^{\epsilon=0} = kT\ln 2 \cdot \log_2 \max_{x'} \sum_x p(x'|x)\,, \tag{6.9}$$

where the sum ranges only over those $x$ that have a non-zero probability of occurring. In the case of deterministic mappings $p(x'|x) \in \{0,1\}$, this corresponds to the maximum number of input states that map to a same output state.

Now let's apply this to the AND gate. Provided all four states **00**, **01**, **10** and **11** have nonzero probability of occurring, there are three input states mapping to the same output state, so (6.9) gives us simply

$$W^{\epsilon=0}_{\text{AND}} = \log_2(3)\cdot kT\ln 2 \approx 1.6\,kT\ln 2\,. \tag{6.10}$$

If we tolerate a small failure probability $\epsilon$, and as long as the input distribution does not have very small eigenvalues, then no eigenvalues will be comparably small to $\epsilon$: all distributions that are $\epsilon$-close to the initial one will have the same rank. Thus the value (6.10) is exact also for an $\epsilon > 0$ which is not too large:

$$W^{\epsilon\approx 0}_{\text{AND}} = \log_2(3)\cdot kT\ln 2 \approx 1.6\,kT\ln 2\,. \tag{6.11}$$

It turns out that this value differs from the value we would obtain from the expression in (6.7): the reason is that the latter was obtained with an additional relaxation of $H^{\epsilon}_{\max,0}(E\,|\,X')$ to $H^{\sqrt{2\epsilon}}_{\max}(E\,|\,X')$ for purposes of presentation.

## 6.4 Dependence on the logical process

Crucially, our result reveals that the minimal work requirement depends in general on the specific logical process, and not only on the input and output states. This contrasts with traditional thermodynamics for large systems, where the minimal work requirement of a state transformation can always be written as a difference of a thermodynamical potential, such as the free energy. For example, the minimal work cost of performing specifically an AND gate may differ from that of another logical process mapping an input distribution $(p_{00}, p_{01}, p_{10}, p_{11})$ (with $\sum_i p_i = 1$) to the distribution $(p'_0, p'_1) = (p_{00} + p_{01} + p_{10}, p_{11})$. (Recall that the classical counterpart of a quantum state is a probability distribution.) To see this, consider the XOR gate, which outputs a **1** exactly when both inputs are different (see Figure 6.1**b**). The minimal work cost requirement of this gate, as given by (6.9), is now only $kT \ln 2$, as in the worst case, only a single bit of information is erased (again supposing that all four input states have non-negligible probability of occurring). Now, suppose that for some reason, the input distribution is such that $p_{01} + p_{10} = p_{11}$, i.e. the input **11** occurs with the same probability as of either **01** or **10** appearing. Then, the XOR gate reproduces the exact same output distribution as the AND gate: in both cases, we have $p'_0 = p_{00} + p_{10} + p_{01} = p_{00} + p_{11}$ and $p'_1 = p_{11} = p_{01} + p_{10}$. In other words, both logical processes have the same input and output state, yet the XOR gate only requires work $kT \ln 2$ compared to the AND gate, which requires $1.6\, kT \ln 2$. In summary, whereas both processes map the distribution (e.g. a quantum state) $(p_{ij})$ to the distribution $(p'_k)$, both the logical processes have the different minimal work costs given by (6.11) and by

$$W_{\text{XOR}}^{\epsilon\approx 0} = kT \ln 2\ . \tag{6.12}$$

[2]Note that there may be several ways to extend a classical logical process to a quantum channel. Observe for example that with this prescription the classical identity map $p(x'|x) = \delta_{x,x'}$ corresponds to a quantum dephasing channel, and not the quantum identity process.

On the one hand, we are by definition interested in the work cost of a given logical process, so one might have expected that this work cost should not only depend on the input and output states. On the other hand, it might seem contradictory that the full logical process matters even though we have fixed an input state $\sigma_X$. However, this makes sense if we consider preparing the input state as part of a pure state on the input system and a reference system. In this case, the logical process which is implemented influences the (in principle detectable) correlations between the output and the reference system, even if the reduced state on the input is the fixed state $\sigma_X$.

We emphasize that the phenomenon observed here is fundamentally different from the notion of thermodynamic irreversibility (we refer in particular to the discussion in Section 2.5.1). Here, we always consider the optimal procedure for implementing the logical process, whereas a thermodynamically irreversible process is in fact a suboptimal physical process which could be replaced by a more efficient, reversible one. In our framework, the thermodynamically irreversibile processes are those physical implementations which do not achieve the bound (6.7).

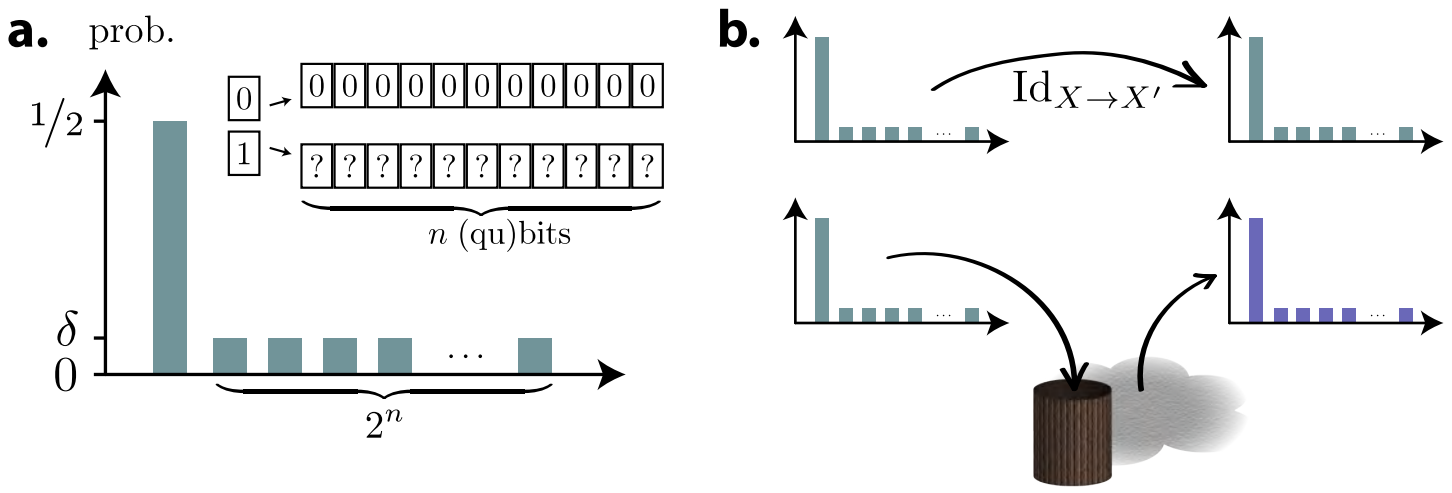

Figure 6.2: Examples of large, non-typical distributions. **a.** The probability distribution (given by the spectrum of the state) of a classical system of one random qubit, along with $n$ other qubits that are all **0** if the first qubit is **0**, or uniformly random otherwise. **b.** Two different operations on this system may have the same input and output state, yet their work cost may differ arbitrarily. The first operation copies its input to its output (identity map), which costs no work. The second destroys the input and reproduces a fresh system at the output.

The difference between the minimal work costs of the AND and XOR gates given by (6.11) and (6.12) may seem small. However, this difference can be arbitrarily large in certain scenarios. In the rest of this section we'll play around with a more drastic example. Consider the state provided in Figure 6.2a, described by the density matrix

$$\rho = \frac{1}{2}\left[ |0\rangle\langle 0| \otimes |0\ldots 0\rangle\langle 0\ldots 0| + |1\rangle\langle 1| \otimes \frac{\mathbb{1}_{2^n}}{2^n} \right] .$$

Consider also the two logical processes depicted schematically in Figure 6.2b. The first logical process $\mathcal{E}_1$ is simply the identity map, $\mathcal{E}_1(\sigma) = \mathrm{id}_{X\to X'}(\sigma) = \sigma$. The second logical process $\mathcal{E}_2$ resets its input and prepares a fresh copy of $\rho$, i.e. $\mathcal{E}_2(\sigma) = \mathrm{tr}(\sigma)\,\rho$.

First note that both computations have exactly the same input and output states. The minimal work requirement of the identity mapping is zero, obviously, because it can be implemented by doing nothing, or also, because it is logically reversible. However, the analysis is different for $\mathcal{E}_2$. If we did nothing as for $\mathcal{E}_1$, then high correlations would remain between the input and the output, and we would not be implementing the computation $\mathcal{E}_2$ but rather $\mathcal{E}_1$. Now the minimal work requirement that will be needed, if we want to be almost certain that the process succeeds, can be intuitively understood as follows: in the worst case, which happens with probability ½, the input is in the state that is almost fully mixed, and one will first have to reset $\sim n$ bits, costing $\sim n\,kT \ln 2$ work. When preparing the output, we can decide randomly on whether to prepare the pure or the mixed state by extracting 1 bit of work from a Szilard box. However, in the worst case with probability ½, we have to prepare the state $|0\dots0\rangle$, and at worst only one bit of work can be extracted. The total worst-case work cost of this strategy is

$$W^{\varepsilon\approx 0}(\mathcal{E}_2, \rho) \approx n\,kT \ln 2 \,. \tag{6.13}$$

This value can be calculated exactly as our example is a special case of Section 6.10.1 below; it turns out to be optimal. The approximation we made above (where we note '$\sim$' and '$\approx$') is simply that $\log(2^n + 1) \approx n$ and $n + 1 \approx n$. Also, we have assumed that $(1 - \epsilon)\,n \approx n$.

Note that the quantity (6.13) can become arbitrarily large, as it scales with the number of qubits $n$.

This distribution might seem very artificially constructed. We however provide here an example of a physical system which exhibits such behavior. Consider a particle detector, which we model in the following way: as long as no particle has shown up, the detector is initialized in a state $|0\rangle$. Once a particle hits the device, the state of the detector is changed to a very disordered state $\tau$, which we may for the sake of the example choose as a uniformly mixed state of rank $d$: $\tau = \frac{1}{d}\mathbb{1}_d$. Suppose we wish to describe the state of the device, not knowing whether a particle has hit it or not. If the probability that a particle was detected is ½, then the state of the detector is precisely $\rho$ (with $d = 2^n$). The first logical processes given in Figure 6.2**b** corresponds to not doing anything to the detector. The second logical process corresponds to resetting the detector, and then again sending a new particle in with probability ½. Note that in this case, by looking at the detector after the process, we may not know the state of the detector at the input of the process.

## 6.5 Work extraction

While erasure requires work, it is well known that in a wide range of frameworks one can in general extract work with the reverse logical process, which corresponds to taking a register of bits which are all in the zero state and making them maximally mixed (Szilard, 1929; Bennett, 1982). Our result intrinsically reproduces this fact: the Stinespring dilation $\mathcal{U}_{X\to X'E}$ of a logical process which generates randomness in fact creates entanglement between the output $X'$ and $E$ (see Figure 6.3). The conditional entropy $H^{\epsilon}_{\max}(E\,|\,X')$ then becomes negative, such that the bound (6.7) allows work to be extracted.

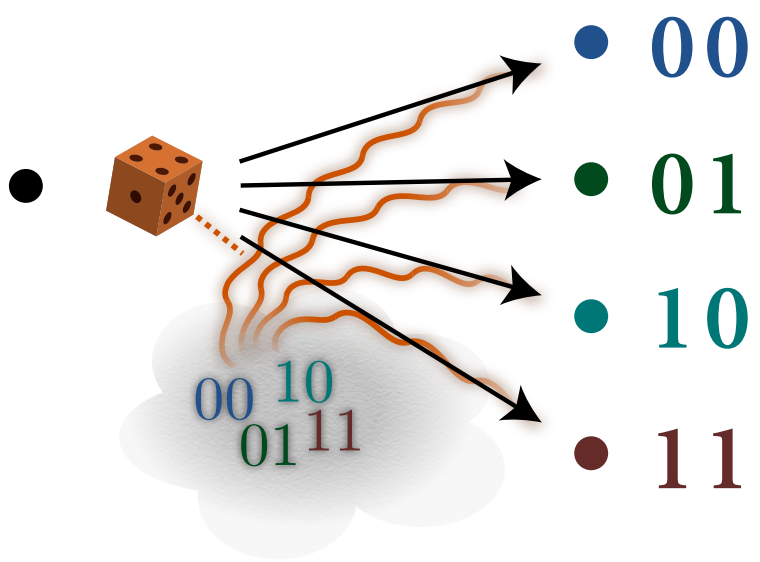

Figure 6.3: Work can be extracted if randomness is being produced: the discarded information is entangled with the output (orange wavy lines), and the conditional entropy on the right hand side of (6.7) is negative.

We remark that, even if the logical process $\mathcal{E}_{X\to X'}$ is classical, the relevant state for the entropic term in (6.7) is entangled, and thus all but classical; this is due to the construction of $E$ as a purifying system for the logical process.

## 6.6 I.i.d. limit

As we have seen in the introduction, considerable previous work has focused on the limit cases where many i.i.d. systems are provided. In such a case, the process $\mathcal{E}^{\otimes n}$ is applied on $n$ independent copies of the input $\sigma^{\otimes n}$, and outputs $\rho^{\otimes n}$. A smoothing parameter $\epsilon > 0$ is chosen freely. We may simply apply our (smoothed) main result to get an expression for our bound on the work cost,

$$W \geqslant kT\ln(2)\cdot H_{\max}^{\bar{\epsilon}}(E^n\,|\,X^n)_{\rho^{\otimes n}}\,, \tag{6.14}$$

however it is known that the smooth entropies converge to the von Neumann entropy in the i.i.d. limit, as explained in Section 3.2.3, which allows us to simplify the expression of the work cost per particle, or per repetition of the process, to

$$H(E\,|\,X)_\rho = H(E\,|\,X)_\rho - H(X)_\rho = H(X)_\sigma - H(X)_\rho\,,$$

where the last equality holds because $\rho_{EX}$ and $\sigma_X$ have the same spectrum being both purifications of the same $\rho_R = \sigma_R$. We conclude that in the asymptotic i.i.d. regime, the work cost is simply given by the difference of entropy between the initial and final state,

$$W \geqslant kT\ln(2)\cdot\left[H(\text{initial state}) - H(\text{final state})\right]. \tag{6.15}$$

Here, $W$ is the average work cost per particle, or per repetition of the process. In the case for example of many independent particles undergoing the same independent process, the total work $W$ required is obtained by considering the entropy of the full system of all particles in both terms in (6.15).

We see that, in the i.i.d. regime, the minimal work cost of the logical process no longer depends on the details of the logical process, and just depends on the input and output state. This is precisely the behavior of macroscopic thermodynamics, and thus allows us to identify the thermodynamic entropy with the von Neumann entropy. The thermodynamic entropy state function is therefore not a property of the microscopic system; it is rather an *emergent quantity* that appears whenever the state of the full system is typical.

This is just a teaser to illustrate the main idea of how to connect our microscopic model to macroscopic thermodynamics. The detailed mechanism according to which macroscopic thermodynamics emerges from our microscopic model is discussed later in Chapter 8.

## 6.7 Erasure with a quantum memory

Recently, del Rio *et al.* (2011) have constructed an explicit procedure capable of resetting a quantum system $S$ to a pure state using an erasure mechanism assisted by a quantum memory $M$ (Figure 6.4). As we shall see, our main result implies that their procedure is nearly optimal.

The setting proposed by del Rio *et al.* (2011) consists in a system $S$ which is correlated with a system $M$ in a joint state $\sigma_{SM}$. The task is to erase $S$ while preserving the reduced state on $M$ and any possible correlations of $M$ with other systems. Formally, given a purification $\sigma_{SMR}$ of $\sigma_{SM}$, we are looking for a process that will bring this state to the state $\rho_{SMR} = |0\rangle\langle 0|_S \otimes \sigma_{MR}$. This includes a requirement to preserve any correlations between $M$ and $R$ contained in the state $\sigma_{MR}$. In del Rio *et al.* (2011) a process is proposed that performs this task at work cost

$$W_{\text{Lídia}} = kT\ln(2)\, H_{\max}^{\epsilon}(S\,|\,M)_{\sigma} + O\big(\log\tfrac{1}{\varepsilon}\big)\,, \tag{6.16}$$

where $H_{\max}^{\epsilon}(S\,|\,M)_{\sigma}$ designates the smooth max-entropy (Section 3.2.3)

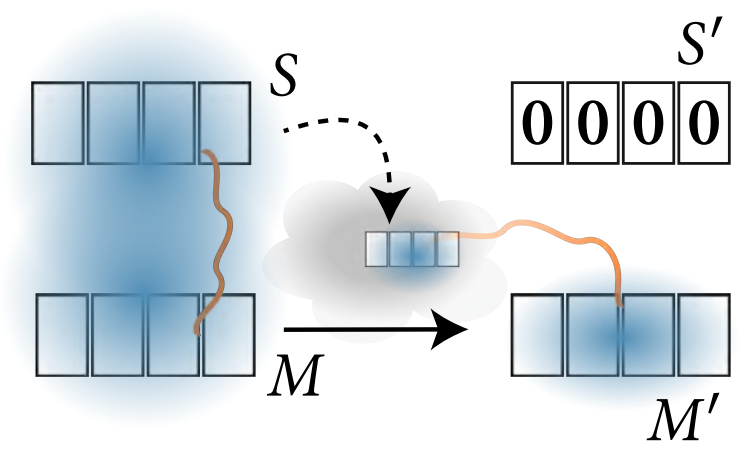

Figure 6.4: The erasure of a quantum system $S$ with access to a quantum memory $M$ must transfer the content of $S$ into the system $E$ containing the discarded information, while preparing $S'$ in a pure state and mapping $M$ to $M'$ identically. The corresponding minimal work cost is $kT\ln(2)\cdot H^{\epsilon}_{\max}(S\,|\,M)$; this can be achieved using the procedure of del Rio *et al.* (2011). If the system is entangled with the memory, this quantity is negative and work may be extracted.

The full process that is eventually performed can be written as

$$\mathcal{E}^{(\text{erasure})}_{SM\to SM}(\sigma) = |0\rangle\langle 0|_S \otimes \operatorname{tr}_S(\sigma)\ . \tag{6.17}$$

It is clear that this process preserves the reduced state $\sigma_{MR}$, because it acts as the identity map on those systems. We can now apply our main result to this particular mapping, simply by considering $X$ to be the joint system of $S$ and the memory $M$, $\mathscr{H}_X = \mathscr{H}_S \otimes \mathscr{H}_M$. Note that we have $\rho_{SMR} = |0\rangle\langle 0|_S \otimes \sigma_{MR}$, purified by $|\rho\rangle_{SMRE} = |0\rangle_S \otimes |\rho\rangle_{MRE}$, where $|\rho\rangle_{MRE} = U_{S\to E}\,|\sigma\rangle_{SMR}$ and $U_{S\to E}$ is an isometry from $S$ to $E$.

Then the bound on the work cost, including a smoothing parameter $\epsilon$, is

$$\begin{aligned} W^{\epsilon} &\geqslant H^{\epsilon}_{\max}(E\,|\,SM)_{\rho}\cdot kT\ln(2) \\ &= H^{\epsilon}_{\max}(E\,|\,M)_{\rho}\cdot kT\ln(2) \\ &= H^{\epsilon}_{\max}(S\,|\,M)_{\sigma}\cdot kT\ln(2)\ , \end{aligned} \tag{6.18}$$

where the first equality follows because $\rho$ is pure on $S$ and the second by reversing the isometry $U$. We can immediately conclude that, within our framework, any process that performs this erasure has to cost at least $kT\ln(2)\,H^{\epsilon}_{\max}(S\,|\,M)_{\sigma}$ work. Thus, the process proposed by del Rio *et al.* is optimal up to logarithmic factors in $\epsilon$. Note that if we take the memory $M$ to be trivial i.e. a pure state, then we are in the standard scenario of Landauer erasure on a single system, and we have $W^{\epsilon}\geqslant H^{\epsilon}_{\max}(S)$ which is achievable, recovering the result of Dahlsten *et al.* (2011).

## 6.8 Achievability with thermal operations

The minimal work cost given in Eq. (6.4) is clearly achievable with Gibbs-preserving maps, as seen in Proposition 6.1. However, we have seen that it is not clear in general how to explicitly implement a Gibbs-preserving map without expending any work. One may ask, therefore, whether it is possible to achieve the same work cost given by (6.4) with an operational framework such as thermal operations.

In fact, the erasure procedure by del Rio *et al.* (2011) can be used to show that for any arbitrary logical process, the minimal amount of work our result associates to it can be in principle achieved to good approximation using an operational framework. The result of del Rio *et al.* is not explicitly written in terms of thermal operations, however it only relies upon few elementary operations: performing a global random unitary, work extraction from a system in a pure state to a fully mixed state, and resetting a system from a completely mixed state to a pure state. These elementary operations can be performed within any reasonable operational framework, including thermal operations with any reasonable form of a work storage system.

Given a logical process $\mathcal{E}$ and an input state $\sigma_X$, we may calculate its Stinespring dilation, which is an isometry $V_{X\to X'E}$ (see Section 3.1.8). Supposing $X \simeq X'$, we may in turn write this as a unitary $U_{XE\to X'E}$ where the $E$ system starts in a pure state $|0\rangle_E$, such that $V_{X\to X'E} = U_{XE\to X'E} \cdot (|0\rangle_E \otimes \mathbb{1}_X)$. For pure convenience, we present the case $X \simeq X'$ in the following. If $X \not\simeq X'$, the same argument applies, although extra care has to be taken in the construction of the unitary: the dimension of the additional system in the unitary may have to be different before and after the process.

Consider now an ancillary system $A_E$ of the same dimension as $E$, initialized in a pure state $|0\rangle_{A_E}$. One can then carry out the unitary $U_{XA_E\to X'A'_E}$ on $X$ and $A_E$, recalling that by construction

$$\mathrm{tr}_{A_E}\Big\{U_{XA_E\to X'A'_E}\big(\sigma_X \otimes |0\rangle\langle 0|_{A_E}\big)U^\dagger_{XA_E\leftarrow X'A'_E}\Big\} = \mathcal{E}_{X\to X'}(\sigma_X)\ . \qquad (6.19)$$

In effect, $A'_E$ impersonates the abstract system $E$ while we perform a unitary corresponding to the Stinespring dilation of $\mathcal{E}$ (see Figure 6.5). This unitary operation can be implemented at no work cost because it is reversible. The aforementioned optimal erasure procedure can then be used to restore the ancilla $A'_E$ to its original pure state, using the output system $X'$ as the quantum memory. The work cost of doing so is approximately $kT\ln(2)\cdot H^\epsilon_{\max}(A'_E\,|\,X')$. As $A'_E$ corresponds to $E$, this matches our bound (6.7) and therefore proves its tightness.

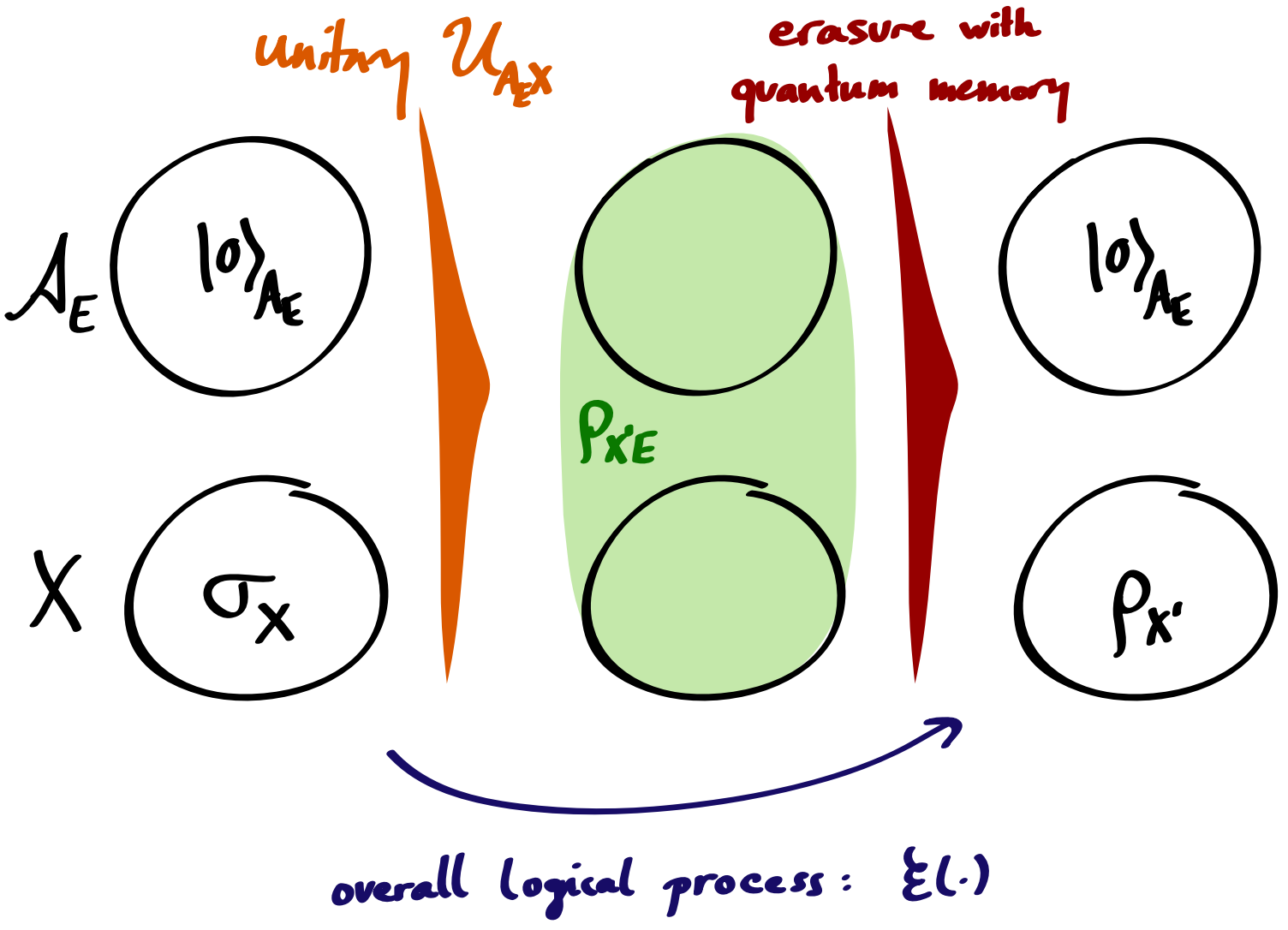

Figure 6.5: Implementation of any logical process $\mathcal{E}$ with thermal operations, at work cost matching our bound. First, we write $\mathcal{E}$ via its Stinespring dilation as a unitary operation $U_{XA_E \to X'A'_E}$ over an additional system $E$ which starts in a pure state $|0\rangle$. We then bring in an ancilla $A_E$ which will play the role of $E$, and initialize it in the pure state $|0\rangle_{A_E}$. Then, we carry out the unitary $U_{XA_E \to X'A'_E}$, which requires no work, and which successfully implements the logical process $\mathcal{E}$ on $X \to X'$. The ancilla $A'_E$ can be restored to its initial state by using the erasure procedure by del Rio *et al.* (2011), by using $X'$ as the quantum memory. The work required to do so matches our bound.

## 6.9 The work requirement of a quantum measurement

The problem of determining the amount of work needed to carry out a quantum measurement has been the subject of much literature (Sagawa and Ueda, 2009; Buscemi *et al.*, 2008; Jacobs, 2012), especially in the context of Maxwell's demon (Bennett, 1982, 2003; Earman and Norton, 1999; Leff and Rex, 2010). A quantum measurement is a logical process (depicted in Figure 6.6a) acting on a system $X$ to be measured and a classical register $C$ initially set to a pure state, and outputting systems $C'$ and $X'$, with $C'$ containing the measurement result and $X'$ the quantum post-measurement state.

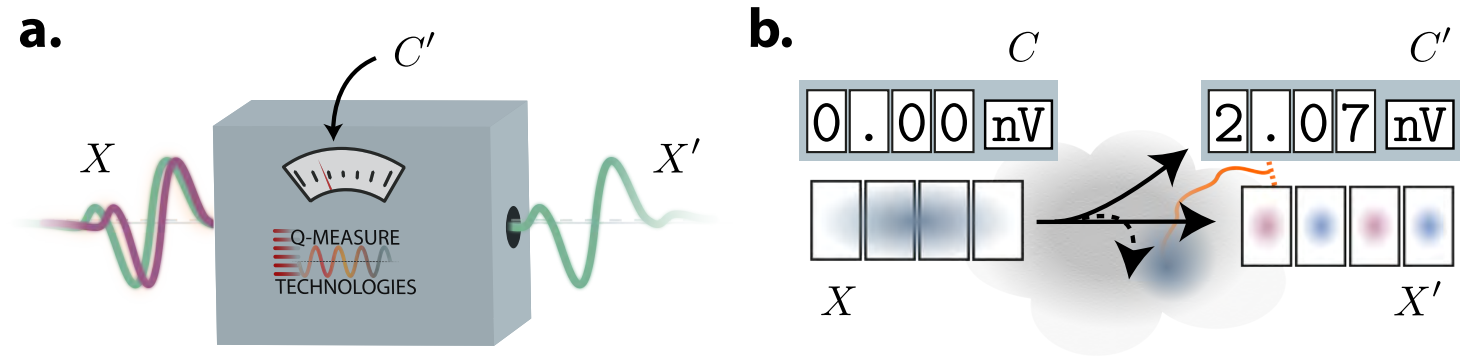

Figure 6.6: The work cost of quantum measurements. **a.** A quantum measurement may be thought of as a device which produces a post-measurement state $X'$ and a classical reading $C'$ from an input state $X$. **b.** The corresponding logical process maps the input system $X$ and classical register $C$ to a classical outcome on the output register $C'$ and a post-measurement state on $X'$. The initial register $C$ is prepared in a pure state. Our main result implies that the measurement costs no work in principle.

Let us first consider a projective measurement. The logical process corresponding to the measurement described by a complete set of projectors $\{P_i\}_i$ takes the form

$$\mathcal{E}_{CX\to C'X'}(\sigma_{CX}) = \sum_i |i\rangle\langle i|_{C'} \otimes (P_i\,\sigma_X P_i)\ . \tag{6.20}$$

Our bound (6.4) for this map is at most zero. Indeed, we have

$$\left\| \mathcal{E}\big(|0\rangle\langle 0|_C \otimes \Pi_X\big)\right\|_\infty \leqslant 1\,, \tag{6.21}$$

implying that the measurement can be carried out in principle at no work cost. This fact was already stated by Bennett (1982), and refutes earlier conjectures based on detailed analyses of particular measurement schemes (Brillouin, 1951; Earman and Norton, 1998, 1999).

It is to be emphasized that we study the minimum required work cost of the device which produces the necessary interaction and evolution for the measurement, whose logical process corresponds to (6.20). No statement is made regarding how the state collapses, or concerning any philosophical assumptions or implications of the collapse.

A related question is the work cost of erasing the information contained in the register $C'$ after the measurement. Doing so would allow us to construct a cycle. The cost of this erasure can be reduced using the post-measurement state as a quantum memory, by employing the procedure presented above, to $kT \ln 2 \cdot H_{\max}^{\epsilon}(C' \,|\, X')$. But because $C'$ and $X'$ may only be classically correlated, no work may be extracted in this way (del Rio *et al.*,

2011). In some cases this work cost may be zero, as we will see below. This might seem to save Maxwell's demon from Bennett's information-theoretic exorcism (Bennett, 1982, 2003) which argues that the demon must pay work to reset its memory (see, e.g., Section 2.4). However the key point is that the demon can't use the post-measurement state to both extract work and to reset its internal memory register.

In the following, we analyse more specifically the minimal work cost of a general quantum measurement and how one can reset the measurement device back to a pure state.

### 6.9.1 The measurement process and its work cost

Suppose that on the system $S$, in the state $\sigma$, we perform a measurement described by a POVM $\{Q_k\}$. Each measurement outcome, labeled by the index $k$, occurs with probability $\operatorname{tr}(Q_k\sigma)$. The completely positive map associated with this measurement is,

$$\mathcal{E}_{S\to S'C}(\sigma) = \sum_k |k\rangle\langle k|_C \otimes \mathcal{E}^{(k)}_{S\to S'}(\sigma) , \tag{6.22}$$

where $C$ is a classical register containing the outcome of the measurement ($C$ is initially in a pure state), and the $\mathcal{E}^{(k)}_{S\to S'}(\cdot) = \sum_i E_i^{(k)}\,(\cdot)\,E_i^{(k)\,\dagger}$ are trace-decreasing completely positive maps that map $\sigma$ to its post-measurement state for the outcome $k$, each of which occurs with probability $\operatorname{tr}(Q_k\sigma) = \operatorname{tr}\mathcal{E}^{(k)}_{S\to S'}(\sigma)$. The Kraus operators $E_i^{(k)}$ are related to the POVM elements $Q_k$ by $\sum_i E_i^{(k)\,\dagger} E_i^{(k)} = Q_k$. Equation (6.22) simply expresses that the output state of the measurement is a mixture of the possible post-measurement states corresponding to the different outcomes $k$. We emphasize that the register $C$ must start off in a pure state; if this is not the case (as in a purity resource framework, for example), it should be initialized first, causing some work cost, before performing the map $\mathcal{E}$ (Jacobs, 2012).

We first need to calculate the Stinespring dilation of the process $\mathcal{E}$, which is given by $\mathcal{E}_{S\to CS'}(\cdot) = \operatorname{tr}_E\big(V_{S\to ECS'}(\cdot)\,V^\dagger\big)$, where the isometry $V_{S\to ECS'}$ can be read off the operator-sum representation of (6.22), that is

$$\mathcal{E}_{S\to CS'} = \sum_{k,i} \left(|k\rangle_C \otimes E_i^{(k)}\right) \sigma \left(\langle k|_C \otimes E_i^{(k)\,\dagger}\right) , \tag{6.23}$$

as $V = \sum_{k,i} |k,i\rangle_E \otimes |k\rangle_C \otimes E_i^{(k)}$. This yields the post-measurement state

including $E$

$$\rho_{ECS'} = V_{S\to ECS'}\ \sigma_S\ V^\dagger\ . \tag{6.24}$$

For convenience, let $R$ be a purifying system for $\sigma_S$, i.e. let $|\sigma\rangle_{SR}$ such that $\mathrm{tr}_R|\sigma\rangle\langle\sigma|_{SR} = \sigma_S$. This allows us to write a full, pure, post-measurement state $|\rho\rangle_{ECS'R}$ as

$$|\rho\rangle_{ECS'R} = \sum_{k\,i} |k,i\rangle_E \otimes |k\rangle_C \otimes \left(E_i^{(k)}|\sigma\rangle_{SR}\right)\ . \tag{6.25}$$

Our main result asserts that the minimal work cost of the measurement described by the mapping (6.22) is simply given by the bound

$$W^\varepsilon \geqslant kT\ln 2\cdot H_{\max}^\epsilon(E\,|\,CS')\ , \tag{6.26}$$

where the entropy measure is evaluated for the state $\rho_{ECS'R}$ given by (6.25), and where the bound can be achieved exactly with Gibbs-preserving maps, or saturated approximately with thermal operations as explained in Section 6.8. In the remainder of this section, all entropy measures are implicitly evaluated on this state, unless indicated otherwise. It is also implied that all work costs are smoothed with a small but finite $\epsilon$ parameter. For convenience, we denote the value of the right-hand side of (6.26) by

$$W_{\text{meas}} = kT\ln 2\cdot H_{\max}^\epsilon(E\,|\,CS')\ . \tag{6.27}$$

As we will see from some simple examples, the quantity (6.27) in its most general form may take any value, from a work cost to a work yield.

Let us consider an important class of measurements: those for which the collapse operators $\mathcal{E}_k$ don't themselves need work. In general, we know that those processes that don't need work are subunital, i.e. they satisfy $\mathcal{E}^{(k)}(\mathbb{1}) \leqslant \mathbb{1}$. This is for example the case for projective measurements, or more generally if the $\mathcal{E}^{(k)}$'s only have a single Kraus operator. (We also note that any general measurement in the form (6.22) can be written as a combination of a measurement with collapse superoperators that have each a single Kraus operator, followed by a partial erasure on the memory register $C$.) Such measurements require no work in principle.

***Proposition 6.2.*** *Let $\mathcal{E}_{S\to CS'}$ be a measurement process of the form (6.22), and assume that for all $k$, $\mathcal{E}^{(k)}(\mathbb{1}) \leqslant \mathbb{1}$. Then $W_{meas}$ as defined by (6.27), satisfies $W_{meas} \leqslant 0$.*

*Proof.* Instead of proving that the entropic quantity (6.27) is negative, we will show that the full measurement process $\mathcal{E}$ itself is a subunital superoperator, which we know from our framework costs no work. This is straightforward to see:

$$\mathcal{E}_{S\to CS'}(\mathbb{1}_S) = \sum_k |k\rangle\langle k|_C \otimes \mathcal{E}^{(k)}_{S\to S'}(\mathbb{1}_S) \leqslant \sum_k |k\rangle\langle k|_C \otimes \mathbb{1}_{S'} \leqslant \mathbb{1}_{CS'} . \qquad \blacksquare$$

### 6.9.2 Resetting the memory register containing the measurement outcome

Let us now consider the task of resetting $C$ to a pure state, after having performed the measurement process above. This resetting can obviously be performed directly, with a cost given by Landauer's principle as $H^{\epsilon}_{\max}(C)_{\rho}$, which in turn depends on the number of possible outcomes the measurement had (if $\epsilon > 0$, we would only consider the measurements that are not extremely unlikely). This procedure, however, is not optimal if we are allowed access for example the information contained in the post-measurement state on $S'$. Indeed, in the latter case, we may use the system $S'$ as a memory as discussed in Section 6.7, and the optimal work cost is then

$$W_{\text{reset } C|S'} = kT\ln 2 \cdot H^{\epsilon}_{\max}(C\,|\,S')_{\rho} . \tag{6.28}$$

This work cost is always positive, or at best, zero, because $\rho_{CS'}$ is classical-quantum (c-q).

We will see that this work cost may be both less and larger than $W_{\text{meas}}$ with some examples. Of course, this does not constitute a violation of the second law, as we will discuss.

Alternatively, we could imagine a scenario where we have kept a purification of the input state $|\sigma\rangle_{SR}$ on an ancilla system $R$, in order to "remember" the state of the initial system $S$. We may then of course use this system also to reduce the work cost of erasing $C$; this work cost is then

$$W_{\text{reset } C|R} = kT\ln 2 \cdot H^{\epsilon}_{\max}(C\,|\,R) . \tag{6.29}$$

It turns out that, if the collapse operators $\mathcal{E}^{(k)}$ all have a single Kraus operator, this work cost is always greater than the work yield of performing the measurement $(-W_{\text{meas}})$, and that the difference between both is precisely the difference between the max and min-entropy of the system $C$ conditioned on $R$.

***Proposition 6.3.*** *Let $\mathcal{E}_{S\to CS'}$ be a measurement process of the form (6.22),*

*with for all k,* $\mathcal{E}^{(k)}(\cdot) = E_k(\cdot)E_k^\dagger$. *Let* $\rho_{ECS'R}$ *be defined as in* (6.25). *Then*

$$W_{\text{meas}} + W_{\text{reset } C|R} = kT\ln 2 \cdot \left[-H_{\min}^{\epsilon}(C\,|\,R) + H_{\max}^{\epsilon}(C\,|\,R)\right]. \tag{6.30}$$

*Additionally, if* $\epsilon$ *is not too large (such that* $H_{\max}^{\epsilon}(\cdot) \geqslant H_{\min}^{\epsilon}(\cdot)$ *(*Vitanov et al., *2013*)), this expression is always positive,*

$$W_{\text{meas}} + W_{\text{reset } C|R} \geqslant 0\,. \tag{6.31}$$

*Proof.* First notice that the state $\rho_{ECS'R}$ takes the form

$$|\rho\rangle_{ECS'R} = \sum_k |k\rangle_E |k\rangle_C \left(E_k|\sigma\rangle_{SR}\right)\,,$$

and in particular, $\rho$ is invariant under interchange of $E$ and $C$ systems. Then, using duality of the smooth entropies (Section 3.2.3), we have $H_{\max}^{\epsilon}(E\,|\,CS') = -H_{\min}^{\epsilon}(E\,|\,R) = -H_{\min}^{\epsilon}(C\,|\,R)$ and thus $W_{\text{meas}} = -kT\ln 2 \cdot H_{\min}^{\epsilon}(C\,|\,R)$. Then recall that $W_{\text{reset } C|R} = kT\ln 2 \cdot H_{\max}^{\epsilon}(C\,|\,R)$ and that the max-entropy is larger than the min-entropy for small $\epsilon$. ■

The final (pure) state on $E$, $C$, $S'$ and $R$, which is the output of the Stinespring isometry $V$ applied on the $S$ system of $|\sigma\rangle_{SR}$, is still given by the expression (6.25).

### 6.9.3 Some examples of measurement processes

Let us now focus on some examples of measurement processes, which are all special cases of (6.22).

*Measurement in the computational basis* $|0\rangle$, $|1\rangle$ *of a single qubit in a maximally mixed state* $\mathbb{1}_2/2$.

The measurement process then simply yields the output state

$$\rho_{CS'} = \frac{1}{2}\,|0\rangle\langle 0|_C \otimes |0\rangle\langle 0|_{S'} + \frac{1}{2}\,|1\rangle\langle 1|_C \otimes |1\rangle\langle 1|_{S'}\,. \tag{6.32}$$

The input state on $S$ is purified by a fully entangled state $|\phi\rangle_{SR}$ on $R$. The system $E$ also purifies the measurement process, and as given by (6.25),

$$\begin{aligned} |\rho\rangle_{ECS'R} &= \sum_k |k\rangle_E |k\rangle_C \left(|k\rangle\langle k|_S |\phi\rangle_{SR}\right) \\ &= \frac{1}{\sqrt{2}}\left[|0\rangle_C|0\rangle_E|00\rangle_{SR} + |1\rangle_C|1\rangle_E|11\rangle_{SR}\right]. \end{aligned} \tag{6.33}$$

It is then evident that

$$W_{\text{meas}} = kT\ln 2 \cdot H^{\epsilon}_{\max}(E\,|\,CS') = 0\;; \tag{6.34-a}$$
$$W_{\text{reset }C|S'} = kT\ln 2 \cdot H^{\epsilon}_{\max}(C\,|\,S') = 0\;; \tag{6.34-b}$$
$$W_{\text{reset }C|R} = kT\ln 2 \cdot H^{\epsilon}_{\max}(C\,|\,R) = 0\;, \tag{6.34-c}$$

as the corresponding reduced states are all classically correlated.

*Measurement of a trivial noisy POVM.*

Consider a POVM in the extreme case where the state is left untouched, but a random outcome is generated according to a distribution $\{p_k\}$. The POVM effects are simply $Q_k = p_k\mathbb{1}$ and the post-measurement operators $\mathcal{E}^{(k)}(\sigma) = p_k\sigma$ are simply the identity superoperator weighted by the probability $p_k$.

Intuitively, this should be no different than rolling a die, or more generally, generating a random outcome with a specific distribution: An implementation of such a logical process can yield work in principle. Indeed, based on the explicit expression of the final state

$$|\rho\rangle_{ECS'R} = \Big(\sum_k \sqrt{p_k}\,|k\rangle_E|k\rangle_C\Big) \otimes |\sigma\rangle_{S'R}\;, \tag{6.35}$$

we may express of the work cost of the measurement using some basic properties of the smooth entropies (see Section 3.2.3). Using the duality of the min- and max-entropies,

$$\begin{aligned} W_{\text{meas}}/(kT\ln 2) &= H^{\epsilon}_{\max}(E\,|\,CS') = -H^{\epsilon}_{\min}(E\,|\,R) \\ &= -H^{\epsilon}_{\min}(E) = -H^{\epsilon}_{\min}(C) \leqslant \log\|\rho_C\|_\infty \leqslant 0\;, \end{aligned} \tag{6.36}$$

because $R$ is not correlated to $E$, and $\rho$ is invariant under exchange of $E$ and $C$ (both these statements can be seen in (6.35)), and by definition the min-entropy of $C$ is given by $H_{\min}(C) = -\log\|\rho_C\|_\infty$. Also, smoothing the min-entropy can only increase the quantity by definition (see Section 3.2.3).

One can also calculate, because $C$ is uncorrelated to both $S'$ and $R$,

$$\begin{aligned} W_{\text{reset }C|S'} &= kT\ln 2 \cdot H^{\epsilon}_{\max}(C\,|\,S') \\ &= kT\ln 2 \cdot H^{\epsilon}_{\max}(C)\;; \end{aligned} \tag{6.37-a}$$
$$\begin{aligned} W_{\text{reset }C|R} &= kT\ln 2 \cdot H^{\epsilon}_{\max}(C\,|\,R) \\ &= kT\ln 2 \cdot H^{\epsilon}_{\max}(C)\;. \end{aligned} \tag{6.37-b}$$

This means that the work we need to invest to reset $C$ is always larger than what we gain from generating the random outcome. In fact, the gap is precisely the difference between the max- and the min-entropy, which is the same kind of irreversibility that is observed between the single-shot entanglement distillation and formation cost between two parties (Hayden *et al.*, 2006).

*Projective measurement of a pure superposition state.*

One may think that intuitively, for the measurement to yield work, the POVM must be noisy. Surprisingly enough, this is not the case. Even projective measurements can yield work for specific input states. For example, consider the state $|\sigma\rangle_S = |+\rangle := \frac{1}{\sqrt{2}}\left(|0\rangle + |1\rangle\right)$. Here $R$ is a trivial system since $\sigma$ is already pure. Now consider the usual projective measurement that measures $\sigma$ in the computational basis $|0\rangle$, $|1\rangle$. The final state is

$$|\rho\rangle_{ECS'} = \frac{1}{\sqrt{2}}\left(|0\rangle_E|0\rangle_C|0\rangle_{S'} + |1\rangle_E|1\rangle_C|1\rangle_{S'}\right) . \tag{6.38}$$

We then evidently have

$$W_{\mathrm{meas}} = kT\ln 2\cdot H^{\epsilon}_{\max}(E\,|\,CS') = -kT\ln 2\,; \tag{6.39-a}$$

$$W_{\mathrm{reset}\ C|S'} = kT\ln 2\cdot H^{\epsilon}_{\max}(C\,|\,S') = 0\,; \tag{6.39-b}$$

$$\begin{aligned} W_{\mathrm{reset}\ C|R} &= kT\ln 2\cdot H^{\epsilon}_{\max}(C\,|\,R) \\ &= kT\ln 2\cdot H^{\epsilon}_{\max}(C) = kT\ln 2\,. \end{aligned} \tag{6.39-c}$$

We conclude that it is possible to extract one bit of work while performing the measurement, and that resetting the memory register can be done at no work cost using $S'$ but needs one bit of work if we use the (trivial) reference system $R$.

Note that resetting the measurement register $C$ using $S'$ costs no work. This is not in violation of the second law of thermodynamics: we have not returned the post-measurement state back to the initial state, but rather we have consumed its purity.

*Measurement with erasure collapse operators.*

It was noted above that if the collapse operators $\mathcal{E}^{(k)}$ were themselves maps that cost work, e.g. erasure channels, then the measurement would also possibly cost work. It is sufficient to consider the following extreme example:

take a single-outcome measurement, i.e. a trivial measurement, with a the single collapse operator $\mathcal{E}_{k=0}(\cdot) = \operatorname{tr}(\cdot)\,|0\rangle\langle 0|$ being an erasure channel. Obviously this operation has to cost work: performing operation $\mathcal{E}_{S\to CS'}$ is exactly the same as performing just the erasure $\mathcal{E}_{k=0}$, which costs work according to our main result (which is of course also in line with Landauer's principle).

*Information gain of a measurement.*

Existing literature (Winter, 2004; Wilde *et al.*, 2012; Buscemi *et al.*, 2008; Berta *et al.*, 2014) has studied and identified the amount of information that a quantum measurement provides about a system being measured. With the notation above, the information gain in the asymptotic, i.i.d. regime is defined in (Buscemi *et al.*, 2008) as

$$\iota(\sigma_S, \mathcal{E}) = I(R:C)_\rho := H(R) - H(R|C)\ . \tag{6.40}$$

In our framework, information contained in a quantum system is represented by how much work we need in order to erase that system. Bearing this in mind, the natural way of defining the amount of information gained about the system using the measurement in the i.i.d. regime is then the difference in work costs of erasing $S$ *before* and *after* the measurement. Since $S$ was consumed by the measurement, this statement doesn't fully make sense, so we will rather consider erasing the system $R$ instead, which is a purification of $S$. This is precisely the quantity described by (6.40).

## 6.10 More properties and examples

### 6.10.1 State transformation while decoupling from the reference system

Let's return to another special case that we can derive as a corollary from our main result. Consider the process that erases its input and prepares the required output independently. This would occur if we required the output state to be completely uncorrelated to the reference system $R$: $\rho_{X'R} = \rho_{X'} \otimes \rho_R$. This corresponds to a replacement map. Any third party $R$ that would have been correlated to the input is now completely uncorrelated to the output.

Again, we may simply apply our main result with the additional condition that $\rho_{X'R} = \rho_{X'} \otimes \rho_R$. In this case, the purification of $\rho_{X'R}$, $\rho_{X'RE}$, takes a special form due to the tensor product structure, with the $E$ system split into two $E_R$ and $E_{X'}$ systems ($E = E_R \otimes E_{X'}$),

$$|\rho\rangle_{X'RE} = |\psi\rangle_{X'E_{X'}} \otimes |\phi\rangle_{RE_R} , \tag{6.41}$$

where $|\psi\rangle_{X'E_{X'}}$ and $|\phi\rangle_{RE_R}$ are purifications of $\rho_{X'}$ and $\rho_R$, respectively.

The (smooth) lower bound on the minimal work cost $W$, given by our main result, then reads

$$\begin{aligned} W &\geqslant kT \ln 2 \cdot H^{\epsilon}_{\max}(E \,|\, X')_{\rho} \\ &= kT \ln 2 \cdot H^{\epsilon}_{\max}(E_R)_{|\phi\rangle} + kT \ln 2 \cdot H^{\epsilon}_{\max}(E_{X'} \,|\, X')_{|\psi\rangle} . \end{aligned} \tag{6.42}$$

Now, the spectrum of $\rho_{E_R}$ is exactly the same as the spectrum of $\rho_R$ by the Schmidt decomposition of $|\phi\rangle$. This in turn has the same spectrum as $\sigma_X$ also by the Schmidt decomposition of $\sigma_{XR}$ and because $\rho_R = \sigma_R$. It follows that $H^{\epsilon}_{\max}(E_R)_{\rho} = H^{\epsilon}_{\max}(X)_{\sigma}$. Also, by duality of the min- and max-entropies, we have $H^{\epsilon}_{\max}(E_{X'} \,|\, X')_{|\psi\rangle} = -H^{\epsilon}_{\min}(E_{X'})_{\rho} = -H^{\epsilon}_{\min}(X')_{\rho}$. In consequence,

$$W \geqslant kT \ln 2 \cdot \left[ H^{\epsilon}_{\max}(X)_{\sigma} - H^{\epsilon}_{\min}(X)_{\rho} \right] . \tag{6.43}$$

That is, to transform a state $\sigma$ to $\rho$ while completely decorrelating $\rho$ from the input, then one has to erase $\sigma$ to a pure state (at cost $H^{\epsilon}_{\max}(X)_{\sigma}$), and then prepare $\rho$ (extracting work $H^{\epsilon}_{\min}(X')_{\rho}$).

### 6.10.2 Coherent preparation of a state on a system with a memory; "reverse" of a logical process

One may also wonder which process is the reverse process of erasure with a quantum memory. Specifically, starting off with a pure system $S$ and some state $\rho_M$ on a memory $M$, one might ask how much work is needed to prepare a given bipartite state $\sigma_{SM}$ on these systems.

The process as such is not clearly defined, as we have not specified which correlations between the initial and final state on $M$ are to be preserved, or, equivalently, which completely positive map $\mathcal{E}$ is to be applied for this preparation.

Let us first study the erasure mapping (6.17) a bit more closely. The output state of the erasure including the reference system $R$ is given by the state $\rho_{X'R} = |0\rangle\langle 0|_S \otimes \rho_{MR}$, where $\mathscr{H}_{X'} = \mathscr{H}_S \otimes \mathscr{H}_M$ is the total output system. As mentioned earlier, the joint state on $X'$ and $R$ may be interpreted as the *process matrix* of the operation $\mathcal{E}$ on $\sigma$: it can be thought of as a joint probability distribution giving the probability that we had $|k\rangle$ at the input and got $|k'\rangle$ at the output; also, $\rho_{X'}$ is the output state and $\rho_R$ is the input state. This consideration gives us a natural way of *reversing* any process: a natural "reverse" process to the process $\rho_{X'R}$ is simply given by swapping the two systems, i.e. considering $R$ as the output and $X'$ as a purification of the input.[3] Let us return to the case of the erasure. First, consider the purification of $\sigma_{SM}$ into a system $\mathscr{H}_R = \mathscr{H}_{R_S} \otimes \mathscr{H}_{R_M}$ explicitly as

$$|\sigma\rangle_{SMR_SR_M} = \sigma_{SM}^{1/2}|\Phi\rangle_{SM|R_SR_M} , \tag{6.44}$$

where $|\Phi\rangle_{SM|R_SR_M} = \sum_{kl}|k\rangle_S|k\rangle_{R_S}|l\rangle_M|l\rangle_{R_M}$. Now applying the erasure process on $SM$ gives us:

$$\rho_{SMR_SR_M} = \mathcal{E}_{SM\to SM}^{(\text{erasure})}\left(\sigma_{SMR_SR_M}\right) = |0\rangle\langle 0|_S \otimes \operatorname{tr}_S\left(\sigma_{SMR_SR_M}\right) . \tag{6.45}$$

It is thus natural, for the preparation scenario, to consider the process matrix

$$\rho_{SMR_SR_M}^{(\text{prep})} = |0\rangle\langle 0|_{R_S} \otimes \operatorname{tr}_{R_S}\left(\sigma_{SMR_SR_M}\right) . \tag{6.46}$$

(Recall that the input state is pure on $S$.)

This is obviously purified by a system $E$ which contains the traced out information, i.e. given an isometry $U_{R_S\to E}$,

$$|\rho^{(\text{prep})}\rangle_{SMR_SR_ME} = |0\rangle_{R_S} \otimes \left(U_{R_S\to E}\,|\sigma\rangle_{SMR_SR_M}\right) . \tag{6.47}$$

If we then calculate the minimal work cost of performing this process according to our main result, we obtain

$$\begin{aligned} W^{(\text{prep})} &= kT\ln 2 \cdot H_{\max}^{\epsilon}(E\,|\,SM)_{\rho^{(\text{prep})}} \\ &= kT\ln 2 \cdot H_{\max}^{\epsilon}(R_S\,|\,SM)_{\sigma} \\ &= -kT\ln 2 \cdot H_{\min}^{\epsilon}(R_S\,|\,R_M)_{\sigma} \\ &= -kT\ln 2 \cdot H_{\min}^{\epsilon}(S\,|\,M)_{\sigma} . \end{aligned} \tag{6.48}$$

(We have used the fact that $\sigma$ and $\rho$ are related by an isometry between $R_S$ and $E$, as well as the duality between min- and max-entropies.)

We notice that $kT\ln 2\cdot H_{\min}^{\epsilon}(S\,|\,M)_{\sigma}$ work can be extracted in the reverse process of the original erasure process, which required $kT\ln 2\cdot H_{\max}^{\epsilon}(S\,|\,M)_{\sigma}$ work. These values can be arbitrarily different; this gap is expected as we require both processes to succeed with high probability. We find that the gap is exactly the difference between the min- and the max-entropy, similarly to the single-shot irreversibility between distillation rate and formation rate of entangled pairs with LOCC operations (Bennett *et al.*, 1996a,b,c; Vidal and Cirac, 2001).

[3] Not surprisingly at all, the more general definition of a *reverse process* in Section 5.4 in terms of a recovery map coincides with the present one for trivial $\Gamma$ operators as considered in this chapter.

### 6.10.3 Example: erasing part of a W state; again, the importance of correlations between the input and the output

Consider the W state (Zeilinger *et al.*, 1991; Dür *et al.*, 2000) on a system $S$, a memory $M$ and a reference system $R$, given by

$$|W\rangle_{SMR} = \frac{1}{\sqrt{3}} \left[|001\rangle + |010\rangle + |100\rangle\right]_{SMR} . \tag{6.49}$$

The reduced states on $SM$ and $M$ are respectively given by $\sigma_{SM} = \frac{1}{3}|00\rangle\langle 00| + \frac{2}{3}|\Psi^+\rangle\langle\Psi^+|$ and $\sigma_M = \frac{2}{3}|0\rangle\langle 0| + \frac{1}{3}|1\rangle\langle 1|$, where $|\Psi^+\rangle = \frac{1}{\sqrt{2}}\left(|01\rangle + |10\rangle\right)$. By symmetry of the W state, the reduced state on any two or one qubit(s) have the same form.

By actions on $S$ and $M$, we would like to erase $S$, leading to the final state on $S$ and $M$ given by $\rho_{SM} = |0\rangle\langle 0| \otimes \sigma_M$. Let us consider two processes that achieve this goal: the first one will preserve correlations with $R$ but will cost work, the second will not cost work but will modify those correlations.

We may directly apply the special case above concerning the erasure of a system conditioned on a memory: the fundamental work cost of such an erasure, if one preserves correlations with a reference system $R$, is given by $H_{\max,0}(S\,|\,M)_\sigma$. In this case we have $H_{\max,0}(S\,|\,M)_\sigma = \log\frac{2}{3} \approx 0.59$ (which we calculate below) and thus this process must cost at least this amount of work. Because of the small system size, we may not assert the achievability of this erasure at this work cost (the error terms discussed in Section 6.8 become overwhelming). However we can safely exclude the possibility of performing this operation at no work cost.

Observe now that both $\sigma_{SM}$ and $\sigma_M$ have the same spectrum $\{2/3, 1/3\}$. This means that there exists a unitary $U$ that performs the erasure simply as $|0\rangle\langle 0| \otimes \sigma_M = U\sigma_{SM}U^\dagger$, and this unitary process by definition does not cost any work. Note though that the correlations with $R$ are not preserved. Indeed, the unitary sends $|00\rangle$ to $|01\rangle$ and $|\Psi^+\rangle$ to $|00\rangle$, so one explicitly calculates that the state after the process is given by $U|\sigma\rangle_{SMR} = \frac{1}{\sqrt{3}}\left[|011\rangle + \sqrt{2}|000\rangle\right] = |0\rangle \otimes \frac{1}{\sqrt{3}}\left[|11\rangle + \sqrt{2}|00\rangle\right]$. We notice that the reduced state on $M$ and $R$ is now

pure and differs from initial one, given by $\sigma_{MR} = \frac{1}{3}|00\rangle\langle 00| + \frac{2}{3}|\Psi^+\rangle\langle\Psi^+|$.

As before, this is an example where one can transform the input state into the output state at no work cost a priori, but if correlations are to be preserved between the memory and a reference system or, equivalently, if the exact erasure process (6.17) is to be performed, then a physical implementation of this operation would require work.

It remains to calculate $H_{\max,0}(S\,|\,M)_\sigma$. Written out explicitly in the basis $\{|0\rangle, |1\rangle\}$, the state $\sigma_{SM}$ and the projector $\Pi_{SM}$ on its support take the form

$$\sigma_{SM} = \begin{pmatrix} ⅓ & & & \\ & ⅓ & ⅓ & \\ & ⅓ & ⅓ & \\ & & & 0 \end{pmatrix} ; \quad \Pi_{SM} = \begin{pmatrix} 1 & & & \\ & ½ & ½ & \\ & ½ & ½ & \\ & & & 0 \end{pmatrix} .$$

(Missing entries are zero.) We then see that

$$\mathrm{tr}_S\, \Pi_{SM} = \begin{pmatrix} 3/2 & \\ & 1/2 \end{pmatrix} ,$$

such that

$$H_{\max,0}(S\,|\,M)_\sigma = \log \left\| \mathrm{tr}_S\, \Pi_{SM} \right\|_\infty = \log \frac{3}{2} . \qquad (6.50)$$

# 7

# A Coherent Relative Entropy Measure

In this chapter, we generalize the results of Chapter 6 to systems described by arbitrary Hamiltonians. The results here are formulated again in the more abstract language of Chapter 5, to highlight their information-theoretic nature and generality.

We consider a system $X$ described by a hidden information $\Gamma_X$. The system $X$ is prepared in an initial state $\sigma_X$, purified as $|\sigma\rangle_{XR}$, on which we implement a logical process $\mathcal{E}_{X\to X'}$ resulting in a process matrix $\rho_{X'R}$ (Section 3.1.9). The output system $X'$ is described by a hidden information operator $\Gamma_{X'}$, which may differ from $\Gamma_X$. The question we ask is, what is the maximal compression yield of the best process which implements the process matrix $\rho_{X'R}$?

The result is a new quantity, which we term the *coherent relative entropy*, and which we denote by $\bar{D}_{R\to X'}(\rho_{X'R} \,\|\, \Gamma_R, \Gamma_{X'})$. This measure depends on the process matrix $\rho_{X'R}$, as well as the hidden information operators $\Gamma_X$, $\Gamma_{X'}$. The coherent relative entropy generalizes the max-entropy, which was obtained in the previous chapter in the special case of finite-size thermodynamics with a trivial Hamiltonian. At the same time, it accurately describes the work cost of erasure and work yield of formation of quantum states in finite-size thermodynamics, known to be given by the min- and max-relative entropies (Åberg, 2013; Horodecki and Oppenheim, 2013a; Brandão *et al.*, 2015). Its interpretation as an information measure is further justified by the fact that it satisfies some essential properties usually enjoyed by such quantities, such as a chain rule and the asymptotic equipartition property.

In order to achieve physically relevant statements, it is important to smooth this quantity with a parameter $\epsilon$ (see Section 2.6.2). The smoothing is achieved as in the smooth entropy framework by optimizing the quantity over states which are close to $\rho_{X'R}$ in a suitable distance measure.

The coherent relative entropy enjoys a number of desirable properties. It can be written as a semidefinite program, can be bound using known entropy measures, and satisfies a chain rule. Furthermore, it converges to the difference of relative entropies $D(\rho_R \,\|\, \Gamma_R) - D(\rho_{X'} \,\|\, \Gamma_{X'})$ in the i.i.d. limit.

In Section 7.1, we introduce the basic definition and essential properties of our new measure. A smooth version is defined in Section 7.2, and some of its properties explored. In Section 7.3, we study special cases where the coherent relative entropy can be related to other known measures. In Section 7.4 we prove bounds to this quantity in terms of known entropy measures. We prove a chain rule in Section 7.5 and an asymptotic equipartition rule in Section 7.6. Finally, we discuss in Section 7.7 where our quantity fits in our entropy zoo (Figure 3.2).

## 7.1 Definition and basic properties

Consider two quantum systems $X$ and $X'$, described by respective hidden information operators $\Gamma_X$ and $\Gamma_{X'}$ (Section 5.2). We would like to perform a logical process from $X$ to $X'$ which is described by the process matrix $\rho_{X'R}$ (Section 3.1.9). This is, as we have seen, equivalent to specifying an input state $\sigma_X$ and a trace-preserving, completely positive map $\mathcal{E}_{X\to X'}$ specified on the support of $\sigma_X$. Recall that $\rho_{X'R} = \mathcal{E}_{X\to X'}\left(\sigma_X^{1/2}\Phi_{X:R}\sigma_X^{1/2}\right)$ where $\Phi_{X:R}$ is the unnormalized maximally entangled state defined in Section 3.1.5. In the following, we write $\Gamma_R := \mathrm{tr}_X(\Phi_{X:R}\Gamma_X) = t_{X\to R}(\Gamma_X)$ recalling that $t_{X\to R}$ is the partial transpose map with respect to the same bases as $\Phi_{X:R}$.

The coherent relative entropy is defined as the maximal compression yield of an implementation of a process matrix $\rho_{X'R}$ relative to hidden information operators $\Gamma_R, \Gamma_{X'}$, in the sense of Chapter 5. By Proposition 5.2, it is possible to perform the requested process matrix $\rho_{X'R}$ with a compression yield of $\lambda_1 - \lambda_2 \equiv \lambda$ if and only if there exists a completely positive, trace non-increasing map $\mathcal{T}_{X\to X'}$ such that $\mathcal{T}_{X\to X'}(\Gamma_X) \leqslant 2^{-\lambda}\Gamma_{X'}$ and $\mathcal{T}_{X\to X'}(\sigma_{XR}) = \rho_{X'R}$. These conditions on $\mathcal{T}_{X\to X'}$ are semidefinite constraints on the Choi matrix $T_{X'R}$ of $\mathcal{T}_{X\to X'}$ (Section 3.1.9). In fact, by introducing a variable $\alpha = 2^{-\lambda}$, the maximization problem over $\lambda$ turns into a minimization problem for $\alpha$ which is also a semidefinite program.

For our definition, we only require that $\rho_{X'R}$ is a subnormalized quantum state. This is first of all a pure technical convenience, as we anyway show that the coherent relative entropy of a subnormalized state is equal to that of its normalized counterpart. Recall also the interpretation of subnormalized states as normalized states projected onto particular events (Section 3.3.1).

***Coherent relative entropy.*** *For a bipartite quantum subnormalized state $\rho_{X'R}$, and two positive semidefinite operators $\Gamma_R$ and $\Gamma_{X'}$ such that $\rho_{X'R}$ lies in the support of $\Gamma_R \otimes \Gamma_{X'}$, we introduce the 'coherent relative entropy' as*

$$\bar{D}_{R\to X'}(\rho_{X'R} \,\|\, \Gamma_R, \Gamma_{X'}) = -\log\alpha\,; \tag{7.1}$$

*where $\alpha$ in (7.1) is the optimal solution to the following semidefinite program in terms of the variables $T_{X'R} \geqslant 0$, $\alpha \geqslant 0$, and dual variables $Z_{X'R} = Z^\dagger_{X'R}$, $\omega_{X'} \geqslant 0$, $X_R \geqslant 0$:*

***Primal problem:***

$$\begin{aligned}
\text{minimize:}\quad & \alpha & & \\
\text{subject to:}\quad & \mathrm{tr}_R\left[T_{X'R}\Gamma_R\right] \leqslant \alpha\Gamma_{X'} & : \omega_{X'} \qquad & \text{(7.2-a)}\\
& \mathrm{tr}_{X'}\left[T_{X'R}\right] \leqslant \mathbb{1}_R & : X_R \qquad & \text{(7.2-b)}\\
& \rho_R^{1/2} T_{X'R} \rho_R^{1/2} = \rho_{X'R} & : Z_{X'R} \qquad & \text{(7.2-c)}
\end{aligned}$$

***Dual problem:***

$$\begin{aligned}
\text{maximize:}\quad & \mathrm{tr}\left[Z_{X'R}\rho_{X'R}\right] - \mathrm{tr}\, X_R & & \\
\text{subject to:}\quad & \mathrm{tr}\left[\omega_{X'}\Gamma_{X'}\right] \leqslant 1 & : \alpha \qquad & \text{(7.3-a)}\\
& \rho_R^{1/2} Z_{X'R} \rho_R^{1/2} \leqslant \Gamma_R \otimes \omega_{X'} + X_R \otimes \mathbb{1}_{X'} & : T_{X'R} \qquad & \text{(7.3-b)}
\end{aligned}$$

In this semidefinite program, the variable $T_{X'R}$ is in fact to be interpreted as the Choi matrix of a trace nonincreasing completely positive map $\mathcal{T}_{X\to X'}$, where the $R$ system is seen as a purifying system for input states on $X$.

Some additional notes about the semidefinite program: the dual is strictly feasible (choose, e.g., $Z_{X'R} = 0$, $\omega_{X'} = 0$ and $X_R = \mathbb{1}_R$), and $T_{X'R} = \rho_R^{-1/2}\rho_{X'R}\rho_R^{-1/2}$ is a feasible primal candidiate, and hence by Slater's sufficiency conditions (Section 3.3.2) we have that strong duality holds and there always exists optimal primal candidates. However, note that the primal problem is not always strictly feasible (indeed, constraint (7.2-c) is very strong and fixes the mapping $T_{X'R}$ on a subspace; it is then not clear if (7.2-b) can be satisfied strictly). This means that there is a possibility that there is no optimal choice of dual variables. However, since we have strong duality, there is always a sequence of choices for dual variables whose attained objective value will converge to the optimal solution of the semidefinite program.

Here are first some basic properties of this quantity.

***Proposition 7.1** (Trivial bounds). We have*

$$\bar{D}_{R\to X'}(\rho_{X'R} \,\|\, \Gamma_R, \Gamma_{X'}) \geqslant -\log \operatorname{tr}\Gamma_R - \log \big\|\Gamma_{X'}^{-1}\big\|_\infty \;; \tag{7.4-a}$$

$$\bar{D}_{R\to X'}(\rho_{X'R} \,\|\, \Gamma_R, \Gamma_{X'}) \leqslant \log \big\|\Gamma_R^{-1}\big\|_\infty + \log \operatorname{tr}\Gamma_{X'} \;. \tag{7.4-b}$$

*(Proof on page 149.)*

These bounds can be understood in terms of work extraction. Say the Γ operators represent systems with Hamiltonians $H_R$ and $H_{X'}$ interacting with a heat bath at inverse temperature $\beta$, as represented by $\Gamma_R = e^{-\beta H_R}$ and $\Gamma_{X'} = e^{-\beta H_{X'}}$. Then $\log \|\Gamma_R^{-1}\|_\infty$ (resp. $\log \|\Gamma_{X'}^{-1}\|_\infty$) is $\beta$ times the maximum energy of $R$ (resp. $X'$), and similarly, $\operatorname{tr}\Gamma_R$ (resp. $\operatorname{tr}\Gamma_{X'}$) is the partition function of $R$ (resp. $X'$). The partition function is directly related to the work cost of erasure (resp. formation) of a thermal state to (resp. from) a pure energy eigenstate of zero energy. In this case, the bounds (7.4) correspond to the ultimate worst and best cases respectively. The ultimate worst case is that we start off in a thermal state and end up in the highest energy level, whereas the absolute best case would be to start in the highest energy eigenstate and finish in the Gibbs state.

Much like the conditional entropy and relative entropy, the coherent relative entropy is invariant under partial isometries of which $\rho_{X'R}$ and Γ operators lie in the support. In particular, the coherent relative entropy is completely oblivious to dimensions of the Hilbert spaces which are not spanned by $\Gamma_R$ and $\Gamma_{X'}$.

***Proposition 7.2** (Invariance under isometries). Let $\tilde{X}'$ and $\tilde{R}$ be new systems. Suppose there exist partial isometries $V_{R\to\tilde{R}}$ and $V'_{X'\to\tilde{X}'}$ such that both $\rho_R$ and $\Gamma_R$ are in the support of $V_{R\to\tilde{R}}$, and both $\rho_{X'}$ and $\Gamma_{X'}$ are in the support of $V'_{X'\to\tilde{X}'}$. Then*

$$\bar{D}_{\tilde{R}\to\tilde{X}'}\big((V'\otimes V)\,\rho_{X'R}\,(V'\otimes V)^\dagger \,\|\, V\Gamma_R V^\dagger, V'\Gamma_{X'}V'^\dagger\big) = \bar{D}_{R\to X'}(\rho_{X'R} \,\|\, \Gamma_R, \Gamma_{X'}) \;. \tag{7.5}$$

*(Proof on page 149.)*

This proposition allows us to embed states in larger dimensions, as well as to show that it is invariant under simultaneous action of unitaries on the states and the Γ operators.

As we've noted before, the operators $\Gamma_R$ and $\Gamma_{X'}$ may have any normal-ization. In the case where $\Gamma = e^{-\beta H}$ for a Hamiltonian $H$ and an inverse temperature $\beta$, we may think of a change in the normalization of $\Gamma$ as a global energy shift at constant $\beta$ (Section 5.2.5). Consequently we expect that changing the normalization of the hidden information operators introduces a constant shift in the coherent relative entropy, which would correspond to providing the required energy to produce the energy shift.

***Proposition 7.3*** *(Scaling the hidden information operators). Let $a, b > 0$. Then*

$$\bar{D}_{R\to X'}(\rho_{X'R} \,\|\, a\Gamma_R, b\Gamma_{X'}) = \bar{D}_{R\to X'}(\rho_{X'R} \,\|\, \Gamma_R, \Gamma_{X'}) + \log\frac{b}{a}\,. \tag{7.6}$$

*(Proof on page 150.)*

The following proposition shows that using subnormalized states doesn't change the value of the coherent relative entropy. This will prove very convenient when smoothing this quantity, as we won't have to bother with normalizing the quantum states we come up with.

***Proposition 7.4*** *(Subnormalized states). Let $\rho_{X'R}$ be any subnormalized quantum state, and let $\Gamma_R, \Gamma_{X'} \geqslant 0$. Then*

$$\bar{D}_{R\to X'}(\rho_{X'R} \,\|\, \Gamma_R, \Gamma_{X'}) = \bar{D}_{R\to X'}\left(\frac{\rho_{X'R}}{\operatorname{tr}\rho_{X'R}} \,\middle\|\, \Gamma_R, \Gamma_{X'}\right). \tag{7.7}$$

*(Proof on page 150.)*

***Proposition 7.5*** *(Superadditivity for tensor products). Let systems $X_1$, $X_1'$, $X_2$, $X_2'$ have respective hidden information operators $\Gamma_{X_1}$, $\Gamma_{X_1'}$, $\Gamma_{X_2}$, $\Gamma_{X_2'}$. Let $\rho_{X_1'R_1}$ and $\zeta_{X_2'R_2}$ be two subnormalized quantum states, where $R_1 \simeq X_1$ and $R_2 \simeq X_2$ as per the construction of the process matrix in Section 3.1.9. Then*

$$\begin{aligned}&\bar{D}_{R_1R_2\to X_1'X_2'}(\rho_{X_1'R_1}\otimes\zeta_{X_2'R_2} \,\|\, \Gamma_{R_1}\otimes\Gamma_{R_2}, \Gamma_{X_1'}\otimes\Gamma_{X_2'})\\ &\qquad\geqslant \bar{D}_{R_1\to X_1'}(\rho_{X_1'R_1} \,\|\, \Gamma_{R_1}, \Gamma_{X_1'}) + \bar{D}_{R_2\to X_2'}(\zeta_{X_2'R_2} \,\|\, \Gamma_{R_2}, \Gamma_{X_2'})\,.\end{aligned} \tag{7.8}$$

*(Proof on page 150.)*

Also, it is possible to restrict the hidden information operators by projecting them onto selected eigenkets, while still having the process matrix lying in their support. Then the coherent relative entropy remains unchanged.

***Proposition 7.6*** *(Restricting the $\Gamma$ operators). Let $P_R$ and $P'_{X'}$ be projectors such that $[P_R, \Gamma_R] = 0$ and $[P'_{X'}, \Gamma_{X'}] = 0$. Define $\Gamma'_R = P_R\Gamma_R P_R$ and $\Gamma'_{X'} = P'_{X'}\Gamma_{X'}P'_{X'}$. Let $\rho_{X'R}$ be any quantum subnormalized state with support inside* *that of* $\Gamma'_{X'} \otimes \Gamma'_R$. *Then*

$$\bar{D}_{R\to X'}(\rho_{X'R} \,\|\, \Gamma_R, \Gamma_{X'}) = \bar{D}_{R\to X'}(\rho_{X'R} \,\|\, \Gamma'_R, \Gamma'_{X'}) \,. \tag{7.9}$$

*(Proof on page 150.)*

Another property relates the coherent relative entropy to that with respect to different $\Gamma$ operators which represent "at least or at most as much hidden information," as represented as an operator inequality.

***Proposition 7.7.*** *Let* $\tilde{\Gamma}_R \geqslant 0$ *and* $\tilde{\Gamma}_{X'} \geqslant 0$ *be such that* $\tilde{\Gamma}_R \leqslant \Gamma_R$ *and* $\Gamma_{X'} \leqslant \tilde{\Gamma}_{X'}$. *Then*

$$\bar{D}_{R\to X'}(\rho_{X'R} \,\|\, \tilde{\Gamma}_R, \tilde{\Gamma}_{X'}) \geqslant \bar{D}_{R\to X'}(\rho_{X'R} \,\|\, \Gamma_R, \Gamma_{X'}) \,. \tag{7.10}$$

*(Proof on page 151.)*

The last of these basic properties is a simple rewriting of the semidefinite program, which allows us to condense somewhat the primal problem. This form is useful for the proofs of the other propositions.

***Proposition 7.8.*** *The optimization problem defining the coherent relative entropy can be rewritten as*

$$2^{-\bar{D}_{R\to X'}(\rho_{X'R} \,\|\, \Gamma_R,\Gamma_{X'})} = \min_{T_{X'R}} \left\| \Gamma_{X'}^{-1/2} \operatorname{tr}_R \left[ T_{X'R} \Gamma_R \right] \Gamma_{X'}^{-1/2} \right\|_\infty \,, \tag{7.11}$$

*where the minimization is taken over all positive semidefinite* $T_{X'R}$ *satisfying both conditions (7.2-b) and (7.2-c), and for which the operator* $\operatorname{tr}_R\left(T_{X'R}\Gamma_R\right)$ *lies within the support of* $\Gamma_{X'}$. *Equivalently,*

$$2^{-\bar{D}_{R\to X'}(\rho_{X'R} \,\|\, \Gamma_R,\Gamma_{X'})} = \min_{\mathcal{T}_{X\to X'}} \left\| \Gamma_{X'}^{-1/2} \mathcal{T}_{X\to X'}\left[\Gamma_X\right] \Gamma_{X'}^{-1/2} \right\|_\infty \,, \tag{7.12}$$

*where the minimization is taken over all trace nonincreasing, completely positive maps* $\mathcal{T}_{X\to X'}$ *which satisfy* $\mathcal{T}_{X\to X'}\left[|\sigma\rangle\langle\sigma|_{XR}\right] = \rho_{X'R}$ *and for which* $\mathcal{T}_{X\to X'}\left(\Gamma_X\right)$ *lies within the support of* $\Gamma_{X'}$. *As above,* $\sigma_{XR}$ *is the purified input state, following the construction in Section 3.1.9.*

We begin by proving Proposition 7.8, which is needed for the other proofs.

*Proof of Proposition 7.8.* Let $T_{X'R}$ be any candidate in the primal problem. If $\operatorname{tr}_R(T_{X'R})$ does not lie within the support of $\Gamma_{X'}$, then condition (7.2-a) is not satisfied and the candidate is not primal feasible; we can hence ignore it in the minimization. Otherwise, by conjugating condition (7.2-a) by $\Gamma_{X'}^{-1/2}$, we see that (7.2-a) is equivalent to

$$\Gamma_{X'}^{-1/2}\,\mathrm{tr}_R\left[T_{X'R}\Gamma_R\right]\Gamma_{X'}^{-1/2} \leqslant \alpha\,\Pi_{X'}^{\Gamma_{X'}}\ , \tag{7.13}$$

which in turn is equivalent to

$$\Gamma_{X'}^{-1/2}\,\mathrm{tr}_R\left[T_{X'R}\Gamma_R\right]\Gamma_{X'}^{-1/2} \leqslant \alpha\,\mathbb{1}\ , \tag{7.14}$$

because the left hand side of (7.13) is entirely within the support of its right hand side. Now, the optimal $\alpha$ which corresponds to this fixed $T_{X'R}$ is given by $\|\Gamma_{X'}^{-1/2}\,\mathrm{tr}_R\left[T_{X'R}\Gamma_R\right]\Gamma_{X'}^{-1/2}\|_\infty$. This chain of equivalences may be followed in reverse order, establishing the equivalence of the minimization problems.

The formulation in terms of channels follows immediately from the translation of one formalism to the other. ∎

*Proof of Proposition 7.1.* Consider the minimization in Proposition 7.8, and let $\mathcal{T}_{X\to X'}$ be any trace nonincreasing completely positive map such that $\mathcal{T}_{X\to X'}(\Gamma_X)$ lies within the support of $\Gamma_{X'}$. Define the normalized state $\gamma_X = \Gamma_X/\,\mathrm{tr}\,\Gamma_X$. Now write

$$\|\Gamma_{X'}^{-1/2}\mathcal{T}\left[\Gamma_X\right]\Gamma_{X'}^{-1/2}\|_\infty = \mathrm{tr}\left[\Gamma_X\right]\|\Gamma_{X'}^{-1/2}\mathcal{T}\left[\gamma_X\right]\Gamma_{X'}^{-1/2}\|_\infty\ . \tag{7.15}$$

As $\mathcal{T}\left[\gamma_X\right] \geqslant 0$ and $\mathrm{tr}\,\mathcal{T}\left[\gamma_X\right] \leqslant \mathrm{tr}\,\gamma_X \leqslant 1$, we have $\mathcal{T}\left[\gamma_X\right] \leqslant \mathbb{1}_{X'}$ and thus

$$(7.15) \leqslant \mathrm{tr}(\Gamma_X)\,\|\Gamma_{X'}^{-1}\|_\infty\ . \tag{7.16}$$

By taking minus the logarithm of both ends of the calculation for the optimal $\mathcal{T}_{X\to X'}$ we obtain (7.4-a).

For bound (7.4-b) we provide feasible candidate dual variables for the semidefinite program. Let $\omega_{X'} = (\mathrm{tr}\,\Gamma_{X'})^{-1}\,\mathbb{1}_{X'}$, $Z_{X'R} = \|\Gamma_R^{-1}\|_\infty^{-1}\omega_{X'}\otimes\mathbb{1}_R$ and $X_R = 0$. We have $\mathrm{tr}(\omega_{X'}\Gamma_{X'}) \leqslant 1$ in accordance with constraint (7.3-a). Observe that $\|\Gamma_R^{-1}\|_\infty^{-1}\rho_R \leqslant \|\Gamma_R^{-1}\|_\infty^{-1}\Pi_R^{\Gamma_R} \leqslant \Gamma_R$ because $\|\Gamma_R^{-1}\|_\infty^{-1}$ is the minimal nonzero eigenvalue of $\Gamma_R$ and $\rho_R$ is assumed to be in the support of $\Gamma_R$. Hence

$$\rho_R^{1/2}Z_{X'R}\rho_R^{1/2} = \|\Gamma_R^{-1}\|_\infty^{-1}\,\omega_{X'}\otimes\rho_R \leqslant \Gamma_R\otimes\omega_{X'}\ . \tag{7.17}$$

All constraints fulfilled, we now calculate the attained value:

$$\mathrm{tr}\left[Z_{X'R}\rho_{X'R}\right] - \mathrm{tr}\,X_R = \|\Gamma_R^{-1}\|_\infty^{-1}\,\mathrm{tr}\left[\omega_{X'}\rho_{X'}\right] = \|\Gamma_R^{-1}\|_\infty^{-1}\left(\mathrm{tr}\left[\Gamma_{X'}\right]\right)^{-1}\ . \tag{7.18}$$

This gives us

$$\bar{D}_{R\to X'}(\rho_{X'R}\,\|\,\Gamma_R,\Gamma_{X'}) \leqslant \log\|\Gamma_R^{-1}\|_\infty + \log\mathrm{tr}\,\Gamma_{X'}\ . \qquad ∎$$

*Proof of Proposition 7.2.* This is clearly the case, because the semidefinite problem lies entirely within the support of the isometries. Formally, any choice of variables for the original problem can be mapped in the new spaces through these partial isometries, and vice versa, and the attained values remain the same. Hence the optimal value of the problem is also the same. ∎

*Proof of Proposition 7.3.* Consider the optimal primal candidiates $T_{X'R}$ and $\alpha$ for the problem defining $2^{-\bar{D}_{R\to X'}(\rho_{X'R}\,\|\,\Gamma_R,\Gamma_{X'})}$. Then $T_{X'R}$ and $\alpha a b^{-1}$ are feasible primal candidates for the semidefinite program with the scaled $\Gamma$ operators. Hence

$$2^{-\bar{D}_{R\to X'}(\rho_{X'R}\,\|\,a\Gamma_R,b\Gamma_{X'})} \leqslant \frac{a}{b}\alpha = \frac{a}{b}\,2^{-\bar{D}_{R\to X'}(\rho_{X'R}\,\|\,\Gamma_R,\Gamma_{X'})}\,. \tag{7.19}$$

The opposite direction follows by applying the same argument to the reverse situation with $\Gamma_R' = a\Gamma_R$ and $\Gamma_{X'}' = b\Gamma_{X'}$. ∎

*Proof of Proposition 7.4.* Let $\bar{\rho}_{X'R} = \rho_{X'R}/\operatorname{tr}\rho_{X'R}$. The only difference between the primal problems defining $\bar{D}_{R\to X'}(\rho_{X'R}\,\|\,\Gamma_R,\Gamma_{X'})$ and $\bar{D}_{R\to X'}(\bar{\rho}_{X'R}\,\|\,\Gamma_R,\Gamma_{X'})$ is condition (7.2-c).

Let $T_{X'R} \geqslant 0$. We have

$$\bar{\rho}_R^{-1/2}T_{X'R}\bar{\rho}_R^{-1/2} = \frac{1}{\operatorname{tr}\rho_{X'R}}\,\rho_R^{1/2}T_{X'R}\rho_R^{1/2}\,; \tag{7.20}$$

hence for any $T_{X'R} \geqslant 0$, $\rho_R^{1/2}T_{X'R}\rho_R^{1/2} = \rho_{X'R}$ if and only if $\bar{\rho}_R^{-1/2}T_{X'R}\bar{\rho}_R^{-1/2} = \bar{\rho}_{X'R}$.

This means the optimization candidates are exactly the same for both problems, and the optimal solution is the same. ∎

*Proof of Proposition 7.5.* Let $T_{X_1'R_1}, \alpha_1$ and $T_{X_2'R_2}, \alpha_2$ be the optimal choice of primal variables for $2^{-\bar{D}_{R_1\to X_1'}(\rho_{X_1'R_1}\,\|\,\Gamma_{R_1},\Gamma_{X_1'})}$ and $2^{-\bar{D}_{R_2\to X_2'}(\zeta_{X_2'R_2}\,\|\,\Gamma_{R_2},\Gamma_{X_2'})}$, respectively.

Now, let $T_{X_1'X_2'R_1R_2} = T_{X_1'R_1}\otimes T_{X_2'R_2}$ and $\alpha = \alpha_1\alpha_2$. Then

$$\operatorname{tr}_{R_1R_2}\!\left[T_{X_1'X_2'R_1R_2}\Gamma_{R_1}\otimes\Gamma_{R_2}\right] \leqslant \alpha_1\alpha_2\Gamma_{X_1'}\otimes\Gamma_{X_2'}\,, \tag{7.21}$$

$$\operatorname{tr}_{X_1'X_2'}\!\left[T_{X_1'X_2'R_1R_2}\right] \leqslant \mathbb{1}_{X_1'}\otimes\mathbb{1}_{X_2'}\,, \tag{7.22}$$

and

$$\left(\rho_{R_1}^{1/2}\otimes\zeta_{R_2}^{1/2}\right)T_{X_1'X_2'R_1R_2}\left(\rho_{R_1}^{1/2}\otimes\zeta_{R_2}^{1/2}\right) = \rho_{X_1'R_1}\otimes\zeta_{X_2'R_2}\,, \tag{7.23}$$

and hence this choice of variables is feasible for the tensor product problem. We then have

$$\begin{aligned}2^{-\bar{D}_{R_1R_2\to X_1'X_2'}(\rho_{X_1'R_1}\otimes\zeta_{X_2'R_2}\,\|\,\Gamma_{R_1}\otimes\Gamma_{R_2},\Gamma_{X_1'}\otimes\Gamma_{X_2'})} &\leqslant \alpha_1\alpha_2\\ &= 2^{-\left(\bar{D}_{R_1\to X_1'}(\rho_{X_1'R_1}\,\|\,\Gamma_{R_1},\Gamma_{X_1'})+\bar{D}_{R_2\to X_2'}(\zeta_{X_2'R_2}\,\|\,\Gamma_{R_2},\Gamma_{X_2'})\right)}\,.\end{aligned}$$ ∎

*Proof of Proposition 7.6.* Let $T_{X'R}$ and $\alpha$ be the optimal feasible candidates for the primal semidefinite problem defining $2^{-\bar{D}_{R\to X'}(\rho_{X'R}\,\|\,\Gamma_R,\Gamma_{X'})}$. Let $T_{X'R}' = (P_{X'}'\otimes P_R)\,T_{X'R}\,(P_{X'}'\otimes P_R)$ and $\alpha' = \alpha$. Then

$$\operatorname{tr}_{X'}T_{X'R}' = P_R\operatorname{tr}_{X'}\left[P_{X'}'T_{X'R}\right]P_R \leqslant P_R\operatorname{tr}_{X'}\left(T_{X'R}\right)P_R \leqslant P_R \leqslant \mathbb{1}_R\,, \tag{7.24}$$

satisfying (7.2-b), and

$$\rho_R^{1/2} T'_{X'R} \rho_R^{1/2} = \left(P'_{X'} \otimes P_R\right) \rho_R^{1/2} T_{X'R} \rho_R^{1/2} \left(P'_{X'} \otimes P_R\right) = \left(P'_{X'} \otimes P_R\right) \rho_{X'R} \left(P'_{X'} \otimes P_R\right) = \rho_{X'R} \,, \tag{7.25}$$

where the first and last inequality hold because $\rho_R$ and $\rho_{X'R}$ lie within the support of $P_R$ and $P'_{X'} \otimes P_R$, respectively. Hence (7.2-c) is fulfilled. Now we have

$$\begin{aligned} \operatorname{tr}_R\left[T'_{X'R}\Gamma'_R\right] &= \operatorname{tr}_R\left[\left(P'_{X'} \otimes P_R\right) T_{X'R} \left(P'_{X'} \otimes P_R\right) \Gamma'_R\right] = \operatorname{tr}_R\left[P'_{X'} T_{X'R} P'_{X'}\, \Gamma'_R\right] \\ &\leqslant P'_{X'} \operatorname{tr}_R\left[T_{X'R}\, \Gamma_R\right] P'_{X'} \leqslant P'_{X'} \left(\alpha \Gamma_{X'}\right) P'_{X'} = \alpha' \Gamma'_{X'} \,, \end{aligned} \tag{7.26}$$

using the fact that $\Gamma'_R \leqslant \Gamma_R$: indeed, $\Gamma'_R = P_R \Gamma_R P_R = P_R \Gamma_R = \Gamma_R^{1/2} P_R \Gamma_R^{1/2} \leqslant \Gamma_R$ because $\left[\Gamma_R, P_R\right] = 0$. Hence

$$2^{-\bar{D}_{R\to X'}(\rho_{X'R} \,\|\, \Gamma'_R, \Gamma'_{X'})} \leqslant 2^{-\bar{D}_{R\to X'}(\rho_{X'R} \,\|\, \Gamma_R, \Gamma_{X'})} \,. \tag{7.27}$$

Let $Z_{X'R}$, $X_R$ and $\omega_{X'}$ be any dual feasible candidates for $2^{-\bar{D}_{R\to X'}(\rho_{X'R} \,\|\, \Gamma_R, \Gamma_{X'})}$. Now let $Z'_{X'R} = \left(P'_{X'} \otimes P_R\right) Z_{X'R} \left(P'_{X'} \otimes P_R\right)$, $X'_R = P_R X_R P_R$, and $\omega'_{X'} = P'_{X'} \omega_{X'} P'_{X'}$. Then $\operatorname{tr}\left(\omega'_{X'} \Gamma'_{X'}\right) = \operatorname{tr}\left(\omega_{X'} \Gamma'_{X'}\right) \leqslant \operatorname{tr}\left[\omega_{X'} \Gamma_{X'}\right] \leqslant 1$ (using the fact that $\Gamma'_{X'} \leqslant \Gamma_{X'}$ since $\left[\Gamma_{X'}, P'_{X'}\right] = 0$), in accordance with (7.3-a). Also,

$$\begin{aligned} \rho_R^{1/2} Z'_{X'R} \rho_R^{1/2} &= \left(P_R \otimes P'_{X'}\right) \rho_R^{1/2} Z_{X'R} \rho_R^{1/2} \left(P_R \otimes P'_{X'}\right) \\ &\leqslant \left(P_R \otimes P'_{X'}\right) \left(\Gamma_R \otimes \omega_{X'} + X_R \otimes \mathbb{1}_{X'}\right) \left(P_R \otimes P'_{X'}\right) \\ &= \Gamma'_R \otimes \omega'_{X'} + X'_R \otimes P'_{X'} \\ &\leqslant \Gamma'_R \otimes \omega'_{X'} + X'_R \otimes \mathbb{1}_{X'} \,, \end{aligned} \tag{7.28}$$

which satisfies (7.3-b). Finally, the attained dual value is

$$\operatorname{tr}\left[Z'_{X'R}\rho_{X'R}\right] - \operatorname{tr} X'_R = \operatorname{tr}\left[Z_{X'R}\rho_{X'R}\right] - \operatorname{tr}\left[P_R X_R\right] \geqslant \operatorname{tr}\left[Z_{X'R}\rho_{X'R}\right] - \operatorname{tr}\left[X_R\right] \,. \tag{7.29}$$

Hence, we now have

$$2^{-\bar{D}_{R\to X'}(\rho_{X'R} \,\|\, \Gamma'_R, \Gamma'_{X'})} \geqslant 2^{-\bar{D}_{R\to X'}(\rho_{X'R} \,\|\, \Gamma_R, \Gamma_{X'})} \,, \tag{7.30}$$

which completes the proof. ∎

*Proof of Proposition 7.7.* Let $T_{X'R}$ and $\alpha$ be the optimal solution to the semidefinite program for $2^{-\bar{D}_{R\to X'}(\rho_{X'R} \,\|\, \Gamma_R, \Gamma_{X'})}$. They are then also feasible candidates for the semidefinite program for $2^{-\bar{D}_{R\to X'}(\rho_{X'R} \,\|\, \tilde{\Gamma}_R, \tilde{\Gamma}_{X'})}$, because the only condition that changes is (7.2-a), which is obviously still satisfied. ∎

## 7.2 Smooth coherent relative entropy

As in the smooth entropy framework, it is necessary to introduce a failure probability which accounts for extremely unlikely events (see Section 2.6.2 and Section 3.2.3).

We ask what is the best compression yield we can achieve by a logical process which implements only approximately the process matrix $\rho_{X'R}$. This is done by optimizing the readily available coherent relative entropy in an ball of states close to $\rho_{X'R}$, controlled by a parameter $\epsilon$.

***Smooth coherent relative entropy***. *For a subnormalized state $\rho_{X'R}$, and positive semidefinite $\Gamma_R, \Gamma_{X'}$, the 'smooth coherent relative entropy' is defined as*

$$\bar{D}^{\epsilon}_{R\to X'}(\rho_{X'R} \parallel \Gamma_R, \Gamma_{X'}) = \max_{\hat{\rho}_{X'R}\approx_\epsilon \rho_{X'R}} \bar{D}_{R\to X'}(\hat{\rho}_{X'R} \parallel \Gamma_R, \Gamma_{X'}) \,, \tag{7.31}$$

*where the maximization in (7.31) is taken over subnormalized quantum states which are in the support of $\Gamma_R \otimes \Gamma_{X'}$ and which satisfy $P(\hat{\rho}_{X'R}, \rho_{X'R}) \leqslant \epsilon$ with the generalized purified distance $P(\cdot,\cdot)$ introduced in Section 3.3.1.*

Because the battery system is a part of the physical implementation of the process, we may ask why it is not included in the definition (7.31) in a way which would allow the physical implementation to fail to produce the appropriate output battery state with a small probability. Remarkably, there would have been no difference had we chosen to smooth the battery states as well. This follows from the following proposition, which asserts that optimization candidates which include smoothing on the battery states are in fact already included in the optimization in the definition above. This holds for the general battery states of the form $P_A\Gamma_A P_A/\operatorname{tr}(P_A\Gamma_A)$, for a projector $P_A$ commuting with the $\Gamma_A$ of the battery (see item (v) of Proposition 5.2).

***Proposition 7.9** (Smoothing battery states). Let $A, A'$ be quantum systems with corresponding $\Gamma_A$, $\Gamma_{A'}$. Let $P_A$, $P'_{A'}$ be projectors such that $[P_A, \Gamma_A] = 0$ and $[P_{A'}, \Gamma_{A'}] = 0$, and let $\Phi_{XA\to X'A'}$ be a trace nonincreasing, completely positive map such that $\Phi_{XA\to X'A'}(\Gamma_X \otimes \Gamma_A) \leqslant \Gamma_{X'} \otimes \Gamma_{A'}$, and such that*

$$P\left[\Phi_{XA\to X'A}\left(\sigma_{XR} \otimes \frac{P_A\Gamma_A P_A}{\operatorname{tr} P_A\Gamma_A}\right),\ \rho_{X'R} \otimes \frac{P'_{A'}\Gamma_{A'}P'_{A'}}{\operatorname{tr} P'_{A'}\Gamma_{A'}}\right] \leqslant \epsilon \,, \tag{7.32}$$

*where $P(\cdot,\cdot)$ is the purified distance (Section 3.1.10). Then there exists a trace-nonincreasing, completely positive map $\mathcal{T}_{X\to X'}$ such both the following condi-*

*tions hold:*

$$P\big(\mathcal{T}_{X\to X'}(\sigma_{XR}),\rho_{X'R}\big)\leqslant\epsilon\,; \tag{7.33-a}$$

$$\mathcal{T}_{X\to X'}(\Gamma_X)\leqslant\frac{\operatorname{tr}(P'_{A'}\Gamma_{A'})}{\operatorname{tr}(P_A\Gamma_A)}\,\Gamma_{X'}\;. \tag{7.33-b}$$

This means that the processes which also allow "fuzziness" on the battery states are *de facto* already included in the optimization defining the smooth coherent relative entropy (7.31). This is formulated explicitly in the following corollary.

***Corollary 7.10.*** *Let $\rho_{X'R}$ be a subnormalized state, let $\Gamma_R,\Gamma_{X'}\geqslant 0$ and let $\epsilon>0$. Then*

$$\bar{D}^{\epsilon}_{R\to X'}(\rho_{X'R}\,\|\,\Gamma_R,\Gamma_{X'})=\max_{A,A',P_A,P'_{A'},\Phi_{XA\to X'A'}}-\log\frac{\operatorname{tr}P'_{A'}\Gamma_{A'}}{\operatorname{tr}P_A\Gamma_A}\;, \tag{7.34}$$

*where the optimization is performed over all systems $A$, $A'$, all operators $\Gamma_A$, $\Gamma_{A'}$, and all projectors $P_A$, $P'_{A'}$ such that $[P_A,\Gamma_A]=0$ and $[P'_{A'},\Gamma_{A'}]=0$, for which there is a trace nonincreasing, completely positive map $\Phi_{XA\to X'A'}$ satisfying $\Phi_{XA\to X'A'}(\Gamma_X\otimes\Gamma_A)\leqslant\Gamma_{X'}\otimes\Gamma_{A'}$ as well as*

$$P\left[\Phi_{XA\to X'A'}\!\left(\sigma_{XR}\otimes\frac{P_A\Gamma_AP_A}{\operatorname{tr}P_A\Gamma_A}\right),\;\rho_{X'R}\otimes\frac{P'_{A'}\Gamma_{A'}P'_{A'}}{\operatorname{tr}P'_{A'}\Gamma_{A'}}\right]\leqslant\epsilon\;. \tag{7.35}$$

*Proof of Proposition 7.9.* Define, for any $\omega_X$,

$$\mathcal{T}_{X\to X'}(\omega_X)=\operatorname{tr}_{A'}\left[P'_{A'}\,\Phi_{XA\to X'A'}\!\left(\omega_X\otimes\frac{P_A\Gamma_AP_A}{\operatorname{tr}P_A\Gamma_A}\right)\right]. \tag{7.36}$$

Then

$$\mathcal{T}_{X\to X'}(\sigma_{XR})=\operatorname{tr}_{A'}\left[P'_{A'}\,\Phi_{XA\to X'A'}\!\left(\sigma_{XR}\otimes\frac{P_A\Gamma_AP_A}{\operatorname{tr}P_A\Gamma_A}\right)\right]=\operatorname{tr}_{A'}\left[P'_{A'}\,\tilde{\rho}_{A'X'R}\right], \tag{7.37}$$

where $\tilde{\rho}_{A'X'R}:=\Phi_{XA\to X'A'}\big(\sigma_{XR}\otimes\frac{P_A\Gamma_AP_A}{\operatorname{tr}P_A\Gamma_A}\big)$ satisfies $P\big(\tilde{\rho}_{A'X'R},\rho_{X'R}\otimes\frac{P'_{A'}\Gamma_{A'}P'_{A'}}{\operatorname{tr}P'_{A'}\Gamma_{A'}}\big)\leqslant\epsilon$ by assumption. Using the monotonicity of the purified distance (Tomamichel, 2012) in particular under the trace-nonincreasing completely positive map $\operatorname{tr}\left[P'_{A'}\,(\cdot)\right]$, we have

$$P\big(\mathcal{T}_{X\to X'}(\sigma_{XR}),\rho_{X'R}\big)\leqslant\epsilon\;. \tag{7.38}$$

We also have

$$\mathcal{T}_{X\to X'}(\Gamma_X) = \operatorname{tr}_{A'}\left[P'_{A'}\,\Phi_{XA\to X'A'}\Big(\Gamma_X\otimes\frac{P_A\Gamma_AP_A}{\operatorname{tr}P_A\Gamma_A}\Big)\right] \leqslant \frac{1}{\operatorname{tr}P_A\Gamma_A}\cdot\operatorname{tr}_{A'}\left[P'_{A'}\,\Gamma_{X'}\otimes\Gamma_{A'}\right], \tag{7.39}$$

using the fact that $P_A\Gamma_AP_A = \Gamma_A^{1/2}P_A\Gamma_A^{1/2} \leqslant \Gamma_A$ (because $[P_A,\Gamma_A]=0$) and also with the fact that $\Phi_{XA\to X'A'}$ is $\Gamma$-sub-preserving. Then

$$\mathcal{T}_{X\to X'}(\Gamma_X) \leqslant \frac{\operatorname{tr}P'_{A'}\Gamma_{A'}}{\operatorname{tr}P_A\Gamma_A}\,\Gamma_X\ , \tag{7.40}$$

which completes the proof. ∎

*Proof of Corollary 7.10.* First, let $A$, $A'$, $P_A$, $P'_{A'}$, $\Gamma_A$, $\Gamma_{A'}$ and $\Phi_{XA\to X'A'}$ satisfy the conditions of the maximization (7.34). Let $\mathcal{T}_{X\to X'}$ the mapping given by Proposition 7.9 Observe that $\big\|\Gamma_{X'}^{-1/2}\mathcal{T}_{X\to X'}(\Gamma_X)\Gamma_{X'}^{-1/2}\big\|_\infty \leqslant (\operatorname{tr}P'_{A'}\Gamma_{A'})/(\operatorname{tr}P_A\Gamma_A)$. Define $\hat\rho_{X'R} = \mathcal{T}_{X\to X'}(\sigma_{XR})$, so that $\mathcal{T}_{X\to X'}$ is a valid candidate in the expression (7.12) for $\bar D_{R\to X'}(\hat\rho_{X'R}\,\|\,\Gamma_R,\Gamma_{X'})$. Hence

$$\bar D^{\epsilon}_{R\to X'}(\rho_{X'R}\,\|\,\Gamma_R,\Gamma_{X'}) \geqslant \bar D_{R\to X'}(\hat\rho_{X'R}\,\|\,\Gamma_R,\Gamma_{X'}) \geqslant -\log\frac{\operatorname{tr}P'_{A'}\Gamma_{A'}}{\operatorname{tr}P_A\Gamma_A}\ , \tag{7.41}$$

as $P(\hat\rho_{X'R},\rho_{X'R}) \leqslant \epsilon$.

To show that equality is achieved in (7.34), let $\hat\rho_{X'R}$ with $P(\hat\rho_{X'R},\rho_{X'R}) \leqslant \epsilon$ and let $\mathcal{T}_{X\to X'}$ be valid optimization candidates in (7.31), as well as in (7.12) for $\bar D_{R\to X'}(\hat\rho_{X'R}\,\|\,\Gamma_R,\Gamma_{X'})$, which achieve the optimal value $y = \bar D^{\epsilon}_{R\to X'}(\rho_{X'R}\,\|\,\Gamma_R,\Gamma_{X'}) = \bar D_{R\to X'}(\hat\rho_{X'R}\,\|\,\Gamma_R,\Gamma_{X'})$, that is such that $-\log\big\|\Gamma_{X'}^{-1/2}\mathcal{T}_{X\to X'}(\Gamma_X)\Gamma_{X'}^{-1/2}\big\|_\infty = y$. Then $\mathcal{T}_{X\to X'}(\Gamma_X) \leqslant 2^{-y}\,\Gamma_{X'}$, and this mapping satisfies the conditions of item (i) of Proposition 5.2. Let $A = A'$ be a qubit system with $P_A = |0\rangle\langle 0|_A$, $P'_{A'} = |1\rangle\langle 1|_{A'}$, and $\Gamma_A = \Gamma_{A'} = g_0\,|0\rangle\langle 0|_A + g_1\,|1\rangle\langle 1|_A$, with $g_0/g_1 = 2^y$. In virtue of item (iii) of Proposition 5.2, there exists a trace-nonincreasing, completely positive map $\Phi_{XA\to X'A'}$ such that $\Phi_{XA\to X'A'}(\Gamma_X\otimes\Gamma_A) \leqslant \Gamma_{X'}\otimes\Gamma_{A'}$ and which satisfies $\Phi_{XA\to X'A'}((\cdot)\otimes|0\rangle\langle 0|_A) = \mathcal{T}_{X\to X'}(\cdot)\otimes|1\rangle\langle 1|_{A'}$. Then

$$\Phi_{XA\to X'A'}(\sigma_{XR}\otimes|0\rangle\langle 0|_A) = \hat\rho_{X'R}\otimes|1\rangle\langle 1|_{A'}\ . \tag{7.42}$$

Recalling that $P(\hat\rho_{X'R},\rho_{X'R}) \leqslant \epsilon$ shows that (7.35) holds, such that all the conditions of the maximization (7.34) are satisfied, and the achieved value is indeed $-\log[(\operatorname{tr}P'_{A'}\Gamma_{A'})/(\operatorname{tr}P_A\Gamma_A)] = -\log(g_2/g_1) = y$. ∎

## 7.3 Special cases; relation to known entropy measures

### 7.3.1 Special states and mappings

First, we show that we can recover the standard max-entropy if we restrict $\rho_{X'R}$ to live inside a subspace where the $\Gamma$ operator is uniform. If the $\Gamma$ operators represent a system with a Hamiltonian in contact with a heat bath, then the constant terms which appear can be simply interpreted as an energy normalization issue as discussed in Section 5.2.5. The main result in Chapter 6 can also be seen as a consequence of this proposition.

***Proposition 7.11** (Mapping within a $\Gamma$ operator eigenspace). Let $P_X$ be a subspace of the eigenspace of $\Gamma_X$ corresponding to the eigenvalue $g$, and let $P'_{X'}$ be a subspace of the eigenspace of $\Gamma_{X'}$ corresponding to the eigenvalue $g'$. Let $\rho_{X'R}$ be any state inside $P'_{X'} \otimes P_R$, i.e. representing a specific mapping of an input state in $P_R$ to a state in $P'_{X'}$. Then*

$$\begin{aligned}\bar{D}_{R\to X'}(\rho_{X'R} \,\|\, \Gamma_R, \Gamma_{X'}) &= \log(g') - \log(g) - \log\left\| \mathrm{tr}_R\, \rho_R^{-1/2} \rho_{X'R} \rho_R^{-1/2} \right\|_\infty \\ &= \log(g') - \log(g) - H_{\max,0}(E\,|\,X')_\rho\,, \end{aligned} \tag{7.43}$$

*where $|\rho\rangle_{X'RE}$ is any purification of $\rho_{X'R}$.* *(Proof on page 157.)*

This proposition in fact has a natural interpretation if $\Gamma_X$ and $\Gamma_{X'}$ represent systems with Hamiltonians $H_X$, $H_{X'}$ which are interacting with a heat bath at inverse temperature $\beta$ a Hamiltonian on $X, X'$ as $\Gamma_X = e^{-\beta H_X}$ and $\Gamma_{X'} = e^{-\beta H_{X'}}$. Then the eigenstates of $\Gamma_X$ and $\Gamma_{X'}$ are energy levels, and a logical process mapping states within one energy eigenspace to another have two contributions from the work cost: one from the difference in energy, the other from the irreversibility of the logical process studied in the last chapter.

The next proposition concerns identity mappings. It is a property that one would expect very naturally: if the process matrix corresponds to the identity mapping, and if the hidden information operators coincide, then the process has zero compression yield (which may be interpreted as requiring zero work). This may seem like a triviality, but it is in fact not so obvious: because the coherent relative entropy is a function of the process matrix only, then the implementation can choose to implement whatever process they like on the complement of the support of the input state. In other words, this proposition tells us that there is no way to extract work by exploiting this complementary subspace when performing the identity map on the support of $\sigma_X$.

***Proposition 7.12** (Identity mapping). Let $\mathcal{I}_{X\to X'}$ be the identity map from a system $X$ to a system $X' \simeq X$. Assume that $\Gamma_{X'} = \mathcal{I}_{X\to X'}(\Gamma_X)$. Let $\sigma_X$ be any state on $X$, $\sigma_{XR}$ any purification on a system $R$, and let $|\rho\rangle_{X'R}$ be the process* *matrix of the identity process applied on $\sigma_X$, i.e. $\rho_{X'R} = \mathcal{I}_{X\to X'}(\sigma_{XR})$. Then*

$$\bar{D}_{R\to X'}(\rho_{X'R} \,\|\, \Gamma_R, \Gamma_{X'}) = 0 \,. \tag{7.44}$$

*(Proof on page 158.)*

The following proposition tells us that one may map the $\Gamma_X / \operatorname{tr}\Gamma_X$ state to the $\Gamma_{X'} / \operatorname{tr}\Gamma_{X'}$ state in however way one wants, i.e. regardless of the logical process, and yet in any case the coherent relative entropy is given by the ratio $\operatorname{tr}\Gamma_{X'} / \operatorname{tr}\Gamma_R$. This is a consequence of allowing any $\Gamma$-preserving maps to be performed for free, and this ratio comes about from the normalization of the respective input and output states.

***Proposition 7.13*** *(Mapping normalized "hidden information states"). Let $\rho_{X'R}$ be a bipartite quantum state with reduced states $\rho_R = \Gamma_R / \operatorname{tr}\Gamma_R$ and $\rho_{X'} = \Gamma_{X'} / \operatorname{tr}\Gamma_{X'}$. The correlations between the two systems may be arbitrary. Then*

$$\bar{D}_{R\to X'}(\rho_{X'R} \,\|\, \Gamma_R, \Gamma_{X'}) = \log\operatorname{tr}\Gamma_{X'} - \log\operatorname{tr}\Gamma_R \,. \tag{7.45}$$

*(Proof on page 159.)*

In general, the coherent relative entropy depends on the precise logical process used to map the input and output states (as already discussed in Section 6.4 for the special case of a trivial Hamiltonian). However, there are some classes of states for which the coherent relative entropy depends only on the input and output state. We may combine Proposition 7.13 with the convenient Proposition 7.6 to construct such a class of states. Furthermore, the coherent relative entropy may be written as the difference of a potential. These states will prove invaluable in the next chapter, when we study the emergence of thermodynamics.

***Corollary 7.14*** *(Mapping Commuting Projections of $\Gamma$ Operators). Let $P_R$ and $P'_{X'}$ be two projectors with $[P_R, \Gamma_R] = 0$ and $[P'_{X'}, \Gamma_{X'}] = 0$. Let $\rho_{X'R}$ be a bipartite quantum state with reduced states $\rho_R = (P_R\Gamma_R P_R) / \operatorname{tr}(P_R\Gamma_R)$ and $\rho_{X'} = (P'_{X'}\Gamma_{X'}P'_{X'}) / \operatorname{tr}(P'_{X'}\Gamma_{X'})$. Then*

$$\bar{D}_{R\to X'}(\rho_{X'R} \,\|\, \Gamma_R, \Gamma_{X'}) = \log\operatorname{tr}(P'_{X'}\Gamma_{X'}) - \log\operatorname{tr}(P_R\Gamma_R) \,. \tag{7.46}$$

*(Proof on page 160.)*

We note that for this special type of states we have the nice expression for their relative entropy to $\Gamma$. If $\Gamma \geqslant 0$ and $P$ is a projector with $[P, \Gamma] = 0$,

then

$$D\left(\frac{P\Gamma}{\operatorname{tr} P\Gamma} \middle\| \Gamma\right) = -\log \operatorname{tr} P\Gamma\ . \tag{7.47}$$

Moreover, these states are precisely those type of states which we allowed on battery systems in Section 5.3.2.

We now proceed to the proofs of the above propositions.

*Proof of Proposition 7.11.* In fact, this is a special case of Proposition 7.6. Here's an independent proof, just for fun. Consider the primal candidate $T_{X'R} = \rho_R^{-1/2}\rho_{X'R}\rho_R^{-1/2}$. Conditions (7.2-b) and (7.2-c) are automatically satisfied. Choosing $\alpha = \|\Gamma_{X'}^{-1/2} \operatorname{tr}_R [T_{X'R}\Gamma_R]\, \Gamma_{X'}^{-1/2}\|_\infty$ satisfies (7.2-a) (because $\rho_{X'}$ is in the support of $\Gamma_{X'}$) and we have

$$\begin{aligned}\alpha &= \left\|\operatorname{tr}_R\left[\left(\Gamma_{X'}^{-1/2}\otimes\Gamma_R^{1/2}\right)\rho_R^{-1/2}\rho_{X'R}\rho_R^{-1/2}\left(\Gamma_{X'}^{-1/2}\otimes\Gamma_R\right)\right]\right\|_\infty\\ &= g'^{-1}g\cdot\left\|\operatorname{tr}_R\left[\rho_R^{-1/2}\rho_{X'R}\rho_R^{-1/2}\right]\right\|_\infty\ ,\end{aligned} \tag{7.48}$$

because $\rho_R^{-1/2}\rho_{X'R}\rho_R^{-1/2}$ is entirely contained in the subspace $P'_{X'}\otimes P_R$, which is an eigenspace of $\Gamma_{X'}^{-1/2}\otimes\Gamma_R^{1/2}$ corresponding to the eigenvalue $g'^{-1/2}g^{1/2}$. Hence

$$\bar{D}_{R\to X'}(\rho_{X'R}\,\|\,\Gamma_R,\Gamma_{X'}) \geqslant -\log\left(g'^{-1}g\cdot\left\|\operatorname{tr}_R\rho_R^{-1/2}\rho_{X'R}\rho_R^{-1/2}\right\|_\infty\right)\ . \tag{7.49}$$

We'll now consider now the dual problem. First, let $\tau_{X'}\geqslant 0$ with $\operatorname{tr}\tau_{X'} = 1$ such that $\operatorname{tr}\left[\tau_{X'}\operatorname{tr}_R\rho_R^{-1/2}\rho_{X'R}\rho_R^{-1/2}\right] = \left\|\operatorname{tr}_R\rho_R^{-1/2}\rho_{X'R}\rho_R^{-1/2}\right\|_\infty$, and note that $\tau_{X'}$ resides inside the subspace $P'_{X'}$. Then, define $\omega_{X'} = g'^{-1}\tau_{X'}$ and note that

$$\operatorname{tr}\omega_{X'}\Gamma_{X'} = g'^{-1}\operatorname{tr}\tau_{X'}\Gamma_{X'} = 1\ , \tag{7.50}$$

because $\tau_{X'}$ resides inside $P'_{X'}$. Thus condition (7.3-a) is satisfied. Now let $X_R = 0$ and $Z_{X'R} = g\rho_R^{-1}\otimes\omega_{X'}$, and see that

$$\rho_R^{1/2}Z_{X'R}\rho_R^{1/2} = g\omega_{X'}\otimes\Pi_R^{\rho_R}\otimes\omega_{X'}\leqslant\Gamma_R\otimes\omega_{X'}\ . \tag{7.51}$$

It remains to what the attained value is,

$$\begin{aligned}\operatorname{tr}\left[Z_{X'R}\rho_{X'R}\right] = g\operatorname{tr}\left[\rho_R^{-1}\otimes\omega_{X'}\cdot\rho_{X'R}\right] &= gg'^{-1}\operatorname{tr}\left[\tau_{X'}\operatorname{tr}_R\rho_R^{-1/2}\rho_{X'R}\rho_R^{-1/2}\right]\\ &= gg'^{-1}\left\|\operatorname{tr}_R\rho_R^{-1/2}\rho_{X'R}\rho_R^{-1/2}\right\|_\infty\ .\end{aligned} \tag{7.52}$$

We now have the converse bound for (7.49) and hence (7.49) becomes an equality.

It remains to see that $\log\left\|\operatorname{tr}_R\rho_R^{-1/2}\rho_{X'R}\rho_R^{-1/2}\right\|_\infty = H_{\max,0}(E\,|\,X')_\rho$ to complete the proof.

This is seen with a simple calculation (Faist *et al.*, 2015a),

$$\begin{aligned}\left\|\mathrm{tr}_R\, \rho_R^{-1/2}\rho_{X'R}\rho_R^{-1/2}\right\|_\infty = \left\|\mathrm{tr}_{RE}\, \rho_R^{-1/2}\rho_{X'RE}\rho_R^{-1/2}\right\|_\infty = \left\|\mathrm{tr}_{RE}\, \rho_{X'E}^{-1/2}\rho_{X'RE}\rho_{X'E}^{-1/2}\right\|_\infty \\ = \left\|\mathrm{tr}_E\, \rho_{X'E}^{-1/2}\rho_{X'E}\rho_{X'E}^{-1/2}\right\|_\infty = \left\|\mathrm{tr}_E\, \Pi_{X'E}^{\rho_{X'E}}\right\|_\infty = 2^{H_{\max,0}(E\,|\,X')_\rho}\ ,\end{aligned} \tag{7.53}$$

where we have used the fact that $|\rho\rangle_{X'RE} = \rho_{X'E}^{1/2}\,|\Phi^\rho_{X'E:R}\rangle = \rho_R^{1/2}\,|\Phi^\rho_{X':E}\rangle$ for the canonical entangled state $|\Phi^\rho_{X'E:R}\rangle$ in the Schmidt basis of $\rho$ with respect to the bipartition $X'E : R$ (Section 3.1.5), and so $\rho_R^{-1/2}|\rho\rangle_{X'RE} = |\Phi^\rho_{X'E:R}\rangle = \rho_{X'E}^{-1/2}|\rho\rangle_{X'RE}$. Also the last step follows because

$$2^{H_{\max,0}(E\,|\,X')_\rho} = \max_{\tau_{X'}} \mathrm{tr}\big[\Pi_{X'E}^{\rho_{X'E}}\ (\mathbb{1}_E \otimes \tau_{X'})\big] = \left\|\mathrm{tr}_E\, \Pi_{X'E}^{\rho_{X'E}}\right\|_\infty\ , \tag{7.54}$$

where the first maximization, by definition of $H_{\max,0}$, is taken over (normalized) density operators. ∎

*Proof of Proposition 7.12.* Let $|\Phi\rangle_{X'R} = \sum_i |i\rangle_{X'}|i\rangle_R$ the unnormalized maximally entangled state on $X'$ and $R$ with an appropriate choice of bases such that $\rho_{X'R} = \rho_R^{1/2}\Phi_{X'R}\rho_R^{1/2}$ and $\mathrm{tr}_{X'}\,\Phi_{X'R} = \mathbb{1}_R$.

First we show that $\bar{D}_{R\to X'}(\rho_{X'R}\,\|\,\Gamma_R, \Gamma_{X'}) \geqslant 0$. Consider the primal candidates $T_{X'R} = \Phi_{X'R}$ and $\alpha = 1$. Note that this $T_{X'R}$ is the Choi matrix of the identity mapping. This clearly satisfies both (7.2-b) and (7.2-c). Also, $\mathrm{tr}_R\left[T_{X'R}\Gamma_R\right] = \Gamma_{X'}$ satisfying (7.2-a). Hence

$$\bar{D}_{R\to X'}(\rho_{X'R}\,\|\,\Gamma_R, \Gamma_{X'}) \geqslant 0\ . \tag{7.55}$$

The reverse inequality is the tricky part. There may not be an optimal choice of dual variables. The best we can do in general is to come up with a sequence of choices for dual candidates whose attained value converges to 1. For any $\mu > 0$, let

$$Z_{X'R} = \mu\rho_R^{-1/2}\Phi_{X'R}\rho_R^{-1/2}\ ; \qquad\qquad \omega_{X'} = \left(\mathrm{tr}\left[\Pi_{X'}^{\rho_{X'}}\Gamma_{X'}\right]\right)^{-1}\Pi_{X'}^{\rho_{X'}}\ . \tag{7.56}$$

Then $\mathrm{tr}\,(\omega_{X'}\Gamma_{X'}) = 1$, satisfying the dual constraint (7.3-a). Let's now study (7.3-b):

$$\rho_R^{1/2}Z_{X'R}\rho_R^{1/2} - \Gamma_R\otimes\omega_{X'} = \mu\Pi_R^{\rho_R}\,\Phi_{X'R}\,\Pi_R^{\rho_R} - \left(\mathrm{tr}\left[\Pi_{X'}^{\rho_{X'}}\Gamma_{X'}\right]\right)^{-1}\Gamma_R\otimes\Pi_{X'}^{\rho_{X'}}\ . \tag{7.57}$$

The operator $\Pi_R^{\rho_R}\Phi_{X'R}\Pi_R^{\rho_R}$ is a rank-1 positive operator with support within $\Pi_R^{\rho_R}\otimes\Pi_{X'}^{\rho_{X'}}$, and its nonzero eigenvalue is given by

$$\mathrm{tr}\left(\Pi_R^{\rho_R}\Phi_{X'R}\Pi_R^{\rho_R}\right) = \mathrm{rank}\,\rho_R\ . \tag{7.58}$$

Let $r = \mathrm{rank}\,\rho_R$. We then have $\Pi_R^{\rho_R}\Phi_{X'R}\Pi_R^{\rho_R} \leqslant r\Pi_R^{\rho_R}\otimes\Pi_{X'}^{\rho_{X'}}$ and we may continue our calculation:

$$(7.57) \leqslant \left(\mu r\Pi_R^{\rho_R} - \left(\mathrm{tr}\left[\Pi_{X'}^{\rho_{X'}}\Gamma_{X'}\right]\right)^{-1}\Gamma_R\right)\otimes\Pi_{X'}^{\rho_{X'}}\ . \tag{7.59}$$

Now, let $P_R$ be the projector onto the eigenspaces associated to the positive (or null) eigen-values of the operator $\left(\mu r\Pi_R^{\rho_R} - \left(\operatorname{tr}\left[\Pi_{X'}^{\rho_{X'}}\Gamma_{X'}\right]\right)^{-1}\Gamma_R\right)$, and let

$$X_R = P_R\left(\mu r\Pi_R^{\rho_R} - \left(\operatorname{tr}\left[\Pi_{X'}^{\rho_{X'}}\Gamma_{X'}\right]\right)^{-1}\Gamma_R\right)P_R\,. \tag{7.60}$$

Then

$$(7.59) \leqslant X_R \otimes \mathbb{1}_{X'}\,. \tag{7.61}$$

Hence, for any $\mu > 0$, this choice of dual variables satisfies the dual constraints. The value attained by this choice of variables is given by

$$\operatorname{tr}\left[Z_{X'R}\rho_{X'R}\right] - \operatorname{tr}X_R = \mu\operatorname{tr}\left[\Pi_R^{\rho_R}\Phi_{X'R}\Pi_R^{\rho_R}\Phi_{X'R}\right] - \operatorname{tr}X_R\,. \tag{7.62}$$

As the object $\Pi_R^{\rho_R}\Phi_{X'R}\Pi_R^{\rho_R}$ is rank-1, we have thanks to (7.58) that $\operatorname{tr}\left[\left(\Pi_R^{\rho_R}\Phi_{X'R}\Pi_R^{\rho_R}\right)^2\right] = \left(\operatorname{tr}\Pi_R^{\rho_R}\Phi_{X'R}\Pi_R^{\rho_R}\right)^2 = r^2$. Then

$$\begin{aligned}
(7.62) &= \mu r^2 - \operatorname{tr}X_R \\
&= \mu r^2 - \mu r\operatorname{tr}\left(P_R\Pi_R^{\rho_R}\right) + \left(\operatorname{tr}\left[\Pi_{X'}^{\rho_{X'}}\Gamma_{X'}\right]\right)^{-1}\operatorname{tr}\left(P_R\Gamma_R\right) \\
&\geqslant \mu r^2 - \mu r\operatorname{tr}\left(\Pi_R^{\rho_R}\right) + \left(\operatorname{tr}\left[\Pi_{X'}^{\rho_{X'}}\Gamma_{X'}\right]\right)^{-1}\operatorname{tr}\left(P_R\Gamma_R\right) \\
&\geqslant \left(\operatorname{tr}\left[\Pi_{X'}^{\rho_{X'}}\Gamma_{X'}\right]\right)^{-1}\operatorname{tr}\left(P_R\Gamma_R\right)\,,
\end{aligned} \tag{7.63}$$

recalling that $\operatorname{tr}\Pi_R^{\rho_R} = \operatorname{rank}\rho_R = r$.

Next episode: the Lemma awakens. Take $A = \mu r\Pi_R^{\rho_R}$ and $B = \left(\operatorname{tr}\left[\Pi_{X'}^{\rho_{X'}}\Gamma_{X'}\right]\right)^{-1}\Gamma_R$; Lemma A.2 then asserts that there exists a constant $c$ independent of $\mu$ such that

$$\Pi_R^{\rho_R} \leqslant P_R + \frac{c}{\mu}\mathbb{1}\,. \tag{7.64}$$

Hence,

$$(7.63) \geqslant \left(\operatorname{tr}\left[\Pi_{X'}^{\rho_{X'}}\Gamma_{X'}\right]\right)^{-1}\left(\operatorname{tr}\left[\Pi_R^{\rho_R}\Gamma_R\right] - \frac{c}{\mu}\operatorname{tr}\Gamma_R\right) = 1 - O\left(1/\mu\right)\,. \tag{7.65}$$

Taking $\mu \to \infty$ yields successive feasible dual candidates with attained objective value converging to 1, hence proving that

$$\bar{D}_{R\to X'}(\rho_{X'R}\,\|\,\Gamma_R, \Gamma_{X'}) \leqslant 0\,. \qquad \blacksquare$$

*Proof of Proposition 7.13.* This is readily seen in terms of our framework and invoking Proposition 7.3; for completeness we give a proof based on the definition of $\bar{D}(\cdot)$. Take any $T_{X'R}$ satisfying $\rho_R^{1/2}T_{X'R}\rho_R^{1/2} = \rho_{X'R}$ and $\operatorname{tr}_{X'}T_{X'R} \leqslant \mathbb{1}_R$. Then since $\rho_R\operatorname{tr}(\Gamma_R) = \Gamma_R$ and $\rho_{X'}\operatorname{tr}(\Gamma_{X'}) = \Gamma_{X'}$, we have

$$\begin{aligned}
\operatorname{tr}_R T_{X'R}\Gamma_R = \operatorname{tr}(\Gamma_R)\operatorname{tr}_R\rho_R^{1/2}T_{X'R}\rho_R^{1/2} &= \operatorname{tr}(\Gamma_R)\operatorname{tr}_R\rho_{X'R} = \operatorname{tr}(\Gamma_R)\rho_{X'} \\
&= \operatorname{tr}(\Gamma_R)\operatorname{tr}(\Gamma_{X'})^{-1}\Gamma_{X'}\,,
\end{aligned} \tag{7.66}$$

and thus $\|\Gamma_{X'}^{-1/2}\,\mathrm{tr}_R\,[T_{X'R}\Gamma_R]\,\Gamma_{X'}^{-1/2}\|_\infty = \mathrm{tr}\,(\Gamma_R)\,\mathrm{tr}\,(\Gamma_{X'})^{-1}$. This argument holds in particular for the optimal solution of the primal problem. ■

*Proof of Corollary 7.14.* This is a straightforward combination of Proposition 7.13 with Proposition 7.6. ■

### 7.3.2 Relation to known entropy measures

In this section, we relate our new measure to other established entropic measures. We recover by studying different special cases the conditional max-entropy and min-entropy, as well as the min- and max-relative entropy. Recall the definitions of these measures given in Section 3.2.3.

We have already seen how to recover the max-entropy in the last chapter. We state it again here for completeness.

***Proposition 7.15** (Conditional max-entropy with trivial hidden information operator). Let $|\rho\rangle_{X'RE}$ be any pure state on systems $R$, $X'$, and $E$. Then*

$$\bar{D}_{R\to X'}(\rho_{X'R}\,\|\,\mathbb{1}_R, \mathbb{1}_{X'}) = -H_{\max,0}(E\,|\,X')_\rho = H_{\min}(E\,|\,R)_{\rho|\rho}\,. \tag{7.67}$$

*Proof of Proposition 7.15.* This follows because the semidefinite program reduces exactly to the situation in Faist *et al.* (2015a) which was covered in Chapter 6. It is also a special case of Proposition 7.11. ■

The min- and max-relative entropies already have known connections to thermodynamics (Åberg, 2013; Horodecki and Oppenheim, 2013a; Brandão *et al.*, 2015) in terms of work cost of erasure and work yield of formation of a state in the presence of a heat bath. These results are recovered here.

***Proposition 7.16** (Recovering the min- and max-relative entropies). The min- and max-relative entropies are recovered with a trivial input or output state:*

$$\bar{D}_{R\to\varnothing}(\rho_R\,\|\,\Gamma_R, 1) = D_{\min,0}(\rho_R\,\|\,\Gamma_R) = D_{\mathrm{H}}^{\eta=1}(\rho_R\,\|\,\Gamma_R)\,; \tag{7.68}$$
$$\bar{D}_{\varnothing\to X'}(\rho_{X'}\,\|\,1, \Gamma_X')= -D_{\max}(\rho_{X'}\,\|\,\Gamma_{X'})\,. \tag{7.69}$$

*Proof of Proposition 7.16.* For the case (7.68), write

$$\bar{D}_{R\to\varnothing}(\rho_R\,\|\,\Gamma_R, 1) = \min\left\{\mathrm{tr}\,[T_R\Gamma_R] : T_R \leqslant \mathbb{1}\ \text{ and }\ \rho_R^{1/2}T_R\rho_R^{1/2} = \rho_R\right\}\,. \tag{7.70}$$

The last condition is equivalent to $\mathrm{tr}\left[T_R\rho_R\right] = 1$, making this minimization exactly the same as the one for $D_{\mathrm{H}}^{\eta=1}(\rho_R \,\|\, \Gamma_R)$. Let's now prove that this is also equal to $D_{\mathrm{min},0}(\rho_R \,\|\, \Gamma_R)$. The choice $T_R = \Pi_R^{\rho}$ provides a feasible primal candidate achieving the value $\mathrm{tr}\left(\Pi_R^{\rho}\Gamma_R\right) = 2^{-D_{\mathrm{min},0}(\rho_R \,\|\, \Gamma_R)}$. In the dual problem, for any $\mu > 0$ let $Z_R = \mu\Pi_R$ and $\omega_{X'} = 1$. Let $P_R$ be the projector onto the eigenspaces associated with the positive (or null) eigenvalues of $(\mu\rho_R - \Gamma_R)$, and let $X_R = P_R\left(\mu\rho_R - \Gamma_R\right)P_R$. Then the dual constraints (7.3-a) and (7.3-b) are clearly satisfied. The attained value is

$$\begin{aligned}\mathrm{tr}\, Z_R\rho_R - \mathrm{tr}\, X_R = \mu\,\mathrm{tr}\,\rho_R - \mu\,\mathrm{tr}\left(P_R\rho_R\right) + \mathrm{tr}\left(P_R\Gamma_R\right) \geqslant \mathrm{tr}\left(P_R\Gamma_R\right)\\ \geqslant \mathrm{tr}\left(\Pi_R\Gamma_R\right) - O\left(1/\mu\right)\,,\end{aligned} \tag{7.71}$$

where we have used Lemma A.2 in the last step. If we take $\mu \to \infty$ we get successive feasible dual candidates whose attained value approaches $\mathrm{tr}\left(\Pi_R\Gamma_R\right) = 2^{-D_{\mathrm{min},0}(\rho_R \,\|\, \Gamma_R)}$; hence this is the optimal value of the semidefinite program.

Let's now prove equality (7.69). The choice $T_{X'} = \rho_{X'}$ and $\alpha = \|\Gamma_{X'}^{-1/2}\rho_{X'}\Gamma_{X'}^{-1/2}\|_\infty = 2^{D_{\mathrm{max}}(\rho_{X'} \,\|\, \Gamma_{X'})}$ clearly satisfies the primal constraints, and thus

$$2^{-\bar{D}_{\varnothing\to X'}(\rho_{X'} \,\|\, 1,\Gamma_{X'})} \leqslant 2^{D_{\mathrm{max}}(\rho_{X'} \,\|\, \Gamma_{X'})}\,. \tag{7.72}$$

By properties of the infinity norm, there exists a $\tau_{X'} \geqslant 0$ with $\mathrm{tr}\,\tau_{X'} = 1$ such that $\|\Gamma_{X'}^{-1/2}\rho_{X'}\Gamma_{X'}^{-1/2}\|_\infty = \mathrm{tr}\big[\tau_{X'}\cdot\Gamma_{X'}^{-1/2}\rho_{X'}\Gamma_{X'}^{-1/2}\big]$. Let $\omega_{X'} = \Gamma_{X'}^{-1/2}\tau_{X'}\Gamma_{X'}^{-1/2}$, $Z_{X'} = \omega_{X'}$ and $X = 0$. Then the dual constraints are trivially satisfied and the attained value is

$$\mathrm{tr}\left[Z_{X'}\rho_{X'}\right] = \mathrm{tr}\left[\Gamma_{X'}^{-1/2}\tau_{X'}\Gamma_{X'}^{-1/2}\rho_{X'}\right] = 2^{D_{\mathrm{max}}(\rho_{X'} \,\|\, \Gamma_{X'})}\,. \tag{7.73}$$

The primal and dual candidates achieve the same value, and hence this is the optimal solution to the semidefinite program. This completes the proof of (7.69). ■

## 7.4 Bounds on the coherent relative entropy

At this point, we further characterize the coherent relative entropy with bounds in terms of simpler quantities depending only on the input and output states.

The first bound applies to product states, and is given in terms of min- and max-relative entropies of input and output. Physically, it asserts that a possible strategy for implementing the product state process matrix is to completely erase the input state (at a cost given by the min-relative entropy), and subsequently prepare the required output state (at a yield given by the max-relative entropy).

***Proposition 7.17*** *(Coherent relative entropy for product states).* *For states* $\sigma_R$

*and $\rho_{X'}$, we have*

$$\bar{D}_{R\to X'}(\sigma_R \otimes \rho_{X'} \,\|\, \Gamma_R, \Gamma_{X'}) \geqslant D_{\mathrm{min},0}(\sigma_R \,\|\, \Gamma_R) - D_{\mathrm{max}}(\rho_{X'} \,\|\, \Gamma_{X'}) \ . \quad (7.74)$$

In order to formulate further bounds on the coherent relative entropy, we introduce a generalization of the *Rob entropy* or *smooth S-entropy* (Vitanov *et al.*, 2013):

$$D_{\mathrm{r}}(\rho \,\|\, \Gamma) = -\log \|\rho^{-1/2}\Gamma\rho^{-1/2}\|_\infty = -\log \min\{\nu : \nu\rho \geqslant \Pi^\rho\Gamma\Pi^\rho\} \ ; \quad (7.75)$$

$$D_{\mathrm{r}}^{\epsilon}(\rho \,\|\, \Gamma) = \max_{\hat\rho \approx_\epsilon \rho} D_{\mathrm{r}}(\hat\rho \,\|\, \Gamma) \ . \quad (7.76)$$

The following series of results are useful mostly for technical reasons, and don't necessarily represent any particular physical situation.

***Proposition 7.18***. *The coherent relative entropy may be upper bounded as*

$$\bar{D}_{R\to X'}(\rho_{X'R} \,\|\, \Gamma_R, \Gamma_{X'}) \leqslant D_{\mathrm{max}}(\rho_R \,\|\, \Gamma_R) - D_{\mathrm{max}}(\rho_{X'} \,\|\, \Gamma_{X'}) \quad (7.77)$$

*(Proof on page 164.)*

***Proposition 7.19***. *We have the upper bound*

$$\bar{D}_{R\to X'}(\rho_{X'R} \,\|\, \Gamma_R, \Gamma_{X'}) \leqslant D(\rho_R \,\|\, \Gamma_R) - D(\rho_{X'} \,\|\, \Gamma_{X'}) \ . \quad (7.78)$$

*(Proof on page 164.)*

***Proposition 7.20***. *We have the lower bound*

$$\bar{D}_{R\to X'}(\rho_{X'R} \,\|\, \Gamma_R, \Gamma_{X'}) \geqslant D_{\mathrm{r}}(\rho_R \,\|\, \Gamma_R) - D_{\mathrm{max}}(\rho_{X'} \,\|\, \Gamma_{X'}) \ . \quad (7.79)$$

*(Proof on page 165.)*

The quantity $D_{\mathrm{r}}(\cdot \,\|\, \cdot)$, when smoothed, is essentially equal to the min-relative entropy: These two differ by a term which is logarithmic in the failure probability. In this way, the smooth quantity $D_{\mathrm{r}}^{\epsilon}(\cdot \,\|\, \cdot)$ may be related to a better known quantity with an operational interpretation.

***Proposition 7.21*** *(Smooth $D_{\mathrm{r}}(\cdot \,\|\, \cdot)$ and $D_{\mathrm{min},0}(\cdot \,\|\, \cdot)$). Let $\epsilon > 0$. Then*

$$D_{\mathrm{r}}^{\epsilon}(\rho \,\|\, \Gamma) \geqslant D_{\mathrm{min},0}(\rho \,\|\, \Gamma) + \log \epsilon' \ , \quad (7.80)$$

*where $\epsilon' = \epsilon^2 / \left(2 + \epsilon^2\right)$, or equivalently, $\epsilon = \sqrt{2\epsilon'/(1-\epsilon')}$.* *(Proof on page 165.)*

The following proposition gives a lower bound to the smooth coherent relative entropy. This will prove crucial to the proof of the asymptotic equipartition theorem.

***Proposition 7.22*** *(Smooth lower bound). Let $\epsilon', \epsilon'' \geqslant 0$ and $\epsilon''' > 0$. Let $\epsilon \geqslant 2\sqrt{2\epsilon'} + 2\sqrt{2(\epsilon'' + \epsilon''')}$. Then*

$$\bar{D}^{\epsilon}_{R\to X'}(\rho_{X'R} \,\|\, \Gamma_R, \Gamma_{X'}) \geqslant D^{\epsilon''}_{\mathrm{min},0}(\rho_R \,\|\, \Gamma_R) - D^{\epsilon'}_{\mathrm{max}}(\rho_{X'} \,\|\, \Gamma_{X'}) + \log \frac{\epsilon'''^2}{2 + \epsilon'''^2} \,. \tag{7.81}$$

*(Proof on page 166.)*

The last bound in this section is devoted to the particular class of states which have been already been singled out as battery states, generalizing Corollary 7.14 to the smooth coherent relative entropy. It is shown, in particular, that the smoothing for these states may only yield a rather small advantage which is essentially proportional to $\epsilon$.

***Proposition 7.23*** *(Smooth coherent relative entropy with commuting projections of $\Gamma$ operators). Let $P_R, P'_{X'}$ be projectors such that $[\Gamma_R, P_R] = 0$ and $[\Gamma_{X'}, P'_{X'}] = 0$. Let $\rho_{X'R}$ be such that $\rho_R = P_R\Gamma_R P_R / \operatorname{tr} P_R\Gamma_R$ and $\rho_{X'} = P'_{X'}\Gamma_{X'}P'_{X'} / \operatorname{tr} P'_{X'}\Gamma_{X'}$. Let $\epsilon \geqslant 0$. Then*

$$\bar{D}^{\epsilon}_{R\to X'}(\rho_{X'R} \,\|\, \Gamma_R, \Gamma_{X'}) = \log \frac{\operatorname{tr} P'_{X'}\Gamma_{X'}}{\operatorname{tr} P_R\Gamma_R} + f(\epsilon, \Gamma_R, \Gamma_{X'}) \,, \tag{7.82}$$

*where the error term $f(\epsilon, \Gamma_R, \Gamma_{X'})$ is bounded as*

$$0 \leqslant f(\epsilon, \Gamma_R, \Gamma_{X'}) \leqslant f_0(\epsilon, \Gamma_R) + f_0(\epsilon, \Gamma_{X'}) \,, \tag{7.83}$$

*where $f_0(\epsilon, \Gamma) = \epsilon \log(\operatorname{rank}\Gamma - 1) + \epsilon \|\log \Gamma\|_\infty + h(\epsilon)$ with the binary entropy $h(\epsilon) = -\epsilon \log \epsilon - (1-\epsilon)\log(1-\epsilon)$; furthermore $f_0(\epsilon, \Gamma) \to 0$ as $\epsilon \to 0$.*

*(Proof on page 167.)*

And now comes this section's worth of proofs.

*Proof of Proposition 7.17.* Choose $T_{X'R} = \Pi^\sigma_R \otimes \rho_{X'}$. This choice trivially satisfies (7.2-b). Also, $\sigma_R^{1/2} T_{X'R} \sigma_R^{1/2} = \sigma_R \otimes \rho_{X'}$ so (7.2-c) is also satisfied. As per Proposition 7.8, we have that $\operatorname{tr}_R T_{X'R}\Gamma_R$ lies in the support of $\Gamma_{X'}$ because $\rho_{X'}$ does so, and the optimal value of $\alpha$ corresponding to this $T_{X'R}$ is given by

$$\begin{aligned} \alpha &= \big\|\Gamma_{X'}^{-1/2} \operatorname{tr}_R\big[T_{X'R}\Gamma_R\big]\Gamma_{X'}^{-1/2}\big\|_\infty = \big\|\Gamma_{X'}^{-1/2} \operatorname{tr}_R\big[\big(\Pi^\sigma_R \otimes \rho_{X'}\big)\Gamma_R\big]\,\Gamma_{X'}^{-1/2}\big\|_\infty \\ &= \operatorname{tr}_R\big[\Pi^\sigma_R \Gamma_R\big]\, \big\|\Gamma_{X'}^{-1/2}\rho_{X'}\Gamma_{X'}^{-1/2}\big\|_\infty = 2^{-D_{\mathrm{min},0}(\sigma_R \,\|\, \Gamma_R)}\, 2^{D_{\mathrm{max}}(\rho_{X'} \,\|\, \Gamma_{X'})} \,. \end{aligned} \tag{7.84}$$

This choice of $\alpha$ and $T_{X'R}$ is feasible for $2^{-\bar{D}_{R\to X'}(\sigma_R\otimes\rho_{X'}\,\|\,\Gamma_R,\Gamma_{X'})}$, hence

$$\bar{D}_{R\to X'}(\sigma_R\otimes\rho_{X'}\,\|\,\Gamma_R,\Gamma_{X'}) \geqslant D_{\min,0}(\sigma_R\,\|\,\Gamma_R) - D_{\max}(\rho_{X'}\,\|\,\Gamma_{X'})\ . \qquad \blacksquare$$

*Proof of Proposition 7.18.* Consider an optimal solution $T_{X'R}$ and $\alpha$ for the primal semidefinite program. Then we have via the semidefinite constraints

$$\rho_{X'} = \mathrm{tr}_R\left[T_{X'R}\rho_R\right] \leqslant 2^{D_{\max}(\rho_R\,\|\,\Gamma_R)}\,\mathrm{tr}_R\left[T_{X'R}\Gamma_R\right] \leqslant \alpha 2^{D_{\max}(\rho_R\,\|\,\Gamma_R)}\Gamma_{X'}\ . \tag{7.85}$$

By definition, we have

$$2^{D_{\max}(\rho_{X'}\,\|\,\Gamma_{X'})} = \min\{\mu : \mu\Gamma_{X'} \geqslant \rho_{X'}\}\ , \tag{7.86}$$

and thus we see that $\alpha 2^{D_{\max}(\rho_R\,\|\,\Gamma_R)}$ is a candidate $\mu$ in this minimization. Hence $2^{D_{\max}(\rho_{X'}\,\|\,\Gamma_{X'})} \leqslant \alpha 2^{D_{\max}(\rho_R\,\|\,\Gamma_R)}$ and

$$\alpha \geqslant 2^{D_{\max}(\rho_{X'}\,\|\,\Gamma_{X'})-D_{\max}(\rho_R\,\|\,\Gamma_R)}\ . \tag{7.87}$$

The claim then follows from $\bar{D}_{R\to X'}(\rho_{X'R}\,\|\,\Gamma_R,\Gamma_{X'}) = -\log\alpha$. $\blacksquare$

*Proof of Proposition 7.19.* Consider the optimal solution $T_{X'R}$ and $\alpha$ to the primal semidefinite program, and let $\mathcal{T}_{X\to X'}$ be the completely positive map corresponding to $T_{X'R}$, i.e. defined by $\mathcal{T}_{X\to X'} = \mathrm{tr}_R\left[T_{X'R}\,(\cdot)^{t_{X\to R}}\right]$ with $(\cdot)^{t_{X\to R}}$ denoting the partial transpose map, as in Section 3.1.9. The mapping defined in this way is completely positive since $T_{X'R} \geqslant 0$ and is trace-nonincreasing thanks to condition (7.2-b).

The map $\mathcal{T}_{X\to X'}$ thus satisfies the conditions of item (i) of Proposition 5.2. Hence, invoking item (ii) of that proposition, let $\tilde{\Phi}_{XA\to X'A'}$ be a trace nonincreasing $\Gamma$-sub-preserving map for large enough $A$, $A'$, with $\Gamma_A = \mathbb{1}_A$, $\Gamma_{A'} = \mathbb{1}_{A'}$, satisfying

$$\tilde{\Phi}_{XA\to X'A'}\left(\sigma_{XR}\otimes\left(2^{-\lambda_1}\mathbb{1}_{2^{\lambda_1}}\right)\right) = \rho_{X'R}\otimes\left(2^{-\lambda_2}\mathbb{1}_{2^{\lambda_2}}\right)\ , \tag{7.88}$$

with $\alpha = 2^{-(\lambda_1-\lambda_2)}$ and $\sigma_{XR}$ a pure state satisfying $\sigma_R = \rho_R$, $\sigma_X = (\rho_X)^{t_{X\to R}}$. (If $\alpha$ is irrational, the following argument may be applied to arbitrary good rational approximations to $\alpha$.)

Now, dilate $\tilde{\Phi}_{XA\to X'A'}$ using Proposition 5.1 to a trace-preserving $\Gamma$-preserving map $\Phi_{XAX'A'Q\to XAX'A'Q}$ with states $|\mathrm{x}\rangle_X$, $|\mathrm{a}\rangle_A$, $|\mathrm{i}\rangle_Q$, $|\mathrm{x}'\rangle_{X'}$, $|\mathrm{a}'\rangle_{A'}$, $|\mathrm{f}\rangle_Q$ (all eigenstates of the respective $\Gamma$ operators), satisfying

$$\Phi_{XAX'A'Q}\left(\Gamma_{XAX'A'Q}\right) = \Gamma_{XAX'A'Q}\ ; \tag{7.89-a}$$

$$\begin{aligned}\Phi_{XAX'A'Q}&\left(\sigma_{XR}\otimes\left(2^{-\lambda_1}\mathbb{1}^{A}_{2^{\lambda_1}}\right)\otimes|\mathrm{x}'\mathrm{a}'\mathrm{i}\rangle\langle\mathrm{x}'\mathrm{a}'\mathrm{i}|_{X'A'Q}\right)\\ &= \rho_{X'R}\otimes\left(2^{-\lambda_2}\mathbb{1}^{A'}_{2^{\lambda_2}}\right)\otimes|\mathrm{xaf}\rangle\langle\mathrm{xaf}|_{XAQ}\ ;\quad\text{and}\end{aligned} \tag{7.89-b}$$

$$\langle\mathrm{xaf}|\,\Gamma_{XAQ}\,|\mathrm{xaf}\rangle_{XAQ} = \langle\mathrm{x}'\mathrm{a}'\mathrm{i}|\,\Gamma_{X'A'Q}\,|\mathrm{x}'\mathrm{a}'\mathrm{i}\rangle_{X'A'Q}\ . \tag{7.89-c}$$

Calculate first (recall $\Gamma_A = \mathbb{1}_A$)

$$D(2^{-\lambda_1}\mathbb{1}^A_{2^{\lambda_1}} \,\|\, \Gamma_A) = \operatorname{tr}\Big[2^{-\lambda_1}\mathbb{1}^A_{2^{\lambda_1}}\Big(\log\Big(2^{-\lambda_1}\mathbb{1}^A_{2^{\lambda_1}}\Big) - \log\mathbb{1}_A\Big)\Big] = \operatorname{tr}\Big[2^{-\lambda_1}\mathbb{1}^A_{2^{\lambda_1}}\big(-\lambda_1\mathbb{1}^A_{2^{\lambda_1}}\big)\Big]$$
$$= -\lambda_1\,, \tag{7.90}$$

and same for $D(2^{-\lambda_2}\mathbb{1}^{A'}_{2^{\lambda_2}} \,\|\, \Gamma_{A'}) = -\lambda_2$. We also see that for any pure state $|y\rangle$ which is an eigenstate of a positive semidefinite operator $\Gamma$, we have

$$D(|y\rangle\langle y| \,\|\, \Gamma) = \operatorname{tr}\big[|y\rangle\langle y|\,(\log|y\rangle\langle y| - \log\Gamma)\big] = -\langle y|\log\Gamma|y\rangle = -\log\langle y|\Gamma|y\rangle\,. \tag{7.91}$$

Then, by the data processing inequality for the relative entropy and with (7.89-b),

$$\begin{aligned}
0 &\leqslant D\Big(\sigma_X \otimes \Big(2^{-\lambda_1}\mathbb{1}^A_{2^{\lambda_1}}\Big) \otimes |x'a'i\rangle\langle x'a'i|_{X'A'Q} \,\|\, \Gamma_{XAX'A'Q}\Big)\\
&\quad - D\Big(\rho_{X'} \otimes \Big(2^{-\lambda_2}\mathbb{1}^{A'}_{2^{\lambda_2}}\Big) \otimes |xaf\rangle\langle xaf|_{XAQ} \,\|\, \Gamma_{XAX'A'Q}\Big)\\
&= D(\sigma_X \,\|\, \Gamma_X) + D(2^{-\lambda_1}\mathbb{1}^A_{2^{\lambda_1}} \,\|\, \Gamma_A) + D(|x'a'i\rangle\langle x'a'i|_{X'A'Q} \,\|\, \Gamma_{X'A'Q})\\
&\quad - D(\rho_{X'} \,\|\, \Gamma_{X'}) - D(2^{-\lambda_2}\mathbb{1}^{A'}_{2^{\lambda_2}} \,\|\, \Gamma_{A'}) - D(|xaf\rangle\langle xaf|_{XAQ} \,\|\, \Gamma_{XAQ})\\
&= D(\sigma_X \,\|\, \Gamma_X) - D(\rho_{X'} \,\|\, \Gamma_{X'}) - \lambda_1 + \lambda_2\\
&\quad - \log\langle x'a'i|\Gamma_{X'A'Q}|x'a'i\rangle + \log\langle xaf|\Gamma_{XAQ}|xaf\rangle\\
&= D(\sigma_X \,\|\, \Gamma_X) - D(\rho_{X'} \,\|\, \Gamma_{X'}) - \lambda_1 + \lambda_2\,,
\end{aligned} \tag{7.92}$$

where we invoked the condition (7.89-c) in the last step. We then have

$$\bar{D}_{R\to X'}(\rho_{X'R} \,\|\, \Gamma_R, \Gamma_{X'}) = \lambda_1 - \lambda_2 \leqslant D(\sigma_X \,\|\, \Gamma_X) - D(\rho_{X'} \,\|\, \Gamma_{X'})\,. \qquad \blacksquare$$

*Proof of Proposition 7.20.* Choose the primal candidate $T_{X'R} = \rho_R^{-1/2}\rho_{X'R}\rho_R^{-1/2}$. We have $\operatorname{tr}_{X'} T_{X'R} = \rho_R^{-1/2}\rho_R\rho_R^{-1/2} = \Pi^\rho_R \leqslant \mathbb{1}_R$ so our candidate satisifes (7.2-b). Also (7.2-c) is satisfied by construction, and $\operatorname{tr}_R(T_{X'R}\Gamma_R)$ is in the support of $\rho_{X'}$ and hence of $\Gamma_{X'}$. According to Proposition 7.8 we choose $\alpha = \|\Gamma_{X'}^{-1/2}\operatorname{tr}_R[T_{X'R}\Gamma_R]\,\Gamma_{X'}^{-1/2}\|_\infty$ and

$$\begin{aligned}
2^{-\bar{D}_{R\to X'}(\rho_{X'R} \,\|\, \Gamma_R,\Gamma_{X'})} \leqslant \alpha &= \|\Gamma_{X'}^{-1/2}\operatorname{tr}_R[T_{X'R}\Gamma_R]\,\Gamma_{X'}^{-1/2}\|_\infty\\
&= \|\Gamma_{X'}^{-1/2}\operatorname{tr}_R\big[T_{X'R}\Pi^\rho_R\Gamma_R\Pi^\rho_R\big]\,\Gamma_{X'}^{-1/2}\|_\infty\\
&\leqslant 2^{-D_{\mathrm{r}}(\rho_R \,\|\, \Gamma_R)}\|\Gamma_{X'}^{-1/2}\operatorname{tr}_R[T_{X'R}\rho_R]\,\Gamma_{X'}^{-1/2}\|_\infty\,,
\end{aligned} \tag{7.93}$$

since by definition $\rho_R^{-1/2}\Gamma_R\rho_R^{-1/2} \leqslant 2^{-D_{\mathrm{r}}(\rho_R \,\|\, \Gamma_R)}\mathbb{1}$ and thus $\Pi^\rho_R\Gamma_R\Pi^\rho_R \leqslant 2^{-D_{\mathrm{r}}(\rho_R \,\|\, \Gamma_R)}\rho_R$. Then

$$(7.93) = 2^{-D_{\mathrm{r}}(\rho_R \,\|\, \Gamma_R)}\|\Gamma_{X'}^{-1/2}\rho_{X'}\Gamma_{X'}^{-1/2}\|_\infty = 2^{-D_{\mathrm{r}}(\rho_R \,\|\, \Gamma_R)}2^{D_{\max}(\rho_{X'} \,\|\, \Gamma_{X'})}\,. \qquad \blacksquare$$

*Proof of Proposition 7.21.* The proof of this proposition proceeds via the hypothesis testing relative entropy, $D^\eta_{\mathrm{H}}(\rho \,\|\, \Gamma)$. Let $\epsilon' = \epsilon^2/(2+\epsilon^2)$ and let $\eta = 1-\epsilon'$. The hypothesis testing relative entropy can be written as the solution of a semidefinite program (Dupuis *et al.*, 2013).

Specifically, there exists $Q \geqslant 0$, $\mu \geqslant 0$ and $X \geqslant 0$ such that

$$2^{-D_{\mathrm{H}}^{\eta}(\rho \,\|\, \Gamma)} = \frac{1}{\eta} \operatorname{tr}\left[Q\Gamma\right] = \mu - \frac{\operatorname{tr} X}{\eta} , \tag{7.94}$$

with $Q$, $\mu$ and $X$ satisfying the conditions

$$Q \leqslant \mathbb{1} ; \tag{7.95-a}$$

$$\operatorname{tr}\left[Q\rho\right] \geqslant \eta ; \tag{7.95-b}$$

$$\mu\rho \leqslant \Gamma + X . \tag{7.95-c}$$

In addition, the complementary slackness relations for these variables read

$$XQ = X ; \tag{7.96-a}$$

$$\operatorname{tr}\left(Q\rho\right) = \eta ; \tag{7.96-b}$$

$$Q\left(\mu\rho - \Gamma - X\right) = 0 . \tag{7.96-c}$$

Define $\bar{\rho} = \Pi^{Q}\rho\Pi^{Q}$, where $\Pi^{Q}$ is the projector onto the support of $Q$. Apply $Q^{-1}\left(\cdot\right)\Pi^{Q}$ onto (7.96-c) to obtain

$$\mu\bar{\rho} = \Pi^{Q}\Gamma\Pi^{Q} + \Pi^{Q}X\Pi^{Q} \geqslant \Pi^{Q}\Gamma\Pi^{Q} . \tag{7.97}$$

In addition, because $\Pi^{Q}\Gamma\Pi^{Q}$ has support on $\Pi^{Q}$, then $\bar{\rho}$ must also have support on the full of $\Pi^{Q}$, i.e. $\Pi^{\bar{\rho}} = \Pi^{Q}$. So, by definition of $D_{\mathrm{r}}(\bar{\rho} \,\|\, \Gamma)$ have that

$$2^{-D_{\mathrm{r}}(\bar{\rho} \,\|\, \Gamma)} \leqslant \mu . \tag{7.98}$$

Also, define $\bar{\rho}' = \bar{\rho}/\operatorname{tr}\bar{\rho}$, and we can see by Lemma A.6 that $P\left(\rho, \bar{\rho}'\right) \leqslant \sqrt{2\epsilon'/\left(1-\epsilon'\right)} = \epsilon$. Also, $2^{-D_{\mathrm{r}}(\bar{\rho}' \,\|\, \Gamma)} \leqslant 2^{-D_{\mathrm{r}}(\bar{\rho} \,\|\, \Gamma)}$ by definition of $D_{\mathrm{r}}(\cdot \,\|\, \cdot)$. Then $\bar{\rho}'$ is a valid optimization candidate in the definition of $D_{\mathrm{r}}^{\bar{\epsilon}}(\rho \,\|\, \Gamma)$ and

$$2^{-D_{\mathrm{r}}^{\bar{\epsilon}}(\rho \,\|\, \Gamma)} \leqslant 2^{-D_{\mathrm{r}}(\bar{\rho}' \,\|\, \Gamma)} \leqslant \mu . \tag{7.99}$$

It thus remains to show that $\mu \leqslant \epsilon'^{-1} 2^{-D_{\mathrm{min},0}(\rho \,\|\, \Gamma)}$. Apply $\operatorname{tr}\left(\Pi^{\rho}\left(\cdot\right)\right)$ onto the constraint (7.95-c) to obtain

$$\mu \leqslant \operatorname{tr}\left(\Pi^{\rho}\Gamma\right) + \operatorname{tr}\left(\Pi^{\rho}X\right) \leqslant \operatorname{tr}\left(\Pi^{\rho}\Gamma\right) + \operatorname{tr}\left(X\right) . \tag{7.100}$$

Now, because of (7.94), we have $0 \leqslant \operatorname{tr}\left[Q\Gamma\right] = \mu\eta - \operatorname{tr} X$, and thus $\operatorname{tr} X \leqslant \mu\eta$. Combining with (7.100) gives

$$\mu\left(1-\eta\right) \leqslant \operatorname{tr}\left(\Pi^{\rho}\Gamma\right) ; \tag{7.101}$$

since $\epsilon' = 1 - \eta$ and $\operatorname{tr}\left(\Pi^{\rho}\Gamma\right) = 2^{-D_{\mathrm{min},0}(\rho \,\|\, \Gamma)}$ we have $\mu \leqslant \left(1/\epsilon'\right) 2^{-D_{\mathrm{min},0}(\rho \,\|\, \Gamma)}$ and the claim follows. ■

*Proof of Proposition 7.22.* Let $\tilde{\rho}_{R}, \tilde{\rho}_{X'}$ be quantum states which are optimal smoothed states

for the quantities

$$D^{\epsilon''}_{\min,0}(\rho_R \,\|\, \Gamma_R) = D_{\min,0}(\tilde{\rho}_R \,\|\, \Gamma_R)\,. \tag{7.102-a}$$

$$D^{\epsilon'}_{\max}(\rho_{X'} \,\|\, \Gamma_{X'}) = D_{\max}(\tilde{\rho}_{X'} \,\|\, \Gamma_{X'})\,. \tag{7.102-b}$$

With $\epsilon''' > 0$ and using Proposition 7.21, we know that

$$D^{\epsilon'''}_{\mathrm{r}}(\tilde{\rho}_R \,\|\, \Gamma_R) \geqslant D_{\min,0}(\tilde{\rho}_R \,\|\, \Gamma_R) + \log\frac{\epsilon'''^2}{2+\epsilon'''^2}\,. \tag{7.103}$$

Let $\tilde{\tilde{\rho}}_R$ be the optimal smoothed state for $D^{\epsilon'''}_{\mathrm{r}}(\tilde{\rho}_R \,\|\, \Gamma_R)$, such that

$$D_{\mathrm{r}}(\tilde{\tilde{\rho}}_R \,\|\, \Gamma_R) = D^{\epsilon'''}_{\mathrm{r}}(\tilde{\rho}_R \,\|\, \Gamma_R)\,. \tag{7.104}$$

At this point, we have

$$D_{\mathrm{r}}(\tilde{\tilde{\rho}}_R \,\|\, \Gamma_R) - D_{\max}(\tilde{\rho} \,\|\, \Gamma_{X'}) \geqslant D^{\epsilon''}_{\min,0}(\rho_R \,\|\, \Gamma_R) - D^{\epsilon'}_{\max}(\rho_{X'} \,\|\, \Gamma_{X'}) + \log\frac{\epsilon'''^2}{2+\epsilon'''^2}\,, \tag{7.105}$$

with

$$P\left(\tilde{\rho}_{X'}, \rho_{X'}\right) \leqslant \epsilon'\,; \qquad P\left(\tilde{\rho}_R, \rho_R\right) \leqslant \epsilon''\,; \qquad P\left(\tilde{\tilde{\rho}}_R, \tilde{\rho}_R\right) \leqslant \epsilon'''\,. \tag{7.106}$$

Now, we'll apply Lemma A.7 twice to construct a state close to $\rho_{X'R}$ which has marginals $\tilde{\rho}_{X'}$ and $\tilde{\tilde{\rho}}_R$ exactly. Let $\tau_{X'R}$ be the quantum state given by Lemma A.7 satisfying

$$\tau_{X'} = \tilde{\rho}_{X'}\,; \qquad \tau_R = \rho_R\,; \qquad P\left(\tau_{X'R}, \rho_{X'R}\right) \leqslant 2\sqrt{2\epsilon'}\,. \tag{7.107}$$

Applying Lemma A.7 again, let $\tau'_{X'R}$ be a quantum state close to $\tau_{X'R}$ such that

$$\tau'_{X'} = \tilde{\rho}_{X'}\,; \qquad \tau'_R = \tilde{\tilde{\rho}}_R\,; \qquad P\left(\tau'_{X'R}, \tau_{X'R}\right) \leqslant 2\sqrt{2\left(\epsilon''+\epsilon'''\right)}\,. \tag{7.108}$$

We thus have by triangle inequality

$$P\left(\tau'_{X'R}, \rho_{X'R}\right) \leqslant 2\sqrt{2\epsilon'} + 2\sqrt{2\left(\epsilon''+\epsilon'''\right)}\,. \tag{7.109}$$

By Proposition 7.20 we can now write

$$\bar{D}_{R\to X'}(\tau'_{X'R} \,\|\, \Gamma_R, \Gamma_{X'}) \geqslant D_{\mathrm{r}}(\tau'_R \,\|\, \Gamma_R) - D_{\max}(\tau'_{X'} \,\|\, \Gamma_{X'}) = D_{\mathrm{r}}(\tilde{\tilde{\rho}}_R \,\|\, \Gamma_R) - D_{\max}(\tilde{\rho}_{X'} \,\|\, \Gamma_{X'})\,. \tag{7.110}$$

Observe now that $\tau'_{X'R}$ is a valid optimization candidate for $\bar{D}^{\epsilon}_{R\to X'}(\rho_{X'R} \,\|\, \Gamma_R, \Gamma_{X'})$. Hence

$$\bar{D}^{\epsilon}_{R\to X'}(\rho_{X'R} \,\|\, \Gamma_R, \Gamma_{X'}) \geqslant \bar{D}_{R\to X'}(\tau'_{X'R} \,\|\, \Gamma_R, \Gamma_{X'})\,. \tag{7.111}$$

Finally, inequality (7.111) followed by (7.110) and (7.105) provides us the seeked lower bound. ∎

*Proof of Proposition 7.23.* The lower bound is given simply as

$$\bar{D}^{\epsilon}_{R\to X'}(\rho_{X'R}\,\|\,\Gamma_R,\Gamma_{X'}) \geqslant \bar{D}^{\epsilon=0}_{R\to X'}(\rho_{X'R}\,\|\,\Gamma_R,\Gamma_{X'}) = \log\frac{\operatorname{tr} P'_{X'}\Gamma_{X'}}{\operatorname{tr} P_R\Gamma_R}\ , \tag{7.112}$$

where the latter expression is provided by Corollary 7.14. For the upper bound, let $\hat{\rho}_{X'R}$ be the optimal state such that $P(\hat{\rho}_{X'R},\rho_{X'R}) \leqslant \epsilon$ and

$$\bar{D}^{\epsilon}_{R\to X'}(\rho_{X'R}\,\|\,\Gamma_R,\Gamma_{X'}) = \bar{D}_{R\to X'}(\hat{\rho}_{X'R}\,\|\,\Gamma_R,\Gamma_{X'})\ , \tag{7.113}$$

and invoke Proposition 7.19 to get

$$(7.113) \leqslant D(\hat{\rho}_R\,\|\,\Gamma_R) - D(\hat{\rho}_{X'}\,\|\,\Gamma_{X'})\ . \tag{7.114}$$

We have $D(\hat{\rho}_R,\rho_R) \leqslant P(\hat{\rho}_R,\rho_R) \leqslant \epsilon$ and analogously $D(\hat{\rho}_{X'},\rho_{X'}) \leqslant \epsilon$. By continuity of the relative entropy given in Proposition A.8, we get

$$\begin{aligned} |D(\hat{\rho}_R\,\|\,\Gamma_R) - D(\rho_R\,\|\,\Gamma_R)| &\leqslant f_0(\epsilon,\Gamma_R)\ ;\\ |D(\hat{\rho}_{X'}\,\|\,\Gamma_{X'}) - D(\rho_{X'}\,\|\,\Gamma_{X'})| &\leqslant f_0(\epsilon,\Gamma_{X'})\ , \end{aligned} \tag{7.115-a}$$

where $f_0(\epsilon,\Gamma)$ is as given in the claim. On the other hand,

$$D(\rho_R\,\|\,\Gamma_R) - D(\rho_{X'}\,\|\,\Gamma_{X'}) = \log\operatorname{tr} P'_{X'}\Gamma_{X'} - \log\operatorname{tr} P_R\Gamma_R\ , \tag{7.116}$$

because $\rho_R = P_R\Gamma_R P_R/\operatorname{tr} P_R\Gamma_R$ and $\rho_{X'} = P'_{X'}\Gamma_{X'}P'_{X'}/\operatorname{tr} P'_{X'}\Gamma_{X'}$, as given by (7.47). This means that

$$(7.114) \leqslant \log\operatorname{tr}\frac{P'_{X'}\Gamma_{X'}}{P_R\Gamma_R} + f_0(\epsilon,\Gamma_R) + f_0(\epsilon,\Gamma_{X'}) \tag{7.117}$$

■

## 7.5 A chain rule

If two individual processes are concatenated, what can be said of the coherent relative entropy of the combined processes? As one would expect, it turns out that the compression yield of a composition of logical maps can only be better than the sum of the compression yields of the individual realizations of each map. To help keep track of the different systems, see Figure 7.1.

***Proposition 7.24** (A chain rule). Let $\tau_{X''RE}$ be any tripartite state. (Interpret $\tau_{X''RE}$ as the process matrix of a mapping $X' \to X''$ on a state $|\rho\rangle_{X'RE}$; the latter is analogously to be interpreted as a purification of the process matrix of a mapping $X \to X'$ with reference system R, see Figure 7.1.) Let $|\rho\rangle_{X'RE}$ be the corresponding "input state," where RE is interpreted as the purifying reference system (technically, let $|\rho\rangle_{X'RE}$ be any purification of $\tau_{RE}$). The*

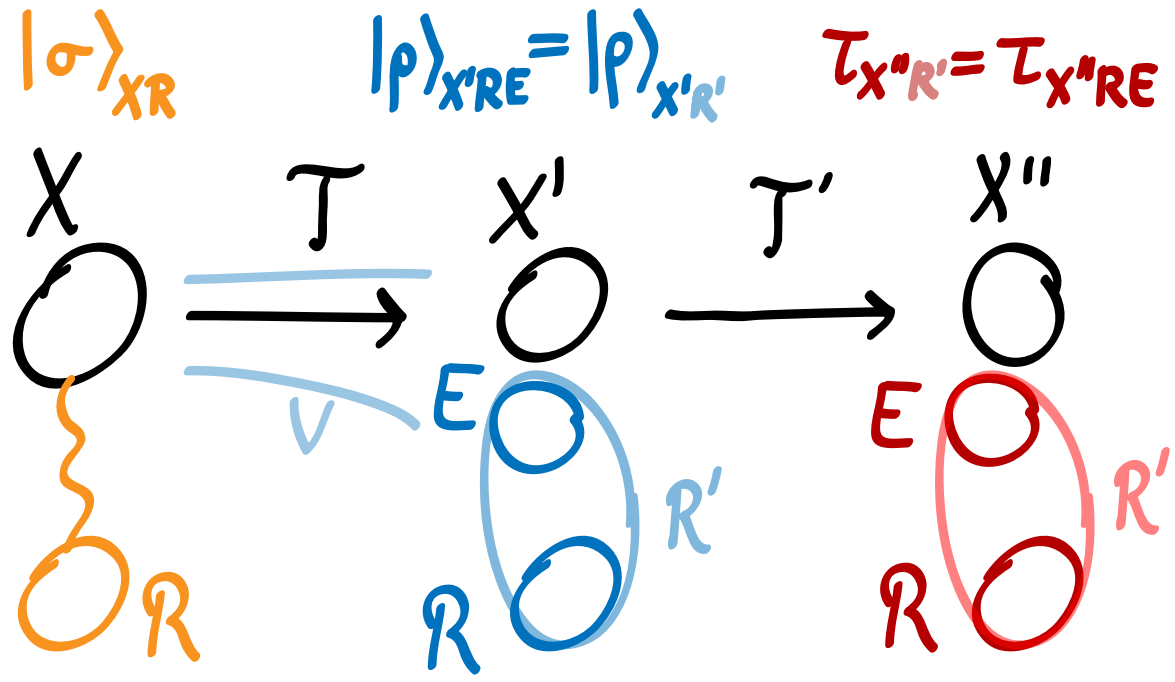

Figure 7.1: Systems involved in the chain rule. Two logical maps $\mathcal{T}$ and $\mathcal{T}'$ are applied in sequence, the first from $X$ to $X'$ and the second from $X'$ to $X''$. These maps are represented by process matrices (Choi matrix weighted by the input state). The input starts with a pure state $|\sigma\rangle_{XR}$ between $X$ and a reference system $R$. After the first map, $|\sigma\rangle_{XR}$ is mapped to another pure state $|\rho\rangle_{X'RE}$ on systems $X'$, $R$ and an Stinespring environment $E$. Now the systems $RE =: R'$ play the role of the reference system for the second mapping. The latter produces the final state $\tau_{X''R'} = \tau_{X''RE}$. The chain rule states that the individual compression yields of these two mappings is worse than the compression yield associated with the total (composite) mapping, which is described by the process matrix $\tau_{X''R}$ ($= \mathrm{tr}_E\, \tau_{X''RE}$). The positive semidefinite operators $\Gamma_X$, $\Gamma_{X'}$ and $\Gamma_{X''}$ are given; from them define $\Gamma_R = \Gamma_X^{t_{X\to R}}$ and $\Gamma_{RE} = \Gamma_{R'} = \Gamma_{X'}^{t_{X'\to R'}}$. As the notation suggests and as the picture with the processes evokes, we should have $\Gamma_R = \mathrm{tr}_E\, \Gamma_{RE} = \mathrm{tr}_E\, \Gamma_{X'}^{t_{X'\to RE}}$.

*positive semidefinite operators $\Gamma_X$, $\Gamma_{X'}$ and $\Gamma_{X''}$ are given. Define $\Gamma_R = \Gamma_X^{t_{X\to R}}$ and $\Gamma_{RE} = \Gamma_{X'}^{t_{X'\to RE}}$. Assume that $\Gamma_{X'}$ satisfies $\Gamma_R = \mathrm{tr}_E\,\Gamma_{RE} = \mathrm{tr}_E\,\Gamma_{X'}^{t_{X'\to RE}}$. Then*

$$\bar{D}_{R\to X'}(\rho_{X'R}\,\|\,\Gamma_R,\Gamma_{X'}) + \bar{D}_{RE\to X''}(\tau_{X''RE}\,\|\,\Gamma_{RE},\Gamma_{X''}) \leqslant \bar{D}_{R\to X''}(\tau_{X''R}\,\|\,\Gamma_R,\Gamma_{X''})\ . \quad (7.118)$$

*Proof of Proposition 7.24.* Let $T_{X'R}$, $\alpha$ be the optimal primal variables for $2^{-\bar{D}_{R\to X'}(\rho_{X'R}\,\|\,\Gamma_R,\Gamma_{X'})}$, and let $T'_{X''RE} = T'_{X''R'}$ and $\alpha'$ be the optimal primal variables for $2^{-\bar{D}_{RE\to X''}(\rho_{X''RE}\,\|\,\Gamma_{RE},\Gamma_{X''})}$.

Now let

$$T''_{X''R} = \mathrm{tr}_{R'}\big(T_{X''R'}\,(T_{X'R})^{t_{X'\to R'}}\big)\ ; \quad (7.119\text{-a})$$

$$\alpha'' = \alpha\cdot\alpha'\ . \quad (7.119\text{-b})$$

(Recall that $T_{X'R}$ and $T'_{X''R'}$ are interpreted as Choi matrices of trace nonincreasing maps $\mathcal{T}_{X\to X'}$ and $\mathcal{T}'_{X'\to X''}$; then $T''_{X''R}$ is the Choi matrix of the composite mapping $\mathcal{T}'_{X'\to X''}\circ\mathcal{T}_{X\to X'}$.)

First, we have $\mathrm{tr}_{X''}\,T''_{X''R} = \mathrm{tr}_{X''R'}\big(T_{X''R'}\,(T_{X'R})^{t_{X'\to R'}}\big) \leqslant \mathrm{tr}_{R'}\big((T_{X'R})^{t_{X'\to R'}}\big) \leqslant \mathbb{1}_R$, satisfying constraint (7.2-b). Then, we have

$$\mathrm{tr}_R\,T''_{X''R}\Gamma_R = \mathrm{tr}_{R'}\Big[T'_{X''R'}\,[\mathrm{tr}_R\,T_{X'R}\Gamma_R]^{t_{X'\to R'}}\Big] \leqslant \alpha\,\mathrm{tr}_{R'}\Big[T'_{X''R'}\,[\Gamma_{X'}]^{t_{X'\to R'}}\Big] = \alpha\,\mathrm{tr}_{R'}\Big[T''_{X''R'}\Gamma_{R'}\Big] \leqslant \alpha\alpha'\Gamma_{X''} = \alpha''\Gamma_{X''}\ . \quad (7.120)$$

Finally (with $\tau_R = \rho_R = \sigma_R$ and $\rho_{RE} = \tau_{RE}$),

$$\tau_R^{1/2}T''_{X''R}\tau_R^{1/2} = \mathrm{tr}_{R'}\,T'_{X''R'}\Big(\rho_R^{1/2}T_{X'R}\rho_R^{1/2}\Big)^{t_{X'\to R'}} = \mathrm{tr}_{R'}\,T'_{X''R'}\rho_{X'R}^{t_{X'\to R'}} = \mathrm{tr}_{R'E}\,T'_{X''R'}\rho_{X'RE}^{t_{X'\to R'}} = \mathrm{tr}_E\,\rho_{RE}^{1/2}T'_{X''RE}\rho_{RE}^{1/2} = \mathrm{tr}_E\,\tau_{X''RE} = \tau_{X''R}\ . \quad (7.121)$$

These primal candidates are hence feasible; they provide an upper bound to the optimal semidefinite program solution. Taking minus the logarithm of this value gives,

$$\bar{D}_{R\to X''}(\tau_{X''R}\,\|\,\Gamma_R,\Gamma_{X''}) \geqslant -\log\alpha'' = -\log\alpha - \log\alpha'\ , \quad (7.122)$$

which completes the proof. ∎

## 7.6 Asymptotic equipartition property

Finally, the coherent relative entropy also obeys an asymptotic equipartition property in the i.i.d. limit. In this limit, the coherent relative entropy converges to the difference of relative entropies of the input and the output to the respective hidden information operators. This behavior is a natural generalization of the i.i.d. limit studied in Section 6.6, where we recovered the difference of the von Neumann entropy of the input and output states.

***Proposition 7.25** (Asymptotic equipartition property). The coherent relative entropy converges to the difference of relative entropies in the i.i.d. limit,*

$$\lim_{\epsilon\to 0}\lim_{n\to\infty}\frac{1}{n}\bar{D}^{\epsilon}_{R^n\to X'^n}(\rho^{\otimes n}_{X'^n R^n}\,\|\,\Gamma^{\otimes n}_R,\Gamma^{\otimes n}_{X'}) = D(\rho_R\,\|\,\Gamma_R) - D(\rho_{X'}\,\|\,\Gamma_{X'})\;. \tag{7.123}$$

*Proof of Proposition 7.25.* We lower bound and upper bound $\bar{D}^{\epsilon}_{R\to X'}(\rho_{X'R}\,\|\,\Gamma_R,\Gamma_{X'})$ by expressions with known asymptotic i.i.d. behavior.

Consider the upper bound (7.78) from Proposition 7.19. Observe that the relative entropy $D(\rho\,\|\,\Gamma)$ is continuous in the first variable: indeed, we have $D(\rho\,\|\,\Gamma) = -H(\rho) - \mathrm{tr}\,[\rho\log\Gamma]$; the von Neumann entropy is continuous (Fannes, 1973) and the second term is linear in $\rho$. It is also known that the relative entropy is additive, that is, $D(\rho^{\otimes n}\,\|\,\Gamma^{\otimes n}) = n\,D(\rho\,\|\,\Gamma)$. Finally, we have

$$\lim_{\epsilon\to 0}\lim_{n\to\infty}\frac{1}{n}\bar{D}^{\epsilon}_{R\to X'}(\rho_{X'R}\,\|\,\Gamma_R,\Gamma_{X'}) \leqslant D(\rho_R\|\Gamma_R) - D(\rho_{X'}\|\Gamma_{X'})\;. \tag{7.124}$$

The lower bound is given by Proposition 7.22. The right hand side of (7.81) is known to converge to the difference of relative entropies thanks to the asymptotic equipartition of the min- and max-relative entropies (Datta, 2009). (The logarithmic term vanishes in the limit $n\to\infty$.) Combined with the upper bound (7.124), this proves the claim. ∎

## 7.7 A guardian for the entropy zoo

As we have now introduced a new entropic quantity, it is a natural question to ask where in the landscape of entropy measures it fits in (Figure 3.2).

The coherent relative entropy is, as we've mentioned before, neither a proper conditional entropy, nor a proper relative entropy. Yet, it contains both aspects. In fact, both the smooth conditional entropies, and the smooth relative entropies are recovered by the coherent relative entropy. (See Proposition 7.15 and Proposition 7.16.) The coherent relative entropy is thus a "guardian" for our entropy zoo (Figure 7.2).

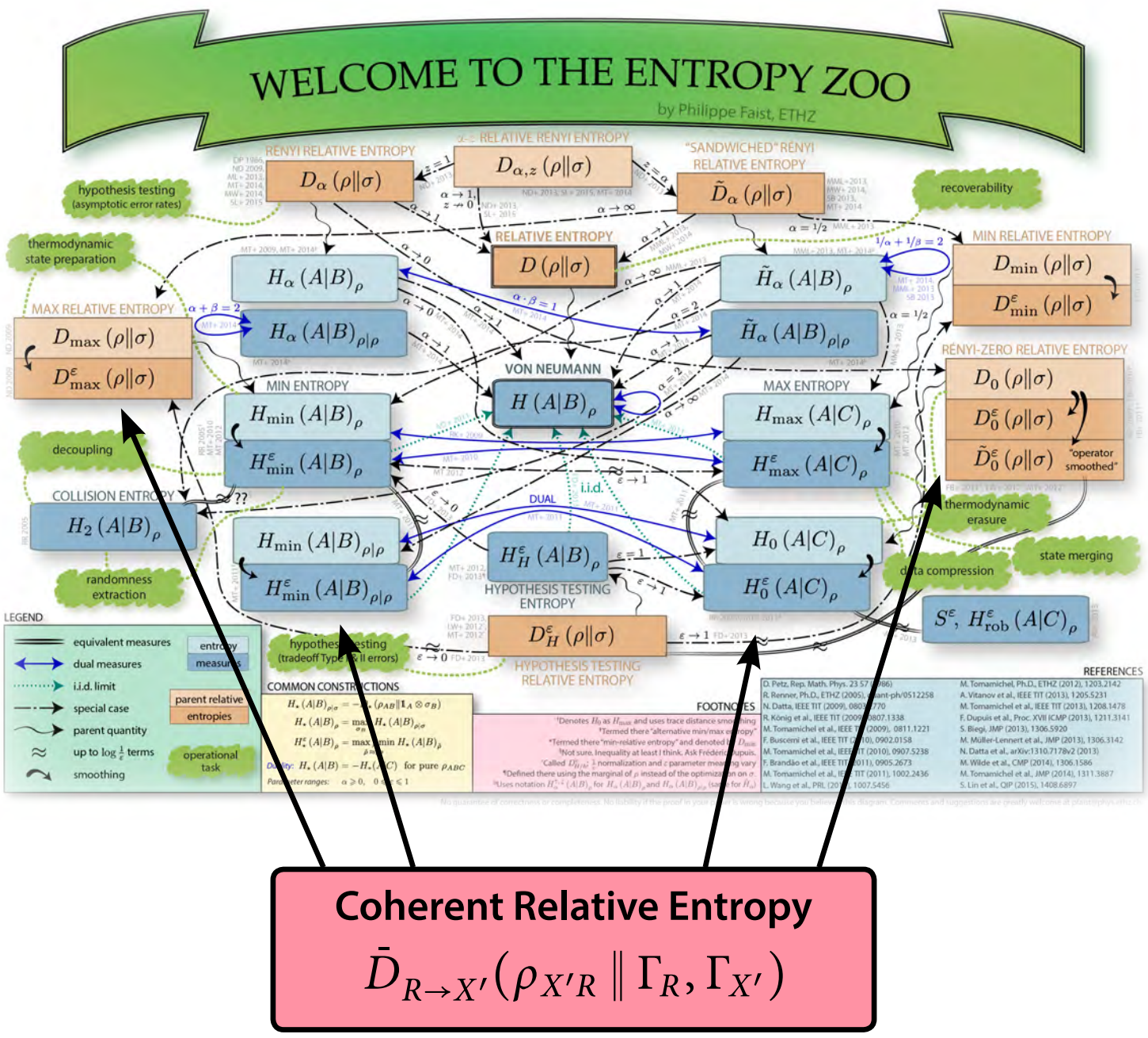

Figure 7.2: A guardian for the entropy zoo in Figure 3.2. The coherent relative entropy is a new type of measure which generalizes both the conditional entropy and the relative entropy.

# 8

# Emergence of Thermodynamics in the Macroscopic Regime

In this chapter, we discuss how to recover macroscopic thermodynamics from the microscopic theory developed in the preceding chapters.

First, we need to clarify which aspects of traditional thermodynamics we expect to recover. In most textbooks, thermodynamics is derived for gases, and then extended—often simply by analogy—to more complicated systems such as electrical systems, spin lattices or even black holes. One may ask, what is the minimal setup of thermodynamics, the common denominator of all these applications? For example, we know that work $\delta W$ may be defined via the pressure and the variation in volume of a gas via $\delta W = -p\, dV$, but in other contexts, such as for magnetic media, a polymer or a black hole, its definition should be adapted appropriately. Also, is it necessary to single out the internal energy, or can any other physical quantity play the same role, such as spin? The question of the minimal assumptions of thermodynamics has been largely addressed in the literature, culminating in self-contained mathematical frameworks for thermodynamics and rigorous definitions of the thermodynamic entropy (Giles, 1964; Gyftopoulos and Beretta, 2005; Beretta and Zanchini, 2011; Lieb and Yngvason, 1999, 2004, 2013, 2014). More recent work (Weilenmann *et al.*, 2015) shows among others that this type of framework can be further simplified, and that it also applies to small quantum systems.

The approach we take here is to study a particular class of states of a system described by the framework defined in the previous chapters. We define *thermodynamic states* as a class of states which can be reversibly interconverted in principle regardless of the details of the microscopic processes.

Because these states may be reversibly interconverted, the compression yield of these transformations depend only on the input and output state and derives from a potential, which we call the *natural thermodynamic potential*. It may only decrease under free operations, giving rise to a general formulation of the macroscopic second law of thermodynamics in the present context. Furthermore, if the thermodynamic states are parameterized by continuous parameters, then we may calculate the differential of the natural thermodynamic potential, since it is a function of these parameters. The differential of this potential describes the physics of the system, as a genuine thermodynamic potential. In our framework, the valuable resource is anything which can increase this natural thermodynamic potential. The standard notion of *thermodynamic work* for a gas is introduced by considering an explicit device, a piston, which can extract this resource, and can be directly related to the natural thermodynamic potential.

In stark contrast to textbook statistical mechanics, our approach does not rely on average quantities. This means that it can be applied to experiments which are performed a single time, and where the quantities of interest are not averaged over many independent repetitions of the experiment. The latter is the case in textbook statistical mechanics, and our approach thus avoids this issue.

An application of our ideas to the textbook example of a gas either considered as an isolated system or in contact with a heat bath illustrates how textbook thermodynamics is recovered, and how the traditional second law formulated in terms of entropy, heat and temperature is recovered as $dS \geqslant \delta Q/T$.

Then, we turn to further, more exotic examples, which illustrate the breadth of the applicability of our emergent thermodynamics picture.

## 8.1 The general emergence mechanism

First, we present what the goal of our approach is, namely, what exactly we are trying to recover. Then, we define *thermodynamic states*, *thermodynamic variables* as well as the *natural thermodynamic potential*. It may be necessary to consider these notions in some *thermodynamic limit*. For continuous thermodynamic variables, the natural thermodynamic potential has a differential structure with generalized chemical potentials, and gives a general formulation of a *macroscopic second law* in the present context.

We then briefly discuss *control systems* as well as a convenient strategy to construct thermodynamic states explicitly.

### 8.1.1 Properties of thermodynamics to recover

Thermodynamics may have various meanings, and is usually explained based on explicit examples of physical systems. Because our microscopic model is entirely information-theoretic, we expect to recover a more general macroscopic theory. Here, we would like to recover the following properties.

(a) The possible states a thermodynamic system can be found in are characterized by continuous parameters called *thermodynamic variables*, which may be controlled externally;

(b) There is a function of the thermodynamic variables, the *natural thermodynamic potential*, which is monotonous under processes may occur spontaneously;

(c) In addition, there may be a notion of *thermodynamic work*, which is a resource provided by an external system enabling otherwise forbidden operations.

The natural thermodynamic potential depends on the thermodynamic variables, which are continuous. We may then consider the differential of the thermodynamic potential with respect to the independent thermodynamic variables. Because the thermodynamic variables correspond to physical quantities a macroscopic observer can control, the coefficients of the differential typically provide information about the physical behavior of the system.

The notion of *thermodynamic equilibrium* is implicit in point (a). We assume the thermodynamic states are singled out for some particular reason, such as thermalizing dynamics. This is not necessary, though, as we will see in additional, more exotic examples which are purely information-theoretic.

Points (b) and (c) correspond to the so-called *second law of thermodynamics* and *first law of thermodynamics*, respectively. The presence or absence of particular kinds of thermodynamic baths simply influences the relevant potential in point (b).

This description of thermodynamics is purposely formulated in extremely abstract terms, without even referring to any particular physical quantities such as energy, volume, or pressure. This is both in line with the observed universality of thermodynamics, and consistent with our approach in the previous chapters.

### 8.1.2 Thermodynamic states and potential

Here, we identify what *thermodynamic states* are. We single them out as a particular class of quantum states on $X$ which satisfy some properties.

Consider a system $X$ with hidden information operator $\Gamma_X$, and suppose that the system and its dynamics are well described by the framework developed in the preceding chapters. In particular, a logical process $\Phi$ corresponding to dynamics which may occur spontaneouly when, e.g., in contact with a heat bath, must satisfy $\Phi(\Gamma_X) \leqslant \Gamma_X$. For convenience, we define also $X' \simeq X$ and $\Gamma_{X'} = \mathrm{id}_{X\to X'}(\Gamma_X)$ so that we can identify logical processes on $X$ by their bipartite process matrix $\rho_{XX'}$ (Section 3.1.9).

We now suppose that there is some kind of physical mechanism which restricts the possible physical states to a certain class of states. This mechanism may be, for example, microscopic mixing dynamics or a particular symmetry which the system must respect. This class of states is then parameterized by the quantities which macroscopic observers have access to, the thermodynamic variables.

First, the class of states needs to be parameterized by some variables $\{z_j\}$, which we also denote together as $\vec{z}$. These are the *thermodynamic variables*. We denote the quantum state corresponding to the thermodynamic variables $\vec{z}$ by $\tau^{(\vec{z})}$. For now, we don't worry about the range of valid values of $\vec{z}$: they may be discrete, confined to an interval, or take any real value. Our results may also hold for more general cases, such as if the object $\vec{z}$ lives on a manifold.

Second, the states need to be reversibly interconvertible. This means the following: consider any $\rho_{XX'}$ such that $\rho_X = \tau^{(\vec{z})}$ and $\rho_{X'} = \tau^{(\vec{z}')}$. Then the coherent relative entropy is given by a potential $\Lambda(\vec{z})$, such that it depends only on the input and output states and not on the details of the logical process:

$$\bar{D}^{\epsilon=0}_{X\to X'}(\rho_{X'X} \,\|\, \Gamma_X, \Gamma_{X'}) = \Lambda(\vec{z}) - \Lambda(\vec{z}') \,, \tag{8.1}$$

The potential $\Lambda(\vec{z})$ is the *natural thermodynamic potential*, and determines which processes are permitted.

***Thermodynamic states.*** *A class $\{\tau^{(\vec{z})}\}$ of quantum states on $X$ are 'thermodynamic states' if the following conditions are satisfied:*

- *the states $\{\tau^{(\vec{z})}\}$ are parameterized by some variables $\vec{z}$, called the 'thermodynamic variables;'*

- *there is a function $\Lambda(\vec{z})$ such that the coherent relative entropy of any log-*

*ical process relating two states of the form $\tau^{(\vec{z})}$ is given by a the difference in $\Lambda(\vec{z})$, as per* (8.1). *The function $\Lambda(\vec{z})$ is the 'natural thermodynamic potential.'*

The $\vec{z}$ parameters are meant to correspond to physical properties of the system which can be measured and manipulated by the macroscopic observer. They may correspond for example directly to eigenspaces of particular observables. They may also represent more general parameters which the macroscopic observer can control, for example the position of a wall enclosing a gas. We will further elaborate on the meaning of these parameters shortly.

An interesting structure emerges when these parameters can take values in a continuous range, at least to good approximation, because we may consider the differential of the natural thermodynamic potential. It may be necessary for this to consider a thermodynamic limit, which we elaborate on below.

The condition (8.1) on the coherent relative entropy with $\epsilon = 0$ is extremely strong, and in fact it may be sufficient to require this condition only approximately, and for $\epsilon$ small enough. This is also fixed by considering a thermodynamic limit, in which $\epsilon$ is taken to go to zero appropriately. We elaborate on this below.

### 8.1.3 Take a thermodynamic limit, if necessary

One of the defining properties of thermodynamics is that systems are described by continuous variables—the thermodynamic variables—rather than discrete ones. This indeed enables the application of differential calculus on the state functions and unveils a collection of useful relations between the physical quantities.

Often, thermodynamic states do not emerge exactly for finite sizes, but in fact do so only in some macroscopic limit. Such a limit may be, for example, one in which the system size grows to infinity. Hence, we will refine our definition of thermodynamic states to allow for the emergence to occur in what we call a *thermodynamic limit*.

Traditionally, the thermodynamic limit refers to considering larger and larger number of particles, or equivalently, looking at larger and larger systems. However, because our framework doesn't explicitly consider a system composed of particles, we seek a more general concept of thermodynamic limit. Here, we define a *thermodynamic limit* as a family of systems, parameterized by a parameter $\xi$ which we can send to infinity; we then require that the structure described above emerges in that limit. That is, in the limit $\xi \to \infty$, there are continuously parameterized thermodynamic states $\rho^{\vec{z}}$ for which the coherent relative entropy of transformations is given by a natural thermodynamic potential.

***Thermodynamic states in a thermodynamic limit.*** *Consider a family of systems $S^\xi$ for a specific sequence of $\xi$ with $\xi \to \infty$, each with a $\Gamma_{S^\xi}$. Assume that for each of these $\xi$, the systems $S^\xi$ has a class of states $\{\tau_{S^\xi}^{(\vec{z})}\}$ parameterized by some parameters $\vec{z}$. Then the $\{\tau_{S^\xi}^{(\vec{z})}\}$ are 'thermodynamic states in the thermodynamic limit $\xi \to \infty$' if:*

- *in the limit $\xi \to \infty$, this parameterization is continuous, i.e. for any real $\vec{z}$ (in some appropriate region), we can find a sequence of valid $\vec{z}_\xi$ which converges to $\vec{z}$; and*
- *there is a potential $\Lambda_\xi(\vec{z})$ parameterized by $\xi$ such that, for any sequences of valid parameterizations $(\vec{z}_\xi)$, $(\vec{z}'_\xi)$, and any bipartite states $\rho_{S^\xi S'^\xi}^{(\xi)}$ with $(S'^\xi \simeq S^\xi$, $\Gamma_{S'^\xi} = \mathrm{id}_{S^\xi \to S'^\xi}(\Gamma_{S^\xi}))$ and with $\rho_{S^\xi}^{(\xi)} = \tau_{S^\xi}^{(\vec{z}_\xi)}$, $\rho_{S'^\xi}^{(\xi)} = \tau_{S'^\xi}^{(\vec{z}'_\xi)}$, we have*

$$\lim_{\epsilon \to 0} \lim_{\xi \to \infty} \frac{\bar{D}^{\epsilon}_{S^\xi \to S'^\xi}(\rho^{(\xi)}_{S'^\xi S^\xi} \,\|\, \Gamma_{S^\xi}, \Gamma_{S'^\xi})}{\Lambda_\xi\big(\vec{z}_\xi\big) - \Lambda_\xi\big(\vec{z}'_\xi\big)} = 1\,. \tag{8.2}$$

This definition is slightly more complicated; one must ensure in particular that the limits are taken in the correct order. As always in information theory, the limit $\epsilon \to 0$ should be taken last (if at all), because the tolerated probability of failure does not depend on the system size. This is important, for example, if we consider many copies of a state and wish to apply the asymptotic equipartition property.

With this definition, we have an emerging structure of thermodynamic states $\vec{z}$ living in a continuous region, to which are associated a limit of thermodynamic states of finite sizes. The regularized natural thermodynamic potential is given in the thermodynamic limit as $\lambda(\vec{z}) = \lim_{\xi \to \infty} c_\xi^{-1}\, \Lambda_\xi\big(\vec{z}_\xi\big)$, where $c_\xi$ is a regularizing factor allowing us to take the limit. It may be the number of systems involved, for example, or more generally $c_\xi = \operatorname{tr} \Gamma_{S(\xi)}$.

Such a regularized form of the natural thermodynamic potential would typically express the same idea as the entropy or energy per particle when similar statements are considered in statistical mechanics. Also in information theory, in the asymptotic of many independent repetitions of an information-theoretic task such as data compression, it is necessary to consider the *rate*, that is the compressed size per repetition, because the amount of information obviously diverges for large numbers of copies.

### 8.1.4 Differential of the natural thermodynamic potential and generalized chemical potentials

Suppose now that we have singled out a class $\mathcal{C}$ of thermodynamic states, and assume additionally that the variables $\vec{z}$ which parameterize this class take values in some continuous range of real values. This means we have a function $\Lambda(\vec{z})$ which is defined over a continuous range of variables, meaning that we can consider its differential. It is given as

$$d\Lambda = \mu_1\, dz_1 + \mu_2\, dz_2 + \cdots \,, \tag{8.3}$$

where

$$\mu_j = \left(\frac{\partial\Lambda}{\partial z_j}\right)_{z_1,\ldots,z_{j-1},z_{j+1},\ldots} \tag{8.4}$$

is the *generalized chemical potential* which is conjugate to the thermodynamic variable $z_j$. Following the standard thermodynamic literature, we denote by $(\partial\Lambda/\partial z_1)_{z_2,\ldots}$ the partial derivative of $\Lambda$ with respect to $z_1$, where the variables $z_2, \ldots$ are kept constant. This notation is very useful if different sets of independent variables are used. In general, $\mu_j$ may loosely be interpreted as a measure of how much effort needs to be provided to increase $z_j$ by an infinitesimal amount while keeping the other $z_j$'s constant.

### 8.1.5 Variational principle for the natural potential and the second law of thermodynamics

In our microscopic framework, the spontaneous processes which can happen are only those which correspond to $\Gamma$-sub-preserving maps.

For any $\Gamma$-sub-preserving map relating two infinitesimally close thermodynamic states $\vec{z}$ and $\vec{z} + d\vec{z}$, the natural thermodynamic potential $\Lambda$ by definition cannot increase. This is because the coherent relative entropy associated to any $\Gamma$-sub-preserving map has to be positive or zero, which according to (8.1) implies that the natural thermodynamic potential must decrease:

$$d\Lambda \leqslant 0 \,. \tag{8.5}$$

This is the formulation of the macroscopic second law of thermodynamics in our language.

Furthermore, if some of the thermodynamic variables are no longer controlled and set at fixed values (e.g. if a separator between two compartments of a gas is removed), and if the microscopic dynamics of the system is sufficiently mixing, then the natural potential $\Lambda$ will tend to its minimum value compatible with whatever constraints have been maintained on the system. This is analogous to the variational principles in thermodynamics, where the equilibrium state of an isolated system is given by maximizing the entropy, or by minimizing the free energy if the system is in contact with a heat bath.

In this respect, the quantities $\mu_j$ gain a physical interpretation analogous to that of temperature, pressure or the chemical potential in thermodynamics. If two systems are combined while exchanging a physical quantity modeled by a thermodynamic variable (and supposing that $\Lambda$ is additive), then the maximization of the total $\Lambda$ subject to the quantity being conserved globally implies that the corresponding $\mu_j$'s must be equal. This is simply the equilibrium condition for temperature, pressure or chemical potential.

### 8.1.6 Control systems

The thermodynamic states, on the microscopic level, do not have to be fully distinguishable. For example, the quantum states corresponding to a gas of particles occupying different volumes overlap and are not fully distinguishable. Macroscopically, the different volumes appear as two clearly different, fully distinguishable states. This is not because of the state of the gas, but rather of the state of the box in which the gas is placed. The box states corresponding to confining the gas into different volumes are orthogonal.

In general, control systems allow to change the $\Gamma_S$ operator of the system controllably. This can be accounted for in our framework by treating the control system explicitly. Suppose the hidden information operator can be tuned to different values $\Gamma_S^{(w)}$ according to a parameter $w$. The control system $C$ remembers which value $w$ the system is tuned to, and we can describe the joint system with the fixed hidden information operator

$$\Gamma_{CS} = \sum |w\rangle\langle w|_C \otimes \Gamma_S^w \ . \tag{8.6}$$

We refer to such a variable $w$ as a *control variable*.

### 8.1.7 Standard construction with projections of the Γ operator

We have until now only presented the required properties of thermodynamic states, and have not specified how to construct them explicitly. Here, we'll concentrate on one particular fashion to construct them which will prove useful for the following.

Consider any system $S$ let $\Gamma_S$ be its hidden information operator. Then a class of thermodynamic states, in the sense of the definition given in Section 8.1.2, can be singled out by states of the form

$$\tau^{(\vec{z})} = \frac{P^{(\vec{z})}\Gamma_S P^{(\vec{z})}}{\operatorname{tr} P^{(\vec{z})}\Gamma_S} , \tag{8.7}$$

for a selection of projectors $P^{\vec{z}}$ which commute with $\Gamma_S$. The fact that these states satisfy the required properties owes to the mystical wonders of Proposition 7.6; the corresponding natural thermodynamic potential is simply given by

$$\Lambda(\vec{z}) = -\log \Omega(\vec{z}) ; \qquad\qquad \Omega(\vec{z}) = \operatorname{tr}\left[P^{(\vec{z})}\Gamma_S\right] . \tag{8.8}$$

(Of course, $\Omega(\vec{z})$ will turn out to be the relevant partition function of the thermodynamic state.)

Additionally, we may have control variables which can tune the hidden information operator of the system. In this case we may model the control system explicitly as explained above and apply the same construction. Suppose we have some control variables $w_1, w_2, \ldots$, which can tune the hidden information operator on $S$ as $\Gamma_S^{\vec{w}}$. The joint hidden information operator is given as per (8.6), as

$$\Gamma_{CS} = \sum |\vec{w}\rangle\langle\vec{w}|_C \otimes \Gamma_S^{\vec{w}} . \tag{8.9}$$

Suppose also that for each such choice, we have variables $y_1, y_2, \ldots$ which further specify the choice of state on the system. These variables taken together form the thermodynamic variables $\vec{z} = (w_1, w_2, \ldots, y_1, y_2, \ldots)$. Then the thermodynamic state is given by

$$\tau^{(\vec{z})} = |\vec{w}\rangle\langle\vec{w}|_C \otimes \frac{P_S^{(\vec{w},\vec{y})}\Gamma_S^{\vec{w}} P_S^{(\vec{w},\vec{y})}}{\operatorname{tr} P_S^{(\vec{w},\vec{y})}\Gamma_S} , \tag{8.10}$$

where $P_S^{(\vec{w},\vec{y})}$ are a collection of projectors which commute with $\Gamma_S^{\vec{w}}$. The corresponding natural thermodynamic potential is again simply given by

$$\Lambda(\vec{z}) = -\log \Omega(\vec{z}) ; \qquad \Omega(\vec{z}) = \mathrm{tr}\left[P^{(\vec{z})}\Gamma_S^{\vec{w}}\right] . \tag{8.11}$$

Depending on the system considered, it may be necessary to consider a thermodynamic limit as explained above. In this case, we consider a parameter $\xi$ of the setup, and send $\xi \to \infty$. Crucially, thanks to Proposition 7.23, the natural potential for fixed $\xi$ and as a function of $\epsilon$ behaves continuously, and the limit $\xi \to \infty$ may thus be taken to determine the (regularized) natural thermodynamic potential.

## 8.2 Example: thermodynamics of a gas

Here we illustrate how to apply the above ideas in order to recover the usual thermodynamics of a gas.

### 8.2.1 Isolated system

We consider a macroscopic gas with a fixed energy, confined inside a well-defined volume and with a fixed number of particles.

Denote by $\hat{N}$ the number operator, and suppose that the Hamiltonian $\hat{H}$ of the gas commutes with the number operator. Assume that the gas can be confined with the help of a piston to a volume $V$. The volume constraint is included in the total Hamiltonian $\hat{H}$ via a confining potential, in such a way that if the gas has $N$ particles and occupies a volume $V$, then it can be effectively described via the Hamiltonian $\hat{H}^{(V)}$. We assume that $\hat{H}^{(V)}$ commutes with $\hat{N}$. Furthermore, we assume that the energy of the gas and the number of particles it is comprised of is fixed to some values $E$ and $N$, meaning that the state of the gas lives in the corresponding eigenspaces of $\hat{H}^{(V)}$ and $\hat{N}$. We denote the projector onto this common eigenspace by $P_S^{E,V,N}$. (Of course, this is simply the microcanonical subspace associated to the variables $E$, $V$ and $N$.)

First we start by specifying the main ingredient to the microscopic framework, that is the $\Gamma$ operator. This operator should encode which microscopic processes must be excluded from happening spontaneously. Any spontaneous process should not touch the state $|E, V, N\rangle_C$ on the control system, and must be a unitary evolution on the system. This implies, in particular, that any spontaneous operation $\Phi_{CS}$ on $CS$ must preserve the subspace compatible with the fixed values of $E$, $V$ and $N$:

$$\Phi_{CS}\Big(|E,V,N\rangle\langle E,V,N|_C \otimes P_S^{(E,V,N)}\Big) \leqslant |E,V,N\rangle\langle E,V,N|_C \otimes P_S^{(E,V,N)} . \tag{8.12}$$

Of course, this is not sufficient, but it turns out it is enough to merely rule out any process which do not satisfy this. The constraint (8.12) defines the hidden information operator $\Gamma_{CS}$ as

$$\Gamma_{CS} = \sum_{E,V,N} |E,V,N\rangle\langle E,V,N|_C \otimes P^{(E,V,N)} . \tag{8.13}$$

We define the thermodynamic states simply as projections of the hidden information operator, using the construction explained in Section 8.1.7. They are given as

$$\tau_{CS}^{(E,V,N)} = |E,V,N\rangle\langle E,V,N|_C \otimes \frac{P^{(E,V,N)}}{\operatorname{tr} P^{(E,V,N)}} . \tag{8.14}$$

Similarly, the corresponding natural thermodynamic potential is simply

$$\Lambda(E,V,N) = -\ln\Omega(E,V,N) ; \qquad \Omega(E,V,N) = \operatorname{tr} P^{(E,V,N)} . \tag{8.15}$$

Here, we choose to represent the potential in units of *nats*, rather than *bits*. (See Section 3.2.1; this is achieved by switching the logarithm from base 2 to base $e$).

We immediately recognize the construction of the Boltzmann entropy of the gas. The latter is defined as

$$S(E,V,N) = k\ln\Omega(E,V,N) , \tag{8.16}$$

where $k$ is Boltzmann's constant. We then have

$$S(E,V,N) = -k\,\Lambda(E,V,N) . \tag{8.17}$$

In other words, the natural thermodynamic potential emergent from our framework is none other than the Boltzmann entropy, up to choices of sign and units.

As seen above, the macroscopic second law emergent from our framework reads

$$d\Lambda \leqslant 0 . \tag{8.18}$$

We then immediately recover the standard formulation of the macroscopic second law of thermodynamics for the thermodynamic entropy of an isolated system,

$$dS \geqslant 0\ . \tag{8.19}$$

Hence, this well-known form of the macroscopic second law is recovered as an emergent quantity in our framework.

### 8.2.2 Contact with a heat bath at fixed temperature

In this second simple example, we consider the gas in contact with a reservoir at a fixed temperature $T$. We proceed as above, by identifying the hidden information operator via the microscopic restriction we would like to impose on the spontaneous processes, and define a class of thermodynamic states. For simplicity, we fix the number of particles to $N$ and thus don't need to define a number operator. The volume constraint is modeled as before, namely at a volume $V$ the Hamiltonian of the system is $\hat{H}^{(V)} = P^{(V)}\hat{H}P^{(V)}$. The temperature $T$ is fixed, and cannot be changed. This models the presence of a heat bath which interacts with the system $S$, but cannot be controlled. There is thus a single controllable thermodynamic variable, the volume $V$.

Now we need to define what the spontaneous processes are and which constraint they satisfy. Because the system is in contact with a heat bath, we may invoke the models developed in Chapter 4, such as thermal operations and Gibbs-preserving maps. Indeed, these were designed to model such a situation. Here, we simply use the constraint that the $\Gamma_S$ operator must be conserved, as argued in Section 5.2.5. At fixed volume, the spontaneous processes which may occur must then preserve the operator $\Gamma_S^{(V)} = e^{-\hat{H}^{(V)}/(kT)}$. Modeling the control system explicitly, we have

$$\Gamma_{CS} = \sum_V |V\rangle\langle V|_C \otimes e^{H^{(V)}/(kT)}\ . \tag{8.20}$$

The thermodynamic states are now given simply as specified with the thermodynamic variables $T, V, N$, as

$$\tau^{(V)} = |V\rangle\langle V|_C \otimes \frac{e^{-\hat{H}^{(V)}/(kT)}}{Z(V)}\ , \tag{8.21}$$

with the *partition function*,

$$Z(V) = \mathrm{tr}\Big(e^{-\hat{H}^{(V)}/kT}\Big)\,. \tag{8.22}$$

As for $\Omega(E, V, N)$ earlier, the partition function $Z(V)$ is introduced merely as a normalization factor enabling us to write the thermodynamic state as a projection of the hidden information operator. However these partition functions coincide are precisely those introduced in the microcanonical and canonical ensembles of statistical mechanics.

The corresponding potential is given by

$$\Lambda(V) = -\ln Z(V)\,, \tag{8.23}$$

and is directly related to the *Helmholtz free energy* $F(V) = -(kT)^{-1} \ln Z(V)$ via

$$F(V) = kT \cdot \Lambda(V)\,. \tag{8.24}$$

It is then possible to recover, via the differential of the thermodynamic potential, all the information about the thermodynamic behavior of the system.

We note that the potential $\Lambda$ corresponds to the quantity introduced as the *free entropy* in Guryanova *et al.* (2016).

Already, we may guess a connection with thermodynamic work: we know that the free energy difference between two thermodynamic states is precisely the minimal amount of work that we need to supply in order to perform the transition.

## 8.3 External devices and thermodynamic work

Now we turn to recovering the notion of thermodynamic work. Of course, we could simply stick to the notions of work defined by the explicit work storage systems presented in Section 4.5. It is however possible to design a work storage battery which is better suited to thermodynamic systems, yields an equivalent notion of work, and is closer to the usual definition of *thermodynamic work* in the literature. The thermodynamic work is defined with the help of this new type of battery, which is also treated with the same emergent thermodynamic framework. The device interacts with the system while satisfying some global physical conservation law.

### 8.3.1 External devices

An external device $W$ is modeled as an explicit system in our emergent thermodynamic framework. It has some thermodynamic variables $\vec{x}$, a corresponding hidden information operator $\Gamma_W = \sum g_W(\vec{x}) \, |\vec{x}\rangle\langle\vec{x}|_W$, thermal states $\tau_W^{(\vec{x})} = |\vec{x}\rangle\langle\vec{x}|_W$, and a corresponding natural potential $\Lambda_W(\vec{x}) = -\ln g_W(\vec{w})$. In the following, we often define the desired $\Lambda_W$ and it is then implied that $\Gamma_W$ must be chosen correspondingly.

The composite system $S \otimes W$ is jointly described by the hidden information operator $\Gamma_S \otimes \Gamma_W$. With the thermodynamic states $\tau_S^{(\vec{z})} \otimes \tau_W^{(\vec{x})}$, we have the natural thermodynamic potential

$$\Lambda_{SW}(\vec{z}, \vec{x}) = \Lambda_S(\vec{z}) + \Lambda_W(\vec{x}) \; . \tag{8.25}$$

According to our framework, infinitesimal processes which can occur naturally on $SW$ must satisfy $d\Lambda_{SW} \leqslant 0$. In particular,

$$d\Lambda_S \leqslant -d\Lambda_W \; . \tag{8.26}$$

We see that it is possible to use the device $W$ in order to provide the necessary resources to increase the potential of $\Lambda_S$ of $S$, at the expense of lowering the potential $\Lambda_W$.

In order to design a work storage device, it may be necessary to tune the device's properties according to a particular thermodynamic process. For example, if we extract work from a gas isothermally with the help of a piston, it is necessary to adjust the force applied on the piston so as to keep it constantly in equilibrium with the gas. Otherwise, the optimal amount of work will not be extracted. Hence, we require the external device not to be universal, but we allow its natural thermodynamic potential to be designed for a particular thermodynamic process.

### 8.3.2 Piston for a gas

We may model a way to supply or extract work from a gas via a piston. (Figure 8.1). The piston has a thermodynamic variable $x$ corresponding to its position, which determines the volume of the gas. The gas also exchanges energy with the piston, because of the particles hitting on the wall of the piston, for example, and thus we must include the energy $e$ of the piston as an explicit thermodynamic variable. The natural thermodynamic potential

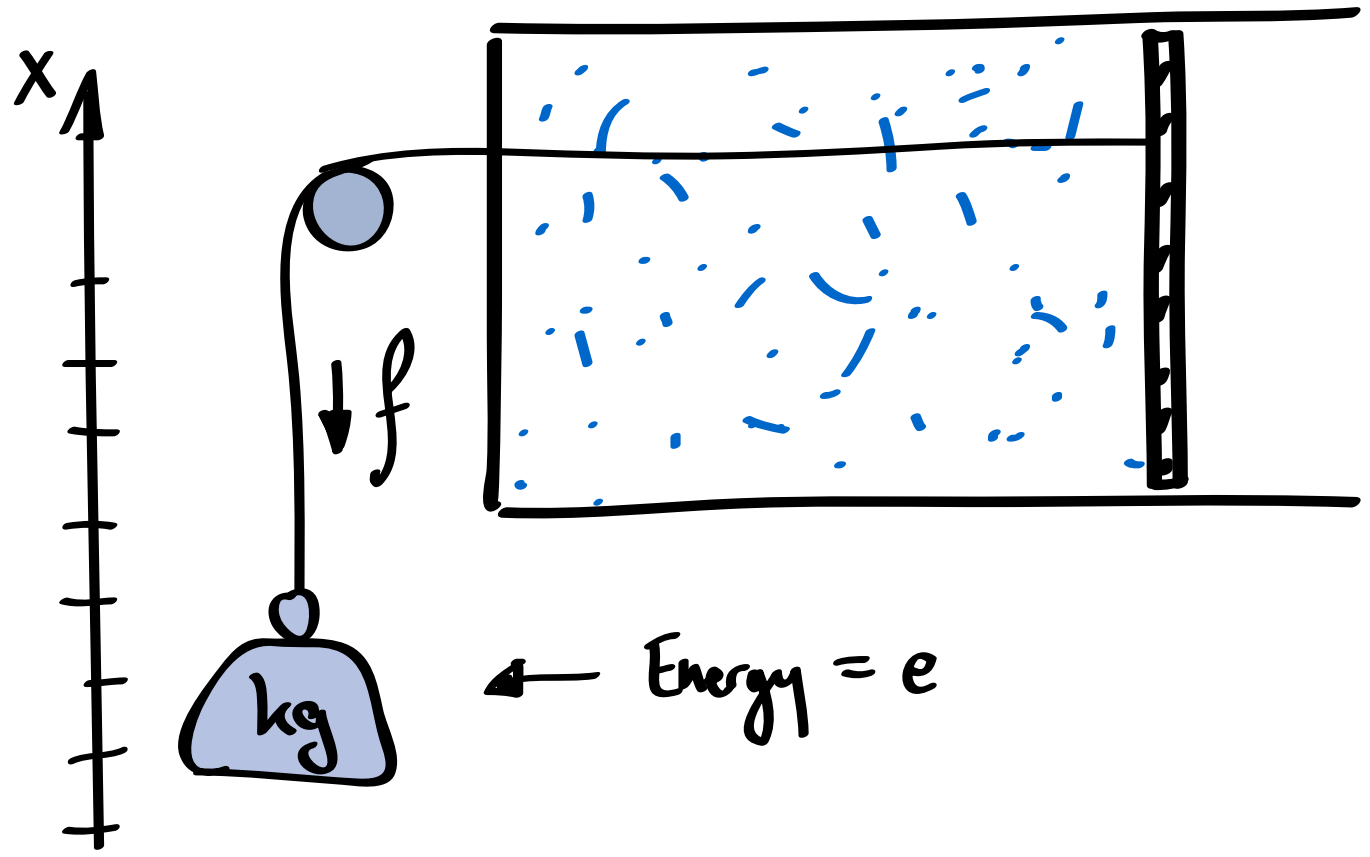

Figure 8.1: A piston can be used to supply or extract work to or from a gas. The piston has a position $x$ and an energy $e$, which are linked to the gas' state by physical conservation laws.

of the piston is

$$\Lambda_W(x, e) = v_e\, de + v_x\, dx\ , \tag{8.27}$$

with $v_e = (\partial\Lambda_W/\partial e)_x$ and $v_x = (\partial\Lambda_W/\partial x)_e$ the generalized chemical potentials as already seen above.

Additionally, the piston's setup has to be tuned such that a force is applied to the gas, in order to counter the gas' action on the piston. The force $f$ which is applied by the piston is defined as the gradient of the energy of the piston, $f = -(\partial e/\partial x)_{\Lambda_W}$. It is determined by the natural thermodynamic potential. Indeed, we have

$$de = \frac{1}{v_e}(d\Lambda_W - v_x\, dx)\ , \tag{8.28}$$

and hence

$$f = -\left(\frac{\partial e}{\partial x}\right)_{\Lambda_W} = \frac{v_x}{v_e}\ . \tag{8.29}$$

The amount of thermodynamic work which is supplied to the gas is defined

as

$$\delta W = f \cdot dx \,. \tag{8.30}$$

The device must interact with the system via conservation laws and coupling of the thermodynamic variables. The system is described by the thermodynamic variables $E, V, N$. First, we require that the total energy be conserved. Second, the piston is coupled to the volume $V$ of the gas via $x = V/A$, where $A$ is the area of the piston. These constraints imply

$$dE = -de \,; \qquad\qquad dV = A \cdot dx \,. \tag{8.31}$$

In our framework, the macroscopic second law of thermodynamics for the joint, isolated system $S \otimes W$ reads $d\Lambda_{SW} \leqslant 0$. Then

$$d\Lambda_S \leqslant -d\Lambda_W = -\nu_e \, de - \nu_x \, dx = \nu_e \left(dE - f \cdot dx\right) = \nu_e \left(dE - \delta W\right) . \tag{8.32}$$

Now, we finally define the *heat* as the contribution to the energy difference which is not due to work, $\delta Q = dE - \delta W$. We then obtain

$$d\Lambda_S \leqslant \nu_e \, \delta Q \,. \tag{8.33}$$

When is the inequality saturated? We get equality when $d\Lambda_S = -d\Lambda_W = \nu_e \, dE - (\nu_x/A) \, dV$. By comparing this differential with the natural representation of $d\Lambda_S$, we obtain

$$\nu_e = \left(\frac{\partial \Lambda_S}{\partial E}\right)_V \,; \qquad\qquad \frac{\nu_x}{A} = \left(\frac{\partial \Lambda_S}{\partial V}\right)_E \,. \tag{8.34}$$

For the case of the gas, we have seen that $\Lambda_S(E, V, N) = -k^{-1} \, S(E, V, N)$ and hence

$$\left(\frac{\partial \Lambda_S}{\partial E}\right)_V = -\frac{1}{kT} \,, \tag{8.35}$$

where $T$ is the temperature of the gas. Finally, we obtain from (8.33) the famous formulation of the second law of thermodynamics,

$$dS \geqslant \frac{\delta Q}{T} \,. \tag{8.36}$$

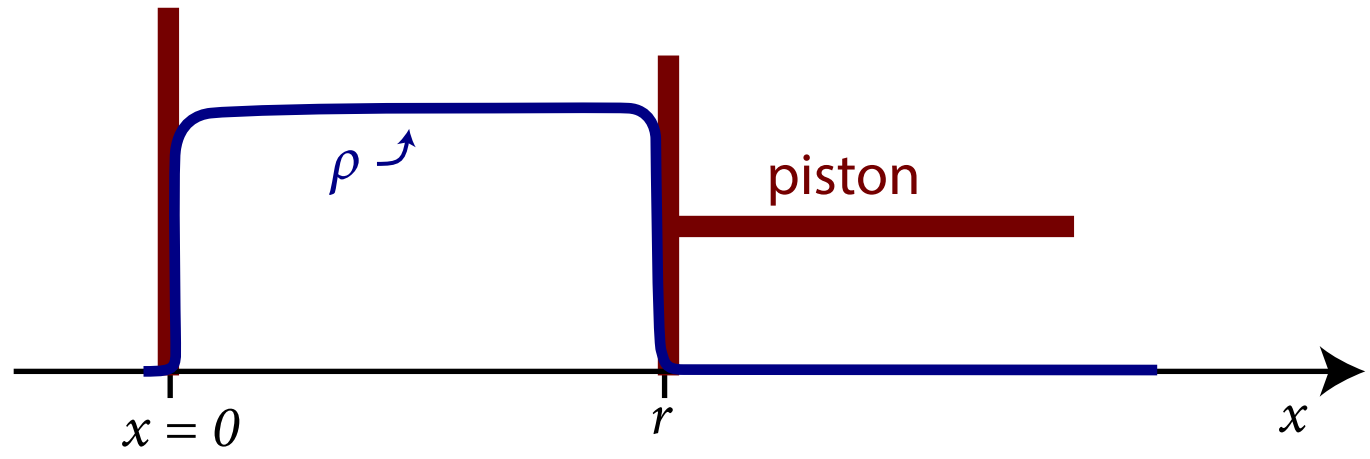

Figure 8.2: Discrete position states chosen at fine enough spacing approximate a particle living on a continuous segment. The length of the segment the particle can be in is controlled externally by a movable wall, and we assume that the particle is described by a maximally mixed state on the given segment. Furthermore, any unitary can be applied to the particle. This means the particle can be prepared in any state described by a density matrix with a flat spectrum.

## 8.4 Example: particle on a segment

We now apply our framework to a slightly different example, illustrating a type of thermodynamic limit which differs from a limit where a large number of systems is considered.

Consider a single particle whose position can be any value in a given segment $[0, r]$ in one dimension (Figure 8.2). For the sake of this thought experiment, we force the Hamiltonian to be completely degenerate, $H = 0$. Although this may not be faithfully representative of a physical particle, we intend this description as a toy model to showcase our arguments in a theoretical situation which is still well described by the formalism of quantum information.

This setting can be approximated by a setup in which the position of the particle can take any set of discrete, evenly-spaced values in that range. We may set the spacing of the available positions to values $1/\xi$ and take $\xi \to \infty$ to reach the continuous setup. This is our thermodynamic limit in the present case. In the following, and as before, we still work in finite dimensions: These considerations are to be applied for a $\xi$ chosen large enough that the finite-dimensional model approximates the continuum case to an acceptable precision.

We now consider free processes consisting in an isolated unitary evolution on the system. We don't necessarily assume the system to have thermalizing dynamics, on the contrary, we may assume that it is well isolated and think of these processes as engineered by the user. For our microscopic model we take the hidden information operator $\Gamma = \mathbb{1}$. In this way the constraint $\Phi(\Gamma) \leqslant \Gamma$ for any free process.

The thermodynamic states depend, of course, on the degree of control of the macroscopic observer. For the sake of the toy example, we assume that the user can very precisely engineer any unitary on the system. Furthermore, we assume that they can impose a constraint on the upper range limit of possible positions the particle can take (for example with a piston). Together, this means effectively that the user can impose such a constraint in any basis: the user may first unitarily rotate the state from the required constraint basis to the position basis, compress the state to the required range, and then undo the rotation to effectively impose the range constraint in any other basis.

A natural choice of thermodynamic states are then those density operators which have a flat spectrum. These states correspond to choosing any projector $P$ and defining a corresponding thermodynamic state with the construction (8.10). The states are parameterized by a rank, $r$, and a basis which we can write as a unitary, $U$. They are given as

$$\tau_{CS}^{(r,U_S)} = |r, U_S\rangle\langle r, U_S|_C \otimes \frac{1}{r} U_S P_S^{(r)} U_S^{\dagger} , \tag{8.37}$$

where $P_S^{(r)}$ is the projector onto the position states $x$ satisfying $0 \leqslant x \leqslant r$, with $r = \operatorname{rank} P^{(r)} = \operatorname{tr} P^{(r)}$.

In fact, there is a redundancy in the representation (8.37), in that full information about $U$ is not needed to identify the state completely. Ultimately, we only need to parameterize a subspace of $S$. However the above specification is more convenient, and does not change our conclusions.

The natural thermodynamic potential associated with these states is given by (8.11) as

$$\Lambda(r, U) = -\ln r , \tag{8.38}$$

which we note coincides with the negative of the state's entropy.

In this example, the thermodynamic variables are not independent real variables. Rather, they live on a manifold consisting in the unitary group combined with an additional real variable. This example is fairly simplistic in that the natural potential depends only on this separate variable $r$. One could imagine more complicated setups, however, they are technically more challenging to analyze as the differential needs to be considered on the manifold.

Finally, we note that these states are precisely the equilibrium states in Weilenmann *et al.* (2015), defined in a different framework based on an order relation for free processes (or, in the language of that reference, *adiabatic processes*). The natural thermodynamic potential also coincides with the thermodynamic entropy derived there (up to a sign convention). The approach of Weilenmann *et al.* (2015) is indeed similar to the present one; the main difference, specifying fully specified logical processes rather than state transitions, becomes irrelevant for thermal states as a consequence of their definition. Furthermore, the method for counting entropy does indeed coincide for these states, whether defined via an explicit information battery or via a notion of 'scaling' of equilibrium states.

## 8.5 Example: information storage with redundancy

In this final example, we turn back to pure classical information theory and coding. We may indeed apply our framework there, too, yielding a thermodynamic-like structure for a purely information-theoretic task. Again, this example is more of a proof of principle, and an illustration of the generality of the concepts introduced here, rather than any kind of useful novel description of information coding. This example is a slightly more involved version of the one presented in Section 5.2.4.

Consider a classical memory register, whose implementation stores the information redundantly internally. For the sake of keeping our example simple, we use a simple repetition code where each symbol is repeated $m$ times in the physical memory (Figure 8.3). For example, when we store the logical bit string $\overline{\underline{\mathbf{0\,1\,1\,1\,0\,1\,0\,1}}}$ in memory, then if the memory device is configured with $m = 3$, then the memory device takes upon itself to store an internal representation of this bit string into its physical space as

$$\overline{\underline{\mathbf{0\,0\,0\,1\,1\,1\,1\,1\,1\,1\,1\,1\,0\,0\,0\,1\,1\,1\,0\,0\,0\,1\,1\,1}}}\,.$$

Apart from the very rudimentary redundancy scheme, this is representative of typical information storage devices such as hard drives, flash devices or CD-ROMs.

Now we introduce a simple model for the memory undergoing exposure to noise from the environment. We assume that the action of the environment results in the operation of an effective channel $\Phi$ on the physical bits of the memory, such that $\Phi$ is subunital, that is, $\Phi(\mathbb{1}) \leqslant \mathbb{1}$. This is the case if the spontaneous processes which occur are noisy operations (Horodecki *et al.*, 2003).

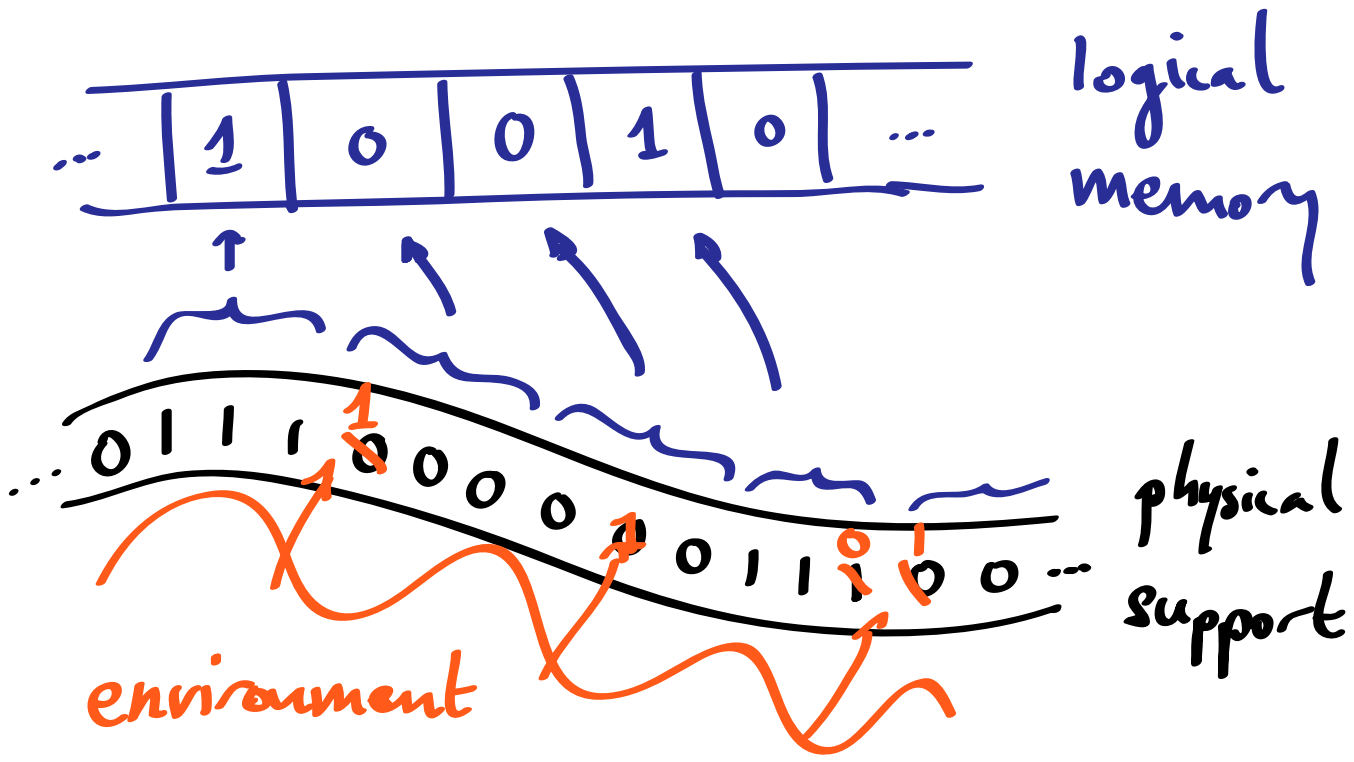

Figure 8.3: Information stored with repetition redundancy code, seen as an instance of our general emergent thermodynamic framework. The environment acts as source of noise for the physical bits, with the effect on the latter modeled as a (sub-)unital map. For large $m$ and $n$, a thermodynamic-like structure emerges, with the thermodynamic states corresponding to the most entropic states which are successfully decoded into a particular logical bit string. The emergent potential describes the amount of empty memory space we need to provide in order to vary the values of $m$ or $n$.

We have now all the elements to apply our finite-size framework. As before, we use an information battery to provide the necessary resources to perform forbidden operations. The battery provides this resource in the form of purity, or pure qubits, or even empty memory space. Here, this is the natural resource in this context and there is no need to convert it into any other form such as energy, as we did for work in finite-size thermodynamics. Logical processes which do not conserve the identity operator, such as an erasure process, are not allowed for free and require available empty memory space to be implemented with free operations. In order to do error correction, for example, a supply of fresh, empty bits must be provided.

The decoding of the information from the memory is done for each logical bit by majority: if there is a majority of zeros in the corresponding physical bits, the decoder assumes the logical bit was a zero, and otherwise, it decodes the logical bit as a one. (Because we are interested in the situation for large $m$, the situation where there is an exact equal number of zeros and ones is unimportant.) For the logical bit number $k$, we denote by $P^{m,k,(0)}$ and $P^{m,k,(1)}$ the POVM effects acting on the $m$ physical bits representing the $k^{\text{th}}$ logical bit, which cause the decoder to report a zero or a one, respectively.

Now we may identify a class of states which are reversibly interconvertible with respect to the coherent relative entropy as per (8.1). Doing so defines a class of thermodynamic states in this context. Consider the class of states

$$\tau^{(n,m,x^n)} = z_{m,n}^{-1} \bigotimes_{k=1}^{n} P^{m,k,(x_k)} \; ; \qquad z_{m,n} = 2^{n\cdot(m-1)} \; , \tag{8.39}$$

where $z_{m,n}$ is such that the state $\tau^{(n,m,x^n)}$ is normalized (since $\operatorname{tr} P^{m,k,(0)} = \operatorname{tr} P^{m,k,(1)} = 2^m/2 = 2^{m-1}$). The state $\tau^{(n,m,x^n)}$ is the most entropic state which is successfully decoded into the bit string $x^n$ with certainty. The natural potential corresponding to these states, satisfying the condition (8.1), is given by

$$\Lambda(n, m, x^n) = -\log z_{m,n} = -n\,(m-1)\;. \tag{8.40}$$

Consider the limit in which the redundancy is very large, $m \gg 1$, and where long logical bit strings are to be stored, $n \gg 1$. In this limit, the variables $m$ and $n$ are essentially continuous compared to the size of the system. What does the natural potential tell us, if seen as a thermodynamic potential? Because the bit string $x^n$ is quite obviously not continuous, we can't consider it in the differential and we simply don't consider any variation of $x^n$ itself. We abide by the convention that if $n$ increases, the new $x^n$ is padded by additional zeros, and if $n$ decreases, the last bits are trashed. The differential of the potential reads

$$d\Lambda = -n\,dm - (m-1)\,dn\;. \tag{8.41}$$

Concretely, this means that the cost of increasing $m$, the redundancy parameter, by a small amount $dm$ is that $n\,dm$ pure qubits need to be supplied so that they can be used as physical storage; this simply reflects the fact that $dm$ physical bits are needed per additional bit in the logical bit string in order to increase the level of redundancy. Also, if the bit string length $n$ is increased by $dn$, then $m-1 \approx m$ bits need to be provided. Again, this simply represents the fact that each additional logical bit needs $m$ physical bits.[1]

This example, albeit simple and rudimentary, serves to illustrate the role of the potential in terms of coarse-graining. Indeed, there is here clearly an explicit coarse-graining procedure between the physical and the logical bits. The coarse-graining is specified by a parameter $m$, which indicates how many bits where forgotten. We could have replaced the redundancy encoding by encoding the bit in a certain state of given energy $E$, and the problem would have been formally exactly the same, by replacing $\log m$ with $E$ in some appropriate units.

[1]In fact, only $m-1$ physical bits are needed instead of $m$. Indeed, the additional physical bit is known to be in the zero state, and we don't need the full redundancy of $m$ bits which is used to encode an unknown bit. However in our example $m$ is considered large, and so in any case $m-1 \approx m$.

# 9

# Observers, Side Information and Reality

In this chapter, we turn to an essential aspect of quantum mechanics: the fact that *measurement outcomes* and *quantum states* are relative to an observer.

The ideas presented here are not novel; they are effectively already understood in various branches of the quantum information community. The content of this chapter is, *de facto*, a reformulation of the ideas underlying *quantum reference frames* (Aharonov and Susskind, 1967; Dowling *et al.*, 2006; Bartlett *et al.*, 2006, 2007), the quantum Bayesian framework *QBism* (Caves *et al.*, 2002a, 2007; Fuchs and Schack, 2010; Fuchs *et al.*, 2014; Fuchs and Schack, 2015; Fuchs, 2016), Everett's *relative state* formulation of quantum mechanics and the *many-worlds interpretation* (Everett III, 1956, 1957; DeWitt, 1970; DeWitt and Everett III, 1973; Tegmark, 2007; Vaidman, 2016), as well as several more recent works including Rovelli (1996); Brassard and Raymond-Robichaud (2013); Brukner (2015); Vedral (2016); Frauchiger and Renner (2016). Furthermore the central thought experiment is a variant on *Wigner's friend* (Wigner, 1961) which is due to Deutsch (1985, 1986) and well summarized in Vedral (2016). It forms the main argument in Frauchiger and Renner (2016), which points out the implications for reality to which we refer to here.

We think it is useful to underscore the fundamental nature of this observer dependence, and to rephrase the existing results and understanding in a natural, quantum information theoretic language which will serve as a basis for the following chapter.

It is to be emphasized that our main motivation is based on a physical requirement of consistency between different observers. Also, we avoid certain questions of interpretation by focusing on understanding how the physical model of quantum mechanics is to be used, rather than attempting to give further meaning to abstract objects than what is required for a successful application of quantum mechanics.

The notion of observer dependence presented in this chapter generalizes the idea of conditioning in classical probability theory. Suppose for example that Alice tosses a classical coin and observes the outcome **0**. If Bob knows that Alice tossed a fair coin but doesn't know the outcome, he will describe the joint state of the coin and Alice's memory as a correlated state, in which either both the coin is **0** and Alice registered a **0** outcome with probability ½, or the coin is **1** and Alice registered a **1** outcome also with probability ½. The state each observer assigns to the coin is different: for Alice, the coin is in the state **0** and for Bob, the coin is in either state **0** or **1** with probability ½. The same happens in quantum mechanics: in the fully quantum case, observers may assign different states to a system.

Our arguments are based on a thought experiment which is in effect a twist around the example often referred to as 'Wigner's friend.' We first present this thought experiment in Section 9.1. In Section 9.2, we introduce the notion of an observer's memory allowing us to model when an observer may experience a particular measurement outcome. For a consistent analysis of such situations, we introduce in Section 9.3 the notion of an *event* in quantum mechanics. We point out that contrary to the classical notion of events, quantum mechanical events may be *incompatible*, meaning that the events cannot be explained by a common underlying reality. In Section 9.4 we elaborate on the connection between the observer-dependent nature of the quantum state and quantum reference frames. Finally, we discuss some conceptual remarks in Section 9.5.

## 9.1 Reversing a quantum measurement

Here, we outline the main argument of the chapter via a thought experiment (Deutsch, 1985, 1986; Vedral, 2016) which is a variant on 'Wigner's friend' (Wigner, 1961). The details of the setup have been slightly adapted to suit our purposes.[1]

Suppose Alice is performing a quantum experiment in her laboratory. She has prepared a quantum state $|\psi\rangle_S$ on a system $S$, and wishes to measure it in a particular basis $\{|k\rangle_S\}_k$. Business as usual: Alice switches on some interaction between the system $S$ and some measurement device, resulting in some joint unitary evolution. This process implements a quantum measurement described by the set of projectors $\{|k\rangle_S\}$, and records a particular measurement outcome $k_{\mathrm{outcome}}$. The particular outcome she obtains is random, as we all know, with probabilities given by the Born rule: the probability of obtaining the outcome $k$ is given by $|\langle k|\psi\rangle|^2$. Alice then describes the system as being in the state $|k_{\mathrm{outcome}}\rangle$, which is the post-measurement state.

[1]This thought experiment, along with the corresponding conclusions, have been recounted to me in the present form by my thesis supervisor, Prof. Renato Renner.

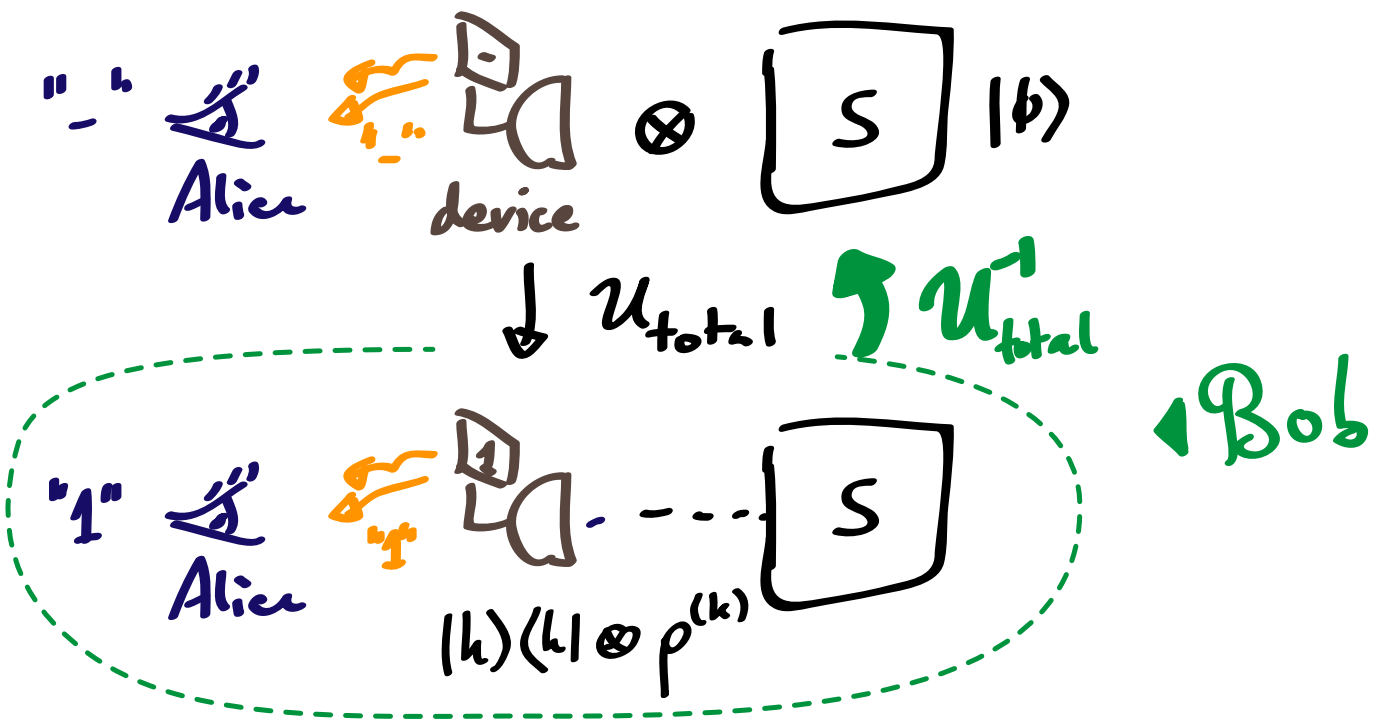

Figure 9.1: Reversing a quantum measurement—why measurement outcomes exist only relative to the observer. A measurement performed by Alice on a system $S$ is represented by a global unitary evolution of the system, a measurement device, the photon field and any other environment system, and Alice herself. Alice observes a definite outcome—this is an empirical fact. If Bob has the necessary technology, he may apply the inverse global unitary reversing the whole evolution of the lab, and thus undoing the measurement. (The inverse unitary also acts on Alice herself.) No record may remain of the observed outcome: All copies of this information, including in the state of the device, the photon field and Alice herself, are coherently restored into the system $S$ by Bob's actions. Requiring that both observers can consistently describe the situation implies that the collapse into a particular measurement outcome was only relative to Alice. Bob, observing a unitary evolution, did not experience any collapse.

Let us analyze what's happening a bit more precisely (Figure 9.1). First, the system interacts with the measurement device. This evolution is unitary, but since the measurement device has an enormous amount of degrees of freedom to which information may get correlated by uncontrolled interactions, the reduced state on the display of the measurement device is decohered in some preferred basis. Actually, once this information has reached the display of the measurement device, it continues to correlate any photons that are emitted from it, all the way to Alice herself. In effect, this evolution taken as a whole is pretty much just a quite elaborate way of performing a C-NOT gate (or corresponding generalization if $S$ is larger than a qubit) between the system and a complicated, highly nonlocalized, subsystem of the measurement device, the photon field and Alice's memory. Alice's memory $M$ is a subsystem of Alice herself which records the information she obtains from the measurement device. Denote this global unitary by $U_{\text{global}}$. Now, suppose that Alice's memory $M$ starts in a state $|\phi\rangle_M$, that the photon field is in a state $\rho^0_{\text{phot}}$, and that the device started in the quantum state $\rho^0_{\text{device}}$. The state of Alice's memory, the photon field, her measurement device, and the system $S$ after the evolution is

$$U_{\text{global}} \left( |\phi\rangle\langle\phi|_M \otimes \rho^0_{\text{phot}} \otimes \rho^0_{\text{device}} \otimes |\psi\rangle\langle\psi|_S \right) U^\dagger_{\text{global}} \tag{9.1}$$

Of course, because Alice is implementing a quantum measurement, the reduced state on her memory and the system $S$ is a classical-quantum state,

$$\sum_k p_k |\text{“}k\text{”}\rangle\langle\text{“}k\text{”}|_M \otimes |k\rangle\langle k|_S \,, \tag{9.2}$$

where $|\text{“}k\text{”}\rangle_M$ denotes the (classical) state of the memory which is used to record the fact that the measurement outcome was $k$, and $p_k = |\langle k|\psi\rangle|^2$ are the corresponding outcome probabilities.

Suppose that we have a second observer, Bob. Futuristic Bob is technologically all-powerful and can engineer any unitary on any particular degree of freedom of macroscopic systems, with extremely high precision. Of course, Bob cannot violate the laws of quantum mechanics. Let Bob have perfect knowledge about the unitary $U_{\text{global}}$. This is because Alice considers Bob a trustworthy friend, and has confided in him the details of the measurement she wishes to carry out.

Now we hit the “play” button, and watch Alice perform the measurement. The unitary $U_{\text{global}}$ unfolds exactly as predicted by Bob. After the unitary evolution is completed, Alice, as we've noted above, observes a measurement outcome $k_{\text{outcome}}$.

Suppose that exactly at the moment when the unitary has completed, super-Bob-the-technologically-all-powerful carries out the inverse unitary—he immediately performs $U^{-1}_{\text{global}}$ on all the systems concerned, reverting them back to their initial state. Of course, this is technologically extremely demanding, because one needs to dig up each electron of each atom of each molecule composing the measurement device, the photon field, Alice's memory, and engineer exactly the reverse interaction which will bring all the information about $|\psi\rangle$ coherently back into $S$. Because Bob has perfect knowledge of $U_{\text{global}}$, this is theoretically possible. Furthermore, it is in general not necessary to perform exactly $U_{\text{global}}^{-1}$; it is only necessary that the unitary which Bob applies coherently reverses the replication of the information extracted from $S$ into the auxiliary systems.

Good. So now, Bob reversed everything, and we're back to square one. But what happened to the measurement outcome $k_{\text{outcome}}$ that Alice observed? After the first unitary evolution was completed, the (classical) information $k_{\text{outcome}}$ was replicated on the display of the measurement device, in the photon field, in Alice's memory, and in tons of other degrees of freedom which interacted with all of these, correlating these systems with that information. But precisely, because Bob reversed the unitary globally, this information was removed from all those locations and no longer exists anywhere! The inverse unitary precisely addresses *all* replicas of this information, otherwise the state on the system $S$ would not be restored in general coherently to $|\psi\rangle$. In other words, after having reversed the measurement and having restored the system back in the state $|\psi\rangle$, *there doesn't exist any system in the universe which has remained correlated to the measurement outcome*—the measurement outcome no longer "exists."

This may seem like a contradiction, because Alice did, after all, observe a particular measurement outcome $k_{\text{outcome}}$, and it is hard to imagine how this fact may have been "erased."[2] We are forced to accept the fact that *the outcome that Alice observed existed only for her, and not for Bob*. This leads us to the conclusion that measurement outcomes are not an absolute concept on which all observers universally agree—they are, indeed, relative to an observer. A revision of our common conception of an absolute reality is necessary: *the events an observer perceives as 'real,' as modeled by measurement outcomes, are only 'real' for that observer, and an observer's notion of 'reality' may differ from that of another observer.*

Of course, the exact nature of *how* and *why* this collapse happens in the first place remains as mysterious as ever. It is also not the goal to address this issue here. For the sake of being precise about what statements we can make, we just propose the following effective model, which we shall accept as a principle: *whenever an observer, described by their memory M, interacts*

[2]As Deutsch (1985, 1986) points out, Alice can be asked to produce a written record that she has observed an outcome—without revealing which one—which still allows the measurement to be undone and at the same time certifies that she has experienced a collapse.

*with a system S such that the joint final state on S and M is classical on M, that is, a state of the form*

$$\rho_{SM} = \sum_k p_k \, |k\rangle\langle k|_M \otimes \rho_S^{(k)} \,, \tag{9.3}$$

*then that observer 'collapses' into one of the k's: namely, they will observe that the 'true' state of S is one of the states* $\rho_S^{(k)}$*, with probability* $p_k$*; however an external observer does not necessarily experience any effect.*

An immediate consequence of this is that different observers may disagree on the state of a system $S$: for example, if Alice's memory $M$ is correlated to a system $S$ causing her to collapse in a state $\rho_S^{(k)}$, then this state can be clearly different from the state that Bob would attribute to $S$, which is the reduced state $\rho_S = \operatorname{tr}_M \rho_{SM}$ of the correlated $S$-$M$ state.

A first question which comes into mind is that we mention in the principle above a state $\rho_{SM}$, but now we've just said that states are relative to observers—so to which observer is the state $\rho_{SM}$ itself relative to? In fact, that state is simply relative to any other observer (such as Bob) who would like to make predictions about what Alice may experience. In other words, the principle above is formulated in terms of an observer who himself can determine the state $\rho_{SM}$ relative to himself, and infer from that what Alice may see. We may, of course, play the game further and introduce a third observer, say Charlie, and represent Bob's description $\rho_{SM}$ as a "collapsed" version of a joint classical-quantum state $\rho_{M_B SM}$, where we've explicitly included Bob's memory $M_B$, and so on.

## 9.2 The observer's memory

The thought experiment above lets us conclude that the reality, or experience, of observing an outcome of a quantum measurement is relative to the observer. But how can we relate claims from different observers? This is done by taking into account the knowledge a particular observer has about the system. We model this knowledge explicitly by means of a *memory*.

An *observer's memory* is a system which may be correlated to a particular system $S$ of interest. The memory's role is to provide information about $S$ to the observer, according to which the observer might act or base decisions on. This system does not have to map one-to-one onto a particular physical system, it is in general a *logical system* (Section 3.4.1). It typically corresponds to a subsystem of the observer's experimental devices, their notebook, their own brain, their computer hard drive, etc.

Let us consider an example scenario. Alice is carrying out an ion trap experiment. The ion is used to implement a logical qubit, and it is prepared in the superposition state $|+\rangle = [|0\rangle + |1\rangle]/\sqrt{2}$. A separate procedure results in a measurement of the ion in the computational basis $\{|0\rangle, |1\rangle\}$. Once the experimental setup is carefully calibrated and procedures understood and established, she observes as expected either output $|0\rangle$ or $|1\rangle$ with probability ½.

Which information should we consider as residing in Alice's memory? Intuitively, there are countless different systems about which Alice has information. An incomprehensive list might include: the result of the measurement, replicated in the measurement device, computer screen and her own brain; but also the decision to perform the experiment with the $|+\rangle$ state instead of the $|-\rangle$ state, all the details of the experimental setup and which components were ordered at which company; and then as well, what she had for lunch, the current weather, where she lives, who her family and friends are, the name of the capital of Switzerland, all aspects of her past history and so on.

All such information which Alice has access to may be modeled as correlations between an explicit memory system and the system which this information refers to. The record of the measurement result is exactly correlated to the actual measurement result she witnessed. The record, either in Alice's brain or her notebook, of the decision to carry out the experiment with the $|+\rangle$ or $|-\rangle$ state is perfectly correlated to the ultimately realized experimental setup, such as the time delay of a pulse during the preparation phase before the measurement which causes either state to be prepared at the time of measurement. Her record of the outside temperature is correlated with the air temperature at the time she was last outside, and so on.

The interest of considering an explicit memory modeling Alice's information about a specific system is to relate statements made by different observers about that system. Alice's colleague, Bob, might not know the outcome of the measurement of her experiment. However, he knows that whatever the outcome, the value she recorded in her notebook is perfectly correlated to the actual result she observed. Hence, Bob best describes the situation by using a correlated state between the ion's measurement outcome and Alice's memory.

This reasoning can be applied at any level. Bob might be working on a different experiment and may not know whether Alice chose to measure a $|+\rangle$ state or a $|-\rangle$ state. Yet Bob knows that, had Alice chosen to implement the experiment for the $|+\rangle$ state, then the time delay would be set to some

specific value, and had it been for $|-\rangle$, the time delay would be set to achieve a relative phase shift of $\pi$. For Bob, Alice's record of her decision of which experiment to pursue is thus correlated to the experimental setup she built based on that decision.

The moral of the example is that *any time Bob wishes to describe which state Alice assigns to a system S, he describes this state as a correlated state between the system S and Alice's memory about that system, which is a subsystem of all systems Alice can use to store information.*

We furthermore note that for all situations which we will consider in the following, the memory of an observer is a *classical memory*, meaning that the information it contains about a system of interest is encoded in a classical-quantum state (see Section 3.1.5). In several situations, it is relevant to consider quantum side information (Berta *et al.*, 2010; del Rio *et al.*, 2011, 2014). In those cases the memory system is always assumed to be a subsystem of a physical device acting as a quantum memory, rather than any part of the observer's brain. We dare not speculate about what a quantum-information-storing brain memory would actually look like, let alone what it would itself experience.

## 9.3 Events, conditioning, and compatibility

Here, we develop a minimal formalism which allows to relate different observer's descriptions of quantum systems, allowing us to make more precise the statements which were deduced in the example in Section 9.1.

Our starting point is the notion of a quantum state for an observer. We adhere to the principle that each observer may apply the formalism of quantum information consistently. Any observer which designs an experiment producing a quantum state $\rho$ may run the experiment many times independently, measuring each state and gathering enough statistics so as to recover the state $\rho$ by performing full tomography. There is hence no ambiguity for this observer about what the quantum state of the system is.

### 9.3.1 Events

A convenient concept to introduce here is the notion of an *event*. An event, in classical probability theory, is some possible outcome or set of outcomes which may be observed. Our notion of a quantum event has a similar interpretation.

***Event.*** *Let $\rho$ be the quantum state of a system S (according to a particular*

*observer). Then a subnormalized state $\sigma$ is a possible 'event' for that observer if $\sigma \leqslant \rho$.*

Recall that a subnormalized state is a positive semidefinite operator with trace less than or equal to one, and that by $\sigma \leqslant \rho$ we mean that $\rho - \sigma$ is positive semidefinite (see Section 3.3.1).

Whenever a state $\rho_S$ is decomposed into events as $\rho_S = \sum_k \sigma_S^{(k)}$ with $0 \leqslant \sigma_S^{(k)} \leqslant \rho_S$, then this can be interpreted as the system being in the state $\sigma_S^{(k)} / \operatorname{tr} \sigma_S^{(k)}$ with probability $\operatorname{tr} \sigma_S^{(k)}$. In this sense, events generalize statistical ensembles of pure states (which are incidentally often used to define the density operator in the first place). We may think in the following way: the value $k$ was chosen at random with probability $\operatorname{tr} \sigma_S^{(k)}$ and stored in a separate register $K$, and subsequently the corresponding state $\sigma_S^{(k)}$ was prepared; the state $\rho_S$ is then the result of forgetting which state was prepared, $\rho_S = \operatorname{tr}_K\big(\sum_k \sigma_S^{(k)} \otimes |k\rangle\langle k|_K\big)$. The converse is also true: for any classical-quantum state $\rho_{SK} = \sum_k |k\rangle\langle k|_K \otimes \tau_S^{(k)}$, then $\tau_S^{(k)} \leqslant \rho_S$.

Events may also be associated with measurements performed on a purification of the state. Such a picture is convenient to represent the state of a system after it has interacted with an environment, such as a measurement device. Suppose that $\rho_S$ is purified by a state $|\rho\rangle_{SR}$ on an additional system $R$. Then any measurement on $R$ described by a POVM $\{Q_R^{(k)}\}$ induces a decomposition of $\rho_S$ into events, given as $\rho_S = \sum_k \sigma_S^{(k)}$ with $\sigma_S^{(k)} = \operatorname{tr}_R\big(Q_R^{(k)} \rho_{SR}\big)$. Observe that indeed $\sigma_S^{(k)} \leqslant \operatorname{tr}_R\big(\rho_{SR}\big) = \rho_S$. The notion of an event is hence useful to describe possible outcomes of a measurement.

### 9.3.2 The conditioning principle

How can an observer Bob make a statement about the way another observer Alice would perceive a system $S$ of interest? Bob models Alice's knowledge about $S$ with the help of a memory system $M_A$. He may then use the *conditioning principle* to describe which events Alice might observe or "collapse onto."

***Conditioning principle.*** *Suppose that Bob describes another observer Alice by their memory as a system $M_A$, and suppose the joint state on a system $S$ and on $M_A$, according to Bob, is a classical-quantum state of the form*

$$\rho_{M_A S}^{\text{Bob}} = \sum_k |k\rangle\langle k|_{M_A} \otimes \rho_S^{(k)} , \tag{9.4}$$

*with* $\operatorname{tr}\rho_S^{(k)} \leqslant 1$*. Then Bob observes the following: Alice will describe the state of* $S$ *as* $\rho_S^{(k)}/\operatorname{tr}\rho_S^{(k)}$ *for some random* $k$ *with probability* $\operatorname{tr}\rho_S^{(k)}$*. Bob himself need not experience any form of collapse.*

The conditioning principle is meant as an effective model, and does not intend to explain how or why the collapse happens for Alice.

As a consequence of the conditioning principle, we can always think of any quantum state on $S$ as a possible event of another state on $S$, which corresponds to having "collapsed" onto a particular outcome described by a classical-quantum state between the system and our own memory:

***Un-conditioning.*** *Suppose that Alice describes the state of a system* $S$ *by a quantum state* $\rho_S^{\text{Alice}}$*. Then, for* $0 < p \leqslant 1$*, one can always think of* $\rho_S^{\text{Alice}}/p$ *as one of the* $\rho_S^{(k)}$ *in the conditioning principle, with any number of other "unobserved" outcomes.*

### 9.3.3 Compatibility of events

The first thing to observe is that different observers may attribute different states to the same system. The simplest case is when two observers have different information about the system $S$. For example, Bob might describe both Alice and Charlie observing the system $S$. Alice has a memory $M_A$ and Charlie a memory $M_C$. If Alice and Charlie are two distinct observers, like we would imagine them intuitively, they have distinct memory systems and Bob would assign a tripartite classical-classical-quantum state of the form

$$\rho^{\text{Bob}}_{M_A M_C S} = \sum |k\rangle\langle k|_{M_A} \otimes |l\rangle\langle l|_{M_C} \otimes \rho_S^{(kl)}\,, \tag{9.5}$$

with subnormalized $\rho_S^{(kl)}$. The possible states which Alice would observe $S$ to occupy are obtained by tracing out $M_C$ from $\rho^{\text{Bob}}_{M_A M_C S}$ in order to obtain a state of the form (9.4) required to apply our conditioning principle. In this case, we get

$$\rho^{\text{Bob}}_{M_A S} = \sum_k |k\rangle\langle k|_{M_A} \otimes \sum_l \rho_S^{(kl)}\,, \tag{9.6}$$

and similarly for Charlie,

$$\rho^{\text{Bob}}_{M_C S} = \sum_l |l\rangle\langle l|_{M_C} \otimes \sum_k \rho_S^{(kl)}\,, \tag{9.7}$$

Then, if Alice observes the outcome $k = 1$ and Charlie the outcome $l = 1$,

then Alice observes $S$ as being in the state $\sum_l \rho_S^{(k=1,l)}$, and Charlie observes $S$ in the state $\sum_k \rho_S^{(k,l=1)}$. The two states do not have to be the same.

While Alice and Charlie may observe different states on $S$, there are some constraints. For example Alice and Charlie will never observe orthogonal states on $S$, or more generally, states whose supports don't overlap at all: distinct observers can never have contradicting statements which they both make with certainty. When two states $\sigma$ and $\sigma'$ may be described as events observed by Alice and Charlie in the way above, we say that $\sigma$ and $\sigma'$ are *compatible events* (Brun *et al.*, 2002)

Intuitively, compatible events are events which can be explained classically by an underlying true state of reality waiting to be uncovered. Consider the following example. Suppose Bob throws a die yielding a uniform random number between 1 and 6, and then copies the information about whether the number is even or odd into Alice's memory, as well as whether the number is small (less than or equal to 3) or large (greater or equal to 4) into Charlie's memory (Figure 9.2a). Bob describes the total joint state with a classical-classical-quantum state of the form (9.5), using the classical basis $\{|1\rangle_S, |2\rangle_S, \ldots |6\rangle_S\}$ on the die $S$, with the events

$$\rho_S^{\text{odd,small}} = \operatorname{diag}(1/6, 0, 1/6, 0, 0, 0)\ ; \quad \rho_S^{\text{even,small}} = \operatorname{diag}(0, 1/6, 0, 0, 0, 0)\ ;$$
$$\rho_S^{\text{odd,large}} = \operatorname{diag}(0, 0, 0, 0, 1/6, 0)\ ; \quad \rho_S^{\text{even,large}} = \operatorname{diag}(0, 0, 0, 1/6, 0, 1/6)\ .$$

If Alice has parity $k$ in her memory, and Charlie has largeness $l$ in his, then Alice and Charlie observe the die with the respective states

$$\rho_S^{\text{Alice: }k} = \rho_S^{k,\text{small}} + \rho_S^{k,\text{large}}\ ; \qquad \rho_S^{\text{Charlie: }l} = \rho_S^{\text{odd},l} + \rho_S^{\text{even},l}\ . \tag{9.8}$$

For example, if the die displays the number '2,' then Alice sees it in the (re-normalized) state $1/3\,|2\rangle\langle 2| + 1/3\,|4\rangle\langle 4| + 1/3\,|6\rangle\langle 6|$ because she only knows that it displays an even number; Charlie, on the other hand, only knows that the number is small, and hence he attributes to the die the state $1/3\,|1\rangle\langle 1| + 1/3\,|2\rangle\langle 2| + 1/3\,|3\rangle\langle 3|$. If Alice and Charlie then meet and compare their results, they will conclude that the state of the die must have been 2, because that would be the only explanation compatible with both their observations. *The fact that Alice and Charlie can meet, discuss their results, and explain their observations by an underlying true state of reality, is an inherent property of compatible events.*

***Compatible events.*** *Let $\rho$ be a quantum state. Let $\sigma, \sigma' \leqslant \rho$ be possible events of $\rho$. The two events $\sigma, \sigma'$ are said to be 'compatible' if there exists an event*

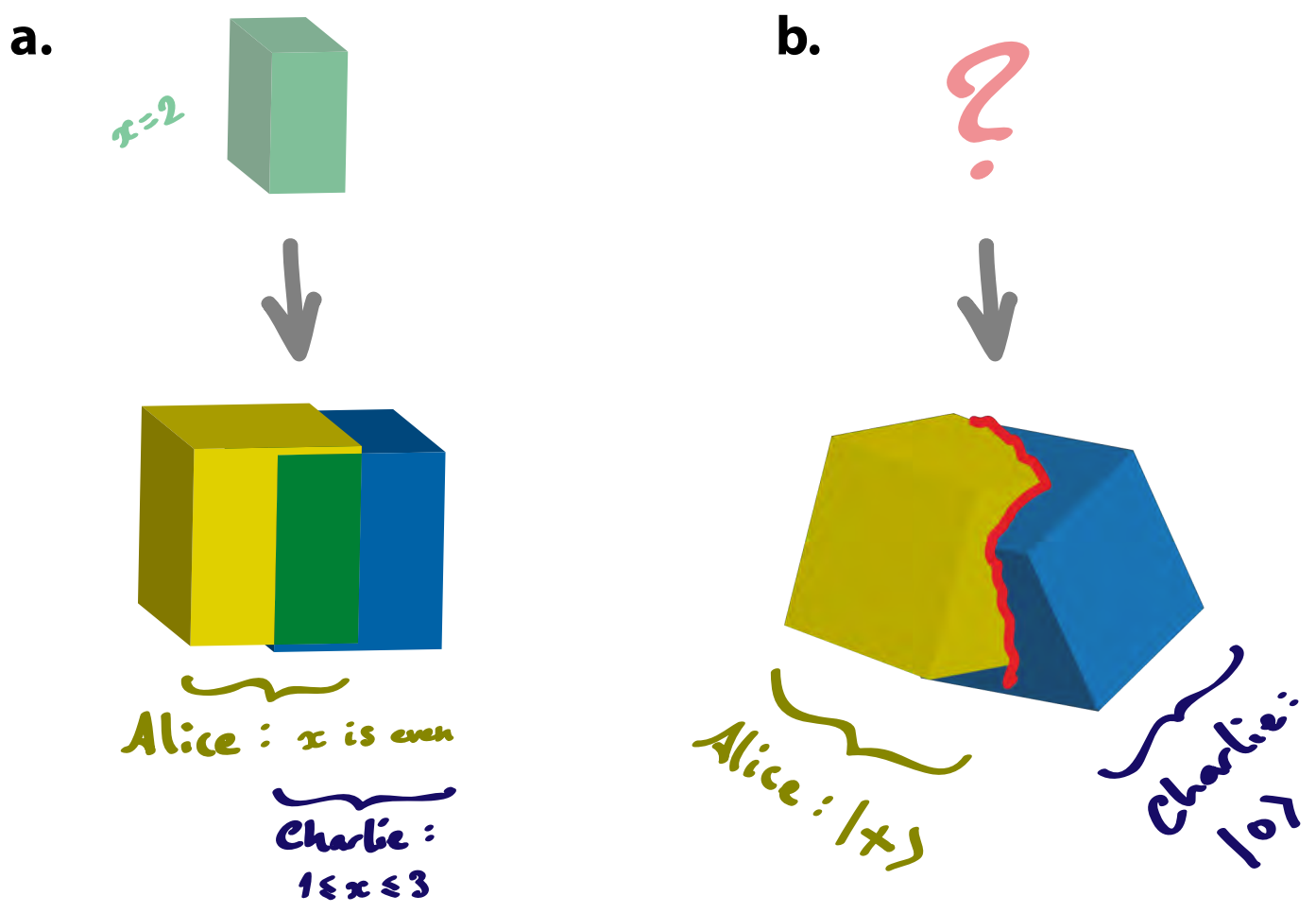

Figure 9.2: Events in quantum mechanics are not necessarily compatible. In this simple geometric analogy, events are represented by rectangular parallelepipeds. **a.** Compatible events have an underlying reality which can be in principle uncovered. In this example of a die throw, Alice's event ("the number is even") and Charlie's event ("the number lies between 1 and 3") are compatible, because both events are explained by the underlying reality that the die shows the value '2.' **b.** In quantum mechanics, events may not always be explained by a common underlying reality. If Alice observes a qubit in the state $|+\rangle$ while Charlie describes it with the state $|0\rangle$, the events are incompatible. For this to happen, Alice and Charlie must share parts of their memory; for example, "Charlie" might represent Alice after a time evolution.

*$\tau$ which is simultaneously an event of $\sigma$ and $\sigma'$, that is, such that $\tau \leqslant \sigma$ and $\tau \leqslant \sigma'$.*

In this definition, $\tau$ is a common event of $\sigma$ and $\sigma'$, and can be interpreted as an underlying true state of reality which the observers can agree on, and which "explains" both $\sigma$ and $\sigma'$. In the example above, $\tau$ corresponds to the event that the die indicates the value '2.' It is clear that if two events $\sigma$ and $\sigma'$ are compatible in the sense of this definition, then we may find a joint classical-classical-quantum state of the form (9.5): One may choose the first $\rho_S^{(k=1,l=1)}$ as being this common event $\tau$, and then complete the state to a normalized state in such a way that $\sum_l \rho_S^{(k=1,l)} = \sigma$ and $\sum_k \rho_S^{(k,l=1)} = \sigma'$.

### 9.3.4 General, incompatible observers

If the events are not compatible, they need not have a common underlying event which simultaneously explains both observed events (Figure 9.2**b**).

In general, the memory systems corresponding to each observer don't necessarily have to be distinct. The two observers are then no longer distinct as we imagine them intuitively, but they may represent for example the same observer at different times. Bob may in such cases describe two memory systems corresponding to different observers as two possibly overlapping subsystems of a common system.

This would be the case, for example, if we want to make statements about what the same observer would say at two different times during which the memory might undergo some quantum evolution. In our formalism, the observer after the time evolution is treated formally as a different observer—it has access to a different quantum memory. Let Bob describe, at time $t = 0$, Alice's memory and the system $S$ in a classical-quantum state $\rho_{M_A S}^{\mathrm{Bob},t=0}$. Suppose a quantum channel $\mathcal{E}_{M_A \to M'_A}$ is applied on the system $M_A$, mapping it to a new system $M'_A$ at time $t = 1$. The Stinespring dilation of $\mathcal{E}_{M_A \to M'_A}$ into an additional system $E$ is given by an isometry $V_{M_A \to M'_A E}$ such that $\mathcal{E}_{M_A \to M'_A}(\cdot) = \mathrm{tr}_E\big(V_{M_A \to M'_A E}\,(\cdot)\,V^\dagger_{M_A \leftarrow M'_A E}\big)$. Then, the two observers (or rather, the same observer Alice at time zero and at time one) can be described simultaneously by Bob by the state

$$\rho'_{M'_A E S} = V_{M_A \to M'_A E}\,\rho_{M_A S}\,V^\dagger_{M_A \leftarrow M'_A E}\,. \tag{9.9}$$

Indeed, the correlated state between $S$ and Alice's memory $M'_A$ at time $t = 1$ is obtained by $\rho'_{M'_A E S}$ by tracing out $E$; similarly the correlated state between $S$ and Alice's memory $M_A$ at time $t = 1$ is obtained by tracing out the com-plementary system of the embedded $M_A$ system inside the bipartite $M'_A E$ via the isometry $V_{M_A \to M'_A E}$. (It is up to the evolution $\mathcal{E}_{M_A \to M'_A}$ to ensure that both bipartite states are classical-quantum states.)

More generally, for any two arbitrary states $|\chi\rangle_S$ and $|\psi\rangle_S$, Bob may imagine a scenario with two observers Alice and Charlie with possibly overlapping memories in which Alice could observe the event $|\chi\rangle_S$, and Charlie could observe the event $|\psi\rangle_S$. We formulate this claim as a proposition.

***Proposition 9.1.*** *Let $|\chi\rangle_S, |\psi\rangle_S$ be any two pure states on S.*[3] *Let $d = \dim \mathscr{H}_S$. Let $\{|k\rangle_S\}$ and $\{|l\rangle_S\}$ be two bases on S such that the respective first elements coincide with $|\chi\rangle_S$ and $|\psi\rangle_S$, respectively. Then there exists a system R, as well as systems $M_A, \bar{M}_A$ and $M_C, \bar{M}_C$ with isometries $V_{R\to M_A \bar{M}_A}$ and $V'_{R\to M_C \bar{M}_C}$, such that there exists a quantum state $\rho_{SR}$ such that both*

$$\mathrm{tr}_{\bar{M}_A}\Big[ V_{R\to M_A\bar{M}_A}\, \rho_{SR}\, V^\dagger_{R\leftarrow M_A\bar{M}_A} \Big] = \frac{1}{d}\sum_k |k\rangle\langle k|_S \otimes |k\rangle\langle k|_{M_A}; \tag{9.10-a}$$

$$\mathrm{tr}_{\bar{M}_C}\Big[ V'_{R\to M_C\bar{M}_C}\, \rho_{SR}\, V'^\dagger_{R\leftarrow M_C\bar{M}_C} \Big] = \frac{1}{d}\sum_l |l\rangle\langle l|_S \otimes |l\rangle\langle l|_{M_C}\,, \tag{9.10-b}$$

*where $\{|k\rangle_{M_A}\}$ and $\{|l\rangle_{M_C}\}$ are bases of $M_A$ and $M_C$.*

It may be, however, that the two events are not compatible, meaning they cannot be explained as lack of knowledge of a finer-grained event. Furthermore, it makes no sense to think of Alice and Charlie as distinct observers if they partially share their memory systems: for instance, the two observers can't meet and compare their results.

*Proof of Proposition 9.1.* Let $R$ be a $d$-dimensional system with a basis $\{|k\rangle_R\}$ and let

$$|\rho\rangle_{SR} = |\psi\rangle_{S:R} = \frac{1}{\sqrt{d}}\sum_k |k\rangle_S |k\rangle_R\,. \tag{9.11}$$

Let $M_A$ and $\bar{M}_A$ each be a $d$-dimensional system with respective bases $\{|k\rangle_{M_A}\}$ and $\{|k\rangle_{\bar{M}_A}\}$, and define the isometry

$$V_{R\to M_A\bar{M}_A} = \sum_k |k\,k\rangle_{M_A\bar{M}_A} \langle k|_R\,. \tag{9.12}$$

Then, (9.10-a) is fulfilled since the resulting state on $M_A\bar{M}_A S$ via this isometry is a GHZ state.

Let $W_S = \sum_k |l = k\rangle\langle k|_S$ be the unitary matrix which relates the bases $\{|k\rangle_S\}$ and $\{|l\rangle_S\}$. Now, let $W_R = t_{S\to R}(W_S)$ be the transpose of this unitary matrix, defined with respect to the

[3]We restrict this argument to pure states for simplicity; in the case of two distinct mixed states, we may reduce the problem to pure states by considering respective purifications.

bases $\{|k\rangle_S\}$ and $\{|k\rangle_R\}$. Using properties of the maximally entangled state $|\psi\rangle_{S:R}$ discussed in Section 3.1.5, we have

$$|\rho\rangle_{SR} = |\psi\rangle_{S:R} = \left(W_S \otimes W_R^\dagger\right)|\psi\rangle_{S:R} \,. \tag{9.13}$$

Define the new basis $\{|l\rangle_R\}$ via $|l\rangle_R = W_R^\dagger|k = l\rangle_R$. Then

$$|\rho\rangle_{SR} = \frac{1}{\sqrt{d}} \sum_l |l\rangle_S |l\rangle_R \,. \tag{9.14}$$

Analogously to before we may define the systems $M_C$, $\bar{M}_C$ each of dimension $d$ and with bases $\{|l\rangle_{M_C}\}$, $\{|l\rangle_{\bar{M}_C}\}$, as well as the isometry

$$V'_{R\to M_C\bar{M}_C} = \sum_l |l\, l\rangle_{M_C\bar{M}_C}\langle l|_R \,, \tag{9.15}$$

and thus Condition (9.10-b) is satisfied as well. ∎

## 9.4 Reference frames and coherence

The idea that a quantum state does not represent an intrinsic property of a system has been evoked in particular in the context of *reference frames* (Bartlett *et al.*, 2007). While studying coherence properties of quantum radiation, a detailed analysis of the workings of a laser by Mølmer concluded that the emitted radiation was in fact not a pure coherent state, but merely a statistical mixture of states (Mølmer, 1997). Quite naturally a controversy ensued (Gea-Banacloche, 1998; Mølmer, 1998; van Enk and Fuchs, 2001; Rudolph and Sanders, 2001; Nemoto and Braunstein, 2003; Wiseman, 2004; Smolin, 2004). The resolution (Bartlett *et al.*, 2006) suggested that both views are in fact correct—they simply refer to statements made in relation to different *reference systems*. If the experimenter has access to a system which acts as a *phase reference*, then they are sensitive to the relative phase of the state of the laser and their reference and may describe the state of a laser with a well-defined phase (Figure 9.3). If, however, they have no phase reference, then for them the laser is in a statistical mixture of all possible phases. This reasoning applies more generally: We may also lift superselection rules due to the conservation of charge (Aharonov and Susskind, 1967), particle number (Dowling *et al.*, 2006) or energy (conservation law) (Brandão *et al.*, 2013; Åberg, 2014). In each of these cases, the superselection rule is lifted by a reference frame which remembers the relative quantum phases between the states of different charge, particle number, or energy.

In essence, this resolution states that different observers may assign different states to a same system, which is in line with our above conclusions.

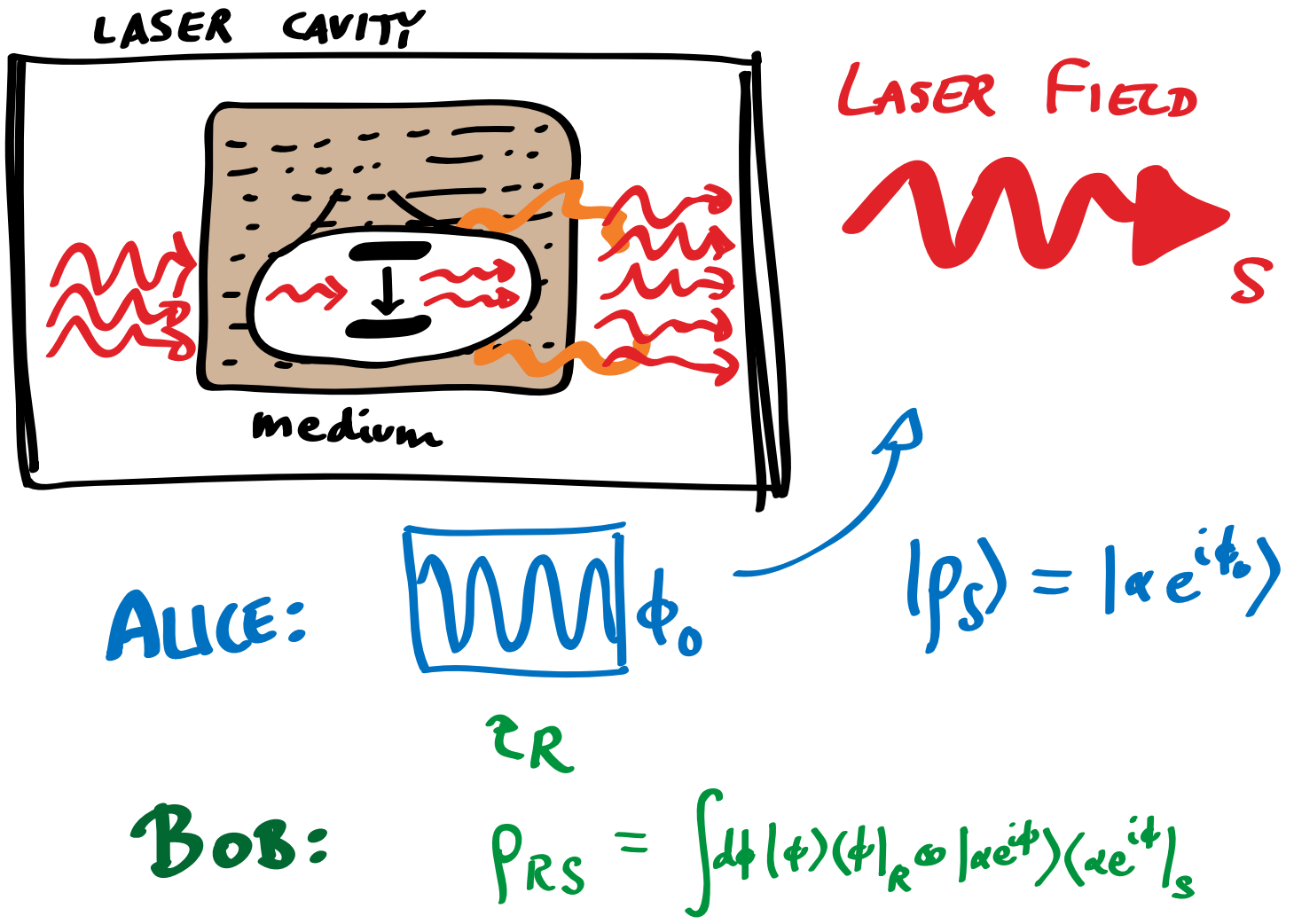

Figure 9.3: Different observers may disagree on whether the state of a laser field is a coherent state (Bartlett *et al.*, 2006). In a laser cavity, atoms of a medium contribute to stimulated emission generating the laser field. The photons in the medium are entangled with the atoms, such that the laser field should be in a mixed state (Mølmer, 1997). However, if Alice has phase-locked the laser, the laser field has a well-defined phase $\phi_0$. She may have done this by running the laser continuously, while comparing and locking its phase to a local oscillator to stabilize it to a well-defined value $\phi_0$. If Bob does not have access to Alice's local oscillator, the state of the field is an incoherent mixture of all phases. He can describe Alice's point of view as a correlated state between Alice's local oscillator and the laser field: the local oscillator acts as a *reference frame*.

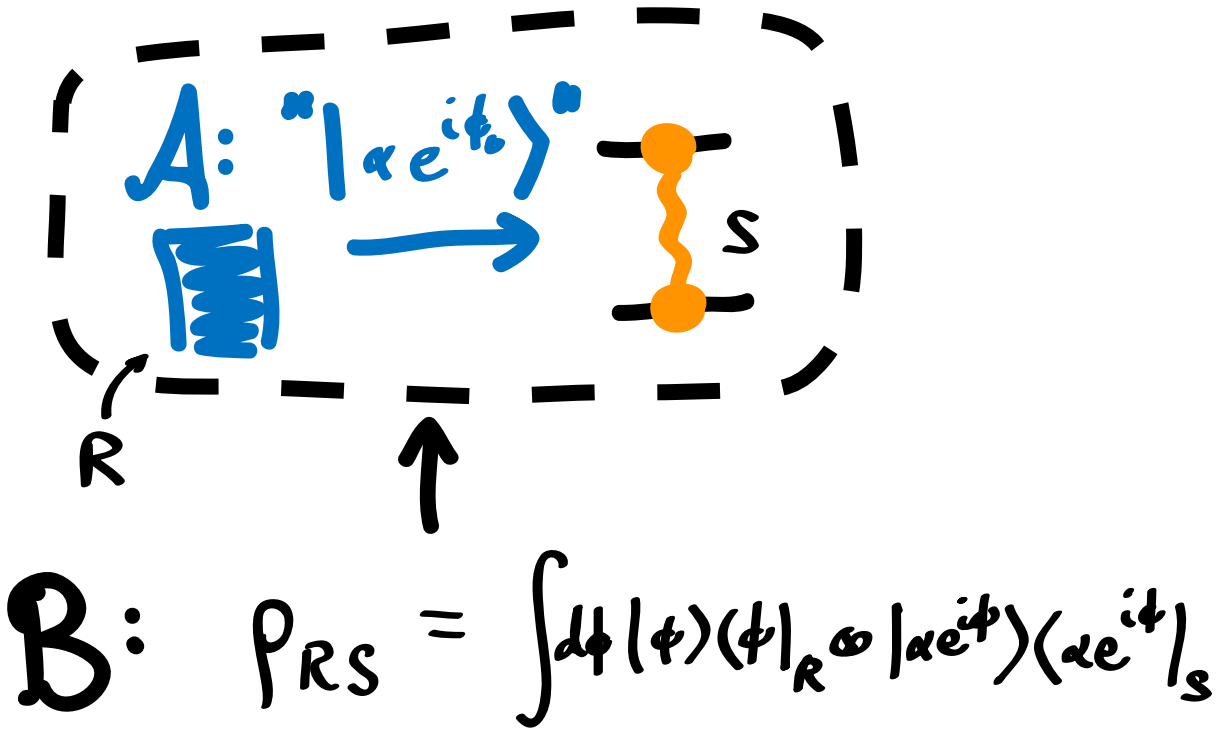

Figure 9.4: A reference frame can be treated as an observer's memory. Alice observes a system in a well-defined phase $\phi_0$, which Bob can explain by a correlated state between Alice's reference frame and the system (Figure 9.3). The reference frame is Alice's memory, and the conditioning principle relates the two descriptions of Alice and Bob, by stating that Alice collapses onto a well-defined value of the phase. In this way, Alice may observe for example coherent superpositions of energy levels on $S$ whereas Bob describes the reduced state on $S$ as an incoherent state.

An observer with a reference frame may see coherent states, one without, may not.

A reference frame may often in fact be treated as the observer's memory in the sense of Section 9.2. If Bob sees that Alice has a phase reference with respect to her laser, Bob may describe the joint state of Alice's memory, the reference frame $R$, and the laser field $S$ as

$$\int d\phi \, |\text{“}\phi\text{”}\rangle\langle\text{“}\phi\text{”}|_R \otimes |\alpha e^{i\phi}\rangle\langle\alpha e^{i\phi}|_S \,. \tag{9.16}$$

Since this is a classical-quantum state, we may apply the collapse principle presented above. According to the collapse principle, Alice observes the system $S$ as being in a state $|\alpha e^{i\phi}\rangle_S$ for some well-defined phase $\phi$, which is chosen randomly according to Bob (Figure 9.4).

States of a laser field with a well-defined phase is only one example of a situation which can be well analyzed in the context of reference frames. The full framework of reference frames has a much wider scope, allowing to analyze any system subject to the representation of a symmetry group. We shall return to reference frames, this time in their full generality, later in the context of coarse-graining.

## 9.5 Conceptual remarks

### 9.5.1 What about tomography?

In quantum information, it is known that the quantum state can be determined by performing quantum tomography. In tomography, statistics are collected over many measurements performed on independent copies of the system in such a way to reconstruct the quantum state of the system. So how can the quantum state be relative to an observer, if the quantum state can be determined *a priori* in this way?

If you're puzzled by this statement, then probably the answer is not the one you're expecting.

Recall that we had the same situation in a completely classical setting. If Alice tosses a coin, and looks at the result without revealing it, then while she observes a particular outcome, say '**0**,' Bob describes the situation as a correlated state between the coin and Alice's memory. For Alice, the state of the coin is '**0**,' and for Bob, it is '**0**' or '**1**' with equal probability.

The key is that the states that Alice and Bob assign to the coin do not describe the same experiment. For Alice, the experiment corresponds to "tossing a coin and observing the outcome '**0**'"—after which, trivially, the probability of seeing the coin in the state '**0**' is 100%. For Bob, the experiment corresponds to just "tossing a coin"—which could then be in the state '**0**' or '**1**' with equal probability. In other words, Alice's experiment includes a post-selection on the '**0**' state which isn't included in Bob's idea of the experiment. So if the coin is tossed many times, and they each perform tomography, Alice will reject half of the outcomes owing to a failed realization of the experiment, only keeping the '**0**' outcomes, while Bob will take into account all outcomes. This is why they arrive at different states.

In practice, it is ubiquitous to post-select on a particular event in an experiment, and this is often done implicitly. For example, if while attempting to toss a coin, I fail to catch it and it falls on the floor, then anyone would expect me to simply pick it up and toss it again, ignoring my failed attempt. Similarly, if a superconducting qubit is to be initialized in the ground state, then a possible strategy to do so is to measure the energy of the qubit in a thermal state, and post-select on it being in the ground state (true story). The remainder of the experiment can then be perfectly carried out by describing the qubit in the ground state, regardless of whether its preparation was subject to post-selection.

In summary, the crucial point is that a quantum state is associated to an *experiment*, as already pointed out in Section 3.4.2. The fact that different observers may associate different states to the same physical system simply means that they refer to different experiment preparations; perhaps one observer is post-selecting on a particular condition being satisfied.

### 9.5.2 Emergent common reality

If reality and events are observer-dependent, how come do we all seem to agree on which events occur or don't occur? Why do we feel like there is a common, unambiguous reality all around us?

The answer is simply that events which we observe are typical states of macroscopic thermodynamic systems which are diagonal in a preferred, classical basis, and thus all events we macroscopic observers see are compatible events. They can always be explained by a common underlying true state of reality in the sense explained above.

Even if experiments are carried out on quantum systems, ultimately the systems we as humans interact with are classical systems, such as the display of a computer or an of an oscilloscope.

### 9.5.3 Lost information

As we noted in Section 9.1, information may be lost, to the point where it may not even make sense to claim that it ever existed. In the example presented there, after having reversed the measurement, there is no system which remains correlated to the value of the outcome $k_{\text{outcome}}$ which Alice had observed: this information is irremediably gone. The way this information is lost is such that it is guaranteed that no other party could be in possession of a copy. The information has ceased to exist entirely.

### 9.5.4 Relation to the relative state and many-worlds interpretation

The view presented above can be seen as a minimal version of the many-worlds interpretation of quantum mechanics. The latter school of thought defends the existence of many worlds, or parallel universes, which endlessly split into many parallel branches (Everett III, 1956, 1957; DeWitt, 1970; DeWitt and Everett III, 1973; Tegmark, 2007; Vaidman, 2016).

If we abandon the notion of an absolute reality, there is no longer a need to decide when or how, in an absolute sense, the branches of the universe split. They split, or do not split, relative to each observer. We might be tempted to call these branches *an observer's parallel realities*, avoiding the words 'universes' or 'worlds' which might bear an "absolute" or "observer independent" connotation. For that matter, one might even put in question the relevance of the 'universal wavefunction' altogether, in favor of a relative description of states between observers which need not fit into any "absolute" picture.

The present view also introduces a kind of "merging" of the different branches of reality. For example, in the setup of Section 9.1, as Alice performs her measurement, Bob might describe her as splitting into branches of parallel realities representing the possible outcomes; after he reversed her measurement, he might interpret this as the branches recombining back into a single reality branch.

While the picture of parallel branches of realities is convenient, we emphasize from a more pragmatic point of view that the question of whether the different 'realities' actually exist in some 'parallel universes' or 'worlds' is more of a metaphysical question, an answer to which is not crucial for the corresponding theory to produce physically sound statements.

Finally, we note the closeness of the present arguments to the original formulation by Everett III (1957), who prefers in that reference the terminology 'relative state formulation' and avoids any mention of 'parallel worlds' or 'universes.' To recover the present view it suffices from there to make explicit the idea that reality is not an absolute concept—which then no longer necessarily requires there being several 'worlds' or 'universes.'

# 10

# Coarse-Graining and Hidden Information

If different observers can assign different states to the same quantum system, why should the states even lie in the same state space? In this chapter, we argue that different observers may have radically different notions of what the fundamental units of information contained in a system are, assigning states from different state spaces to the same system. Yet, we will see that this is still completely compatible with quantum mechanics.

We propose to take the framework of quantum information seriously for each individual observer. Namely, an observer should be able to specify what their fundamental unit of information is, and then apply consistently the framework of quantum information on that state space. Crucially, because quantum states are relative to the observer and not an absolute concept, different observers may have different descriptions of the same physical system.

We develop a *coarse-graining procedure* which allows to connect the statements of two different observers which may not have access to the same information. The coarse-graining operation is simply defined as forgetting part of the information contained in a quantum system, as represented by the partial trace of some logical subsystem of the system of interest as described by an observer with access to the fine-grained information.

This is precisely what happens in thermodynamics, for example. For a macroscopic observer, the fundamental units of information are those macroscopic quantities such as temperature, volume and pressure, and they only care about physical statements involving these quantities. The macroscopic observer may even be completely ignorant of even the existence of microscopic degrees of freedom. Yet another observer might have knowledge and control over the microscopic degrees of freedom of the thermodynamic system, and they would well know that what the macroscopic observer ignorantly calls 'temperature,' 'volume' and 'pressure' corresponds in fact to a particular well-defined quantum state on the microscopic state space.

For the coarse-grained observer there might remain some signs that there are in fact microscopic states hiding behind the coarse notions of temperature, volume and pressure. One important sign is that the macroscopic observer can derive a *thermodynamic potential*, the *thermodynamic entropy*, that governs which processes are possible.[1] On the other hand one might suppose that there are microscopic particles which form the thermodynamic system, and that the macroscopic quantities correspond to coarse-graining microscopic states together. Each macroscopic state may correspond to a different number of microscopic states: this number is precisely the entropy.[2]

This is represented in our general analysis by the *hidden information*. We argue that the hidden information operator assigned to a system can be interpreted *as if* the system were a coarse-grained version of additional inaccessible degrees of freedom, finally justifying our choice of terminology.

Furthermore, this picture is still completely compatible with quantum mechanics. Indeed, we may take the formalism of quantum information seriously, and apply it for each individual observer using their individual idea of what the state space of the system is. This can be seen by writing out the coarse-graining procedure explicitly as a partial trace of the system containing the inaccessible information: the remaining system containing the macroscopic information is still a quantum system, and nothing prevents the macroscopic observer *a priori* from creating coherent superpositions of these states, provided they possess a suitable reference frame.

The idea of applying the formalism of quantum information to coarse-grained states might seem contradicting because in thermodynamics one never observer "quantum superpositions of different thermodynamic states." We will argue, however, that this is because in this specific case the macroscopic observer does not have access to an appropriate reference frame. To illustrate our claim, we exhibit a rudimentary experimental setup which may legitimately be interpreted as preparing a quantum superposition of coarse-grained states, in the above sense.

[1] This is indeed how the concept of thermodynamic entropy was originally put forward by Clausius (1854, 1865) …

[2] … and this is indeed how the thermodynamic entropy was later interpreted by Boltzmann, Gibbs, Maxwell, and the rest of that crowd.

## 10.1 Coarse-grained observers

Here we proceed to define a *coarse-grained observer* with respect to another observer. The coarse-graining refers to the fact that the coarse-grained observer doesn't have access to part of the information the original observer has. The partial information about the system is represented by embedding the state of the system into logical systems (see Section 3.4.1).

***Coarse-Grained Observer**. Let a physical system Σ be described by an observer with quantum states in the state space $\mathscr{H}^{\mathcal{M}}$. For any way to embed $\mathscr{H}^{\mathcal{M}}$ into two logical systems $C \otimes M$, a 'coarse-grained observer' is an observer who has access to $C$ only, and who has no access to $M$. The coarse-grained observer will then effectively describe the original system with quantum states in $\mathscr{H}^{\mathcal{A}} = \mathscr{H}_C$.*

All along this chapter, we'll call the original observer Mike, and the coarse-grained observer Alice. The situation is depicted in Figure 10.1. From the point of view of Alice, the system is described with quantum states on $\mathscr{H}_C$. If she suspects the existence of a space $M$ with additional information, she would call that space the "microstates" of the system Σ.

To avoid confusion, we distinguish wherever might be ambiguous the *physical system* being described from the *quantum system* used to describe it. The physical system may be for example an atom, a gas of particles, or any system in the usual physical sense. A quantum system refers to a particular Hilbert space which acts as an abstract storage space for quantum information, and which is used by a particular observer to represent the possible states of a physical system.

## 10.2 The hidden information

Alice, the macroscopic observer, will describe the physical system Σ using the quantum system $C$. That is because every experiment, observation, evolution, or measurement she can conceive can be perfectly well explained by a quantum state, an evolution or measurements on $\mathscr{H}_C$. This is precisely because any of her actions may not access the degrees of freedom in $M$.

There may be signs left by the microscopic degrees of freedom, however, if there is some form of uncontrolled interaction between $C$ and $M$. For example, if $M$ may be a heat bath thermalizing $C$. In this case, different kets in $M$ may behave differently: For example, the equilibrium state may have different weights on different kets. This is the case of the Gibbs state.

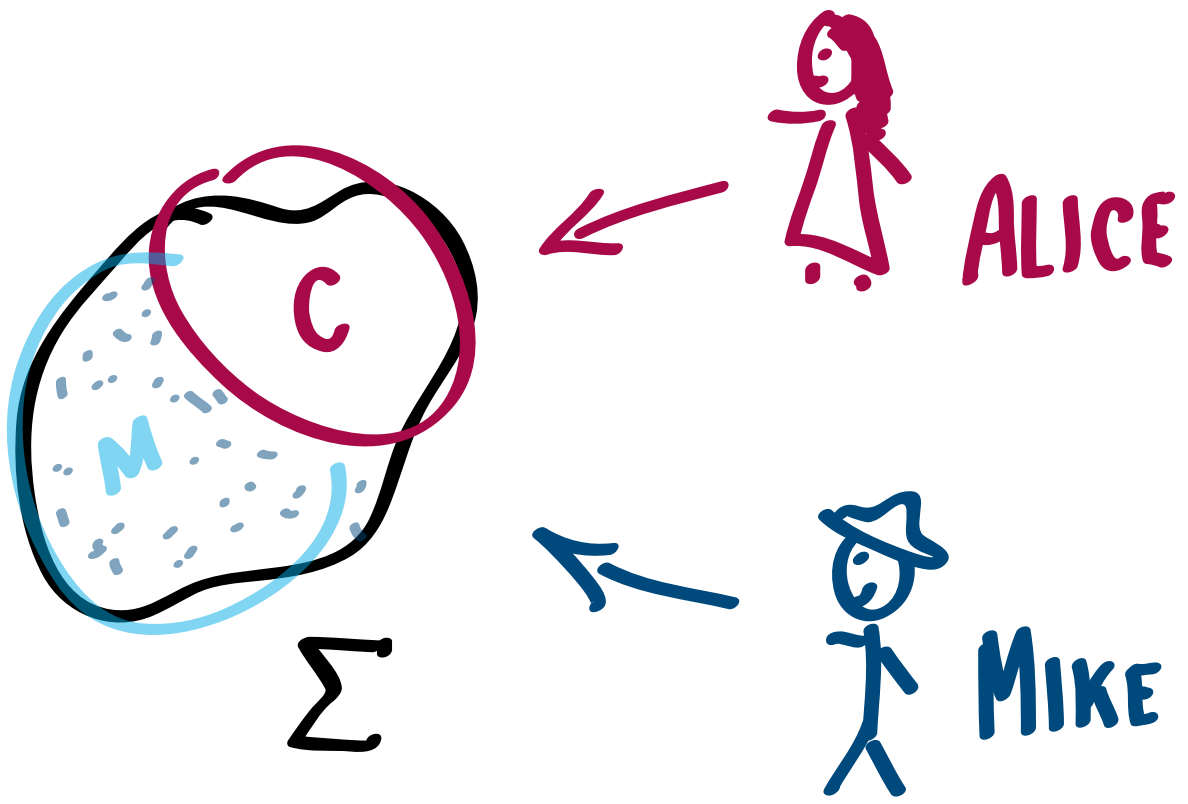

Figure 10.1: Coarse-Grained Observer. Mike describes the physical system $\Sigma$ as taking states in the state space $\mathscr{H}^{\mathcal{M}}$. Alice, on the other hand, only sees a part of the system and then only uses the state space $\mathscr{H}^{\mathcal{A}} = \mathscr{H}_C$ to describe the states of the system. The complementary system $M$ contains the degrees of freedom which are inaccessible to Alice. The systems $C$ and $M$ are typically logical systems, i.e. not parts of the system which can be separated physically, but rather different types of information such as the position of a particle versus its spin.

The weights on each ket may be interpreted as a witness for the inaccessible degrees of information, the *hidden information*.

First, we briefly study the example of coarse-graining hydrogen states to illustrate our point. We then focus on the general question of relating different observers' descriptions, and argue that the evolutions as seen by the microscopic observer are related to those of the coarse-grained observer via a recovery map. This completes the physical interpretation of the hidden information as witness of possible coarse-grained states. The classical situation was exposed in Section 5.1.

### 10.2.1 Example: hydrogen states

There might be some constraints due to the structure of the embedding of $\mathscr{H}^{\mathcal{M}}$ into the logical systems, causing some states to be forbidden. For example, if hydrogen states $|k\ l\ m\rangle_S$ are embedded into systems $S_K \otimes S_L \otimes S_M$ as $|k, l, m\rangle \rightarrow |k\rangle_{S_K}|l\rangle_{S_L}|m\rangle_{S_M}$, then states $|k\rangle_{S_K}|l\rangle_{S_L}|m\rangle_{S_M}$ with for example $m > l$ are forbidden and will never be observed. Suppose Alice has access to $C = S_K \otimes S_L$ but is oblivious to the information in $M = S_M$. This means that each state $|k\ l\rangle_{S_K S_L}$ which Alice has access to effectively corresponds to a different numbers of microstates in $M$. This is relevant if the register $C$ is not isolated and we have microscopic dynamics going on: for example, the atom may be in contact with a heat bath causing its state to equilibrate to a Gibbs state. The Gibbs state is then a witness for the existence of the microstates: for example, if the atom is in zero field, and fine and hyperfine corrections are neglected to assume that all the block with $k = 0$ is degenerate, and that Alice lets the atom thermalize while keeping $k$ at a definite fixed value, then she will observe that the values of $l$ are not all equiprobable: higher values of $l$ will appear more likely than others, even though their energy are all equal. This is because of the different number of $m$ states which each $l$ state corresponds to.

### 10.2.2 Coarse-graining the spontaneous processes

Suppose the microscopic observer Mike himself describes spontaneous logical processes on $C \otimes M$ with a trace-nonincreasing, $\Gamma$-sub-preserving map $\Phi$ for a given $\Gamma_{CM}$,

$$\Phi(\Gamma_{CM}) \leqslant \Gamma_{CM} . \tag{10.1}$$

In the case of the hydrogen atom above, we might choose for example

$$\Gamma_{CM} = \Gamma_{S_K S_L S_M} = \sum_{\text{valid } k,l,m} |k,l,m\rangle\langle k,l,m|_{S_K S_L S_M} \,. \tag{10.2}$$

How are then the spontaneous dynamics best described by Alice? Alice does not know the state of the register $M$, so we may not simply calculate the reduced map on $C$. Rather, we would like to express the fact that Alice is maximally ignorant about the state of $M$. One could be tempted to assign a Gibbs state to $M$, say $\Gamma_M / \operatorname{tr} \Gamma_M$, which would correspond to the case in thermodynamics with a heat bath. However, this would not correctly account for correlations between $C$ and $M$ in the $\Gamma_{CM}$ operator. (For example with the hydrogen states, for a given $\rho_{S_K S_L}$, if we try to write down a state of the form $\rho_{S_K S_L} \otimes \gamma_M$ for some Gibbs state $\gamma_M$, then we would most likely overlap with forbidden states.)

The answer is to turn to the *reverse process* of the partial trace discussed in Section 5.4. The reverse mapping of the logical process which traces out the $M$ system is given by the recovery map (5.25),

$$\mathcal{R}_{C\to CM}(\cdot) = \Gamma_{CM}^{1/2} \left( \left( \Gamma_C^{-1/2} \, (\cdot) \, \Gamma_C^{-1/2} \right) \otimes \mathbb{1}_M \right) \Gamma_{CM}^{1/2} \,. \tag{10.3}$$

Indeed, and as argued in Section 5.4, the recovery map is the reverse process which accurately describes one's lack of information about the $M$ system.

We now return to the question about the reduced dynamics for Alice. Whenever Mike describes the dynamics via the microscopic map $\Phi_{CM}$, then Alice should describe her coarse-grained dynamics as

$$\Phi_C^{\mathcal{A}}(\cdot) = \operatorname{tr}_M(\Phi_{CM}(\mathcal{R}_{C\to CM}(\cdot))) \,. \tag{10.4}$$

The recovery map fills in the state on $M$ while accurately describing Alice's lack of information about it; $\Phi_{CM}$ is the actual microscopic dynamics, and the partial trace again refers to Alice's lack of access to $M$.

Remarkably, $\Phi_C^{\mathcal{A}}$ is a trace-nonincreasing $\Gamma_C$-sub-preserving map. It is trace-nonincreasing as a composition of such maps. Also,

$$\begin{aligned}\Phi_C^{\mathcal{A}}(\Gamma_C) = \operatorname{tr}_M(\Phi_{CM}(\mathcal{R}_{C\to CM}(\Gamma_C))) &\leqslant \operatorname{tr}_M(\Phi_{CM}(\Gamma_{CM})) \\ &\leqslant \operatorname{tr}_M(\Gamma_{CM}) = \Gamma_C \,. \end{aligned} \tag{10.5}$$

We emphasize that the reasoning above holds for any arbitrary $\Gamma_{CM}$. Hence, the coarse-grained dynamics of $\Gamma_{CM}$-sub-preserving maps for any $\Gamma_{CM}$ gives $\Gamma_C$-sub-preserving maps.

### 10.2.3 Physical interpretation of the hidden information

Finally, we may give a well justified interpretation to the hidden information operator $\Gamma$ also in the fully quantum case.

First, it is defined as an abstract operator which spontaneous processes must sub-preserve.

However, we now see that the hidden information $\Gamma_C = \sum_k g_k |k\rangle\langle k|_C$ on a system $C$ can always be seen as a coarse-grained version of a bipartite system $CM$ with a global constraint on the possible states, and where to each ket $|k\rangle_C$ on $C$ may correspond exactly a number of states in $M$ proportional to $g_k$.

There are many ways this information can be coarse-grained, and in fact, $\Gamma_C$ can be explained by the coarse-graining of any extension $\Gamma_{CM}$ of $\Gamma_C$ with $\Gamma_C = \mathrm{tr}_M \Gamma_{CM}$ on the system $M$. As seen in the previous section, these extensions are fully compatible with the definitions of $\Gamma$ as an operator which the spontaneous dynamics must sub-preserve.

Of course, the additional system $M$ doesn't have to actually *exist*, but the formalism behaves exactly *as if* they existed.

## 10.3 Thermodynamic states as coarse-grained states

Recall our *modus operandi* in Chapter 8, where we showed how to recover macroscopic thermodynamics from our generic microscopic information-theoretic framework. On a system $S$, a class of *thermodynamic states* was singled out, corresponding to states which we have access to and which can be interconverted reversibly according to the coherent relative entropy. The class of thermal states need not exhibit any particular structure such as spanning a subspace; they are merely parameterized by some variables $\{z_j\}$.

The thermodynamic variables $\{z_j\}$, by definition, are states which can be controlled by the macroscopic observer. Let's keep them in an explicit memory register $C$, the *control system*, in the form of orthogonal states $\{|\vec{z}\rangle\}$ (as we did in Section 8.1.6). The register $C$ may be thought of as describing the state of the device manipulating the system $S$, for example storing the positions of the walls of a box enclosing a gas. The thermal states $\{\vec{z}\}$, in this picture, are represented by microstates of the form $|\vec{z}\rangle\langle\vec{z}|_C \otimes \tau_S^{(\vec{z})}$. In the most general case, the hidden information of the bipartite $CS$ is given by an operator $\Gamma_{CS} = \sum_{\vec{z}} |\vec{z}\rangle\langle\vec{z}|_C \otimes \Gamma_S^{(\vec{z})}$.

A macroscopic observer does not have access to the quantum system $S$, which contains the microscopic details of the prepared state, but rather has access to $C$, the register which remembers which thermal state was prepared. This is a natural context in which our coarse-graining procedure applies.

## 10.4 Coarse-graining by symmetry

A natural situation of interest is if the coarse-graining is due to a physical symmetry of the system. In this case, the possible states which a system can assume is not the same from the perspective of an observer which is oblivious to the symmetry, than from the point of view of an observer who has a reference frame.

The symmetry is embodied by the representation of a group on the Hilbert space. A representation of a group is a *symmetry* if an observer is oblivious to transformations corresponding to the representation of a group.

Examples of symmetry group are the rotation group $\mathrm{SO}(3)$, for example, representing rotations of our three-dimensional space, or $\mathrm{U}(1)$, representing time evolution, or also $\mathrm{S}_N$, the symmetric group of $N$ particles.

### 10.4.1 Symmetry group

The situation in which the quantum system is subject to a particular symmetry has been extensively studied, and the formalism pertaining to it has been developed in the context of reference frames (Bartlett *et al.*, 2007). Suppose there is a symmetry group $G$ with a unitary representation $U(g)$ on Mike's quantum system. Concretely, this means that one associates to an abstract transformation element $g \in G$ a concrete rule of how to transform the kets in the quantum system according to the abstract transformation $g$.

Suppose that these transformations form a symmetry of the system. This means that if such a transformation is carried out, then one cannot tell the difference before or after. Imagine, for example, that the state was prepared in a laboratory whose orthonormal coordinate system has unknown orientation with respect to Mike's. In this case, the quantum system has to be represented by a state which is symmetrized according to all possible rotations. In general, Mike must assign to the system a state which is *symmetrized* using the *symmetrization superoperator* or *twirling superoperator*

$$\mathcal{G}(\cdot) = \int dg\, U(g)\,(\cdot)\,U^\dagger(g)\,. \tag{10.6}$$

This can be seen, for example, as simply averaging over all possible relative orientations of the reference coordinate systems used for the state preparation and for the measurement. (In the case of a discrete group, the integral is just a sum over the group elements. For a continuous group, we need an appropriate Haar measure.)

### 10.4.2 Charge sectors and superselection rule

Representations of groups (and especially compact groups) present a lot of structure we can exploit. Again, this has been used in the context of reference frames; we'll quickly review here the main steps (Bartlett *et al.*, 2007). The state space $\mathscr{H}_S$ of the full quantum system $S$ splits into *charge sectors* labeled by *charges* $q$:

$$\mathscr{H}_S = \bigoplus_q \mathscr{H}_S^q \,. \tag{10.7}$$

These charge sectors further decompose into tensor products as

$$\mathscr{H}_S^q = \mathscr{M}_S^q \otimes \mathscr{N}_S^q \,, \tag{10.8}$$

where $\mathscr{M}_S^q$ carries an irreducible representation of $G$, and where $\mathscr{N}_S^q$ acts as the multiplicity space and does not transform under $G$. We will further denote by $d_q = \dim \mathscr{M}_S^q$ the dimension of a particular irreducible representation corresponding to the charge $q$, and by $n_q = \dim \mathscr{N}_S^q$ its multiplicity. So we can write

$$\mathscr{H}_S = \bigoplus_q \mathscr{M}_S^q \otimes \mathscr{N}_S^q \,, \tag{10.9}$$

while the representation takes the form

$$U_S(g) = \bigoplus_q U^q_{\mathscr{M}_S^q}(g) \otimes \mathbb{1}_{\mathscr{N}_S^q} \,. \tag{10.10}$$

As a consequence of this decomposition, any operator $A_S$ which is symmetric, that is which is left invariant by the symmetry or equivalently $\mathcal{G}_S(A_S) = A_S$, must take the block-diagonal form

$$A_S = \bigoplus_q \mathbb{1}_{\mathscr{M}_S^q} \otimes A^q_{\mathscr{N}_S^q} \,. \tag{10.11}$$

In particular, a quantum state which respects the symmetry can be written

as

$$\rho_S = \bigoplus_q \left( d_q^{-1} \mathbb{1}_{\mathscr{M}_S^q} \otimes \rho^q_{\mathscr{N}_S^q} \right) , \tag{10.12}$$

or, visually,

$$\rho = \begin{pmatrix} \boxed{d_1^{-1}\mathbb{1}_{d_1} \otimes \rho^1} & 0 & \cdots \\ 0 & \boxed{d_2^{-1}\mathbb{1}_{d_2} \otimes \rho^2} & \\ \vdots & & \ddots \end{pmatrix} , \tag{10.13}$$

where $\rho^q$ are positive semidefinite operators such that $\sum_q \operatorname{tr} \rho^q = 1$.

The state is forced to exhibit a block-diagonal form, forbidding coherent superpositions between charge sectors. This is known as a *superselection rule*.

### 10.4.3 Coarse-graining

Suppose now that another observer, Alice, is completely ignorant even of the existence of this symmetry and of what the underlying quantum state space is. Rather, Alice performs experiments, and observes that she can create, measure, and in general play around with states which may be described using the quantum formalism as block-diagonal,

$$\rho^{\mathcal{A}} = \begin{pmatrix} \rho^1 & 0 & \cdots \\ 0 & \rho^2 & \\ \vdots & & \ddots \end{pmatrix} , \tag{10.14}$$

where each of the $\rho^q$ happen to live in some space of some empirical dimensionality $n_q$. Alice concludes the quantum Hilbert space associated to her system is of dimension $\sum_q n_q$, with an apparent physical restriction forbidding her to create any superposition of states between different blocks in (10.14). She deduces that there is an observable corresponding to a physical charge which must be conserved; this is her way of formulating the superselection rule.

Unless Alice can put her hand on an experiment which probes the microscopic degrees of freedom, or any reference system which is not symmetric, then only by guesswork could she interpret this superselection rule as a restriction imposed by a symmetry on a larger quantum space.

In order to connect this with the coarse-graining procedure defined above, define the basis $|q, m, n\rangle_S$ compatible with the decomposition (10.9), such that the quantum number $q$ designates which charge sector the ket lives in, $m$ the basis vector in $\mathscr{M}_S^q$ and $n$ the basis vector in $\mathscr{N}_S^q$. Now embed this state into the (logical) quantum systems $Q \otimes M \otimes N$ as $|q, m, n\rangle_S \to |q\rangle_Q |m\rangle_M |n\rangle_N$. The spaces $Q$, $M$ and $N$ have to be chosen large enough such that all valid combinations of quantum numbers may be represented. This leads also to "forbidden" states, namely those basis states corresponding to an invalid combination of $(q, m, n)$ quantum numbers. As long as our considerations are restricted to the subspace consisting of valid quantum numbers, however, we do not have to worry about this.

The coarse-grained state of the system can be related to the microscopic version by tracing out the system $M$:

$$\rho^{\mathcal{A}} = \mathrm{tr}_M\big[\rho^{\mathcal{M}}_{QMN}\big] . \tag{10.15}$$

In particular, we can check that if the state $\rho$ is symmetric with respect to the action of $G$, then $\rho^{\mathcal{A}}$ must be of the form

$$\rho^{\mathcal{A}} = \sum_q |q\rangle\langle q|_Q \otimes \rho_N^q , \tag{10.16}$$

that is, in accordance with (10.14), a classical-quantum state on the $QN$ systems.

On the other hand, if $\rho^{\mathcal{A}}$ is of the form (10.16), then the corresponding symmetric $\rho$ is simply given as

$$\rho^{\mathcal{M}} = \sum_q |q\rangle\langle q|_Q \otimes d_q^{-1} \mathbb{1}_M^{d_q} \otimes \rho_N^q \tag{10.17}$$

by adding on an mixed state on $M$ with support on the allowed values of $m$ for this charge sector, denoted by $\mathbb{1}_M^{d_q}$. The state $\rho^{\mathcal{M}}$ given in this way is symmetric because the group action for each $q$ is precisely nontrivial only on the $M$ system with precisely the support of dimension $d_q$. Equivalently, by knowing only the state on $QN$, Alice is maximally ignorant about $M$, as represented with a maximally mixed state, except for the states which are anyway forbidden and unphysical—as represented by the mixed state over a certain rank $d_q$ only.

Equation (10.17) represents Alice's knowledge as perceived from the microscopic point of view. Another way to write this microscopic description is to imagine that the microscopic observer initializes $M$ in a state $|0\rangle_M$ which is compatible with all values of $Q$, effectively embedding Alice's state as

$$\rho^{\mathcal{A}} \to \rho^{\mathcal{A}}_{QN} \otimes |0\rangle\langle 0|_M \,, \tag{10.18}$$

which corresponds to assuming that the microscopic observer has full knowledge of the microscopic details, i.e. they possess a reference frame. The choice of a particular state $|0\rangle_M$ is without loss of generality, as a different choice could be absorbed in the choice of the embedding $S \to QMN$. If we have a reference frame, we may represent the situation as if we knew which transformation $g$ has been applied (as this information is stored in the reference frame), and we may represent the embedding of Alice's state via the isometry

$$W_{QN\to QMN}(g) = \sum_{qn} |qn\rangle\langle qn|_Q \otimes \left(U^q_M(g)|0\rangle_M\right) , \tag{10.19}$$

as

$$\rho^{\mathcal{A}} \quad \to \quad W_{QN\to QMN}(g)\, \rho^{\mathcal{A}}\, W^\dagger_{QN\leftarrow QMN}(g) \,. \tag{10.20}$$

If no reference frame is present, we need to average the resulting state over all possible transformations. This is equivalent to applying the symmetrization superoperator $\mathcal{G}(\cdot)$ on the state (10.18), which gives us exactly the state (10.17).

We note that the embedding (10.17) is exactly the one given by the recovery map (10.3). Indeed, with $C = QN$ and $\Gamma_{QNM} = \sum_q |q\rangle\langle q| \otimes \Gamma^q_N \otimes \mathbb{1}^{d_q}_M$, we have

$$\begin{aligned} \Gamma^{1/2}_{QNM}\Gamma^{-1/2}_{QN} &= \sum_q \left(|q\rangle\langle q|_Q \otimes \Gamma^q_N \otimes \mathbb{1}^{d_q}_M\right) \cdot d_q^{-1/2}\left(|q\rangle\langle q|_Q \otimes \Gamma^q_N\right)^{-1/2} \\ &= \sum_q d_q^{-1/2} \left(|q\rangle\langle q|_Q \otimes \Pi^{\Gamma^q_N}_N \otimes \mathbb{1}^{d_q}_M\right) , \end{aligned} \tag{10.21}$$

and hence for $\rho^{\mathcal{A}} = \sum |q\rangle\langle q|_Q \otimes \rho^q_N$, and assuming that $\Gamma^q_N$ has full support for simplicity,

$$\mathcal{R}_{QN\to QNM}\left(\rho^{\mathcal{A}}\right) = \sum_q |q\rangle\langle q|_Q \otimes \rho^q_N \otimes \left(d_q^{-1}\mathbb{1}^{d_q}_M\right) , \tag{10.22}$$

as given in (10.17).

### 10.4.4 Coarse-grained dynamics and hidden information

Assume, now, that there are some microscopic dynamics going on, which can be described by the formalism presented in Chapter 5 on the microscopic level. This may be complicated unitary dynamics of movements of particles for example, which may result in the system appearing to thermalize. Following Chapter 5, the microscopic dynamics, as a quantum channel, are simply assumed to conserve a given operator $\Gamma_S$.

We assume that the system and its dynamics are invariant under the symmety, we may assume that $\Gamma_S$ is also symmetric. In particular, the operator is diagonal in the $|q, m, n\rangle$ basis as it must also have the block-diagonal form (10.11). So define $\Gamma^q_{\mathscr{N}^q_S}$ such that we can write

$$\Gamma_S = \bigoplus_q \mathbb{1}_{\mathscr{M}^q_S} \otimes \Gamma^q_{\mathscr{N}^q_S} , \tag{10.23}$$

and its embedded version into $QMN$,

$$\Gamma_{QMN} = \sum_q |q\rangle\langle q|_Q \otimes \mathbb{1}^{d_q}_M \otimes \Gamma^q_N , \tag{10.24}$$

where $\mathbb{1}^{d_q}_M$ is a projector of rank $d_q$ and where $\Gamma^q_N$ is the embedded version of $\Gamma^q_{\mathscr{N}^q_S}$ into $N$. These are given by the embedding $|q, m, n\rangle_S \to |q\rangle_Q |m\rangle_M |n\rangle_N$.

The microscopic dynamics of the system are given in terms of a trace-nonincreasing completely positive map $\mathcal{E}^{\mathcal{M}}_{QMN}$ which satisfies

$$\mathcal{E}^{\mathcal{M}}_{QMN}(\Gamma_{QMN}) \leqslant \Gamma_{QMN} . \tag{10.25}$$

How are the dynamics described for Alice? Her description of states on the systems $QN$ should be represented microscopically via the isometry (10.19) depending on the group element $g$ which has been applied. Hence, the effective dynamics for Alice are given by the completely positive, trace-preserving map

$$\mathcal{E}^{\mathcal{A}}_{QN}\big(\rho^{\mathcal{A}}_{QN}\big) = \int dg \; \mathrm{tr}_M \, \mathcal{E}^{\mathcal{M}}\big(W_{QN\to QMN}(g) \, \rho^{\mathcal{A}}_{QN} \, W^\dagger_{QN\leftarrow QMN}(g)\big) . \tag{10.26}$$

By defining $\Gamma_{QN} = \mathrm{tr}_M \, \Gamma_{QNM}$, we have using (10.24) that

$$\Gamma_{QN} = \sum_q d_q \, |q\rangle\langle q|_Q \otimes \Gamma^q_N , \tag{10.27}$$

and that

$$\int dg\, W_{QN\to QMN}(g)\,\Gamma_{QN}\, W^{\dagger}_{QN\leftarrow QMN}(g) \\ = \mathcal{G}(\Gamma_{QN}\otimes |0\rangle\langle 0|_M) = \Gamma_{QMN}\;. \quad (10.28)$$

So, because $\mathcal{E}^{\mathcal{M}}_{QMN}$ satisfies (10.25), we see from (10.26) that

$$\mathcal{E}^{\mathcal{A}}_{QN}(\Gamma_{QN}) \leqslant \Gamma_{QN}\;. \quad (10.29)$$

This means that *microscopic dynamics which are* $\Gamma$*-sub-preserving for a symmetric* $\Gamma_{QMN}$*, when represented in the coarse-grained picture, are again* $\Gamma$*-sub-preserving for* $\Gamma_{QN} = \mathrm{tr}_M\,\Gamma_{QMN}$.

Of course, we reach the same conclusions by applying directly the recovery map to infer the reduced dynamics as in Section 10.2.2.

Consider the example where the microscopic Hamiltonian is completely degenerate, with $\Gamma_S = \mathbb{1}_S$. Then the coarse-grained $\Gamma_{QN}$ given by (10.27) takes the form

$$\Gamma^{\mathcal{A}}_{QN} = \sum_{q,n} d_q\, |q,n\rangle\langle q,n|_{QN}\;, \quad (10.30)$$

and we now see that the eigenvalues of $\Gamma^{\mathcal{A}}_{QN}$ simply count how many degrees of freedom have been "hidden" in the coarse-grained state $|q,n\rangle$. This is what we had concluded already in Section 10.2.

### 10.4.5 Lifting the superselection rule

The superselection rule in (10.14) may be lifted if the observer has access to a reference system which is in a non-symmetric state. This is a *reference frame*. The symmetry acts jointly on the system and the reference frame. If we condition on the state of the reference frame, treating it as an observer's memory, then it is then possible to observe states which are not symmetric on the system, even though the global bipartite state on the system and reference frame is symmetric.

Concretely, from the microscopic picture, Alice's state $\rho^{\mathcal{A}}_{QN}$ is represented on $QNM$ by applying the isometry $W_{QN\to QMN}(g)$ for some transformation element $g$, as per (10.20). If Mike possesses a reference frame $R$ which records the group element $g$ used to transform the system, the state is represented as

$$\int dg\, |g\rangle\langle g|_R \otimes \big(W_{QN\to QMN}(g)\,\rho^{\mathcal{A}}_{QN}\, W^{\dagger}_{QN\leftarrow QMN}(g)\big)\;. \quad (10.31)$$

Because this is an isometric embedding of $\rho^{A}_{QN}$ into a subspace of $QNMR$ which is invariant under the symmetry, the states $\rho^{A}_{QN}$ may display coherent superpositions between different charge sectors, and this state will be represented by a symmetric state including the reference frame.

So if Alice now also has access to this reference frame, which records the transformation element $g$ which has been applied to the state, then she may observe coherent superpositions between states of different charge values.

Giving Alice access to the full reference frame sounds somewhat like cheating, or at least defeating the purpose, because we originally assumed that Alice didn't even know about the symmetry. Yet the reference frame has to be carefully engineered and initialized with knowledge of the symmetry group.

In the last part of this section, we show that even if we also apply the same type of coarse-graining to the reference frame, tracing out the register carrying the irreducible representation, then Alice may still observe some coherence between different charge sectors. We decompose $R$ into registers $R_Q R_M R_N$ as for the system, and show that by coarse-graining the register $R_M$, an initial state $|g = 1\rangle_R$ displays coherent superpositions between different charge values. This is sufficient for $R_Q R_N$ to act as a reference frame for the Alice's charge superselection, although we will see that this reference frame is noisy. If Alice desired a good reference frame, she might need many copies of an $R$ system.

Now we demonstrate this claim. The reference frame $R$ above can be constructed in the following way (Bartlett *et al.*, 2007). Define the system $R$ to possess an orthonormal basis $\{|g\rangle\}$ consisting in one basis ket for each group element.[3] The vectors in this space transform simply as $U_R(g)\,|g'\rangle_R = |gg'\rangle_R$. Now, for any $\rho^0_S$, consider the state

$$\mathcal{G}_{RS}\big(|g = 1\rangle\langle g = 1|_R \otimes \rho^0_S\big) = \int dg\; |g\rangle\langle g|_R \otimes \big(U_S(g)\,\rho^0_S\,U^\dagger_S(g)\big)\,, \quad (10.32)$$

where the symmetrization superoperator $\mathcal{G}(\cdot)$ acts jointly on $R$ and $S$. As in Section 9.4, if $R$ is considered as the observer's memory, then that observer

[3] Mathematically, we need to consider the space $L^2(G)$ of all square-integrable functions on $G$. The square-integrability condition is relevant only if $G$ is infinite, and restricts the wavefunctions to those we can handle with the quantum formalism. In the following we write $|g\rangle$ also for infinite spaces, symbolically meaning "a pure wavefunction appropriately sharply peaked at $g$." The representation we consider on this space is the *left-regular representation.*

collapses in the state $g$, leaving the system in the state

$$U_S(g)\,\rho_S^0\,U_S^\dagger(g)\,. \tag{10.33}$$

They can then undo the transformation $U_S(g)$ and observe the state $\rho_S^0$. This state may be arbitrary, in particular it may exhibit coherent superpositions between charge sectors. Hence, the system $R$ has acted as a reference frame allowing the observer to see states which violate the superselection rule on $S$.

Now, the question is, how is $R$ decomposed into irreducible representations? We need this decomposition in order to embed $R$ into registers $R_Q R_M R_N$ as before, so as to coarse-grain $R$ by dropping the $R_M$ register. We need to determine how the basis states $|g\rangle_R$ are embedded into $R_Q R_M R_N$. A ket $|q\,m\,n\rangle_{R_Q R_M R_N}$ transforms under the symmetry as

$$U_{R_Q R_M R_N}(g)\,|q\,m\,n\rangle_{R_Q R_M R_N} = |q\rangle_{R_Q}\big(U^q_{R_M}(g)|m\rangle_{R_M}\big)\,|n\rangle_{R_N}\,, \tag{10.34}$$

where $U^q_{R_M}(g)$ is the unitary corresponding to the irreducible representation of charge $q$ but projected onto the image of the embedding of the irreducible representation into $R_M$ (making $U^q_{R_M}(g)$ technically a partial isometry). We may invoke the Peter-Weyl theorem (Carter *et al.*, 1995), which gives the irreducible representation structure of $R$, and which has been used in the context of the quantum Fourier transform (Harrow, 2005). This result states that the ket $|g\rangle_R$ is represented with $R_N \simeq R_M$ as

$$|g\rangle_{QMN} = \mathcal{N}^{-1/2}\sum_{q\,m\,m'}\sqrt{d_q}\,u^q_{mm'}(g)\cdot|q\rangle_{R_Q}|m\rangle_{R_M}|n=m'\rangle_{R_N}\,, \tag{10.35}$$

where $u^q_{mm'}(g) = \langle m|\,U^q_{R_M}(g)\,|m'\rangle_{R_M}$ are the matrix elements of $U^q_{R_M}(g)$, and where $\mathcal{N}$ is a constant normalization factor; for finite groups, $\mathcal{N} = |G|^2$, and for infinite groups, we may take $\mathcal{N} = 1$ along with an appropriate normalization of the measure $dq$ used to replace the sum over $q$ by an integral. We may further write (10.35) as

$$|g\rangle_{QMN} = \mathcal{N}^{-1/2}\sum_q d_q\,|q\rangle_{R_Q}\otimes|\psi_q(g)\rangle_{R_M R_N}\,, \tag{10.36}$$

with

$$|\psi_q(g)\rangle_{R_M R_N} = \frac{1}{\sqrt{d_q}}\big(U^q_{R_M}(g)\otimes\mathbb{1}_{R_N}\big)\,|\Phi\rangle_{M:N}\,, \tag{10.37}$$

where $|\Phi\rangle_{M:N} = \sum_m |m\rangle_{R_M} |n = m\rangle_{R_N}$ is the unnormalized maximally entangled state between $M$ and $N$ in their respective bases, and recalling that we've defined $U^q_{R_M}(g)$ to include a projector of rank $d_q$ onto the support of the irreducible representation of charge $q$ into $R_M$. This means that $|\psi_q(g)\rangle$ is a maximally entangled state between two subspaces in $R_M$ and $R_N$ of rank $d_q$, with a local unitary applied on one side according to the transformation $g$.

Let's now coarse-grain the reference frame $R$. Following the prescription for the system $S$, we trace out the $R_M$ system from $R$. Furthermore we suppose that the $R$ system was initially prepared as above in the state $|g = 1\rangle_R$. This gives us

$$\rho^{\mathcal{A}}_{R_Q R_N} = \mathcal{N}^{-1} \sum_{q\,q'} d_q d_{q'}\, |q\rangle\langle q'|_{R_Q} \otimes \mathrm{tr}_{R_M} |\psi_q(1)\rangle\langle\psi_{q'}(1)|_{R_M R_N}\,, \tag{10.38}$$

with, using (10.37) and $U^q_{R_M}(1) = \mathbb{1}^{d_q}_{R_M}$,

$$\mathrm{tr}_{R_M} |\psi_q(1)\rangle\langle\psi_{q'}(1)|_{R_M R_N} = \frac{1}{\sqrt{d_q\, d_{q'}}} \cdot \left(\mathbb{1}^{d_q}_{R_M} \mathbb{1}^{d_{q'}}_{R_M}\right)\,, \tag{10.39}$$

which depends on how the irreducible representations are embedded into $R_M$ for each charge $q$. Importantly, this operator need not be zero for $q \neq q'$. Hence we obtain for $\rho^{\mathcal{A}}_{R_Q R_N}$,

$$\rho^{\mathcal{A}}_{R_Q R_N} = \mathcal{N}^{-1} \sum_{q\,q'} \sqrt{d_q d_{q'}}\, |q\rangle\langle q'|_{R_Q} \otimes \left(\mathbb{1}^{d_q}_{R_M} \mathbb{1}^{d_{q'}}_{R_M}\right)\,. \tag{10.40}$$

Because $\mathbb{1}^{d_q}_{R_M} \mathbb{1}^{d_{q'}}_{R_M}$ need not be zero for different charge values $q \neq q'$, this state may display coherent superpositions of different charge states.

This means that Alice may use the coarse-grained version of $R$ to resolve—to some extent which is not optimal—coherent superpositions on a system $S$. Should she require a better reference frame, she may request several copies of this one.

## 10.5 Quantum superposition of thermodynamic states: a sketch

We have argued above that the thermodynamic state associated to the thermodynamic variables $\vec{z}$, for the macroscopic observer, may be associated with a quantum state $|\vec{z}\rangle$—that's their unit of information. Doing so suggests that it is possible in principle to observe a coherent superposition of different

such states. Here, we illustrate how this may happen.

Consider for our illustration an isolated gas with thermodynamic variables $E, V, N$ corresponding to energy, volume and number of particles. (Our example can of course be adapted to more general systems.) The thermodynamic state is specified by the values of $E, V, N$, and the corresponding state of the gas is the Boltzmann state or microcanonical state

$$\tau_S^{(E,V,N)} = \frac{1}{\Omega(E, V, N)}\, \Pi_S^{(E,V,N)} \,, \tag{10.41}$$

where $\Pi_S^{(E,V,N)}$ is the projector onto the microcanonical subspace of fixed $(E, V, N)$ and where $\Omega = \operatorname{tr} \Pi^{(E,V,N)}$ is the microcanonical partition function. This microcanonical subspace is denoted by $\mathcal{M}_{E,V,N}$. Furthermore, we have seen above that the macroscopic observer may represent the different possible states as kets $|E, V, N\rangle_C$ living in a quantum state space $\mathscr{H}_C$.

From equilibration arguments and under suitable assumptions on the internal dynamics of the system, the thermal state $\tau_S^{(E,V,N)}$ of the gas may also be written as the time average of the evolution from a fixed initial pure state, over a large time $T \to \infty$:

$$\tau_S^{(E,V,N)} = \frac{1}{T} \int_0^T dt\, U_S(t)\, |\psi_0^{E,V,N}\rangle\langle\psi_0^{E,V,N}|_S\, U_S^\dagger(t) \,, \tag{10.42}$$

where $U_S(t)$ is the unitary time evolution of the system, and where $|\psi_0^{E,V,N}\rangle_S$ is any chosen pure state on $S$ in the microcanonical subspace $\mathcal{M}_{E,V,N}$ (Gemmer *et al.*, 2009). Physically, for any $E, V, N$, if we see a thermal state $\tau_S^{(E,V,N)}$ on the gas, we can always assume that it has been prepared in the following way. First, the thermodynamic state $(E, V, N)$ is embedded into the microscopic state space of the gas as

$$|E, V, N\rangle_C \quad \longrightarrow \quad |\psi_0^{E,V,N}\rangle_S \,, \tag{10.43}$$

for a fixed set of $\{|\psi_0^{E,V,N}\rangle_S\}$ whose choice is irrelevant as long as each ket lies in the corresponding $\mathcal{M}_{E,V,N}$. Second, the gas evolves for a time $t$ which is unknown. In other words, the state of the gas can be thought of as if it were really in a pure state which is evolving unitarily, but we didn't know for how much time it has been evolving (Figure 10.2).

Consider two copies of this system, which we call $S$ and $S'$. They are each initialized in some initial state $|\psi_0^{E,V,N}\rangle_S$ and $|\phi_0^{E,V,N}\rangle_{S'}$ respectively, and they have each been evolving for some unknown times $t$ and $t'$. Clearly

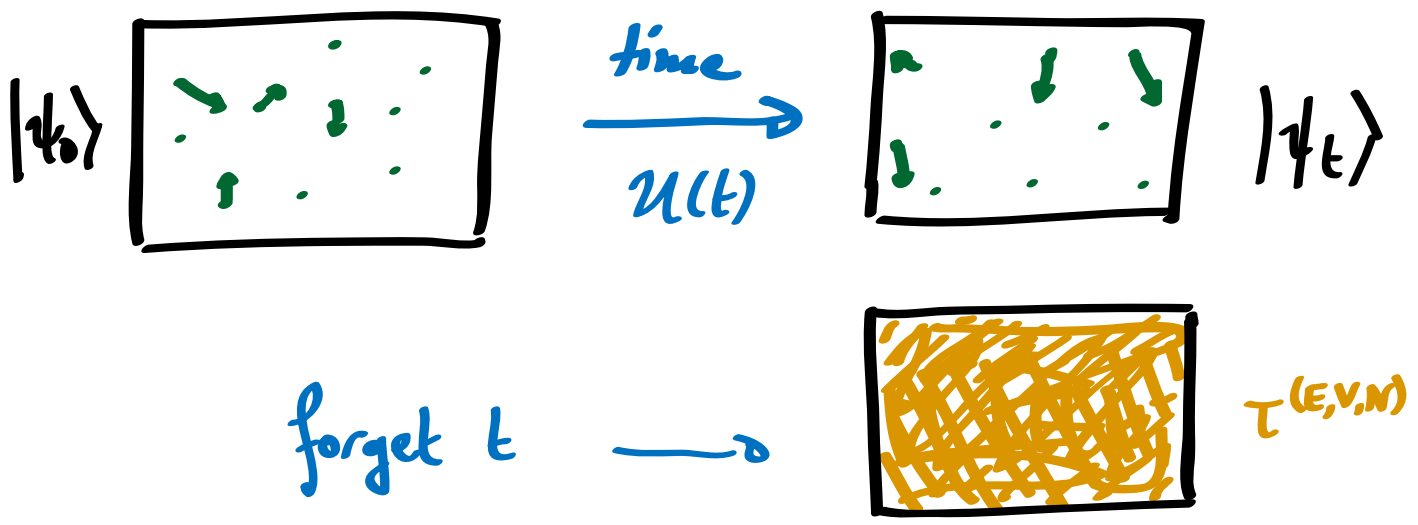

Figure 10.2: The thermal state of a gas can be interpreted as the result of forgetting how much time the gas has evolved from an initial state.

the state of the systems, according to us, are a tensor product of two thermal states

$$\tau_S^{(E,V,N)} \otimes \tau_S^{(E,V,N)} \,. \tag{10.44}$$

Assume now that the two systems are prepared in some arbitrary pure state $|\psi_0\rangle$ which is the same on both copies of the system, and suppose that the two systems are synchronized, meaning that they have been evolving for the same amount of time, $t = t'$, which is known to us. Both gases then remain in a pure state and evolve according to

$$|\psi_t\rangle = U(t)\,|\psi_0\rangle \,. \tag{10.45}$$

There is no thermalization happening here according to us, because we now have the time reference. Hence, depending on the initial state $|\psi_0\rangle$, we may observe arbitrary pure states on the system, such as coherent superpositions of states of different energy. (With thermalizing dynamics, such a pure state would instantly decohere.) But physically speaking, we have given ourselves all the information about the microscopic details of the system, and this i not desirable in order to describe macroscopic observers.

It turns out that we don't need to know what the initial state is, nor how much time it has evolved. We only need to assume that the two systems are synchronized, meaning that they have evolved for the same time starting from the same pure initial state. Suppose for example that both systems are initialized in the state

$$|\psi_0\rangle = \left[|\psi_0^{E,V,N}\rangle + |\psi_0^{E',V',N'}\rangle\right]/\sqrt{2}\,. \tag{10.46}$$

Such a state is simply the embedding according to (10.43) of the coarse-

grained state

$$\left[|E,V,N\rangle_C + |E',V',N'\rangle_C\right]/\sqrt{2}\,, \tag{10.47}$$

which displays coherent superpositions of different thermodynamic states. For simplicity, assume that these systems are not left to evolve ($t = t' = 0$). If two copies of this state are given, then we may use one copy as a reference frame to resolve coherent superpositions of states on the other copy in the coarse-grained picture (Bartlett *et al.*, 2007). (Such a bounded reference frame is clearly not as useful as the universal reference frame as presented above.) Observe how the choice of $|\psi_0^{E,V,N}\rangle$ is irrelevant, as long as both systems are prepared in the same state.

# 11

# Conclusion and Outlook

The present thesis revisits the fundamental connection between information theory and thermodynamics from a purely information-theoretic angle. The result is a general, multipurpose theory of thermodynamics and information, consisting in quantum information processing under the simple restriction of preserving the hidden information.

Our results further expose various features of thermodynamics in the microscopic regime that are not present in the standard setting of large systems. In particular, the minimum work cost of a logical process cannot be given in terms of a state function, such as the entropy or the free energy in thermodynamics. Rather, the latter emerges by typicality after singling out a class of states, the thermodynamic states.

The framework of generalized thermodynamics we have developed can be applied at any level of knowledge, from the macroscopic observer who has access to macroscopic variables to the microscopic observer with precise knowledge of the individual position and momentum of each particle. Our framework applies to more complicated setups with partial information, such as the case of classical computing devices and memory registers working in a thermodynamic regime. There, information is stored in the form of thermodynamic states of many electrons, but where the degree of control is sufficient to reliably store and process logical information. Such setups are at the origin of several Maxwell's-demon-like thought experiments, for instance the Szilárd engine (Szilard, 1929; Bennett, 1982; Earman and Norton, 1998, 1999).

From a more abstract and fundamental perspective, our results can be applied at any level of coarse-graining. Any observer may specify their own idea of what the fundamental unit of information is (for example macroscopic thermodynamic variables, or the individual position and momentum of each and every particle). The units of information are promoted to quantum kets, yielding a system which can be described using the formalism of quantum information. Intuitively, these kets carry a weight which betrays the number of underlying states that they hide, that is, how many states have been coarse-grained into a single macroscopic unit of information: These weights are the hidden information. Furthermore, because quantum states are relative to the observer, the coarse-grained observer may in principle observe quantum superpositions of such states, provided they have access to a suitable reference frame.

*A general, finite-size information-theoretic model for restricted processes.*

One could think that thermodynamics, as a physical theory by essence, would require physical concepts such as energy to be built into the framework. Our results consolidate the opposite picture, where thermodynamics is a generic structure which can be applied to different physical situations, as previously proposed (Giles, 1964; Gyftopoulos and Beretta, 2005; Lieb and Yngvason, 1999, 2004; Beretta and Zanchini, 2011). The physical properties of the system, such as its Hamiltonian, in our case only serve to determine the hidden information operator.

Logical processes are often most associated to quantum or classical computing and information processing. In this respect, our results provide ultimate limits to the thermodynamic resource costs for any quantum or classical computation. Our formula is given in terms of a well-characterized formula which can be computed efficiently as a semidefinite program. However, the logical processes we consider are precisely any evolution which is allowed by quantum mechanics. As such, our result provides a fundamental characterization of quantum mechanical processes.

The resource cost of a logical process is counted at its most fundamental level as purity, or more precisely, as a number of pure qubits. In turn, this may be converted to or from energy if one has access to a heat bath at a given temperature, but also to any other conserved physical quantity such as spin (Barnett and Vaccaro, 2013; Weilenmann *et al.*, 2015). Again, this underscores the fundamentally information-theoretic nature of the resource cost of implementing logical processes, which need not be formulated in terms of mechanical work.

We further note that our framework is based on defining those operations which are explicitly forbidden, rather than specifying those which are expressly allowed. In the simpler case of thermodynamics with a completely degenerate Hamiltonian, we have seen that the operations in our framework can be achieved by an operational framework such as thermal operations to good approximation. Whether this is the case in the most general scenario remains an open question.

*Quantifying information with entropy measures.*

Our results establish a direct connection between the resource costs of implementing a logical process and known measures of information. That the resource cost is given by information-theoretic properties of the logical process is expected, since the model is entirely information-theoretic. Our contribution is to single out the exact measure which is relevant.

In the simple case where the hidden information is the identity operator, this measure is the max-entropy known from quantum information theory. This is the case, for example, if the Hamiltonian of the system is completely degenerate, such as for an idealized memory register. The max-entropy is known as being the relevant entropy measure for data compression or information reconciliation. Hence we may understand our result as measuring the smallest size to which the information discarded by a process can be compressed, in order for it to be disposed of.

The result in the general case is expressed in terms of a new entropy measure, the coherent relative entropy. This entropy measure generalizes both the max-entropy from the special case above, as well as the relative entropy whose relevance is known in thermodynamics. The coherent relative entropy thus acts as a parent quantity from which several known entropy measures can be recovered as special cases.

Our approach provides an additional operational meaning to these entropy measures, as counting the amount of purity consumed by a logical process. It may even be used to *define* information entropy (Feynman, 1996).

*The emergence of thermodynamics by typicality.*

Our results provide a fresh view on macroscopic thermodynamics. The latter can be seen as a general framework, in which the second law postulates the existence of a state function, the thermodynamic entropy, which relates to the heat flow in processes. Many standard results of thermodynamics follow from that starting point. It is now the role of a microscopic theory to construct a state function with this property, based on the microscopic dynamics of the particular system. In textbook statistical mechanics, this construction is given for several physical setups, such as gases or lattices; one usually considers, for example, the configuration entropy, or an appropriately normalized Shannon or von Neumann entropy of the density of the statistical ensemble. Our result generalizes this construction and sheds light on when it is justified: The state function, in general, appears whenever the inherent fluctuations due to the microscopic stochastic nature of the process vanish by typicality. The existence of an entropy state function is therefore not a property of the microscopic system; it is rather an emergent quantity that appears whenever the full system is typical, such as in the limit of macroscopic processes.

We note that the system in consideration need not be large for the typicality arguments to apply. For example, if one considers the work requirement of performing many independent repetitions of a single given logical process (seen as one big joint process), then the work requirement $W^\epsilon$ per repetition converges to the average work requirement as calculated via statistical mechanics, even if the individual system is small: In this case, the entropy function emerges. This further justifies the usage of the von Neumann entropy in statistical mechanics even for small systems. Conversely, a large system does not necessarily display typicality; such is the case for systems out of thermodynamic equilibrium.

The present emergence paradigm opens the door to the question of finding further classes of states which may be considered as thermodynamic states, beyond those constructed in our examples. For example, this is clearly the case for i.i.d. states which need not be of the form of our examples.

Textbook thermodynamics also offers many more features than those recovered here—the third law, equations of state, entropic forces, and so on. We expect that these could also be recovered in our picture, a task we leave open for future work.

*Information is relative to the observer.*

A contribution of this thesis is to bring the status and applicability of quantum information to a new level. We argue that the quantum state is relative to the observer in a strong sense: Different observers may even disagree about which state space a particular physical system should be described with.

A gas appears to a macroscopic observer as being perfectly well described by thermodynamic variables such as temperature, volume and pressure. However, an observer which has knowledge of the microscopic degrees of freedom (or even just of their existence) describes the gas using a distribution over possible particle positions and momenta. These two observers are usually treated in a fundamentally different way. Our framework, however, applies consistently to either observer, and thus treats both points of view on an equal footing. We then take this idea a step further, and point out that the macroscopic observer can consistently apply the formalism of quantum information to their level of knowledge. For example, in our picture, the deterministic and reversible nature of the resource cost for transforming two thermodynamic states into one another, as predicted by thermodynamics for example in an isothermal reversible compression, is treated exactly like transforming a microscopic quantum state from one energy eigenstate to another.

It is also known in thermodynamics that different observers may or may not be able to distinguish two thermodynamic states, and that this affects physical statements such as the amount of extractable work. This argument is probably most elegantly explained in an example by Jaynes (1992, Sections 5 and 6) involving two different hypothetical Argon kinds in the context of the Gibbs paradox. Our contribution may be interpreted as extending this argument to both finite-size, as well as fully quantum, systems.

*Relevance for experiments.*

While our contribution is not directly targeted at discovering new physical effects, phenomena, or improving the understanding of current physical experiments, we still anticipate that it might provide useful higher-level concepts in terms of which more complex experiments could be analyzed. We build upon the concepts of quantum information which have already brought many useful tools for the analysis of experiments (no-cloning theorem, entanglement, information entropy, etc.).

Furthermore, we expect it is possible to develop our rudimentary sketch on how to observe "quantum superpositions of thermodynamic states" into a full-fledged experimental setup. Current technology such as optomechanical systems may provide the necessary building blocks to this end.

*Implications for physics.*

The point of view that the quantum state is relative to the observer might have consequences both for the symmetrization postulate for identical particles in quantum mechanics, as well as for a possible theory of emergent space-time and gravity. More generally, one can speculate about the role of information in a future, more universal theory of physics.

Since our early courses on quantum mechanics, we know that when particles are identical, a physical postulate enforces the total wave function to be either symmetric (for bosons) or antisymmetric (for fermions). From an information-theoretic perspective, this principle is puzzling because distinguishability should be a matter of the observer, and not of physics laws, much like in the case of the Gibbs paradox (Jaynes, 1992). With our results, the symmetrization principle acquires a new interpretation: For us standard, macroscopic observers, systems composed of many bosons or fermions which appear identical to us should be described in a symmetric or antisymmetric state. Nothing prevents, however, other observers from being able to distinguish in principle the particles. For some reason, we appear to have the necessary reference frame to describe such systems with a well-defined phase between the different permutation states, and not with a mixture of all possible particle permutations. A future theory of fundamental particle physics might explain this explicitly.

Recently, there have been proposals that gravity should be viewed as a entropic force in a thermodynamically emergent space-time (Verlinde, 2011), building up on earlier ideas (Jacobson, 1995; Padmanabhan, 2010) as well as the holographic principle (Bousso, 2002). The specific proposal by Verlinde suggests that the state of a particle at a particular position in space is in fact a thermodynamic state, and that the gravitational force felt by the particle is of entropic nature, owing to different amounts of entropy coarse-grained into different position states. This view has drawn criticism (Chaichian *et al.*, 2011; Kobakhidze, 2011a,b) because neutron beam experiments have demonstrated interference between different position states in a gravitational field (Colella *et al.*, 1975; Nesvizhevsky *et al.*, 2002, 2003), from which it has been deduced that position states cannot be of thermodynamic nature. Our results may put in question this criticism: Because quantum states are relative to an observer, and need not represent an absolute, fundamental unit of information, there is *a priori* no issue in observing superpositions of position states, even if those correspond to coarse-grained thermodynamic states. One may speculate the superposition can be explained by the experiment having access to a suitable reference frame, as illustrated in our sketch suggesting the possibility to prepare quantum superpositions of thermodynamic states.

More generally, one may ask how big of a role information theory should play in fundamental laws of physics. In the context of our framework, observers are treated on an equal footing from our information-theoretic perspective. However, at this point, we do not take into account other physical aspects of the system, such as time evolution. For example, a superposition of two energy eigenstates evolves in time with an oscillating relative phase between the two states; for coarse-grained states corresponding to two different thermodynamic states with same energy, this need not necessarily be the case. So, at this point, one might be able to experimentally determine whether two states correspond to genuine energy eigenstates, or whether they are coarse-grained versions of thermodynamic states. In other words, our different observers obey the same "information-theoretic laws," but this does not imply that they obey the same physical laws. Given the growing signs of the fundamental nature of information for physics, we might speculate that the physical laws will, at some point, also be formulated so as to make of this coarse-graining a fundamental aspect of physics. This is in line with a growing community believing that space-time is emergent (Seiberg, 2006; Hamma *et al.*, 2010; Padmanabhan, 2010; Verlinde, 2011; Pastawski *et al.*, 2015), as well as the coarse-graining or renormalization group interpretation of the additional bulk dimension in AdS/CFT (de Haro *et al.*, 2001; McGreevy, 2010; Sachdev, 2011).

APPENDIX

# A

# Technical Utilities

## A.1 Working with positive semidefinite matrices

### A.1.1 Checking for positive semidefiniteness

A useful trick to detect when a matrix is not positive semidefinite is that if a diagonal entry is zero, then there may be no nonzero entries in that particular row and on that particular column of the matrix to which the diagonal entry corresponds.

A less brutal trick is that for any two diagonal entries of the matrix which have values $x$ and $y$, then the two off-diagonal elements which are located in the row of $x$ and column of $y$ and vice versa must both be less in magnitude than $\sqrt{xy}$. Mathematically, for any basis elements $|i\rangle$, $|j\rangle$, if we have $|\langle i|A|j\rangle|^2 \nleqslant \langle i|A|i\rangle\langle j|A|j\rangle$ then $A$ is not positive semidefinite.

Finally, it is also useful for a large matrix to try to permute the basis vectors (permute rows and the corresponding columns) in order to bring it in block-diagonal form; then the full matrix is positive semidefinite if and only if each block is.

### A.1.2 Identities with positive semidefinite matrices

***Proposition.*** *For any positive semidefinite matrix $A$, we have*

$$A \leqslant a\mathbb{1} \Leftrightarrow \|A\|_\infty \leqslant a \; . \tag{A.1}$$

***Proposition.*** *For all $0 \leqslant A \leqslant B$, and for all $D \geqslant 0$, for any arbitrary matrix $C$*

*and for any completely positive map $\mathcal{E}$,*

$$CAC \leqslant CBC\,; \tag{A.2}$$

$$A\otimes D \leqslant B\otimes D\,; \tag{A.3}$$

$$\mathcal{E}(A) \leqslant \mathcal{E}(B)\,; \quad \textit{and} \tag{A.4}$$

$$\|A\|_\infty \leqslant \|B\|_\infty\,. \tag{A.5}$$

### A.1.3 Additional helpful theorems

***Proposition A.1.*** *Let $A$ and $B$ be positive semidefinite operators, and assume that $A \leqslant B$. Then there exists a positive semidefinite operator $C$ satisfying $C \leqslant \mathbb{1}$ such that*

$$A = CBC\,. \tag{A.6}$$

Observe that $0 \leqslant C^2 \leqslant \mathbb{1}$: this is useful to define a POVM effect from $C$ via $Q^{1/2} = C$.

*Proof of Proposition A.1.* The explicit $C$ for this to work is given by

$$C = B^{-1/2}\left(B^{1/2}AB^{1/2}\right)^{1/2}B^{-1/2}\,. \tag{A.7}$$

This expression has been introduced before as the *geometric mean* or more generally the *α-power mean* of $B^{-1}$ with $A$ (Pusz and Woronowicz, 1975; Kubo and Ando, 1980; Hiai and Petz, 1993; Ando and Hiai, 1994).

A straightforward calculation gives $CBC = A$, recalling that $A$ must lie in the support of $B$ because of $A \leqslant B$. Furthermore, $C \geqslant 0$ by construction. Also, since $A \leqslant B$,

$$C \leqslant B^{-1/2}\left(B^{1/2}BB^{1/2}\right)^{1/2}B^{-1/2} \leqslant \Pi^B \leqslant \mathbb{1}\,. \qquad \blacksquare$$

***Lemma A.2.*** *Let $A \geqslant 0$, $B \geqslant 0$ and let $\Pi$ be the projector onto the support of $A$. Let $\mu > 0$. Define $P$ as the projector onto the eigenspaces associated to nonnegative eigenvalues of the operator $(\mu A - B)$. Then there exists a constant* *c which is independent of $\mu$ such that*

$$\|\Pi - P\Pi P\|_\infty \leqslant \frac{c}{\mu} \,. \tag{A.8}$$

*In particular,*

$$\Pi \leqslant P + \frac{c}{\mu}\mathbb{1} \,. \tag{A.9}$$

*Proof of Lemma A.2.* This lemma follows from a result of perturbation of matrix eigenspaces (Bhatia, 1997). We'll consider the operators $A - \frac{1}{\mu}B$ and $A$. Let $Q = \mathbb{1} - P$ be the projector on the eigenspaces associated to the strictly negative eigenvalues of $A - \frac{1}{\mu}B$. Let $a_{\min} = \|A^{-1}\|_\infty^{-1}$ be the smallest nonzero eigenvalue of $A$. Recall that $\Pi$ projects onto the eigenspaces of $A$ associated to eigenvalues larger or equal to $a_{\min}$. We may now invoke Bhatia (1997, Theorem VII.3.1), which asserts that for any unitarily invariant norm $\|\cdot\|_\bullet$,

$$\|Q\Pi\|_\bullet \leqslant \frac{1}{\mu a_{\min}}\|QB\Pi\|_\bullet \leqslant \frac{1}{\mu a_{\min}}\|B\|_\bullet \,. \tag{A.10}$$

(The gap $\delta$ in Bhatia (1997, Theorem VII.3.1) is here the gap between 0 and $a_{\min}$.) In particular, we have $\|Q\Pi\|_\infty \leqslant (\mu a_{\min})^{-1}\|B\|_\infty$. We then have

$$\begin{aligned}\|\Pi - P\Pi P\|_\infty \leqslant \|\Pi - P\Pi\|_\infty + \|P\Pi - P\Pi P\|_\infty \leqslant \|\Pi - P\Pi\|_\infty + \|P\|_\infty\|\Pi - \Pi P\|_\infty \\ = 2\|\Pi - P\Pi\|_\infty = 2\|Q\Pi\|_\infty \leqslant \frac{c}{\mu} \,,\end{aligned} \tag{A.11}$$

with $c = 2\,(a_{\min})^{-1}\|B\|_\infty$. This implies (A.9) because

$$\Pi - P\Pi P \leqslant \frac{c}{\mu}\mathbb{1} \qquad \Rightarrow \qquad \Pi \leqslant P\Pi P + \frac{c}{\mu}\mathbb{1} \leqslant P + \frac{c}{\mu}\mathbb{1} \,. \qquad \blacksquare$$

## A.2 Semidefinite programs for the one norm and the infinity norm

***Proposition A.3.*** *For $A \geqslant 0$, we have that*

$$\|A\|_\infty = \max_{\substack{\omega \geqslant 0 \\ \mathrm{tr}\,\omega = 1}} \mathrm{tr}\left(\omega A\right) \; . \tag{A.12}$$

*(The condition $\mathrm{tr}\,\omega = 1$ may also be replaced by $\mathrm{tr}\,\omega \leqslant 1$.) In particular, if $\omega_A$ is a density operator and $A_{AB} \geqslant 0$,*

$$\mathrm{tr}(\omega_A A_{AB}) \leqslant \|\mathrm{tr}_B\, A_{AB}\|_\infty \; . \tag{A.13}$$

*Proof of Proposition A.3.* Use

$$\|A\|_\infty = \max_{|\psi\rangle:\,\langle\psi|\psi\rangle=1} \langle\psi|A|\psi\rangle = \max_{|\psi\rangle:\,\langle\psi|\psi\rangle=1} \mathrm{tr}(|\psi\rangle\langle\psi|\, A) = \max_{\substack{\omega \geqslant 0 \\ \mathrm{tr}\,\omega = 1}} \mathrm{tr}\left(\omega A\right) \; . \tag{A.14}$$

The last equality holds because by linearity of the optimized expression; indeed there is always a pure state which achieves the maximum. ∎

***Proposition A.4.*** *Let $A \geqslant 0$. The 1-norm $\|A\|_1$ can be written as the semidefinite program in terms of the variables $\Delta^\pm \geqslant 0$:*

$$\text{minimize}: \quad \mathrm{tr}\left(\Delta^+ + \Delta^-\right) \tag{A.15-a}$$
$$\text{subject to}: \quad A = \Delta^+ - \Delta^- \; . \tag{A.15-b}$$

*The dual to this program is an alternate expression of the same quantity, in terms of the Hermitian variable $Z$:*

$$\text{maximize}: \quad \mathrm{tr}\left(Z A\right) \tag{A.16-a}$$
$$\text{subject to}: \quad -\mathbb{1} \leqslant Z \leqslant \mathbb{1} \; . \tag{A.16-b}$$

*Proof of Proposition A.4.* The first constraint forces $\Delta^+$ to be the positive part of $A$ and $\Delta^-$ to be its negative part. Then the trace in the objective function calculates the trace of the absolute value of $A$, which is the sum of its singular values counted with multiplicity. This is the definition of the 1-norm. The second problem follows from elementary operations on the semidefinite program. ∎

## A.3 Trace Distance and Useful Lemmas for Smoothing

First we provide an expression for the trace distance derived from the semidefinite expression for the one-norm given by Proposition A.4.

***Proposition A.5.*** *Let $\rho$ and $\sigma$ be quantum states. The trace distance $D(\rho, \sigma)$ between $\rho$ and $\sigma$ can be written as the semidefinite program in terms of the variables $\Delta^{\pm} \geqslant 0$:*

$$\text{minimize}: \quad \frac{1}{2}\operatorname{tr}\left(\Delta^{+} + \Delta^{-}\right) \tag{A.17-a}$$

$$\text{subject to}: \quad \sigma = \rho + \Delta^{+} - \Delta^{-}\,. \tag{A.17-b}$$

*Furthermore, $\operatorname{tr}\Delta^{+} = \operatorname{tr}\Delta^{-} = D(\rho, \sigma)$ for the optimal solution. The dual to this program is an alternate expression of the same quantity, in terms of the Hermitian variable $Z$:*

$$\text{maximize}: \quad \frac{1}{2}\operatorname{tr}\left(Z\left(\rho - \sigma\right)\right) \tag{A.18-a}$$

$$\text{subject to}: \quad -\mathbb{1} \leqslant Z \leqslant \mathbb{1}\,. \tag{A.18-b}$$

*Proof of Proposition A.5.* See, e.g. (Wilde, 2013). Basically this just follows from $D(\rho, \sigma) = \frac{1}{2}\|\rho - \sigma\|_1$ and from Proposition A.4. For the optimal $\Delta^{\pm}$, the trace of (A.17-b) gives $\operatorname{tr}\Delta^{+} = \operatorname{tr}\Delta^{-}$. ∎

***Lemma A.6*** *(Distance to projected state). Let $\rho \geqslant 0$ with $\operatorname{tr}\rho = 1$. Let $\epsilon \geqslant 0$. Let $\Pi$ be a projector such that $\operatorname{tr}(\Pi\rho) \geqslant 1 - \epsilon$. Then*

$$P\left(\rho, \frac{\Pi\rho\Pi}{\operatorname{tr}(\Pi\rho)}\right) \leqslant \frac{\sqrt{2\epsilon}}{\sqrt{1-\epsilon}}\,. \tag{A.19}$$

*Proof of Lemma A.6.* Calculate

$$\begin{aligned} P\left(\rho, \frac{\Pi\rho\Pi}{\operatorname{tr}(\Pi\rho)}\right) &= \sqrt{1 - F^2\left(\rho, \frac{\Pi\rho\Pi}{\operatorname{tr}(\Pi\rho)}\right)} = \frac{1}{\sqrt{\operatorname{tr}(\Pi\rho)}}\sqrt{\operatorname{tr}(\Pi\rho) - F^2(\rho, \Pi\rho\Pi)} \\ &\leqslant \frac{1}{\sqrt{\operatorname{tr}(\Pi\rho)}}\sqrt{1 - F^2(\rho, \Pi\rho\Pi)} \leqslant \frac{1}{\sqrt{1-\epsilon}}\sqrt{1 - F^2(\rho, \Pi\rho\Pi)} \\ &= \frac{1}{\sqrt{1-\epsilon}} P(\rho, \Pi\rho\Pi)\,. \end{aligned} \tag{A.20}$$

Now, applying Berta *et al.* (2010, Lemma 7), we have

$$(A.20) \leqslant \frac{\sqrt{2\epsilon - \epsilon^2}}{\sqrt{1-\epsilon}} = \sqrt{\epsilon}\frac{\sqrt{2-\epsilon}}{\sqrt{1-\epsilon}} \leqslant \sqrt{\epsilon}\frac{\sqrt{2}}{\sqrt{1-\epsilon}} \qquad \blacksquare$$

***Lemma A.7** (Smoothing "part of" a state). Let $\rho_{AB}$ be a bipartite normalized quantum state and let $\tilde{\rho}_A$ be a normalized quantum state such that $D(\tilde{\rho}_A, \rho_A) \leqslant \delta$. Then there exists a normalized quantum state $\hat{\rho}_{AB}$ such that $\mathrm{tr}_B\, \hat{\rho}_{AB} = \tilde{\rho}_A$, $\mathrm{tr}_A\, \hat{\rho}_{AB} = \rho_B$ and $P(\hat{\rho}_{AB}, \rho_{AB}) \leqslant 2\sqrt{2\delta}$.*

*Proof of Lemma A.7.* Because $\tilde{\rho}_A$ and $\rho_A$ are $\delta$-close in trace distance, by Proposition A.5 there exists $\Delta_A^{\pm} \geqslant 0$ such that $\mathrm{tr}\,\Delta_A^- = \mathrm{tr}\,\Delta_A^+ = D\left(\tilde{\rho}_A, \rho_A\right)$ and

$$\tilde{\rho}_A = \rho_A + \Delta_A^+ - \Delta_A^- \,. \tag{A.21}$$

Let $A = \rho_A - \Delta_A^-$. Then we have $\mathrm{tr}\,A = 1 - D\left(\tilde{\rho}_A, \rho_A\right) \geqslant 1 - \delta$ and

$$\tilde{\rho}_A = A + \Delta_A^+ \geqslant A\,; \qquad\qquad \rho_A = A + \Delta_A^- \geqslant A\,. \tag{A.22}$$

Define $M_A = A^{1/2}\rho_A^{-1/2}$. Note that

$$M_A^\dagger M_A = \rho_A^{-1/2} A \rho_A^{-1/2} \leqslant \rho_A^{-1/2}\left(A + \Delta_A^-\right)\rho_A^{-1/2} = \Pi_A^{\rho_A} \leqslant \mathbb{1}_A\,. \tag{A.23}$$

Consider now the completely positive map

$$\mathcal{M}_{A\to A}\left[\cdot\right] = M_A\left(\cdot\right)M_A^\dagger + \mathrm{tr}_A\left(\left[\mathbb{1}_A - M_A^\dagger M_A\right]\left(\cdot\right)\right)\frac{\Delta_A^+}{\mathrm{tr}\,\Delta_A^+}\,. \tag{A.24}$$

The mapping is trace-preserving:

$$\mathcal{M}_{A\leftarrow A}^\dagger\left[\mathbb{1}_A\right] = M_A^\dagger\, M_A + \mathrm{tr}\left(\frac{\Delta_A^+}{\mathrm{tr}\,\Delta_A^+}\right)\left[\mathbb{1}_A - M_A^\dagger M_A\right] = \mathbb{1}_A\,. \tag{A.25}$$

Define now the state $\hat{\rho}_{AB}$ as

$$\hat{\rho}_{AB} = \mathcal{M}_{A\to A}\left[\rho_{AB}\right] = M_A\rho_{AB}M_A^\dagger + \mathrm{tr}\left(\left[\mathbb{1}_A - M_A^\dagger M_A\right]\rho_{AB}\right)\frac{\Delta_A^+}{\mathrm{tr}\,\Delta_A^+}\,, \tag{A.26}$$

where the identity mapping is understood on system $B$. Then by properties of quantum channels the state on $B$ must be preserved,

$$\mathrm{tr}_A\,\hat{\rho}_{AB} = \rho_B\,. \tag{A.27}$$

Also, the state on $A$ is

$$\mathrm{tr}_B\,\hat{\rho}_{AB} = M_A\rho_A M_A^\dagger + \mathrm{tr}\left(\left[\mathbb{1}_A - M_A^\dagger M_A\right]\rho_A\right)\frac{\Delta_A^+}{\mathrm{tr}\,\Delta_A^+} = A + \left(1 - \mathrm{tr}\left[A\right]\right)\frac{\Delta_A^+}{\mathrm{tr}\,\Delta_A^+}\,, \tag{A.28}$$

and since $\operatorname{tr} A = 1 - D(\tilde{\rho}_A, \rho_A) = 1 - \operatorname{tr} \Delta_A^+$,

$$(A.28) = A + \Delta_A^+ = \tilde{\rho}_A \,. \tag{A.29}$$

It remains to see that $\hat{\rho}_{AB}$ and $\rho_{AB}$ are close in purified distance. Let $|\rho\rangle_{ABC}$ be a purification of $\rho_{AB}$. Apply (Dupuis *et al.*, 2013, Lemma A.4)—itself a reformulation of (Tomamichel *et al.*, 2009, Lemma 15)—with $G = M_A = A^{1/2}\left(A + \Delta_A^-\right)^{-1/2}$ and $|\psi\rangle = |\rho\rangle_{ABC}$ to obtain

$$P\left(M_A \rho_{ABC} M_A^\dagger, \rho_{ABC}\right) \leqslant \sqrt{(2 - \operatorname{tr} \Delta_A^-) \operatorname{tr} \Delta_A^-} \leqslant \sqrt{2 \operatorname{tr} \Delta_A^-} \leqslant \sqrt{2\delta} \,. \tag{A.30}$$

This distance can only decrease if we trace out the system $C$, and thus $P\left(M_A \rho_{AB} M_A^\dagger, \rho_{AB}\right) \leqslant \sqrt{2\delta}$. Then, we can see from (A.26) that $\hat{\rho}_{AB} = M_A \rho_{AB} M_A^\dagger + \Delta'_{AB}$ for some operator $\Delta'_{AB} \geqslant 0$. We can thus bound the generalized trace distance (Tomamichel *et al.*, 2010) between $M_A \rho_{AB} M_A^\dagger$ and $\hat{\rho}_{AB}$ with the trace of $\Delta'_{AB}$,

$$\bar{D}\left(M_A \rho_{AB} M_A^\dagger, \hat{\rho}_{AB}\right) \leqslant \operatorname{tr} \Delta'_{AB} = 1 - \operatorname{tr} A = D(\rho_A, \tilde{\rho}_A) \leqslant \delta \,. \tag{A.31}$$

The generalized trace distance bounds the purified distance as $P(\rho, \rho') \leqslant \sqrt{2\bar{D}(\rho, \rho')}$, and we finally get by triangle inequality

$$P(\rho_{AB}, \hat{\rho}_{AB}) \leqslant P\left(\rho_{AB}, M_A \rho_{AB} M_A^\dagger\right) + P\left(M_A \rho_{AB} M_A^\dagger, \hat{\rho}_{AB}\right) \leqslant 2\sqrt{2\delta} \,.$$ ∎

## A.4 Continuity of the Relative Entropy in its First Argument

***Proposition A.8.*** *Let $\Gamma \geqslant 0$. Let $\rho, \sigma$ lie within the support of $\Gamma$. Assume that $D(\rho, \sigma) \leqslant \epsilon$. Then*

$$|D(\rho \,\|\, \Gamma) - D(\sigma \,\|\, \Gamma)| \leqslant \epsilon \log(\operatorname{rank} \Gamma - 1) + h(\epsilon) + \epsilon \left\|\log \Gamma\right\|_\infty \,, \tag{A.32}$$

*where $h(\epsilon) = -\epsilon \log \epsilon - (1 - \epsilon)\log(1 - \epsilon)$ is the binary entropy.*

*Proof of Proposition A.8.* First, write

$$D(\rho \,\|\, \Gamma) = \operatorname{tr}[\rho \log \rho - \rho \log \Gamma] = -H(\rho) - \operatorname{tr}[\rho \log \Gamma] \,, \tag{A.33}$$

and so

$$|D(\rho \,\|\, \Gamma) - D(\sigma \,\|\, \Gamma)| \leqslant |H(\sigma) - H(\rho)| + |\operatorname{tr}[\sigma \log \Gamma] - \operatorname{tr}[\rho \log \Gamma]| \,. \tag{A.34}$$

Using the continuity bound of Audenaert (2007), we have

$$|H(\rho) - H(\sigma)| \leqslant \epsilon \log(\operatorname{rank} \Gamma - 1) + h(\epsilon) \,, \tag{A.35}$$

where the states $\rho$ and $\sigma$ can be seen as living in a subspace of the full Hilbert space of dimension at most $\Gamma$ (because they must both lie within the support of $\Gamma$), and where $h(\epsilon) = -\epsilon \ln \epsilon - (1-\epsilon) \ln(1-\epsilon)$ is the binary entropy. On the other hand,

$$\operatorname{tr} \rho \log \Gamma - \operatorname{tr} \sigma \log \Gamma = \|\log \Gamma\|_\infty \operatorname{tr}\left[(\rho - \sigma) \frac{\log \Gamma}{\|\log \Gamma\|_\infty}\right] \leqslant \|\log \Gamma\|_\infty D(\rho, \sigma) , \qquad \text{(A.36)}$$

as $\log \Gamma / \|\log \Gamma\|_\infty$ is a valid candidate for $Z$ in Proposition A.5. Inverting the roles of $\rho$ and $\sigma$ in the equation above we finally obtain:

$$|\operatorname{tr} \rho \log \Gamma - \operatorname{tr} \sigma \log \Gamma| \leqslant \|\log \Gamma\|_\infty D(\rho, \sigma) \leqslant \|\log \Gamma\|_\infty \cdot \epsilon . \qquad \blacksquare$$

# Summary of Notation

Throughout this thesis, Hilbert states are of finite dimension, bases are orthonormal, and projectors are always Hermitian.

| Symbol | Meaning | Reference |
|---|---|---|
| $\mathscr{H}_S$ | Hilbert space associated with a quantum system $S$ | Section 3.1.2 |
| $\Pi_S^{\rho_S}$ | Projector onto the support of $\rho_S$ (subscripts $S$ may be omitted if unambiguous) | Section 3.1.1 |
| $\lvert\Phi\rangle_{A:A'}$ | Maximally entangled, unnormalized vector between two spaces $A \simeq A'$ with respect to some fixed bases, $\lvert\Phi\rangle_{A:A'} = \sum_i \lvert i\rangle_A \lvert i\rangle_{A'}$. | Section 3.1.5 |
| $\Phi^\rho_{A:B}$ ; $\lvert\Phi^\rho\rangle_{A:B}$ | Entangled, unnormalized vector in the Schmidt bases of $\lvert\rho\rangle_{AB}$ such that $\rho_{AB} = \rho_A^{1/2}\Phi^\rho_{A:B}\rho_A^{1/2} = \rho_B^{1/2}\Phi^\rho_{A:B}\rho_B^{1/2}$ | Section 3.1.5 |
| $\mathrm{id}_{A\to A'}$ | the identity process between two spaces $A \simeq A'$, with respect to some fixed bases of $A$ and $A'$ | Section 3.1.6 |
| $t_{A\to A'}$ | partial transpose operation with respect to some fixed bases of $A$ and $A'$ | Section 3.1.6 |
| $\mathcal{E}_{X\to X'}$ | a generic logical process, i.e. a completely positive, trace preserving map; sometimes trace-nonincreasing when specified | Chapter 5, Chapter 6 |
| $\Gamma_S$ | hidden information operator on system $S$ | Chapter 5 |
| $\Phi_{X\to X'}$ | a generic admissible operation, i.e. a completely positive, trace nonincreasing map satisfying $\Phi_{X\to X'}(\Gamma_X) \leqslant \Gamma_{X'}$ | Chapter 5 |
| $\{z_j\}$ ; $\vec{z}$ | generic thermodynamic variables | Chapter 8 |
| $\tau^{(\vec{z})}$ | thermodynamic state associated with the thermodynamic variables $\vec{z}$ | Chapter 8 |

# Index